\documentclass[a4paper,11pt]{article}

\usepackage{amsmath,amsfonts}
\usepackage{hyperref,makecell}
\usepackage{graphicx,tikz}
\usetikzlibrary{shapes}
\usetikzlibrary{decorations.pathmorphing}
\usetikzlibrary{decorations.markings}

\tikzset{
    partial ellipse/.style args={#1:#2:#3}{
      insert path={+ (#1:#3) arc (#1:#2:#3)}
    }
}

\def\be{\begin{equation}}
\def\ee{\end{equation}}
\def\bea{\begin{eqnarray}}
\def\eea{\end{eqnarray}}

\newcommand{\NN}{{\cal N}}

\newcommand{\tr}{\mbox{tr}}

\newcommand{\Urm}{\mathrm{U}}
\newcommand{\U}{\mathrm{U}}
\newcommand{\SU}{\mathrm{SU}}
\newcommand{\SO}{\mathrm{SO}}

\newcommand{\Dcal}{\mathcal{D}}
\newcommand{\Hcal}{\mathcal{H}}
\newcommand{\Wcal}{\mathcal{W}}
\newcommand{\Mcal}{\mathcal{M}}
\newcommand{\Ncal}{\mathcal{N}}
\newcommand{\Scal}{\mathcal{S}}

\newcommand{\Qcal}{\mathcal{Q}}
\newcommand{\Zset}{\mathbb{Z}}

\newcommand{\Cset}{{\,\,{{{^{_{\pmb{\mid}}}}\kern-.47em{\mathrm C}}}}}

\newcommand{\doublet}[2]{\left(\begin{array}{c}#1\\#2\end{array}\right)}
\newcommand{\triplet}[3]{\left(\begin{array}{c}#1\\#2\\#3\end{array}\right)}

\newcommand{\twobytwo}[4]{\left(\begin{array}{cc} #1&#2\\#3&#4\end{array}\right)}
\newcommand{\threebythree}[9]{\left(\begin{array}{ccc} #1&#2&#3\\#4&#5&#6\\#7&#8&#9\\\end{array}\right)}

\newcommand{\ket}[1]{\left|#1\right.\rangle}
\newcommand{\bra}[1]{\left.\langle#1\right|}

\newcommand{\ra}{\rightarrow}
\newcommand{\half}{\frac12}

\newcommand{\Tr}{\mathrm{Tr}}
\newcommand{\Dfour}{\hat{D}_4}
\newcommand{\Xb}{\bar{X}}
\newcommand{\Yb}{\bar{Y}}
\newcommand{\Zb}{\bar{Z}}
\newcommand{\Phib}{\bar{\Phi}}
\newcommand{\Qb}{\bar{Q}}

\newcommand{\p}{\partial}
\newcommand{\mparagraph}[1]{\paragraph{#1}\mbox{}}

\newcommand{\codif}{\text{D}}
\newcommand{\bcodif}{\Bar{\text{D}}}

\newcommand{\spcbr}{\text{ }}
\newcommand{\YM}{{\text{YM}}}

\newcommand{\ver}{\text{V}}

\newcommand{\ghost}{\text{ghost}}
\newcommand{\chiral}{\text{chiral}}

\newcommand{\phan}{\phantom{}}
\newcommand{\Msing}{\Mcal^{(\mathbf{1})}}
\newcommand{\Mtrip}{\Mcal^{(\mathbf{3})}}
\newcommand{\expval}[1]{\left\langle#1\right\rangle}

\newcommand{\gym}{g_\text{YM}}
\newcommand{\gymi}{g_{\text{YM},i}}

\newcommand{\Ocal}{\mathcal{O}}

\begin{document}

\numberwithin{equation}{section}

\mbox{}

\vspace{40pt}

\begin{center}

 {\Large \bf Spin chains for ADE quiver theories}\\
\vspace{43pt}

{\large {\mbox{{\bf Jarryd Bath$\,{}^a$} \hspace{.2cm} and \hspace{.2cm} {\bf Konstantinos Zoubos$\,{}^{a,b}$}}}}%

\vspace{.5cm}

${}^a$Department of Physics, University of Pretoria\\
Private Bag X20, Hatfield 0028, South Africa

\mbox{}

and

\mbox{}

${}^b$National Institute for Theoretical and Computational Sciences (NITheCS) \\
Gauteng, South Africa 

\end{center}

\vspace{40pt}

\begin{abstract}
  The spectral problem of four-dimensional superconformal quiver gauge theories can be mapped to one-dimensional spin chains with restricted Hilbert spaces, where the composition of neighbouring spins follows the path algebra of the quiver. To better understand such spin chains, we compute the one-loop planar dilatation operator for the 4d $\Ncal=2$ ADE quiver gauge theories obtained by orbifolding the $\Ncal=4$ Super-Yang-Mills theory and marginally deforming by independently varying the gauge couplings. This extends previous work which was mainly focused on the $\Zset_2$ quiver. We characterise the general features of the resulting ADE spin-chain models and construct the 2-magnon Bethe ansatz for holomorphic states. We also evaluate, at large $N$, the $\Ncal=2$ superconformal index of these gauge theories and use it to study their protected spectrum in specific sectors. 
\end{abstract}

\vspace{.3cm}

\large

\noindent

\normalsize

\noindent

\vspace{3.6cm}

\emph{Email:} jarryd.bath@tuks.co.za, kzoubos@up.ac.za

\vspace{0.5cm}

\setcounter{page}{0}
\thispagestyle{empty}
\newpage

\setcounter{tocdepth}{2}
\tableofcontents

\section{Introduction}

Quantum field theories with a large amount of supersymmetry are an ideal setting for obtaining exact results in Quantum Field Theory. This is particularly true if they exhibit the property of planar integrability, which allows for the computation of non-protected quantities both as regards to the spectrum and correlation functions. The best-understood example is the $\Ncal=4$ super-Yang-Mills theory (SYM) where the planar dilatation operator was shown to map, at one loop, to the Hamiltonian of a nearest-neighbour Heisenberg spin chain \cite{Minahan:2002ve}. The study of higher loops showed that the spin chain remains integrable, with the Hamiltonian becoming increasingly long-range, which, together with insights from the dual $\mathrm{AdS}_5\times \mathrm{S}^5$ perspective, led to the eventual solution of the $\Ncal=4$ SYM spectral problem. We refer to the reviews \cite{Beisert:2004ry,Serban:2010sr,Beisert:2010jr,Bombardelli:2016rwb} for details and references.   

It is important to understand how much of the structure behind planar integrability in $\Ncal=4$ SYM is still present in theories with reduced supersymmetry. This question has been studied extensively for the $\Ncal=1$ theories arising through marginal superpotential deformations of $\Ncal=4$ SYM, as well as theories obtained via orbifolding (see e.g. \cite{Zoubos:2010kh} for a review). The orbifolding process has been shown to respect the integrable structures of $\Ncal=4$ SYM \cite{Wang:2003cu,Ideguchi:2004wm,Beisert:2005he,Solovyov:2007pw} including at higher-loop \cite{Beccaria:2011qd,deLeeuw:2012hp,Skrzypek:2022cgg}. However, given an orbifold theory, one can marginally deform by varying the gauge couplings of each gauge group independently. The case of the marginally deformed $\Zset_2$ orbifold was first studied in \cite{Gadde:2009dj,Gadde:2010zi,Gadde:2010ku,Gadde:2012rv}. Apart from its intrinsic interest, as explained in those works, an important motivation for the study of this theory was that, as one of the deformed couplings tends to zero, it limits to Superconformal QCD in the Veneziano limit \cite{Veneziano:1976wm}. Although $\Ncal=2$ supersymmetry of course allows for the calculation of an abundance of protected quantities in the marginally deformed theories, the results of these early works were not encouraging with respect to integrability. In particular, it quickly became clear that standard structures such as the Yang-Baxter equations were absent and therefore the story of integrability, if present, would necessarily be more subtle.\footnote{These statements refer to the scalar sector which would be the analogue of the $\SO(6)$ sector in $\Ncal=4$ SYM, on which we will focus in this work. The $\SU(2,1|2)$ sector of generic $\Ncal=2$ SCFT, consisting of vector-multiplet fields, has been argued to inherit the integrability properties of $\Ncal=4$ SYM, up to a redefinition of the coupling constant \cite{Gadde:2012rv,Pomoni:2013poa,Mitev:2014yba,Mitev:2015oty}. For an early study of integrability for (non-conformal) pure $\Ncal=2$ SYM, see \cite{DiVecchia:2004jw}.}

Recently, the marginally deformed $\Ncal=2$ orbifold theories were revisited with the goal of uncovering any such additional structures and understanding what their implications would be. The work \cite{Pomoni:2021pbj} studied the spin chains that arise in the holomorphic sector of the deformed $\Zset_2$ orbifold theory, and argued that the appropriate setting to understand them is that of \emph{dynamical} $R$-matrices, similar to those appearing in the study of elliptic quantum groups \cite{Felder:1994pb,Felder:1994be}. Extending studies of the 2-magnon problem in \cite{Gadde:2010zi,Pomoni:2021pbj}, the 3- and 4- magnon problems in a specific sector of these theories were considered in \cite{Bozkurt:2024tpz,Bozkurt:2025exl} and shown to be solvable by taking long-range contributions into account. Meanwhile, \cite{Bertle:2024djm} took a step back and considered the underlying symmetries of the $\Zset_2$ theory, showing that the $\SU(4)$ generators which are broken by the orbifolding procedure can be usefully recovered by working in a Lie algebroid, rather than a Lie algebra setting. The implications of these hidden symmetries and their possible relevance to integrability are still being worked out. 

Given the progress made in the $\Zset_2$ theory, it is relevant to explore the most general class of $\Ncal=2$ theories that can be obtained by orbifolding $\Ncal=4$ SYM and marginally deforming. These correspond to the finite subgroups of $\SU(2)$, which have an ADE classification. As mentioned, it is known that integrability persists for such theories at the orbifold point \cite{Beisert:2005he,Solovyov:2007pw}. However, beyond the case of $\Zset_2$, the marginally deformed theories and their corresponding spin chains have not received much attention so far. 

In this work, we derive the one-loop dilatation operator for generic marginally-deformed ADE quiver theories and discuss the main features of the corresponding spin chains. We will be particularly interested in the non-abelian orbifold theories, which have some fundamental differences to the $\Zset_k$ cases. To illustrate these differences, we treat some concrete examples, in particular the $\Zset_3$, $\hat{D}_4$ and $\hat{E}_6$ cases, in more detail. For the protected spectrum, we check our results by comparing with superconformal index and Molien series computations.

We find it useful to perform our derivation of the dilatation operator fully in superspace. Since most such computations in the $\Ncal=4$ SYM context are performed in components (for reviews, see e.g. \cite{Minahan:2010js, deLeeuw:2017cop}), or focus only on the holomorphic sector \cite{Sieg:2010jt, Sieg:2010tz}, we provide some details of these computations. Apart from ensuring that our computations explicitly preserve supersymmetry, working in superspace is likely to be essential in extending our computations to higher loops, as was done for the $\Zset_2$ quiver in \cite{Pomoni:2011jj}.

The outline of this paper is as follows: In Section \ref{sec:Orbifolding} we review the procedure of orbifolding $\Ncal=4$ SYM by a subgroup of $\SU(2)$ to obtain the ADE quiver theories. For completeness, we include some elements from the theory of finite groups. In Section \ref{sec:ADEDilatationOperator}, we discuss the superspace derivation of the ADE dilatation operator, and summarise the corresponding Hamiltonian in spin-chain language in Section \ref{sec:ADEHamiltonian}. In Section \ref{sec:Protected} we then switch gears to discuss the protected spectrum, and in particular evaluate the superconformal index and its various limits for the ADE theories. In Section \ref{sec:ADEspinchains} we summarise some generic features of the ADE spin chains. After this, we are finally ready to consider some concrete examples. We study the $\Zset_3$, $\Dfour$ and $\hat{E}_6$ theories in Sections \ref{sec:Z3}, \ref{sec:D4} and \ref{sec:E6}, respectively. For each example, we write out the Hamiltonian for that specific case, consider the protected states and, where possible, compare with the explicit diagonalisation, study the spectrum for short closed chains, and finally construct the 2-magnon Bethe ansatz and compare the resulting energies and states with those found by diagonalisation of the Hamiltonian. 

We have included several appendices with additional details on the finite subgroups of $\SU(2)$ (Appendices \ref{sec:CharacterTables} and \ref{sec:MattercontentN=2}), our superspace conventions and Feynman rules (Appendix \ref{sec:SuperspaceFeynman}) (including, as a check, a verification of the vanishing of the beta function in Appendix \ref{sec:betafunction}), more details on the index for some relevant multiplets (Appendix \ref{sec:Indexshort}), the on-shell $\Ncal=2$ supersymmetry transformations (Appendix \ref{sec:ExtendedSusyTransformations}) and the explicit forms of the descendants of the Konishi operator (Appendix \ref{sec:Konishi}).

\section{Orbifolding $\Ncal=4$ SYM} \label{sec:Orbifolding}

In this section we will review the orbifolding procedure as applied to the $\Ncal=4$ SYM theory. This subject has a long history, beginning with \cite{Douglas:1996sw,Johnson:1996py,Kachru:1998ys,Lawrence:1998ja,Oz:1998hr,Gukov:1998kk}. Taking a IIB string theory perspective, recall the UV description of $\Ncal=4$ SYM with gauge group $\SU(|\Gamma|N)$ as a stack of $|\Gamma| N$ D3-branes in 10-dimensional flat space, with transverse space $\Cset^3$. Considering instead the orbifolded geometry $\Cset^3/\Gamma$, with $\Gamma$ a finite group of order $|\Gamma|$, one obtains $|\Gamma|$ stacks of $N$ D3 Branes, with $\Gamma$ acting by permuting these stacks of branes. Orbifolding results in a certain collection of stacks being identified as a single stack and multiple stacks being identified with each other. The result is $M$ stacks of $n_iN$ D3 branes each $(i=1,\ldots,M)$, with $M$ being the number of conjugacy classes of $\Gamma$. From the field theory perspective, one has constructed a gauge theory with $\prod_{i=1}^M\SU(n_iN)$ gauge group and fields in the adjoint or bifundamental representations of these groups, depending on the precise action of $\Gamma$.   

\begin{table}[h]
    \begin{center}
    \begin{tabular}{|c|c|c|c|c|c|c|c|c|c|c|}
    \hline
         &$X^0$&$X^1$&$X^2$&$X^3$&$X^4$&$X^5$&$X^6$&$X^7$&$X^8$&$X^9$ \tabularnewline\hline   
        $N$ D3&$-$&$-$&$-$&$-$&$\cdot$&$\cdot$&$\cdot$&$\cdot$&$\cdot$&$\cdot$ \tabularnewline\hline 
        $\Gamma$-orbifold &$\cdot$&$\cdot$&$\cdot$&$\cdot$&$\times$&$\times$&$\times$&$\times$&$\cdot$&$\cdot$ \tabularnewline\hline 
    \end{tabular}\end{center}
    \caption{Type IIB engineering of $\NN=2$ orbifold daughters of $\NN=4$ SYM.}
    \label{tab:EngineeringN=2}
\end{table}

In practice, one can just work at the level of the field theory, by asking that the fields of $\SU(|\Gamma|N)$ $\Ncal=4$ SYM are invariant under the combined action of the finite group on the gauge as well as the $R$-symmetry indices. 
\be
\Phi^I=R^I_{\;J}(g) \gamma_{\text{reg}}(g) \Phi^J \gamma^{-1}_{\text{reg}}(g) \;\;,\;\; g\in \Gamma\;.
\ee
Here $g$ are the elements of the finite group and $\gamma_{\text{reg}}(g)$ is a specific representation of $g$ on the $\SU(|\Gamma|N)$ indices, to be elaborated below. The matrix $R^I_{\;J}(g)$ is the induced representation of $g$ on the $R$-symmetry indices. If one is interested in preserving $\Ncal=2$ supersymmetry, $\Gamma$ is required to be a finite subgroup of $\SU(2)$. In the next section we will review these groups and their representations, in order to be able to impose the above condition.

\subsection{Some finite group theory}

In this section we will review some background material on finite groups. For a more detailed exposition, we refer to textbooks such as \cite{Lomont1959,Sternberg94}. Recall that for a finite group the order denotes the number of elements. Let $\Gamma$ be a finite group of order $|\Gamma|$ with $M$ unitary irreducible representations (irreps) $\mathbf{r}_i=(\rho_i,\mathcal{V}_i)$, for $i=1,\ldots,M$, where $\mathcal{V}_i$ is a vector space and $\rho_i:\Gamma\mapsto\mathrm{GL}(\mathcal{V}_i)$ is the action of the group on $\mathcal{V}_i$. For a finite group $M$ equals the number of conjugacy classes. Let $n_i\equiv\text{dim}(\mathbf{r}_i)$ and $\{(e_i)_1,\ldots,(e_i)_{n_i}\}$ be a basis for the corresponding vector space $\mathcal{V}_i$. We can then identify $\rho_i$ with its matrix elements: 
\be
\rho_i(g)^m\phan_n(e_i)_m=(e_i)_n \;\;\; \text{for} \;\; g\in \Gamma\;.
\ee
The order of the group can be expressed in terms of the representation space dimensions $n_i$ as  
\be\label{eq:sumoverdimsquared}
|\Gamma|=\sum_{i=1}^Mn_i^2\;.
\ee
The unitary irreps satisfy the unitarity relation
\be
\frac{1}{|\Gamma|}\sum_{g\in\Gamma}\rho_i(g)^m\phan_n\bar{\rho}_j(g)_k\phan^l=\frac{1}{n_i}\delta_{ij}\delta^m\phan_k\delta^l\phan_n \;.
\ee
The product representation of two irreps $\mathbf{r}_k$ and $\mathbf{r}_i$ can be decomposed as a sum over irreps:
\be
\mathbf{r}_k\otimes\mathbf{r}_i=\bigoplus_{j=1}^Ma^k_{ij}\mathbf{r}_j\;,
\ee
where $a^k_{ij}$ denotes the multiplicity and is given by
\be\label{eq:multiplicityprod}
a^k_{ij}=\frac{1}{|\Gamma|}\sum_{g\in\Gamma}\chi_k(g)\chi_i(g)\bar{\chi}_j(g)\;,
\ee
where $\chi_i(g)$ is the character of the element $g$ in the representation $\mathbf{r}_i$. The equation \eqref{eq:multiplicityprod} follows from the orthogonality relation for characters:
\be
\sum_{g\in\Gamma}\chi_i(g)\bar{\chi}_j(g)=|\Gamma|\delta_{ij} \;.
\ee 
We also have 
\be
\sum_{j=1}^Ma^k_{ij}n_j=n_kn_i\;.
\ee
For simply reducible groups (which is the case for all the finite subgroups of $\SU(2)$ apart from $\Zset_2$), we have either $a^k_{ij}=0$ or $a^k_{ij}=1$. The basis vectors of the product representations and the irreps of simply reducible groups are related by the Clebsch-Gordan coefficients $K^{jn}_{kl,\,im}$ as follows
\be\label{eq:Clebsch-Gordan}
(e_k)_l\otimes(e_i)_m=\sum_{j=1}^{M}\sum_{n=1}^{n_j}K^{jn}_{kl,\,im}(e_j)_n\;.
\ee 
From unitarity, we have 
\be
(e_j)_n=\sum_{i,k=1}^M\sum_{m=1}^{n_i}\sum_{l=1}^{n_k}\bar{K}^{jn}_{kl,\,im}(e_i)_m\otimes(e_k)_l.
\ee
The Clebsch-Gordan coefficients satisfy 
\be
   \sum_{m=1}^{n_i}\sum_{l=1}^{n_k}\bar{K}^{j'n'}_{kl,\,im}K^{jn}_{kl,\,im}=\delta^{j'j}\delta^{n'n}\;\;\text{and}\;\;
        \sum_{j=1}^M\sum_{n=1}^{n_j}\bar{K}^{jn}_{kl',\,im'}K^{jn}_{kl,\,im}=\delta_{l'l}\delta_{m'm}\;.
        \ee
Let us now consider the invariant subspaces of product representations. The $\Gamma$-invariant subspace of $\mathcal{V}_i\otimes\mathcal{V}_j^*$ is 
\be\label{eq:invariantsubr_rdual}
(\mathcal{V}_i\otimes\mathcal{V}_j^*)^\Gamma=\{\delta_{ij}\delta^m\phan_{n}(e_i)_m\otimes(e^*_j)^n\}.
\ee
Hence from \eqref{eq:Clebsch-Gordan} and \eqref{eq:invariantsubr_rdual} we can find the invariant subspace of $\mathcal{V}_k\otimes\mathcal{V}_i\otimes\mathcal{V}_j^*$:
\be
(\mathcal{V}_k\otimes\mathcal{V}_i\otimes\mathcal{V}_j^*)^\Gamma=\{\bar{K}^{jn}_{kl,\,im}(e_k)_l\otimes(e_i)_m\otimes (e^*_j)^n\}.
\ee
Let us consider an important (reducible) representation: the regular representation $\mathbf{r}_\text{reg}$. The regular representation is the representation defined by $\text{dim}(\mathbf{r}_\text{reg})=|\Gamma|$. We can interpret the regular representation as the representation acting on the ring over $\Gamma$, $\Cset[\Gamma]$. Following \cite{Solovyov:2007pw}, we refer to this as the ``orbit basis" and denote it as $\mathbf{r}_\text{reg}=(\tau,\Cset[\Gamma])$. The matrix elements of $\tau$ can be found as follows:
\be \label{orbitmatrices}
\tau(g_i)^m\phan_n=\begin{cases}
        1\;,\;\;\text{if}\; g_mg_n^{-1}=g_i\;,\\
        0\;,\;\;\text{otherwise}\;.
    \end{cases}
\ee
Essentially, this means that the matrix elements of $\tau(g_i)$ can be found by setting $1$ where the element $g_i$ appears in the Cayley table of $\Gamma$. 

We can also decompose the regular representation as a sum over over the irreps, which, again following \cite{Solovyov:2007pw} we refer to as the ``quiver basis". We will denote it as $\mathbf{r}_\text{reg}=(\gamma,\mathcal{V}_\text{reg})$. Let $m_i$ denote the multiplicity of $\mathbf{r}_i$ in $\mathbf{r}_\text{reg}$:
\be
\mathbf{r}_\text{reg}=\bigoplus_{i=1}^Mm_i\mathbf{r}_i\;.
\ee
The quiver basis takes the form of a block diagonal matrix. Now notice that the order of the group is expressed in terms of the dimensions $n_i$ and the multiplicities as
\be
|\Gamma|=\chi_\text{reg}(e)=\sum_{i=1}^Mm_i\chi_i(e)=\sum_{i=1}^Mm_in_i\;,
\ee
where $e$ is the identity element of $\Gamma$. From \eqref{eq:sumoverdimsquared}, we have $m_i=n_i$. Hence, we can write 
\be \label{gammareg}
 \gamma(g_i)=\bigoplus_{i=1}^M\rho_i(g_i)\otimes I_{n_i\times n_i}\;,\quad\mathcal{V}_\text{reg}=\bigoplus_{i=1}^M\mathcal{V}_i^{\oplus n_i}\;.
\ee
In the following sections, we will use the quiver basis matrices to impose invariance under the finite group action.

\subsection{The finite subgroups of $\SU(2)$}

In this section we review some well-known aspects of the finite subgroups of $\SU(2)$  \cite{Klein1884}. They are the binary polyhedral groups, specifically the binary cyclic, dihedral, tetrahedral, octahedral and icosahedral groups. Let us briefly summarise their definition and structure. We refer to Appendix \ref{sec:CharacterTables} for the character tables.

\subsubsection{The cyclic group $\Zset_k$}

The cyclic group $\mathbb{Z}_k$ is defined as
\begin{equation}
    \lbrace{a|a^k=e\rbrace}.
\end{equation}
where $e$ is the identity element. It has order $k$ and has $k$ conjugacy classes, hence it also has $k$ irreducible representations, which are all one-dimensional. Denoting by $\omega_k$ the $k$-th root of unity, these representations are simply the 1-dimensional matrices
\begin{equation}
    \rho_n(a)= \omega_k^n\;,
\end{equation}
where $n=0,\ldots,k-1$ and where $n=0$ is the trivial representation.

\subsubsection{The binary dihedral group $\hat{\mathrm{D}}_k$}

The binary dihedral group $\hat{\mathrm{D}}_k$, also referred to as the dicyclic group Dic$_k$ (and also denoted by 2D$_k$) is a group of order $4k-8$. It is given by the following presentation 
\begin{equation}
    \lbrace{a,b|a^{2n}=e,b^2=a^{n}=z,b^{-1}ab=a^{-1}\rbrace},
\end{equation}
where $k=n+2$ and $z$ is the central element of order 2. For $k\ge4$, the centre of the group is the two-element subgroup $\expval{z}$, and the quotient $\hat{\text{D}}_{k}/\expval{z}$ is the standard dihedral group D$_{k}$ of order $2k-4$. $\hat{\text{D}}_{k}$ has $k+1$ conjugacy classes:
\be
 \{e\}\;,\; \{a^{n}\}\;,\;\{a^m,a^{-m}\}\;, \; \{b,a^2b,\ldots,a^{2n-2}b\}\;, \; \{ab,a^3b,\ldots,a^{2n-1}b\}\;.
  \ee
where $m=1,\ldots,n-1$. Correspondingly, there are $k+1$ irreducible representations. Since the characters are simply the traces of the representations, the one-dimensional representations can be read off from the character tables in Tables \ref{tab:Dnevencharactertable} and \ref{tab:Dnoddcharactertable}. Elements in the same conjugacy class have the same 1-dimensional representations, but differ in the 2-dimensional ones. 
  
The two-dimensional irreps are parametrised by the odd integers $n=1,\ldots,k-1$:
\begin{equation}
      \rho(a)=\begin{pmatrix}
            \omega_{2n}^m&0\\
            0&\omega_{2n}^{-m}
        \end{pmatrix}\;\;\text{and}\;\;
        \rho(b)=\begin{pmatrix}
            0&-1\\
            1&0
        \end{pmatrix}\;.
\end{equation}

\subsubsection{The binary tetrahedral group 2T}

This is a non-abelian group of order 24, defined as

\begin{equation} \label{2Tdefinition}
  \lbrace{r,s,t|r^2=s^3=t^3=rst=z\rbrace},
\end{equation}
where $z$ is the central element of order 2. Equivalently, we can write $r=zt^{-1}s^{-1}$, hence, $r^2=(t^{-1}s^{-1}z)^2=(st)^{-2}=z$. Now since $z^{-1}=z$, we have that $z=(st)^2$. Thus, we can write the binary tetrahedral group as
\begin{equation}
    \lbrace{s,t|(st)^2=s^3=t^3=z\rbrace}.
\end{equation}
In terms of quaternions, the generators can be written as 
\begin{equation}
    r=\hat{i},\spcbr s=\frac{1}{2}(1+\hat{i}+\hat{j}-\hat{k}),\spcbr t=\frac{1}{2}(1+\hat{i}+\hat{j}+\hat{k}).
\end{equation}
2T has seven conjugacy classes and hence seven irreducible representations. They are presented in Section \ref{sec:E6}.

\subsubsection{The binary octahedral group 2O}

This is a non-abelian group of order 48, with the following presentation

\begin{equation}
    \lbrace{r,s,t|r^2=s^3=t^4=rst\rbrace},
\end{equation}
equivalently
\begin{equation}
    \lbrace{s,t|(st)^2=s^3=t^4\rbrace}.
\end{equation}
It has 8 conjugacy classes/irreducible representations. We refer to e.g. \cite{Benvenuti:2006qr} for its representations.

\subsubsection{The binary icosahedral group 2I}

This is an non-abelian subgroup of order 120, with the following presentation 
\begin{equation}
    \lbrace{r,s,t|r^2=s^3=t^5=rst\rbrace},
\end{equation} 
equivalently 
\begin{equation}
    \lbrace{s,t|(st)^2=s^3=t^5\rbrace}.
\end{equation}
It has 8 conjugacy classes/irreducible representations. We again refer to e.g. \cite{Benvenuti:2006qr} for more details.

\mbox{}

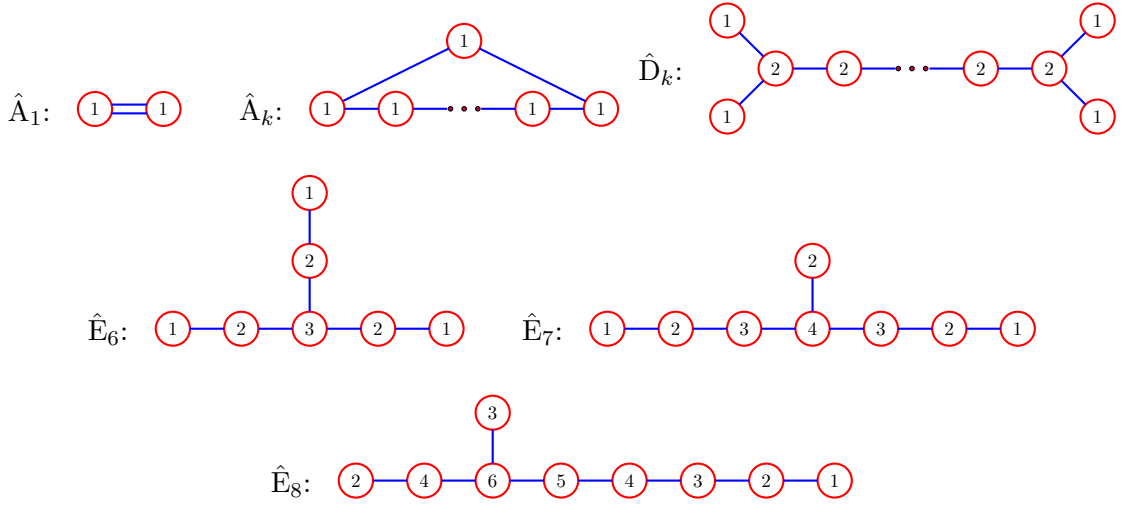
\begin{figure}
    \begin{center}
\begin{tikzpicture}[scale=0.9]
            \node[left] at (-0.5,0) {$\hat{\text{A}}_1$:};
            \draw[red,thick] (0,0) circle (0.25);
            \draw[red,thick] (1,0) circle (0.25);
            \draw[blue,thick] (0.24148,0.064704)--(0.75851,0.064704);
            \draw[blue,thick] (0.24148,-0.064704)--(0.75851,-0.064704);
            \node[scale=0.65] at (0,0) {$1$};
            \node[scale=0.65] at (1,0) {$1$};
        \end{tikzpicture}\hspace{0.5cm}
        \begin{tikzpicture}[scale=0.9]
            \node[left] at (-0.5,0) {$\hat{\text{A}}_k$:};
            \draw[red,thick] (0,0) circle (0.25);
            \draw[red,thick] (1,0) circle (0.25);
            \draw[fill=purple] (1.8,0) circle (.2ex);
            \draw[fill=purple] (2,0) circle (.2ex);
            \draw[fill=purple] (2.2,0) circle (.2ex);
            \draw[red,thick] (3,0) circle (0.25);
            \draw[red,thick] (4,0) circle (0.25);
            \draw[red,thick] (2,1) circle (0.25);
            \draw[blue,thick] (0.25,0)--(0.75,0);
            \draw[blue,thick] (1.25,0)--(1.75,0);
            \draw[blue,thick] (2.25,0)--(2.75,0);
            \draw[blue,thick] (3.25,0)--(3.75,0);
            \draw[blue,thick] (0.2236,0.1118)--(1.77639,0.88819);
            \draw[blue,thick] (3.77639,0.1118)--(2.2236,0.88819);
            \node[scale=0.65] at (0,0) {$1$}; \node[scale=0.65] at (1,0) {$1$}; \node[scale=0.65] at (3,0) {$1$}; \node[scale=0.65] at (4,0) {$1$}; \node[scale=0.65] at (2,1) {$1$};
        \end{tikzpicture}
        \begin{tikzpicture}[scale=0.9]
            \node[left] at (-0.5,0) {$\hat{\text{D}}_k$:};
            \draw[red,thick] (0,0.70716) circle (0.25);
            \draw[red,thick] (0,-0.70716) circle (0.25);
            \draw[red,thick] (0.70716,0) circle (0.25);
            \draw[red,thick] (1.70716,0) circle (0.25);
            \draw[fill=purple] (2.5,0) circle (.2ex);
            \draw[fill=purple] (2.7,0) circle (.2ex);
            \draw[fill=purple] (2.9,0) circle (.2ex);
            \draw[red,thick] (3.70716,0) circle (0.25);
            \draw[red,thick] (4.70716,0) circle (0.25);
            \draw[red,thick] (5.41421,0.70716) circle (0.25);
            \draw[red,thick] (5.41421,-0.70716) circle (0.25);
            \draw[blue,thick] (0.9571,0)--(1.4571,0);
            \draw[blue,thick] (3.9571,0)--(4.4571,0);
            \draw[blue,thick] (0.5303,0.1768)--(0.1768,0.5303);
            \draw[blue,thick] (0.5303,-0.1768)--(0.1768,-0.5303);
            \draw[blue,thick] (4.88391,0.1768)--(5.23746,0.5303);
            \draw[blue,thick] (4.88391,-0.1768)--(5.23746,-0.5303);
            \draw[blue,thick] (1.95716,0)--(2.45716,0);
            \draw[blue,thick] (2.95716,0)--(3.45716,0);
            \node[scale=0.65] at (0,0.70716) {$1$}; \node[scale=0.65] at (0,-0.70716) {$1$}; \node[scale=0.65] at (0.70716,0) {$2$}; 
            \node[scale=0.65] at (1.70716,0) {$2$}; \node[scale=0.65] at (3.70716,0) {$2$}; \node[scale=0.65] at (4.70716,0) {$2$}; \node[scale=0.65] at (5.41421,0.70716) {$1$};
            \node[scale=0.65] at (5.41421,-0.70716) {$1$};
        \end{tikzpicture}\\\vspace{0.5cm}
        \begin{tikzpicture}[scale=0.9]
            \node[left] at (-0.5,0) {$\hat{\text{E}}_6$:};
            \draw[red,thick] (0,0) circle (0.25);
            \draw[red,thick] (1,0) circle (0.25);
            \draw[red,thick] (2,0) circle (0.25);
            \draw[red,thick] (3,0) circle (0.25);
            \draw[red,thick] (4,0) circle (0.25);
            \draw[red,thick] (2,1) circle (0.25);
            \draw[red,thick] (2,2) circle (0.25);
            \draw[blue,thick] (0.25,0)--(0.75,0);
            \draw[blue,thick] (1.25,0)--(1.75,0);
            \draw[blue,thick] (2.25,0)--(2.75,0);
            \draw[blue,thick] (3.25,0)--(3.75,0);
            \draw[blue,thick] (2,0.25)--(2,0.75);
            \draw[blue,thick] (2,1.25)--(2,1.75);
            \node[scale=0.65] at (0,0) {$1$};
            \node[scale=0.65] at (1,0) {$2$};
            \node[scale=0.65] at (2,0) {$3$};
            \node[scale=0.65] at (3,0) {$2$};
            \node[scale=0.65] at (4,0) {$1$};
            \node[scale=0.65] at (2,1) {$2$};
            \node[scale=0.65] at (2,2) {$1$};
        \end{tikzpicture}\hspace{0.5cm}
        \begin{tikzpicture}[scale=0.9]
            \node[left] at (-0.5,0) {$\hat{\text{E}}_7$:};
            \draw[red,thick] (0,0) circle (0.25);
            \draw[red,thick] (1,0) circle (0.25);
            \draw[red,thick] (2,0) circle (0.25);
            \draw[red,thick] (3,0) circle (0.25);
            \draw[red,thick] (4,0) circle (0.25);
            \draw[red,thick] (5,0) circle (0.25);
            \draw[red,thick] (6,0) circle (0.25);
            \draw[red,thick] (3,1) circle (0.25);
            \draw[blue,thick] (0.25,0)--(0.75,0);
            \draw[blue,thick] (1.25,0)--(1.75,0);
            \draw[blue,thick] (2.25,0)--(2.75,0);
            \draw[blue,thick] (3.25,0)--(3.75,0);
            \draw[blue,thick] (4.25,0)--(4.75,0);
            \draw[blue,thick] (5.25,0)--(5.75,0);
            \draw[blue,thick] (3,0.25)--(3,0.75);
            \node[scale=0.65] at (0,0) {$1$};
            \node[scale=0.65] at (1,0) {$2$};
            \node[scale=0.65] at (2,0) {$3$};
            \node[scale=0.65] at (3,0) {$4$};
            \node[scale=0.65] at (4,0) {$3$};
            \node[scale=0.65] at (5,0) {$2$};
            \node[scale=0.65] at (6,0) {$1$};
            \node[scale=0.65] at (3,1) {$2$};
        \end{tikzpicture}\\ \vspace{0.5cm}
        \begin{tikzpicture}[scale=0.9]
            \node[left] at (-0.5,0) {$\hat{\text{E}}_8$:};
            \draw[red,thick] (0,0) circle (0.25);
            \draw[red,thick] (1,0) circle (0.25);
            \draw[red,thick] (2,0) circle (0.25);
            \draw[red,thick] (3,0) circle (0.25);
            \draw[red,thick] (4,0) circle (0.25);
            \draw[red,thick] (5,0) circle (0.25);
            \draw[red,thick] (6,0) circle (0.25);
            \draw[red,thick] (7,0) circle (0.25);
            \draw[red,thick] (2,1) circle (0.25);
            \draw[blue,thick] (0.25,0)--(0.75,0);
            \draw[blue,thick] (1.25,0)--(1.75,0);
            \draw[blue,thick] (2.25,0)--(2.75,0);
            \draw[blue,thick] (3.25,0)--(3.75,0);
            \draw[blue,thick] (4.25,0)--(4.75,0);
            \draw[blue,thick] (5.25,0)--(5.75,0);
            \draw[blue,thick] (6.25,0)--(6.75,0);
            \draw[blue,thick] (2,0.25)--(2,0.75);
            \node[scale=0.65] at (0,0) {$2$};
            \node[scale=0.65] at (1,0) {$4$};
            \node[scale=0.65] at (2,0) {$6$};
            \node[scale=0.65] at (3,0) {$5$};
            \node[scale=0.65] at (4,0) {$4$};
            \node[scale=0.65] at (5,0) {$3$};
            \node[scale=0.65] at (6,0) {$2$};
            \node[scale=0.65] at (7,0) {$1$};
            \node[scale=0.65] at (2,1) {$3$};
        \end{tikzpicture}\end{center}
    \caption{The Dynkin Diagrams for the affine $\hat{A}, \hat{D}$ and $\hat{E}$ series. Each node has an associated Coxeter label $n_i$, which denotes the corresponding vector space dimension. When orbifolding by the corresponding subgroup of $\SU(2)$, these Dynkin diagrams become the quiver diagrams of the orbifolded theories, with each node associated to a $\SU(n_iN)$ gauge group.}
    \label{fig:ADEDynkin}
\end{figure}

Through the McKay correspondence \cite{McKay1980}, these groups are related to the affine $\hat{A}_k$, $\hat{D}_k$, $\hat{E}_6$, $\hat{E}_7$ and $\hat{E}_8$ Lie groups (see e.g. \cite{Fuchs:1992nq,DiFrancesco}), respectively. Concretely, the Dynkin diagrams of these affine groups, shown in Fig. \ref{fig:ADEDynkin},  provide the adjacency diagrams of the binary polyhedral groups. 

The adjacency matrices for all ADE cases are listed in Appendix \ref{sec:MattercontentN=2}. We note that $\Zset_2$ is the only non-simply laced case, indicated by a double line in the Dynkin diagram and correspondingly a 2 in the adjacency matrix. This leads to a degeneracy in the orbifolding procedure which requires the use of an additional label. As the $\Zset_2$ orbifold theory and its deformations is well studied \cite{Gadde:2009dj,Gadde:2010zi,Pomoni:2021pbj,Bertle:2024djm}, in the following we will focus on the case of $\Zset_k$ with $k>2$, and will therefore not include this additional degeneracy label in our notation.

\subsection{The orbifolding procedure}

 Orbifolding by a finite subgroup of $\SU(2)$ breaks the $\SU(4)_R$ $R$-symmetry group of $\Ncal=4$ SYM to $\SU(2)_L/\Gamma\times\SU(2)_R\times\Urm(1)_r$. The action of the unbroken R-symmetry group on the fields is summarised in Table \ref{tab:Raction}.
\begin{table}[h]
    \centering
        \renewcommand{\arraystretch}{1.3}
    \begin{tabular}{|c|c|c|c|c|}\hline
         $\varphi^{ab}$& $\sigma_R^+$&$\sigma_R^-$&$\sigma_R^3$&$\sigma_r$\tabularnewline\hline
         $X$&$0$&$\Yb$&$\half X$&$0$\tabularnewline\hline
         $Y$&$0$&$-\Xb$&$\half Y$&$0$\tabularnewline\hline
         $Z$&$0$&$0$&$0$&$-Z$\tabularnewline\hline
         $\Xb$&$-Y$&$0$&$-\frac{1}{2}\Xb$&$0$\tabularnewline\hline
         $\Yb$&$X$&$0$&$-\frac{1}{2}\Yb$&$0$
         \tabularnewline\hline
         $\Zb$&$0$&$0$&$0$&$\Zb$\tabularnewline\hline
    \end{tabular}
    \caption{The action of $\mathfrak{su}(2)_R\times\mathfrak{u}(1)_r$ on the fields of $\NN=4$ SYM.}
    \label{tab:Raction}
\end{table}
Given a finite subgroup of $\SU(2)$, orbifolding requires us to impose invariance of the fields under $\Gamma$. As mentioned, $\Gamma$ also acts on the $R$-symmetry index of the fields. However, in our $\Ncal=2$ case we have the decomposition  $\mathbf{3}=\mathbf{2}_\text{ind}\oplus\mathbf{1}_\text{triv}$, where $\mathbf{2}_\text{ind}$ is the induced representation that arises from $\Gamma$ being a subgroup of $\SU(2)$ and $\mathbf{1}_\text{triv}$ is the trivial representation. The action can therefore be written as
\be
R=R^{(\mathbf{2})}\oplus 1\;.
\ee
Therefore $\Gamma$ acts on the vector field of $\Ncal=4$ SYM, as well as on one of the chiral multiplets (which we call $\Phi^3=Z$) simply by conjugation, and invariance implies:
\be \label{invarianceVZ}
V=\gamma_{\text{reg}}(g) V \gamma^{-1}_{\text{reg}}(g)\;\;,\;\; Z=\gamma_{\text{reg}}(g) Z \gamma^{-1}_{\text{reg}}(g)\;\;,\;\; g\in\Gamma\;.
\ee
The other two chiral fields of $\Ncal=4$ SYM, $\Phi^1=X$ and $\Phi^2=Y$, transform under the $\SU(2)$, so they are also acted upon by the induced representation:
\be \label{invarianceXY}
\Phi^I=R^I_{\;J}(g) \gamma_{\text{reg}}(g) \Phi^J \gamma^{-1}_{\text{reg}}(g) \;\;,\;\; g\in \Gamma\;.
\ee
The fields of the mother theory decompose as
\be
\begin{split}
    [V]^{im}\phan_{jn}=&\delta^i\phan_j\delta^m\phan_nV_i\;,\\
    [\Phi^I]^{im}\phan_{jn}=&K^{im}_{\mathbf{3}I,\,jn}Q_{ij},\quad\text{for }\;I=1,2\;,\\
    [\Phi^3]^{im}\phan_{jn}=&\delta^i\phan_j\delta^m\phan_n Z_i\;,\\
    [\Phib_I]^{im}\phan_{jn}=&\bar{K}^{jn}_{\mathbf{3}I,\,im}\Qb_{ij},\quad\text{for }\;I=1,2\;,\\
    [\Phib_3]^{im}\phan_{jn}=&\delta^i\phan_j\delta^m\phan_n\Zb_i\;.
\end{split}
\ee
As is well known, this procedure leads to $\Urm(N)$ groups, with the $\Urm(1)$ factors having non-trivial $\beta$ functions and thus spoiling the conformal invariance. As discussed in \cite{Dymarsky:2005uh,Dymarsky:2005nc}, integrating out these $\Urm(1)$ factors leads to $\SU(N)$ gauge groups and restores conformality, but also results in the addition of quartic double-trace terms in the component Lagrangian of the theory, arising by integrating out the $\Urm(1)$ parts of the adjoint $F_Z$ and $D$ auxiliary fields. We are therefore left with a quiver gauge theory with product gauge group
\be
\prod_i \SU(n_i N)\;.
\ee
The matter content is expressed in terms of the adjacency matrix 
\be
a^\mathbf{3}_{ij}=\frac{1}{|\Gamma|}\sum_{g\in\Gamma}\chi_\mathbf{3}(g)\chi_i(g)\bar{\chi}_j(g) \;,
\ee
where a nonzero element with $i=j$ denotes an adjoint field while a nonzero element with $i\neq j$ a bifundamental. These matrices are tabulated in Appendix \ref{sec:MattercontentN=2}. For $\Ncal=2$ theories, all gauge groups have a corresponding adjoint field, forming an $\Ncal=2$ vector multiplet. Therefore $a^{\bf 3}_{ii}=1$ for any $i$. So to focus on the bifundamental fields between nodes $i$ and $j$, we can subtract the diagonal part of the adjacency matrix:
\be
a^\mathbf{2}_{ij}\equiv a^\mathbf{3}_{ij}-\delta_{ij}\;.
\ee
Given the non-chiral nature of the $\Ncal=2$ theory, if the multiplet $Q_{ij}$ is in the $\Box\times \bar{\Box}$ representation of the $\SU(N_i)\times \SU(N_j)$ gauge groups, the $Q_{ji}$ multiplet will be in the conjugate $\bar{\Box}\times \Box$ representation.

The $i^{\mathrm{th}}$ $\Ncal=1$ vector multiplet $V_i$ combines with the $i^{\mathrm{th}}$ adjoint multiplet $Z_i$ to form an $\Ncal=2$ vector multiplet, while the two $\Ncal=1$ chiral multiplets $Q_{ij}$ and $Q_{ji}$ combine to form an $\Ncal=2$ hypermultiplet. Let us denote by $M$ the number of gauge groups/nodes of the quiver, and by $H$ the number of hypermultiplets. Table \ref{tab:numberhypermultiplets} summarises these numbers for the ADE groups.  
\begin{table}[h]
     \centering\small
         \renewcommand{\arraystretch}{1.3}
     \begin{tabular}{|c|c|c|c|c|c|}\hline
         Quiver &  $\hat{A}_{k}$& $\hat{D}_{k}$&$\hat{E}_6$&$\hat{E}_7$&$\hat{E}_8$\tabularnewline\hline
         $\Gamma$&$\Zset_{k+1}$&$\hat{\text{D}}_{k}$&2T&2O&2I\tabularnewline\hline
         $|\Gamma|$& $k+1$ & $4k-8$& $24$ & $48$ & $120$ \tabularnewline\hline
         $M$&$k+1$&$k+1$&$7$&$8$&$9$\tabularnewline\hline 
          $H$&$k+1$&$k$&$6$&$7$&$8$\tabularnewline\hline 
     \end{tabular}     \caption{The orbifold group, the corresponding $\hat{A}\hat{D}\hat{E}$ classification, number of nodes and number of hypermultiplets for the $\Ncal=2$ orbifolds.}
 \end{table}    \label{tab:numberhypermultiplets}

This information can be conveniently expressed in a quiver diagram, which is based on the corresponding affine Dynkin diagram (see Fig. \ref{fig:ADEDynkin}), with each node corresponding to a $\SU(n_i N)$ gauge group and  with the number of lines between nodes determined by the $a^{\bf 3}_{ij}$ matrix. Gauge theories with these product gauge groups, called ADE quiver theories, were first studied in \cite{Kronheimer:1989zs} with a focus on their Higgs-branch geometry, which was understood in a mirror-symmetry context in \cite{Intriligator:1996ex}. 

In the following it will be convenient to define the weighted (by the size of the gauge group) adjacency matrix
\be
d_{ij}\equiv\sum_{I,J=1}^2\sum_{m,n=1}^{n_i,n_j}\varepsilon_{IJ3}K_{\mathbf{3} I,\,im}^{jn}K_{\mathbf{3}J,\,jn}^{im}\;.
\ee
As this matrix will appear in the superpotential, we call it the superpotential coefficient matrix. It satisfies
\be
d_{ji}=-d_{ij} \;,
\ee
and
\be\label{eq:modulusd}
\bar{d}_{ij}d_{ij}=n_i^2n_j^2\,.
\ee
From \eqref{eq:modulusd}, we have $|d_{ij}|^2=n_i^2n_j^2$, hence (taking $d_{ij}$ to be real) we have 
\be\label{dij}
d_{ij}=\pm a^{\mathbf{2}}_{ij}n_in_j\;.
\ee
The choice of sign in (\ref{dij}) is conventional\footnote{The ambiguity in the sign of $d_{ij}$ follows from the fact that the Clebsch-Gordan coefficients $K_{\mathbf{3} I,\,im}^{jn}$ are unique up to a choice in phase.}. In the concrete cases that we will study, we will be writing $Q_{ij}\to X_{ij}/Y_{ij}$, in such a way that  $Q_{ij}\to X_{ij}$ if $d_{ij}=-1$ and $Q_{ij}\to Y_{ij}$ if $d_{ij}=+1$.

The quivers clearly have discrete symmetries related to relabelling the nodes (permuting the branes in the string theory construction). For a complete list of the outer automorphisms of the ADE quivers, we refer to \cite{DiFrancesco}. The symmetry group of the $\Zset_k$ quivers is also $\Zset_k$, and is implemented by the $\tau(g)$ matrices defined above acting by conjugation. These matrices (or at least a suitable subset of them) are expected to play a similar role in the non-abelian quivers, as we will see in our examples. 

As we will come back to in Section \ref{sec:ADEspinchains}, the orbifold theories allow for \emph{twisted sectors}, which can be obtained by inserting the $\gamma(g)$ matrix in the gauge trace of single-trace operators. 
\be
\mathcal{O}_g=\Tr\left(\gamma(g)\Phi^{I_1}\Phi^{I_2}\ldots\Phi^{I_{L-1}}\Phi^{I_L}\right)\;.
\ee
Therefore, the spectrum of the theory at the orbifold point will organise itself in separate twisted sectors. We will indicate the states belonging to each sector in our specific examples.

\subsection{The ADE Quiver Lagrangian}

Having described the field content of the general $\Ncal=2$ ADE quiver theory in the previous section, we are ready to write down the $\Ncal=1$ superspace Lagrangian that one obtains by the orbifolding procedure. As the $\Zset_2$ case has been extensively treated in previous works (e.g. \cite{Gadde:2009dj, Gadde:2010zi}) we will focus on the simply reducible case where there are no multiplicities in the products of irreducible representations. We will also let the couplings of each gauge node vary independently:
\be \label{kappamarginal}
\gym^i=\kappa_i \gym\;.
\ee
The Lagrangian one obtains is given by  $S_{\NN=2,\,\Gamma}=S_\text{vector}+S_\ghost+S_\chiral$, where:

\begin{equation}\label{eq:NN=2generalaction}
    \begin{split}
        S_{\text{vector}}&=\sum_{i=1}^M\frac{{\color{red}n_i}}{\gym^2{\color{blue}\kappa_i^2}}\int d^4x\Tr_i\left(\frac{1}{4}\left(\int d^2\theta(W^{\alpha})_i(W_{\alpha})_i+\text{h.c.}\right)-\frac{1}{\xi_i}\int d^4\theta~\codif^2V_i\bcodif^2V_i\right),\\
        S_\ghost&=\sum_{i=1}^M{\color{red}n_i}\int d^4xd^4\theta~\Tr_i\left(\left(c_i'+\bar{c}'_i\right)\text{L}_{\frac{\gym{\color{blue}\kappa_i}}{2}V_i}\left(c_i+\bar{c}_i+\coth\text{L}_{\frac{\gym{\color{blue}\kappa_i}}{2}V_i}\left(c_i-\bar{c}_i\right)\right)\right),\\
        S_\chiral&=\sum_{i=1}^M{\color{red}n_i}\int d^4xd^4\theta~\Tr_i\biggl(e^{-\gym{\color{blue}\kappa_i}V_i}\bar{Z}_{i}e^{\gym{\color{blue}\kappa_i}V_i}Z_i\biggr)\\
        &+\sum_{i,j=1}^M{\color{red}a^\mathbf{2}_{ij}n_in_j}\int d^4xd^4\theta~\Tr_{j}\biggl(e^{-\gym{\color{blue}\kappa_j} V_j}\bar{Q}_{ji}e^{\gym{\color{blue}\kappa_i}V_i}Q_{ij}\biggr)\\
        &+\left(\int d^4x d^2\theta~ \Wcal_\Gamma + h.c.\right)
    \end{split}
\end{equation}
where 
\begin{equation}
    (W_{\alpha})_i\equiv i\bcodif^2\left(e^{-\gym{\color{blue}\kappa_i}V_i}\codif_{\alpha}e^{\gym{\color{blue}\kappa_i}V_i}\right),
\end{equation}
and the superpotential is given by 
\begin{equation}\label{eq:ADEsuperpotential}
    \mathcal{W}_{\Gamma}=i\gym \sum_{i,j=1}^M{\color{blue}\kappa_i}{\color{red}d_{ji}}\Tr_j\left(Q_{ji}Z_iQ_{ji}\right).
\end{equation}
We have highlighted in red the differences to the $\Ncal=2$ theory due to the orbifolding procedure, and in blue the marginal deformation parameters $\kappa_i$ which take the theory away from the orbifold point.
The superspace Feynman rules arising from the above Lagrangian are summarised in Appendix \ref{sec:SuperspaceFeynman}.  

As is well known \cite{Kachru:1998ys,Lawrence:1998ja}, the orbifolded theories inherit the finiteness properties of $\Ncal=4$ SYM. As also discussed in those works, changing the gauge couplings away from their orbifold-point values as in (\ref{kappamarginal}) is a marginal deformation preserving $\Ncal=2$ superconformal invariance. Although this can be argued in general, it is also straightforward to directly check that the vanishing of the beta functions is unaffected by the deformation. For completeness, we show this in Appendix \ref{sec:betafunction}. 

As we will eventually want to study the spectrum of these theories in terms of component fields, it is crucial to know how the fields are related by supersymmetry transformations. Therefore, we list the unbroken on-shell $\Ncal=2$ supercharges and show their actions on the component fields in Appendix \ref{sec:ExtendedSusyTransformations}.

\section{The planar ADE Dilatation Operator} \label{sec:ADEDilatationOperator}

In this section we provide some details of the computation of the one-loop dilatation operator of the ADE quiver theories, in the full scalar sector. In $\Ncal=4$ SYM, this computation was performed in \cite{Minahan:2002ve}, leading to the $\SO(6)$ XXX Hamiltonian. For reviews and additional details, see \cite{Minahan:2010js,deLeeuw:2017cop}. We will work in an $\Ncal=1$ supersymmetric setting. See \cite{Sieg:2010tz, Sieg:2010jt} for previous superspace approaches to the $\Ncal=4$ dilatation operator and \cite{Pomoni:2011jj} for the $\Ncal=2$ $\Zset_2$ orbifold case. Similar techniques were also employed in \cite{Minahan:2011dd} in the study of a specific $\Ncal=1$ marginal deformation of $\Ncal=4$ SYM. In the above works, the focus was on higher-loop computations where the use of superspace was essential. However, even in our one-loop case we find that superspace simplifies some of the computations and of course guarantees that our result is fully supersymmetric. Furthermore, working in components often entails fixing the supersymmetric gauge to be the Wess-Zumino gauge, in order to reduce the number of auxiliary fields. However, the Wess-Zumino gauge explicitly breaks supersymmetry, with the result that the fundamental fields of the theory acquire non-zero anomalous dimensions, naively making the theory appear to be non-finite. Of course, as we verify in Appendix \ref{sec:betafunction}, working in superspace the finiteness of the theory is evident, as the anomalous dimensions of all fundamental fields vanish.     

For a recent review of superspace techniques as applied to $\Ncal=2$ supersymmetric gauge theory, see \cite{Pomoni:2019oib}. The superspace notation and Feynman rules that we use largely follow \cite{Gates:1983nr} and are summarised in Appendix \ref{sec:SuperspaceFeynman}.

We are interested in computing the one-loop, planar contribution to the two point function
\be
\langle O_\Delta(x)\bar{O}_\Delta(y)\rangle=\frac{1}{(x-y)^{2\Delta}}\;,
\ee
where $O_\Delta(x)$ are gauge-invariant operators in the quiver gauge theory. We focus on the scalar sector, where the operators are made up of the lowest components of the superfields $Q_{ij},\bar{Q}_{ij},Z_i,\bar{Z}_i$. Due to the planar limit, the one-loop contributions are a sum of interactions between neighbouring fields, which allows us to reduce the problem to finding all (divergent) one-loop contributions to all possible pairs of fields. 

Rather than evaluating the two-point function directly, We will take an effective vertex approach \cite{Beisert:2002bb, Beisert:2003tq} where one does not consider the $\bar{O}(y)$ operator but only the outgoing fields from the one-loop corrected $O(x)$ operator. Renormalising each operator as $\Ocal^I_{\text{ren}}=\mathcal{Z}^{I}_{\;J}(\lambda,\epsilon) \Ocal^J_{\text{bare}}$, the action of the dilatation operator will be given by
\be
\mathfrak{D}^{I}_{\;J}=2\text{Res}_{\epsilon=0}\left[\lambda\frac{d}{d\lambda}\log \mathcal{Z}^I_{\;J}(\lambda,\epsilon)\right]\;.
\ee
where the 't Hooft coupling is defined as $\lambda=\gym^2 N$. As this action mixes different monomials in the fields, constructing the operators with definite scaling dimensions $\Delta$ amounts to diagonalising the dilatation operator.

\subsection{Reduction to $\Ncal=4$ SYM diagrams}\label{sec:Reduction}

Given the structure of the ADE Lagrangian (\ref{eq:NN=2generalaction}), it is clear that all the Feynman diagrams will be the same as in $\Ncal=4$ SYM, but now decorated with different choices of gauge groups as well as a dependence on the $\kappa_i$ coefficients of the marginally deformed theory. In this section, we will therefore first keep track of these differences to $\Ncal=4$ SYM. The Feynman integrals themselves will be the same as in $\Ncal=4$ SYM, and will be evaluated in the next subsection. 

As a first example of how the ADE theory diagrams map to those of $\Ncal=4$ SYM, consider the following diagram giving the correction to $\cdots Q_{ij}Q_{ji} \cdots$. We read the diagram from the bottom up, and the dotted line indicates that in computing the effective vertex we only consider the diagram up to this point, and in particular the top index loop is not counted.  
\be \label{FD:QQhol}
\begin{tikzpicture}[scale=1,baseline=1.1cm]

  \draw[red,thick](0.2,0)--(1.8,0);\draw[blue,thick](0,0)--(-0.2,0);\draw[gray,thick](-0.2,-0.2)--(2.2,-0.2);\draw[blue,thick] (2.2,0)--(2,0);
  \draw[red,thick](1.8,0) arc (0:180:0.8);
  \draw[blue,thick](2,0) arc (0:85:1);
  \draw[blue,thick](0,0) arc (180:95:1);
  \draw[blue,thick](0.91,0.99)--(0.9,2);
  \draw[blue,thick](1.09,0.99)--(1.09,2);
  \draw[green,thick](1.8,3) arc (0:-180:0.8);
  \draw[blue,thick](2,3) arc (0:-85:1);
  \draw[blue,thick](0,3) arc (-180:-95:1);
  \draw[green,thick](0.2,3)--(1.8,3);\draw[blue,thick](0,3)--(-0.2,3);\draw[gray,thick] (-0.2,3.2)--(2.2,3.2);\draw[blue,thick](2.2,3)--(2,3);
  \node at (0.2,-0.6) {$Q_{ij}$};\node at (2,-0.6) {$Q_{ji}$};
  \node at (0.2,3.6) {$\bar{Q}_{ik}$};\node at (2,3.6) {$\bar{Q}_{ki}$};
  \node[right,scale=0.7] at (2.2,1.5) {\LARGE $=\kappa_i^2\frac{\bar{d}_{ji}d_{ki}}{n_jn_i^3}\;\times$};
\draw[-,dotted,gray,thick] (-0.2,2.6)--(2.2,2.6);
\end{tikzpicture}\hspace{0.1cm}
\begin{tikzpicture}[scale=1,baseline=1.1cm]

  \draw[violet,thick](0.2,0)--(1.8,0);\draw[violet,thick](0,0)--(-0.2,0);\draw[gray,thick](-0.2,-0.2)--(2.2,-0.2);\draw[violet,thick] (2.2,0)--(2,0);
  \draw[violet,thick](1.8,0) arc (0:180:0.8);
  \draw[violet,thick](2,0) arc (0:85:1);
  \draw[violet,thick](0,0) arc (180:95:1);
  \draw[violet,thick](0.91,0.99)--(0.9,2);
  \draw[violet,thick](1.09,0.99)--(1.09,2);
  \draw[violet,thick](1.8,3) arc (0:-180:0.8);
  \draw[violet,thick](2,3) arc (0:-85:1);
  \draw[violet,thick](0,3) arc (-180:-95:1);
  \draw[violet,thick](0.2,3)--(1.8,3);\draw[violet,thick](0,3)--(-0.2,3);\draw[gray,thick] (-0.2,3.2)--(2.2,3.2);\draw[violet,thick](2.2,3)--(2,3);
  \node at (0.2,-0.6) {$X$};\node at (2,-0.6) {$Y$};
  \node at (0.2,3.6) {$\bar{X}$};\node at (2,3.6) {$\bar{Y}$};
  \node at (1.3,1.2) {$\Zb$};\node at (1.3,1.8) {$Z$};
\draw[-,dotted,gray,thick] (-0.2,2.6)--(2.2,2.6);
\end{tikzpicture}
\ee
The colouring of the double lines in the $\Ncal=2$ diagram represents the different nodes under which the corresponding field is charged. The convention is that different colours on the same propagator are exclusive, so in the above case $i\neq j$ and $i \neq k$, but it could be that $k=j$. The planar diagram on the right represents the $\Ncal=4$ SYM case, where there is only one node and therefore only one type of gauge group index. The notation is slightly schematic in that in the $\Ncal=2$ diagram, the choice of $X$ vs. $Y$ fields is determined by the specific choice made for $Q_{ij}$ for a given quiver, so it actually corresponds to four $\Ncal=4$ SYM diagrams. These diagrams have different signs (schematically, $XY\ra XY$ vs. $XY\ra -YX$) which in the $\Ncal=2$ case are taken care of by the $d_{ij}$ coefficients. 

Note that the powers of $n_i$ come both from the index loops as well as the propagators as listed in Appendix \ref{sec:SuperspaceFeynman}. Suppressing the dotted line in the following, we can similarly relate 
\be\label{FD:QPhihol}
\begin{tikzpicture}[scale=1,baseline=1.1cm]
  \draw[red,thick](0.2,0)--(1.8,0);\draw[blue,thick](0,0)--(-0.2,0);\draw[gray,thick](-0.2,-0.2)--(2.2,-0.2);\draw[red,thick]
  (2.2,0)--(2,0); \draw[red,thick](1.8,0) arc (0:180:0.8);
  \draw[red,thick](2,0) arc (0:85:1); \draw[blue,thick](0,0) arc
  (180:95:1); \draw[blue,thick](0.91,0.99)--(0.9,2);
  \draw[red,thick](1.09,0.99)--(1.09,2); \draw[red,thick](1.8,3) arc
  (0:-180:0.8); \draw[red,thick](2,3) arc (0:-85:1);
  \draw[blue,thick](0,3) arc (-180:-95:1);
  \draw[red,thick](0.2,3)--(1.8,3);\draw[blue,thick](0,3)--(-0.2,3);\draw[gray,thick]
  (-0.2,3.2)--(2.2,3.2);\draw[red,thick](2.2,3)--(2,3); \node at
  (0.2,-0.6) {$Q_{ij}$};\node at (2,-0.6) {$Z_{j}$}; \node at
  (0.2,3.6) {$\bar{Q}_{ij}$};\node at (2,3.6) {$\bar{Z}_{j}$};
  \node[right,scale=0.7] at (2.2,1.5)
       {\LARGE$=\kappa_j^2\bar{d}_{ij}d_{ij}=\kappa_j^2$};
\end{tikzpicture}
\begin{tikzpicture}[scale=1,baseline=1.1cm]

  \draw[violet,thick](0.2,0)--(1.8,0);\draw[violet,thick](0,0)--(-0.2,0);\draw[gray,thick](-0.2,-0.2)--(2.2,-0.2);\draw[violet,thick]
  (2.2,0)--(2,0); \draw[violet,thick](1.8,0) arc (0:180:0.8);
  \draw[violet,thick](2,0) arc (0:85:1); \draw[violet,thick](0,0) arc
  (180:95:1); \draw[violet,thick](0.91,0.99)--(0.9,2);
  \draw[violet,thick](1.09,0.99)--(1.09,2); \draw[violet,thick](1.8,3) arc
  (0:-180:0.8); \draw[violet,thick](2,3) arc (0:-85:1);
  \draw[violet,thick](0,3) arc (-180:-95:1);
  \draw[violet,thick](0.2,3)--(1.8,3);\draw[violet,thick](0,3)--(-0.2,3);\draw[gray,thick]
  (-0.2,3.2)--(2.2,3.2);\draw[violet,thick](2.2,3)--(2,3); \node at
  (0.2,-0.6) {$X$};\node at (2,-0.6) {$Z$}; \node at
  (0.2,3.6) {$\bar{X}$};\node at (2,3.6) {$\bar{Z}$};
    \node at (1.3,1.2) {$\bar{Y}$};\node at (1.3,1.8) {$Y$};
\end{tikzpicture}
\ee
and 
\be \label{FD:QPhihol1}
\begin{tikzpicture}[scale=1,baseline=1.1cm]

  \draw[red,thick](0.2,0)--(1.8,0);\draw[blue,thick](0,0)--(-0.2,0);\draw[gray,thick](-0.2,-0.2)--(2.2,-0.2);\draw[red,thick] (2.2,0)--(2,0);
  \draw[red,thick](1.8,0) arc (0:180:0.8);
  \draw[red,thick](2,0) arc (0:85:1);
  \draw[blue,thick](0,0) arc (180:95:1);
  \draw[blue,thick](0.91,0.99)--(0.9,2);
  \draw[red,thick](1.09,0.99)--(1.09,2);
  \draw[blue,thick](1.8,3) arc (0:-180:0.8);
  \draw[red,thick](2,3) arc (0:-85:1);
  \draw[blue,thick](0,3) arc (-180:-95:1);
  \draw[blue,thick](0.2,3)--(1.8,3);\draw[blue,thick](0,3)--(-0.2,3);\draw[gray,thick] (-0.2,3.2)--(2.2,3.2);\draw[red,thick](2.2,3)--(2,3);
  \node at (0.2,-0.6) {$Q_{ij}$};\node at (2,-0.6) {$Z_{j}$};
   \node at (0.2,3.6) {$\bar{Z}_{i}$};\node at (2,3.6) {$\bar{Q}_{ij}$};
   \node[right,scale=0.7] at (2.2,1.5) {\LARGE$=\kappa_i\kappa_j\bar{d}_{ij}d_{ji}=-\kappa_i\kappa_j$};
\end{tikzpicture}\begin{tikzpicture}[scale=1,baseline=1.1cm]

  \draw[violet,thick](0.2,0)--(1.8,0);\draw[violet,thick](0,0)--(-0.2,0);\draw[gray,thick](-0.2,-0.2)--(2.2,-0.2);\draw[violet,thick]
  (2.2,0)--(2,0); \draw[violet,thick](1.8,0) arc (0:180:0.8);
  \draw[violet,thick](2,0) arc (0:85:1); \draw[violet,thick](0,0) arc
  (180:95:1); \draw[violet,thick](0.91,0.99)--(0.9,2);
  \draw[violet,thick](1.09,0.99)--(1.09,2); \draw[violet,thick](1.8,3) arc
  (0:-180:0.8); \draw[violet,thick](2,3) arc (0:-85:1);
  \draw[violet,thick](0,3) arc (-180:-95:1);
  \draw[violet,thick](0.2,3)--(1.8,3);\draw[violet,thick](0,3)--(-0.2,3);\draw[gray,thick]
  (-0.2,3.2)--(2.2,3.2);\draw[violet,thick](2.2,3)--(2,3); \node at
  (0.2,-0.6) {$X$};\node at (2,-0.6) {$Z$}; \node at
  (0.2,3.6) {$\bar{X}$};\node at (2,3.6) {$\bar{Z}$};
     \node[right,scale=0.7] at (2.2,1.5) {\LARGE$=\kappa_i\kappa_j$};
    \node at (1.3,1.2) {$\bar{Y}$};\node at (1.3,1.8) {$Y$};
\end{tikzpicture}
\begin{tikzpicture}[scale=1,baseline=1.1cm]

  \draw[violet,thick](0.2,0)--(1.8,0);\draw[violet,thick](0,0)--(-0.2,0);\draw[gray,thick](-0.2,-0.2)--(2.2,-0.2);\draw[violet,thick]
  (2.2,0)--(2,0); \draw[violet,thick](1.8,0) arc (0:180:0.8);
  \draw[violet,thick](2,0) arc (0:85:1); \draw[violet,thick](0,0) arc
  (180:95:1); \draw[violet,thick](0.91,0.99)--(0.9,2);
  \draw[violet,thick](1.09,0.99)--(1.09,2); \draw[violet,thick](1.8,3) arc
  (0:-180:0.8); \draw[violet,thick](2,3) arc (0:-85:1);
  \draw[violet,thick](0,3) arc (-180:-95:1);
  \draw[violet,thick](0.2,3)--(1.8,3);\draw[violet,thick](0,3)--(-0.2,3);\draw[gray,thick]
  (-0.2,3.2)--(2.2,3.2);\draw[violet,thick](2.2,3)--(2,3); \node at
  (0.2,-0.6) {$X$};\node at (2,-0.6) {$Z$}; \node at
  (0.2,3.6) {$\bar{Z}$};\node at (2,3.6) {$\bar{X}$};
      \node at (1.3,1.2) {$\bar{Y}$};\node at (1.3,1.8) {$Y$};
\end{tikzpicture}
\ee
The transposed diagrams with $Z Q$ at the bottom are equal to the above. 
As we will discuss in the next section, diagrams with gauge boson exchange in the holomorphic sector are finite by power counting.
So the above are all the diagrams in the holomorphic sector which are UV divergent and contribute to the dilatation operator.

The diagrams for the antiholomorphic sector are the conjugates of the above and will take equal values. So we now need to consider the diagrams in the mixed sector, i.e. where one field is holomorphic and the other antiholomorphic. To save space, we will only indicate the factors that differ from the corresponding $\Ncal=4$ SYM diagrams, which are straightforward to reproduce. Let us first write the diagrams with vertical gauge fields, which are 
\be \label{FD:QQbgauge}
  \begin{tikzpicture}[scale=0.8,baseline=1.1cm]
  \draw[red,thick](0.2,0)--(1.8,0);  \draw[blue,thick](0,0)--(-0.2,0);\draw[gray,thick](-0.2,-0.2)--(2.2,-0.2);\draw[blue,thick] (2.2,0)--(2,0);
  \draw[red,thick](1.8,0) arc (0:180:0.8);
  \draw[blue,thick](2,0) arc (0:85:1);
  \draw[blue,thick](0,0) arc (180:95.5:1);
  \draw[blue,thick,decorate, decoration={snake, amplitude=-1.1, segment length=8.7}](0.9,0.99)--(0.9,2);
  \draw[blue,thick,decorate, decoration={snake, amplitude=-1.1, segment length=8.7}](1.09,0.99)--(1.09,2);
  \draw[green,thick](1.8,3) arc (0:-180:0.8);
  \draw[blue,thick](2,3) arc (0:-85.5:1);
  \draw[blue,thick](0,3) arc (-180:-95:1);
  \draw[green,thick](0.2,3)--(1.8,3);\draw[blue,thick](0,3)--(-0.2,3);\draw[gray,thick] (-0.2,3.2)--(2.2,3.2);\draw[blue,thick](2.2,3)--(2,3);
  \node at (0.2,-0.6) {$Q_{ij}$};\node at (2,-0.6) {$\Qb_{ji}$};
    \node at (0.2,3.6) {$Q_{ik}$};\node at (2,3.6) {$\Qb_{ki}$};
     \node[right,scale=0.7] at (2.2,1.5) {\LARGE$=\half\kappa_i^2\frac{n_k}{n_i}=-$};
\end{tikzpicture}\begin{tikzpicture}[scale=0.8,baseline=1.1cm]

  \draw[red,thick](0.2,0)--(1.8,0);  \draw[blue,thick](0,0)--(-0.2,0);\draw[gray,thick](-0.2,-0.2)--(2.2,-0.2);\draw[blue,thick] (2.2,0)--(2,0);
  \draw[red,thick](1.8,0) arc (0:180:0.8);
  \draw[blue,thick](2,0) arc (0:85:1);
  \draw[blue,thick](0,0) arc (180:95.5:1);
  \draw[blue,thick,decorate, decoration={snake, amplitude=-1.1, segment length=8.7}](0.9,0.99)--(0.9,2);
  \draw[blue,thick,decorate, decoration={snake, amplitude=-1.1, segment length=8.7}](1.09,0.99)--(1.09,2);
  \draw[green,thick](1.8,3) arc (0:-180:0.8);
  \draw[blue,thick](2,3) arc (0:-85.5:1);
  \draw[blue,thick](0,3) arc (-180:-95:1);
  \draw[green,thick](0.2,3)--(1.8,3);\draw[blue,thick](0,3)--(-0.2,3);\draw[gray,thick] (-0.2,3.2)--(2.2,3.2);\draw[blue,thick](2.2,3)--(2,3);
  \node at (0.2,-0.6) {$Q_{ij}$};\node at (2,-0.6) {$\bar{Q}_{ji}$};
  \node at (0.2,3.6) {$\bar{Q}_{ik}$};\node at (2,3.6) {$Q_{ki}$};
  \node[right,scale=0.7] at (2.6,1.5) {\LARGE ,};
  \end{tikzpicture}\hspace{0.5cm}
    \begin{tikzpicture}[scale=0.8,baseline=1.1cm]

  \draw[red,thick](0.2,0)--(1.8,0);  \draw[blue,thick](0,0)--(-0.2,0);\draw[gray,thick](-0.2,-0.2)--(2.2,-0.2);\draw[blue,thick] (2.2,0)--(2,0);
  \draw[red,thick](1.8,0) arc (0:180:0.8);
  \draw[blue,thick](2,0) arc (0:85:1);
  \draw[blue,thick](0,0) arc (180:95.5:1);
  \draw[blue,thick,decorate, decoration={snake, amplitude=-1.1, segment length=8.7}](0.9,0.99)--(0.9,2);
  \draw[blue,thick,decorate, decoration={snake, amplitude=-1.1, segment length=8.7}](1.09,0.99)--(1.09,2);
  \draw[blue,thick](1.8,3) arc (0:-180:0.8);
  \draw[blue,thick](2,3) arc (0:-85.5:1);
  \draw[blue,thick](0,3) arc (-180:-95:1);
  \draw[blue,thick](0.2,3)--(1.8,3);\draw[blue,thick](0,3)--(-0.2,3);\draw[gray,thick] (-0.2,3.2)--(2.2,3.2);\draw[blue,thick](2.2,3)--(2,3);
  \node at (0.2,-0.6) {$Q_{ij}$};\node at (2,-0.6) {$\Qb_{ji}$};
    \node at (0.2,3.6) {$Z_{i}$};\node at (2,3.6) {$\Zb_{i}$};
     \node[right,scale=0.7] at (2.2,1.5) {\LARGE$=\half\frac{\kappa_i^2}{n_i}=-$};
\end{tikzpicture}
\begin{tikzpicture}[scale=0.8,baseline=1.1cm]

  \draw[red,thick](0.2,0)--(1.8,0);  \draw[blue,thick](0,0)--(-0.2,0);\draw[gray,thick](-0.2,-0.2)--(2.2,-0.2);\draw[blue,thick] (2.2,0)--(2,0);
  \draw[red,thick](1.8,0) arc (0:180:0.8);
  \draw[blue,thick](2,0) arc (0:85:1);
  \draw[blue,thick](0,0) arc (180:95.5:1);
  \draw[blue,thick,decorate, decoration={snake, amplitude=-1.1, segment length=8.7}](0.9,0.99)--(0.9,2);
  \draw[blue,thick,decorate, decoration={snake, amplitude=-1.1, segment length=8.7}](1.09,0.99)--(1.09,2);
  \draw[blue,thick](1.8,3) arc (0:-180:0.8);
  \draw[blue,thick](2,3) arc (0:-85.5:1);
  \draw[blue,thick](0,3) arc (-180:-95:1);
  \draw[blue,thick](0.2,3)--(1.8,3);\draw[blue,thick](0,3)--(-0.2,3);\draw[gray,thick] (-0.2,3.2)--(2.2,3.2);\draw[blue,thick](2.2,3)--(2,3);
  \node at (0.2,-0.6) {$Q_{ij}$};\node at (2,-0.6) {$\bar{Q}_{ji}$};
    \node at (0.2,3.6) {$\Zb_{i}$};\node at (2,3.6) {$Z_{i}$};
     \node[right,scale=0.7] at (2.2,1.5) {.};
\end{tikzpicture}\;,
  \ee
as well as 
  \be \label{FD:PhiPhibgauge}
\begin{tikzpicture}[scale=0.8,baseline=1.1cm]

  \draw[blue,thick](0.2,0)--(1.8,0);  \draw[blue,thick](0,0)--(-0.2,0);\draw[gray,thick](-0.2,-0.2)--(2.2,-0.2);\draw[blue,thick] (2.2,0)--(2,0);
  \draw[blue,thick](1.8,0) arc (0:180:0.8);
  \draw[blue,thick](2,0) arc (0:85:1);
  \draw[blue,thick](0,0) arc (180:95.5:1);
  \draw[blue,thick,decorate, decoration={snake, amplitude=-1.1, segment length=8.7}](0.9,0.99)--(0.9,2);
  \draw[blue,thick,decorate, decoration={snake, amplitude=-1.1, segment length=8.7}](1.09,0.99)--(1.09,2);
  \draw[blue,thick](1.8,3) arc (0:-180:0.8);
  \draw[blue,thick](2,3) arc (0:-85.5:1);
  \draw[blue,thick](0,3) arc (-180:-95:1);
  \draw[blue,thick](0.2,3)--(1.8,3);\draw[blue,thick](0,3)--(-0.2,3);\draw[gray,thick] (-0.2,3.2)--(2.2,3.2);\draw[blue,thick](2.2,3)--(2,3);
  \node at (0.2,-0.6) {$Z_{i}$};\node at (2,-0.6) {$\Zb_{i}$};
    \node at (0.2,3.6) {$Z_{i}$};\node at (2,3.6) {$\Zb_{i}$};
    \node[right,scale=0.7] at (2.2,1.5) {\LARGE$=\half\kappa_i^2=-$};
\end{tikzpicture}
\begin{tikzpicture}[scale=0.8,baseline=1.1cm]

  \draw[blue,thick](0.2,0)--(1.8,0);  \draw[blue,thick](0,0)--(-0.2,0);\draw[gray,thick](-0.2,-0.2)--(2.2,-0.2);\draw[blue,thick] (2.2,0)--(2,0);
  \draw[blue,thick](1.8,0) arc (0:180:0.8);
  \draw[blue,thick](2,0) arc (0:85:1);
  \draw[blue,thick](0,0) arc (180:95.5:1);
  \draw[blue,thick,decorate, decoration={snake, amplitude=-1.1, segment length=8.7}](0.9,0.99)--(0.9,2);
  \draw[blue,thick,decorate, decoration={snake, amplitude=-1.1, segment length=8.7}](1.09,0.99)--(1.09,2);
  \draw[blue,thick](1.8,3) arc (0:-180:0.8);
  \draw[blue,thick](2,3) arc (0:-85.5:1);
  \draw[blue,thick](0,3) arc (-180:-95:1);
  \draw[blue,thick](0.2,3)--(1.8,3);\draw[blue,thick](0,3)--(-0.2,3);\draw[gray,thick] (-0.2,3.2)--(2.2,3.2);\draw[blue,thick](2.2,3)--(2,3);
  \node at (0.2,-0.6) {$Z_{i}$};\node at (2,-0.6) {$\bar{Z}_{i}$};
    \node at (0.2,3.6) {$\bar{Z}_{i}$};\node at (2,3.6) {$Z_{i}$};
    \node[right,scale=0.7] at (2.6,1.5) {\LARGE ,};
\end{tikzpicture}\hspace{0.5cm}
\begin{tikzpicture}[scale=0.8,baseline=1.1cm]

  \draw[blue,thick](0.2,0)--(1.8,0);  \draw[blue,thick](0,0)--(-0.2,0);\draw[gray,thick](-0.2,-0.2)--(2.2,-0.2);\draw[blue,thick] (2.2,0)--(2,0);
  \draw[blue,thick](1.8,0) arc (0:180:0.8);
  \draw[blue,thick](2,0) arc (0:85:1);
  \draw[blue,thick](0,0) arc (180:95.5:1);
  \draw[blue,thick,decorate, decoration={snake, amplitude=-1.1, segment length=8.7}](0.9,0.99)--(0.9,2);
  \draw[blue,thick,decorate, decoration={snake, amplitude=-1.1, segment length=8.7}](1.09,0.99)--(1.09,2);
  \draw[red,thick](1.8,3) arc (0:-180:0.8);
  \draw[blue,thick](2,3) arc (0:-85.5:1);
  \draw[blue,thick](0,3) arc (-180:-95:1);
  \draw[red,thick](0.2,3)--(1.8,3);\draw[blue,thick](0,3)--(-0.2,3);\draw[gray,thick] (-0.2,3.2)--(2.2,3.2);\draw[blue,thick](2.2,3)--(2,3);
  \node at (0.2,-0.6) {$Z_{i}$};\node at (2,-0.6) {$\bar{Z}_{i}$};
    \node at (0.2,3.6) {$Q_{ij}$};\node at (2,3.6) {$\Qb_{ji}$};
    \node[right,scale=0.7] at (2.2,1.5) {\LARGE$=\half\kappa_i^2n_j=-$};
\end{tikzpicture}
\begin{tikzpicture}[scale=0.8,baseline=1.1cm]

  \draw[blue,thick](0.2,0)--(1.8,0);  \draw[blue,thick](0,0)--(-0.2,0);\draw[gray,thick](-0.2,-0.2)--(2.2,-0.2);\draw[blue,thick] (2.2,0)--(2,0);
  \draw[blue,thick](1.8,0) arc (0:180:0.8);
  \draw[blue,thick](2,0) arc (0:85:1);
  \draw[blue,thick](0,0) arc (180:95.5:1);
  \draw[blue,thick,decorate, decoration={snake, amplitude=-1.1, segment length=8.7}](0.9,0.99)--(0.9,2);
  \draw[blue,thick,decorate, decoration={snake, amplitude=-1.1, segment length=8.7}](1.09,0.99)--(1.09,2);
  \draw[red,thick](1.8,3) arc (0:-180:0.8);
  \draw[blue,thick](2,3) arc (0:-85.5:1);
  \draw[blue,thick](0,3) arc (-180:-95:1);
  \draw[red,thick](0.2,3)--(1.8,3);\draw[blue,thick](0,3)--(-0.2,3);\draw[gray,thick] (-0.2,3.2)--(2.2,3.2);\draw[blue,thick](2.2,3)--(2,3);
  \node at (0.2,-0.6) {$Z_{i}$};\node at (2,-0.6) {$\bar{Z}_{i}$};
    \node at (0.2,3.6) {$\bar{Q}_{ij}$};\node at (2,3.6) {$Q_{ji}$};
    \node[right,scale=0.7] at (2.2,1.5) {\LARGE .};
\end{tikzpicture}\;.
\ee
Then we have identity-type diagrams with horizontal gauge field exchange:
\be\label{FD:gaugehorizontal}
\begin{tikzpicture}[scale=0.8,baseline=1.1cm]

  \draw[blue,thick](0,0)--(-0.2,0);\draw[gray,thick](-0.2,-0.2)--(2.2,-0.2);\draw[green,thick] (2.2,0)--(2,0);\draw[blue,thick](0,0)--(0,3);\draw[green,thick](2,0)--(2,3);
  \draw[red,thick](0.2,1.4)--(0.2,0)--(1.8,0)--(1.8,1.4);
  \draw[red,thick](0.2,1.6)--(0.2,3)--(1.8,3)--(1.8,1.6);\draw[blue,thick](0,3)--(-0.2,3);\draw[gray,thick] (-0.2,3.2)--(2.2,3.2);\draw[green,thick](2.2,3)--(2,3);
  \draw[red,thick,decorate, decoration={snake, amplitude=-1.1, segment length=9.6}](0.2,1.4)--(1.8,1.4);
  \draw[red,thick,decorate, decoration={snake, amplitude=-1.1, segment length=9.6}](0.2,1.6)--(1.8,1.6);
\node at (0.2,-0.6) {$Q_{ij}$};\node at (2,-0.6) {$\bar{Q}_{jk}$};
    \node at (0.2,3.6) {$\bar{Q}_{ij}$};\node at (2,3.6) {$Q_{jk}$};
    \node[right,scale=0.7] at (2.2,1.5) {\large$=-\kappa_j^2\;,$};
\end{tikzpicture}\hspace{0.7cm}
\begin{tikzpicture}[scale=0.8,baseline=1.1cm]

  \draw[blue,thick](0,0)--(-0.2,0);\draw[gray,thick](-0.2,-0.2)--(2.2,-0.2);\draw[blue,thick] (2.2,0)--(2,0);\draw[blue,thick](0,0)--(0,3);\draw[blue,thick](2,0)--(2,3);
  \draw[blue,thick](0.2,1.4)--(0.2,0)--(1.8,0)--(1.8,1.4);
  \draw[blue,thick](0.2,1.6)--(0.2,3)--(1.8,3)--(1.8,1.6);\draw[blue,thick](0,3)--(-0.2,3);\draw[gray,thick] (-0.2,3.2)--(2.2,3.2);\draw[blue,thick](2.2,3)--(2,3);
  \draw[blue,thick,decorate, decoration={snake, amplitude=-1.1, segment length=9.6}](0.2,1.4)--(1.8,1.4);
  \draw[blue,thick,decorate, decoration={snake, amplitude=-1.1, segment length=9.6}](0.2,1.6)--(1.8,1.6);
\node at (0.2,-0.6) {$Z_{i}$};\node at (2,-0.6) {$\Zb_{i}$};
    \node at (0.2,3.6) {$\Zb_{i}$};\node at (2,3.6) {$Z_{i}$};
    \node[right,scale=0.7] at (2.2,1.5) {\large$=-\kappa_i^2\;,$};
\end{tikzpicture}\hspace{0.7cm}
\begin{tikzpicture}[scale=0.8,baseline=1.1cm]
  \draw[blue,thick](0,0)--(-0.2,0);\draw[gray,thick](-0.2,-0.2)--(2.2,-0.2);\draw[red,thick] (2.2,0)--(2,0);\draw[blue,thick](0,0)--(0,3);\draw[red,thick](2,0)--(2,3);
  \draw[red,thick](0.2,1.4)--(0.2,0)--(1.8,0)--(1.8,1.4);
  \draw[red,thick](0.2,1.6)--(0.2,3)--(1.8,3)--(1.8,1.6);\draw[blue,thick](0,3)--(-0.2,3);\draw[gray,thick] (-0.2,3.2)--(2.2,3.2);\draw[red,thick](2.2,3)--(2,3);
  \draw[red,thick,decorate, decoration={snake, amplitude=-1.1, segment length=9.6}](0.2,1.4)--(1.8,1.4);
  \draw[red,thick,decorate, decoration={snake, amplitude=-1.1, segment length=9.6}](0.2,1.6)--(1.8,1.6);
  \node at (0.2,-0.6) {$Q_{ij}$};\node at (2,-0.6) {$\bar{Z}_{j}$};
   \node at (0.2,3.6) {$\bar{Q}_{ij}$};\node at (2,3.6) {$Z_{j}$};
   \node[right,scale=0.7] at (2.2,1.5) {\large$=-\kappa_j^2$};
\end{tikzpicture}\;.\hspace{0.5cm}
\ee
Finally, we have diagrams with horizontal chiral field exchange, 
\be\label{FD:chiralhorizontal1}
\begin{tikzpicture}[scale=0.8,baseline=1.1cm]
  \draw[blue,thick](0,0)--(-0.2,0);\draw[gray,thick](-0.2,-0.2)--(2.2,-0.2);\draw[green,thick] (2.2,0)--(2,0);\draw[blue,thick](0,0)--(0,3);\draw[green,thick](2,0)--(2,3);
  \draw[red,thick](0.2,1.4)--(0.2,0)--(1.8,0)--(1.8,1.4);
  \draw[red,thick](0.2,1.6)--(0.2,3)--(1.8,3)--(1.8,1.6);\draw[blue,thick](0,3)--(-0.2,3);\draw[gray,thick] (-0.2,3.2)--(2.2,3.2);\draw[green,thick](2.2,3)--(2,3);
  \draw[red,thick](0.2,1.4)--(1.8,1.4);  \draw[red,thick](0.2,1.6)--(1.8,1.6);
  \node at (0.2,-0.6) {$Q_{ij}$};\node at (2,-0.6) {$\Qb_{jk}$};
      \node at (0.2,3.6) {$Q_{ij}$};\node at (2,3.6) {$\Qb_{jk}$};
      \node[right,scale=0.7] at (2.2,1.5) {\LARGE$=\kappa_j^2\frac{\bar{d}_{ij}d_{kj}}{n_in_kn_j^2}\;,$};
\end{tikzpicture}\hspace{0.7cm}
\begin{tikzpicture}[scale=0.8,baseline=1.1cm]

  \draw[blue,thick](0,0)--(-0.2,0);\draw[gray,thick](-0.2,-0.2)--(2.2,-0.2);\draw[blue,thick] (2.2,0)--(2,0);\draw[blue,thick](0,0)--(0,3);\draw[blue,thick](2,0)--(2,3);
  \draw[red,thick](0.2,1.4)--(0.2,0)--(1.8,0)--(1.8,1.4);
  \draw[blue,thick](0.2,1.6)--(0.2,3)--(1.8,3)--(1.8,1.6);\draw[blue,thick](0,3)--(-0.2,3);\draw[gray,thick] (-0.2,3.2)--(2.2,3.2);\draw[blue,thick](2.2,3)--(2,3);
  \draw[red,thick](0.2,1.4)--(1.8,1.4);  \draw[blue,thick](0.2,1.6)--(1.8,1.6);
  \node at (0.2,-0.6) {$Q_{ij}$};\node at (2,-0.6) {$\bar{Q}_{ji}$};
      \node at (0.2,3.6) {$Z_{i}$};\node at (2,3.6) {$\Zb_{i}$};
      \node[right,scale=0.7] at (2.2,1.5) {\LARGE$=\kappa_i^2\frac{\bar{d}_{ji}d_{ji}}{n_i}=\frac{\kappa_i^2}{n_i}\;,$};
\end{tikzpicture}
\ee
as well as 
\be\label{FD:chiralhorizontal2}
\begin{tikzpicture}[scale=0.8,baseline=1.1cm]

  \draw[blue,thick](0,0)--(-0.2,0);\draw[gray,thick](-0.2,-0.2)--(2.2,-0.2);\draw[blue,thick] (2.2,0)--(2,0);\draw[blue,thick](0,0)--(0,3);\draw[blue,thick](2,0)--(2,3);
  \draw[blue,thick](0.2,1.4)--(0.2,0)--(1.8,0)--(1.8,1.4);
  \draw[red,thick](0.2,1.6)--(0.2,3)--(1.8,3)--(1.8,1.6);\draw[blue,thick](0,3)--(-0.2,3);\draw[gray,thick] (-0.2,3.2)--(2.2,3.2);\draw[blue,thick](2.2,3)--(2,3);
  \draw[blue,thick](0.2,1.4)--(1.8,1.4);  \draw[red,thick](0.2,1.6)--(1.8,1.6);
  \node at (0.2,-0.6) {$Z_{i}$};\node at (2,-0.6) {$\Zb_{i}$};
      \node at (0.2,3.6) {$Q_{ij}$};\node at (2,3.6) {$\Qb_{ji}$};
      \node[right,scale=0.7] at (2.2,1.5) {\LARGE$=\kappa_i^2\bar{d}_{ji}d_{ji}n_j=\kappa_i^2n_j\;\;,$};
\end{tikzpicture}\hspace{0.5cm}
\begin{tikzpicture}[scale=0.8,baseline=1.1cm]

  \draw[blue,thick](0,0)--(-0.2,0);\draw[gray,thick](-0.2,-0.2)--(2.2,-0.2);\draw[red,thick] (2.2,0)--(2,0);\draw[blue,thick](0,0)--(0,3);\draw[red,thick](2,0)--(2,3);
  \draw[red,thick](0.2,1.4)--(0.2,0)--(1.8,0)--(1.8,1.4);
  \draw[blue,thick](0.2,1.6)--(0.2,3)--(1.8,3)--(1.8,1.6);\draw[blue,thick](0,3)--(-0.2,3);\draw[gray,thick] (-0.2,3.2)--(2.2,3.2);\draw[red,thick](2.2,3)--(2,3);
  \draw[red,thick](0.2,1.4)--(1.8,1.4);  \draw[blue,thick](0.2,1.6)--(1.8,1.6);
  \node at (0.2,-0.6) {$Q_{ij}$};\node at (2,-0.6) {$\bar{Z}_{j}$};
      \node at (0.2,3.6) {$Z_{i}$};\node at (2,3.6) {$\bar{Q}_{ij}$};
      \node[right,scale=0.7] at (2.2,1.5) {\LARGE$=\kappa_i\kappa_j\bar{d}_{ji}d_{ij}=-\kappa_i\kappa_j\;\;.$};
\end{tikzpicture}
\ee
In the next section we will see that all the above Feynman diagrams are proportional to a single one-loop integral.

\subsection{Evaluation of the Feynman diagrams}

Having expressed all the contributions to the planar dilatation operator in terms of their $\Ncal=4$ SYM counterparts, the next step is to evaluate the corresponding integrals. As we have already taken the gauge index structure into account, in this section we will drop the double-line notation. 

Let us use the following shorthand notation for a 1-loop Feynman integral:
\be
    I(\lambda,\mu,\epsilon)\equiv\lambda\mu^{2\epsilon}\;\;
\begin{tikzpicture}[rotate=90,baseline=-0.1cm]
    \fill[blue!20, opacity=0.5]
        (-0.5,0)
        .. controls (-0.25,0.25) and (0.25,0.25) .. (0.5,0)
        .. controls (0.25,-0.25) and (-0.25,-0.25) .. (-0.5,0)
        -- cycle;

    \draw (-0.5,0).. controls (-0.25,0.25) and (0.25,0.25) .. (0.5,0);
    \draw (-0.5,0).. controls (-0.25,-0.25) and (0.25,-0.25) .. (0.5,0);
\end{tikzpicture}=\lambda\mu^{2\epsilon}\int\frac{d^{4-2\epsilon}k}{(2\pi)^{4-2\epsilon}}\frac{1}{k^2\left(k^2-p^2\right)} \;.
\ee
As we will see, all the 1-loop diagrams contributing to the dilatation operator will turn out to be proportional to this integral. A standard computation gives
\be\begin{split}
    I(\lambda,\mu,\epsilon)=&\,\frac{\lambda\Gamma(\epsilon)\Gamma(1-\epsilon)^2}{16\pi^2\Gamma(2-2\epsilon)}\left[\frac{4\pi\mu^2}{p^2}\right]^{\epsilon}\\
    =&\,\frac{\lambda}{16\pi^2}\left[\frac{1}{\epsilon}-\gamma_\text{E}+\log\left(\frac{4\pi\mu^2}{p^2}\right)+\Ocal(\epsilon)\right],
\end{split}
\ee
where $\Gamma(\epsilon)$ is the standard Gamma function and $\gamma_\text{E}$ is the Euler-Mascheroni constant.
Then
\be \label{Iresidue}
2\text{Res}_{\epsilon=0}\left[\lambda\frac{d}{d\lambda}I(\lambda,\mu,\epsilon)\right]=\frac{2\lambda}{16\pi^2}\;.
\ee
To evaluate the diagrams of the previous section, let us quickly revise some supergraph rules (for more details, see \cite{Gates:1983nr,Wess:1992cp}, as well as the review \cite{Sieg:2010jt}): Each vertex introduces an integral over the fermionic coordinates $d^4\theta$ while each loop introduces an integral over the unconstrained loop momentum $d^{4-2\epsilon}p$. We can slide covariant derivatives $\codif_\alpha$ (or $\bcodif_{\dot{\alpha}}$) across propagators,  picking up a minus sign each time:
\be
\begin{tikzpicture}[baseline={(0,0)}]
    \draw[violet,thick] (0,0)--(3,0);
    \node[above,scale=0.85] at (2.8,0) {$\codif_\alpha$};
    \node[right] at (3.2,0) {$=$};
\end{tikzpicture}
\begin{tikzpicture}[baseline={(0,0)}]
    \node[left] at (-0.2,0) {$-$};
    \draw[violet,thick] (0,0)--(3,0);
    \node[above,scale=0.85] at (0.2,0) {$\codif_\alpha$};
\end{tikzpicture}
\ee
Note that we must keep the ordering of the $\codif$'s the same as we slide them
\be
\begin{tikzpicture}[baseline={(0,0)}]
    \draw[violet,thick] (0,0)--(3,0);
    \node[above,scale=0.85] at (2.6,0) {$\bcodif_{\dot{\alpha}}\codif_{\alpha}$};
    \node[right] at (3.2,0) {$=$};
\end{tikzpicture}
\begin{tikzpicture}[baseline={(0,0)}]

    \draw[violet,thick] (0,0)--(3,0);
    \node[above,scale=0.85] at (0.4,0) {$\bcodif_{\dot{\alpha}}\codif_{\alpha}$};
    \node[right] at (3.2,0) {$\ne$};
\end{tikzpicture}
\begin{tikzpicture}[baseline={(0,0)}]

    \draw[violet,thick] (0,0)--(3,0);
    \node[above,scale=0.85] at (0.4,0) {$\codif_{\alpha}\bcodif_{\dot{\alpha}}$};

\end{tikzpicture}
\ee
Integration by parts on the supergraph is depicted as 
\be
\begin{tikzpicture}[baseline={(0,0)}]
    \draw[violet,thick] (-1,0)--(0,0);
    \node[above,scale=0.7] at (-0.25,0){$\codif_\alpha$};  
    \draw[violet,thick] (0,0)--(0.5,0.866025);
    \draw[violet,thick] (0,0)--(0.5,-0.866025);
    \node[right] at (0.7,0) {$=$};
\end{tikzpicture}
\begin{tikzpicture}[baseline={(0,0)}]
    \draw[violet,thick] (-1,0)--(0,0);
    \node[above,scale=0.7,rotate=60] at (0.15,0.259807621135) {$\codif_\alpha$}; 
    \draw[violet,thick] (0,0)--(0.5,0.866025);
    \draw[violet,thick] (0,0)--(0.5,-0.866025);
    \node[right] at (0.7,0) {$+$};
\end{tikzpicture}
\begin{tikzpicture}[baseline={(0,0)}]
    \draw[violet,thick] (-1,0)--(0,0);
    \node[below,scale=0.7,rotate=300] at (0.125,-0.259807621135) {$\codif_\alpha$}; 
    \draw[violet,thick] (0,0)--(0.5,0.866025);
    \draw[violet,thick] (0,0)--(0.5,-0.866025);
\end{tikzpicture}
\ee
We also have to take into account that every time we perform an integration by parts on a bosonic function, we get a minus sign, and every time we perform an integration by parts on a fermionic function, we get a plus sign. Note that when we perform the integration by parts, we always integrate so that the covariant derivative furthest away from the vertex is moved first: 
\be
\begin{tikzpicture}[baseline={(0,0)},scale=0.95]
    \draw[violet,thick] (-1,0)--(0,0);
    \node[above,scale=0.7] at (-0.3,0){$\codif_\beta\codif_\alpha$};  
    \draw[violet,thick] (0,0)--(0.5,0.866025);
    \draw[violet,thick] (0,0)--(0.5,-0.866025);
    \node[right] at (0.7,0) {$=$};
\end{tikzpicture}
\begin{tikzpicture}[baseline={(0,0)},scale=0.95]
    \draw[violet,thick] (-1,0)--(0,0);
    \node[above,scale=0.7] at (-0.3,0){$\codif_\alpha$};
    \node[above,scale=0.7,rotate=60] at (0.175,0.303108891325) {$\codif_\beta$}; 
    \draw[violet,thick] (0,0)--(0.5,0.866025);
    \draw[violet,thick] (0,0)--(0.5,-0.866025);
    \node[right] at (0.7,0) {$+$};
\end{tikzpicture}
\begin{tikzpicture}[baseline={(0,0)},scale=0.95]
    \node[above,scale=0.7] at (-0.3,0){$\codif_\alpha$};
    \draw[violet,thick] (-1,0)--(0,0);
    \node[below,scale=0.8,rotate=300] at (0.175,-0.303108891325) {$\codif_\beta$}; 
    \draw[violet,thick] (0,0)--(0.5,0.866025);
    \draw[violet,thick] (0,0)--(0.5,-0.866025);
    \node[right] at (0.7,0) {$=$};
\end{tikzpicture}
\begin{tikzpicture}[baseline={(0,0)},scale=0.95]
    \draw[violet,thick] (-1,0)--(0,0);
    \node[above,scale=0.7,rotate=60] at (0.175,0.303108891325) {$\codif_\alpha\codif_\beta$}; 
    \draw[violet,thick] (0,0)--(0.5,0.866025);
    \draw[violet,thick] (0,0)--(0.5,-0.866025);
    \node[right] at (0.7,0) {$+$};
\end{tikzpicture}
\begin{tikzpicture}[baseline={(0,0)},scale=0.95]
    \draw[violet,thick] (-1,0)--(0,0);
    \node[above,scale=0.7,rotate=60] at (0.175,0.303108891325) {$\codif_\alpha$};
    \node[below,scale=0.8,rotate=300] at (0.175,-0.303108891325) {$\codif_\beta$}; 
    \draw[violet,thick] (0,0)--(0.5,0.866025);
    \draw[violet,thick] (0,0)--(0.5,-0.866025);
    \node[right] at (0.7,0) {$+$};
\end{tikzpicture}
\begin{tikzpicture}[baseline={(0,0)},scale=0.95]
    \draw[violet,thick] (-1,0)--(0,0);
    \node[above,scale=0.7,rotate=60] at (0.175,0.303108891325) {$\codif_\alpha\codif_\beta$}; 
    \draw[violet,thick] (0,0)--(0.5,0.866025);
    \draw[violet,thick] (0,0)--(0.5,-0.866025);
    \node[right] at (0.7,0) {$+$};
\end{tikzpicture}
\begin{tikzpicture}[baseline={(0,0)},scale=0.95]
    \draw[violet,thick] (-1,0)--(0,0);
    \node[below,scale=0.8,rotate=300] at (0.175,-0.303108891325) {$\codif_\alpha\codif_\beta$}; 
    \draw[violet,thick] (0,0)--(0.5,0.866025);
    \draw[violet,thick] (0,0)--(0.5,-0.866025);
\end{tikzpicture}
\ee
We also recall that $\codif^3=\bcodif^3=0$, so we can only have a maximum of two $\codif$'s and two $\bcodif$'s sequentially on a line. Furthermore we use the identities
\be
\codif_\alpha\codif^\alpha=-\codif^2\,,\quad\bcodif^{\dot{\alpha}}\bcodif_{\dot{\alpha}}=-\bcodif^2\;,
\ee
and also that 
\be
\codif^2\bcodif^2\codif^2=\square\codif^2\;\;\text{and}\;\;\bcodif^2\codif^2\bcodif^2=\square\bcodif^2\,.
\ee
In momentum space, a factor of $\square$ purely introduces a factor of $-p^2$, where $p$ is the momentum of the propagator that $\square$ is attached to. Finally, we have
\be\label{eq:codiff2bcodiff2delta}
\int d^4\theta\bcodif^2\codif^2\delta^{(4)}(\theta)=\int d^4\theta\codif^2\bcodif^2\delta^{(4)}(\theta)=1\,.
\ee
We make use of \eqref{eq:codiff2bcodiff2delta} to collapse integrals over momentum space and of the fermionic coordinates over products of delta functions in fermionic coordinates to integrals over only momentum space: if we have a loop where a single propagator has either $\bcodif^2\codif^2$ or $\codif^2\bcodif^2$ attached to it, and all other propagators in that loop have no covariant derivatives attached to it, then that integral is localised in superspace and is converted to a regular integral over momentum space. As in e.g. \cite{Sieg:2010tz,Sieg:2010jt,Pomoni:2011jj}, we will indicate this localisation to a regular momentum space loop by shading in the loop. 

We first consider the chiral diagrams. Recall that we are treating the composite operators as a vertex inserted at a point. From superspace rules, due to how functional derivatives with respect to chiral fields are defined, each chiral field in a diagram always introduces a factor of $\codif^2$. However, in order to convert the superspace integral over a chiral vertex from one being over half of superspace $d^2\theta$, we convert one of the superspace derivatives $\bcodif^2$ into the measure $d^2\bar{\theta}$ to get a total measure of $d^4\theta$. This means that whenever we have a chiral vertex consisting of $m$ fields, we always have $m-1$ covariant derivatives $\bcodif^2$ at that vertex. Since we treat the operator insertions as a vertex, if the composite operator is chiral and comprised of $L$ fields, we will always have $L-1$ covariant derivatives $\bcodif^2$ at this insertion. Let us also recall that the superficial degree of divergence of a supergraph is given by \cite{Gates:1983nr}
\begin{equation}
    \text{D}_\infty=2-C-E_c,
\end{equation}
where $C$ is the number of $\Phi\Phi$ or $\Bar{\Phi}\Bar{\Phi}$ propagators (which are only relevant for massive theories) and $E_c$ is the number of external chiral or anti-chiral lines. In considering the superficial degree of divergence, we will be cutting the diagram above the loop and will be treating the operator as an external line. So we can immediately conclude that 
\be
\begin{tikzpicture}[baseline=1.1cm]
        \node[above, rotate=90, scale=0.7] at (-0.5,0.2) {$\bcodif^2$};
        \draw[line width=2,violet] (-0.5,0)--(0.5,0); 
        \draw[violet,thick] (-0.5,0)--(-0.5,3);
        \draw[violet,thick] (0.5,0)--(0.5,3);
        \node[above, rotate=90, scale=0.7] at (-0.5,2.8) {$\bcodif^2$};
        \node[below, rotate=90, scale=0.7] at (0.5,2.8) {$\bcodif^2$};
        \draw[decorate, decoration={snake, amplitude=1, segment length=4.5},violet,thick] (-0.5,1.5)--(0.5,1.5);
        \node[above, rotate=90, scale=0.7] at (-0.5,1.3) {$\codif^2$};
        \node[below, rotate=90, scale=0.7] at (0.5,1.3) {$\codif^2$};
        \node[right,scale=0.8] at (1,1.5) {$\to\text{D}_\infty=-1\to$ finite};
    \end{tikzpicture}
\ee
i.e.  when both fields are holomorphic, the gauge boson exchange diagrams are finite and do not contribute. For the holomorphic sector, we therefore only consider the chiral diagrams (\ref{FD:QQhol}),(\ref{FD:QPhihol}) and (\ref{FD:QPhihol1}).
To evaluate the corresponding $\Ncal=4$ SYM diagrams, we apply the D-algebra until we obtain a loop containing a single $\codif^2$ and a single $\bcodif^2$. As explained, the loop integral is then converted into a regular loop:
\begin{equation}\label{SFD:VertChiral}
    \begin{split}
    \begin{tikzpicture}
        \node[above, rotate=90, scale=0.7] at (-0.5,0.2) {$\bcodif^2$};
        \draw[line width=2,violet] (-0.5,0)--(0.5,0); 
        \draw[violet,thick] (-0.5,0).. controls (-0.5,1) and (0,1).. (0,1);
        \node[above, rotate=45, scale=0.7] at (-0.3,0.8) {$\codif^2$};
        \draw[violet,thick] (0.5,0).. controls (0.5,1) and (0,1).. (0,1);
        \draw[violet,thick] (0,1)--(0,2); \node[below, rotate=90, scale=0.7] at (0,1.3) {$\codif^2$};
        \draw[violet,thick] (-0.5,3).. controls (-0.5,2) and (0,2).. (0,2);
        \draw[violet,thick] (0.5,3).. controls (0.5,2) and (0,2).. (0,2);
        \node[above, rotate=90, scale=0.7] at (-0.5,2.8) {$\bcodif^2$};
        \node[below, rotate=90, scale=0.7] at (0.5,2.8) {$\bcodif^2$};
        \node[right] at (1,1.5) {$=$};
    \end{tikzpicture}
    \begin{tikzpicture}
    \fill[violet!20, opacity=0.5]
        (-0.5,0)
        .. controls (-0.5,1) and (0,1).. (0,1)
        .. controls (0,1) and (0.5,1).. (0.5,0)
        -- cycle;
    \draw[line width=2,violet] (-0.5,0)--(0.5,0); 
    \draw[violet,thick] (-0.5,0).. controls (-0.5,1) and (0,1).. (0,1);
    \draw[violet,thick] (0.5,0).. controls (0.5,1) and (0,1).. (0,1);
    \draw[violet,thick] (0,1)--(0,2); 
    \node[below, rotate=90, scale=0.7] at (0,1.5) {$\codif^2\bcodif^2$};
    \draw[violet,thick] (-0.5,3).. controls (-0.5,2) and (0,2).. (0,2);
    \draw[violet,thick] (0.5,3).. controls (0.5,2) and (0,2).. (0,2);
    \node[above, rotate=90, scale=0.7] at (-0.5,2.8) {$\bcodif^2$};
    \node[below, rotate=90, scale=0.7] at (0.5,2.8) {$\bcodif^2$};
    \node[right,scale=0.8] at (1,1.5) {$\to -I(\lambda,\mu,\epsilon)$};
\end{tikzpicture}
    \end{split}
\end{equation}
In the final step we converted the $d^4\theta$ integral at the top vertex into $d^2\theta\bcodif^2$ to fully localise the fermionic coordinates and then the remaining $\codif^2$ and $\bcodif^2$ combine as a $\square$, cancelling the propagator between the chiral and anti-chiral vertex and we are left with an overall factor of $-1$ as the propagator of the chiral field is of the form $-\square^{-1}$.

Let us now move on to the mixed sector. Since the composite operators comprise both chiral and anti-chiral fields, the composite operator cannot be chiral and we can no longer apply the logic of removing one covariant derivative $\bcodif^2$ as we argued above in the case of chiral composite operators.

The evaluation of the diagrams of type (\ref{FD:QQbgauge}) and (\ref{FD:PhiPhibgauge}) goes as 

\begin{equation}\label{SFD:VertVector}
    \begin{tikzpicture}
        \node[above, rotate=90, scale=0.7] at (-0.5,0.2) {$\bcodif^2$};
        \draw[line width=2,color=violet] (-0.5,0)--(0.5,0); 
        \node[below, rotate=90, scale=0.7] at (0.5,0.2) {$\codif^2$};
        \draw[violet,thick] (-0.5,0).. controls (-0.5,1) and (0,1).. (0,1);
        \node[above, rotate=45, scale=0.7] at (-0.3,0.8) {$\codif^2$};
        \draw[violet,thick] (0.5,0).. controls (0.5,1) and (0,1).. (0,1);
        \node[below, rotate=135, scale=0.7] at (0.3,0.8) {$\bcodif^2$};
        \draw[violet,thick] (0.5,0).. controls (0.5,1) and (0,1).. (0,1);
        \draw[decorate, decoration={snake, amplitude=1, segment length=4.5},violet,thick] (0,1)--(0,2) ;
        \draw[violet,thick] (-0.5,3).. controls (-0.5,2) and (0,2).. (0,2);
        \draw[violet,thick] (0.5,3).. controls (0.5,2) and (0,2).. (0,2);
        \node[right] at (0.7,1.5) {$=$};
    \end{tikzpicture}
    \begin{tikzpicture}
        \draw[line width=2,violet] (-0.5,0)--(0.5,0); 
        \node[below, rotate=90, scale=0.7] at (0.5,0.2) {$\codif^2$};
        \draw[violet,thick] (-0.5,0).. controls (-0.5,1) and (0,1).. (0,1);
        \node[above, rotate=45, scale=0.7] at (-0.3,0.8) {$\bcodif^2\codif^2$};
        \draw[violet,thick] (0.5,0).. controls (0.5,1) and (0,1).. (0,1);
        \node[below, rotate=135, scale=0.7] at (0.3,0.8) {$\bcodif^2$};
        \draw[violet,thick] (0.5,0).. controls (0.5,1) and (0,1).. (0,1);
        \draw[decorate, decoration={snake, amplitude=1, segment length=4.5},violet,thick] (0,1)--(0,2) ;
        \draw[violet,thick] (-0.5,3).. controls (-0.5,2) and (0,2).. (0,2);
        \draw[violet,thick] (0.5,3).. controls (0.5,2) and (0,2).. (0,2);
        \node[right] at (0.7,1.5) {$=$};
    \end{tikzpicture}
    \begin{tikzpicture}
        \draw[line width=2,violet] (-0.5,0)--(0.5,0); 
        \node[below, rotate=90, scale=0.7] at (0.5,0.2) {$\codif^2$};
        \draw[violet,thick] (-0.5,0).. controls (-0.5,1) and (0,1).. (0,1);
        \node[above, rotate=45, scale=0.7] at (-0.3,0.8) {$\bcodif^2\codif^2$};
        \draw[violet,thick] (0.5,0).. controls (0.5,1) and (0,1).. (0,1);
        \draw[violet,thick] (0.5,0).. controls (0.5,1) and (0,1).. (0,1);
        \draw[decorate, decoration={snake, amplitude=1, segment length=4.5},violet,thick] (0,1)--(0,2);
        \node[below, rotate=90, scale=0.7] at (0,1.5) {$\bcodif^2$};
        \draw[violet,thick] (-0.5,3).. controls (-0.5,2) and (0,2).. (0,2);
        \draw[violet,thick] (0.5,3).. controls (0.5,2) and (0,2).. (0,2);
        \node[right] at (0.7,1.5) {$=$};
    \end{tikzpicture}
    \begin{tikzpicture}
        \node[above, rotate=90, scale=0.7] at (-0.5,0.2) {$\bcodif^2$};
        \draw[line width=2,violet] (-0.5,0)--(0.5,0); 
        \draw[violet,thick] (-0.5,0).. controls (-0.5,1) and (0,1).. (0,1);
        \node[above, rotate=45, scale=0.7] at (-0.3,0.8) {$\codif^2$};
        \draw[violet,thick] (0.5,0).. controls (0.5,1) and (0,1).. (0,1);
        \node[below, rotate=135, scale=0.7] at (0.3,0.8) {$\codif^2$};
        \draw[violet,thick] (0.5,0).. controls (0.5,1) and (0,1).. (0,1);
        \draw[decorate, decoration={snake, amplitude=1, segment length=4.5},violet,thick] (0,1)--(0,2);
        \node[below, rotate=90, scale=0.7] at (0,1.5) {$\bcodif^2$};
        \draw[violet,thick] (-0.5,3).. controls (-0.5,2) and (0,2).. (0,2);
        \draw[violet,thick] (0.5,3).. controls (0.5,2) and (0,2).. (0,2);
        \node[right] at (0.7,1.5) {$=$};
    \end{tikzpicture}
    \begin{tikzpicture}
        \node[above, rotate=90, scale=0.7] at (-0.5,0.2) {$\bcodif^2$};
        \draw[line width=2,violet] (-0.5,0)--(0.5,0); 
        \draw[violet,thick] (-0.5,0).. controls (-0.5,1) and (0,1).. (0,1);
        \node[above, rotate=45, scale=0.7] at (-0.3,0.8) {$\codif^2$};
        \draw[violet,thick] (0.5,0).. controls (0.5,1) and (0,1).. (0,1);
        \draw[violet,thick] (0.5,0).. controls (0.5,1) and (0,1).. (0,1);
        \draw[decorate, decoration={snake, amplitude=1, segment length=4.5},violet,thick] (0,1)--(0,2);
        \node[below, rotate=90, scale=0.7] at (0,1.5) {$\codif^2\bcodif^2$};
        \draw[violet,thick] (-0.5,3).. controls (-0.5,2) and (0,2).. (0,2);
        \draw[violet,thick] (0.5,3).. controls (0.5,2) and (0,2).. (0,2);
        \node[right] at (0.7,1.5) {$=$};
    \end{tikzpicture}
    \begin{tikzpicture}
    \fill[violet!20, opacity=0.5]
        (-0.5,0)
        .. controls (-0.5,1) and (0,1).. (0,1)
        .. controls (0,1) and (0.5,1).. (0.5,0)
        -- cycle;
        \draw[line width=2,violet] (-0.5,0)--(0.5,0); 
        \draw[violet,thick] (-0.5,0).. controls (-0.5,1) and (0,1).. (0,1);
        \draw[violet,thick] (0.5,0).. controls (0.5,1) and (0,1).. (0,1);
        \draw[violet,thick] (0.5,0).. controls (0.5,1) and (0,1).. (0,1);
        \draw[decorate, decoration={snake, amplitude=1, segment length=4.5},violet,thick] (0,1)--(0,2);
        \node[below, rotate=90, scale=0.7] at (0,1.5) {$\codif^2\bcodif^2$};
        \draw[violet,thick] (-0.5,3).. controls (-0.5,2) and (0,2).. (0,2);
        \draw[violet,thick] (0.5,3).. controls (0.5,2) and (0,2).. (0,2);
        \node[right,scale=0.8] at (0.7,1.5) {$\to I(\lambda,\mu,\epsilon)$};
    \end{tikzpicture}
\end{equation}
Note that the factor $\codif^2\bcodif^2$ again contributes a $\square$ to the diagram and fully localises the fermionic coordinates, however, now the vector superfield's propagator is of the form $\square^{-1}$, thus the relative sign difference between (\ref{SFD:VertChiral}) and (\ref{SFD:VertVector}).   
For the mixed diagrams with gluon exchange (\ref{FD:gaugehorizontal}) we find 
\begin{equation}
    \begin{split}
        &\begin{tikzpicture}
        \node[above, rotate=90, scale=0.7] at (-0.5,0.2) {$\bcodif^2$};
        \node[below, rotate=90, scale=0.7] at (0.5,0.2) {$\codif^2$};
        \draw[line width=2,violet] (-0.5,0)--(0.5,0); 
        \draw[violet,thick] (-0.5,0)--(-0.5,3);
        \draw[violet,thick] (0.5,0)--(0.5,3);
        \node[above, rotate=90, scale=0.7] at (-0.5,2.8) {$\codif^2$};
        \node[below, rotate=90, scale=0.7] at (0.5,2.8) {$\bcodif^2$};
        \draw[decorate, decoration={snake, amplitude=1, segment length=4.5},violet,thick] (-0.5,1.5)--(0.5,1.5);
        \node[above, rotate=90, scale=0.7] at (-0.5,1.3) {$\codif^2$};
        \node[above, rotate=90, scale=0.7] at (-0.5,1.8) {$\bcodif^2$};
        \node[below, rotate=90, scale=0.7] at (0.5,1.3) {$\bcodif^2$};
        \node[below, rotate=90, scale=0.7] at (0.5,1.8) {$\codif^2$};
        \node[right] at (1,1.5) {$=$};
    \end{tikzpicture}
    \begin{tikzpicture}
        \node[above, rotate=90, scale=0.7] at (-0.5,0.2) {$\bcodif^2$};
        \node[below, rotate=90, scale=0.7] at (0.5,0.2) {$\codif^2$};
        \draw[line width=2,violet] (-0.5,0)--(0.5,0); 
        \draw[violet,thick] (-0.5,0)--(-0.5,3);
        \draw[violet,thick] (0.5,0)--(0.5,3);
        \node[above, rotate=90, scale=0.7] at (-0.5,2.8) {$\codif^2$};
        \node[below, rotate=90, scale=0.7] at (0.5,2.8) {$\bcodif^2$};
        \draw[decorate, decoration={snake, amplitude=1, segment length=4.5},violet,thick] (-0.5,1.5)--(0.5,1.5);
        \node[above, rotate=90, scale=0.7] at (-0.5,1.3) {$\codif^2$};
        \node[right, scale=0.6] at (-0.5,1.3) {$\bcodif^2$};
        \node[below, rotate=90, scale=0.7] at (0.5,1.3) {$\bcodif^2$};
        \node[left, scale=0.6] at (0.5,1.3) {$\codif^2$};
        \node[right] at (1,1.5) {$=$};
    \end{tikzpicture}
    \begin{tikzpicture}
        \node[above, rotate=90, scale=0.7] at (-0.5,0.2) {$\bcodif^2$};
        \node[below, rotate=90, scale=0.7] at (0.5,0.2) {$\codif^2$};
        \draw[line width=2,violet] (-0.5,0)--(0.5,0); 
        \draw[violet,thick] (-0.5,0)--(-0.5,3);
        \draw[violet,thick] (0.5,0)--(0.5,3);
        \node[above, rotate=90, scale=0.7] at (-0.5,2.8) {$\codif^2$};
        \node[below, rotate=90, scale=0.7] at (0.5,2.8) {$\bcodif^2$};
        \draw[decorate, decoration={snake, amplitude=1, segment length=4.5},violet,thick] (-0.5,1.5)--(0.5,1.5);
        \node[below, rotate=90, scale=0.7] at (0.5,1.3) {$\bcodif^2$};
        \node[left, scale=0.6] at (0.5,1.3) {$\codif^2\bcodif^2\codif^2$};
        \node[right] at (1,1.5) {$=$};
    \end{tikzpicture}
    \begin{tikzpicture}
        \node[above, rotate=90, scale=0.7] at (-0.5,0.2) {$\bcodif^2$};
        \node[below, rotate=90, scale=0.7] at (0.5,0.2) {$\codif^2$};
        \draw[line width=2,violet] (-0.5,0)--(0.5,0); 
        \draw[violet,thick] (-0.5,0)--(-0.5,3);
        \draw[violet,thick] (0.5,0)--(0.5,3);
        \node[above, rotate=90, scale=0.7] at (-0.5,2.8) {$\codif^2$};
        \node[below, rotate=90, scale=0.7] at (0.5,2.8) {$\bcodif^2$};
        \draw[decorate, decoration={snake, amplitude=1, segment length=4.5},violet,thick] (-0.5,1.5)--(0.5,1.5);
        \node[below, rotate=90, scale=0.7] at (0.5,1.3) {$\bcodif^2$};
        \node[left, scale=0.6] at (0.5,1.3) {$\square\codif^2$};
        \node[right] at (1,1.5) {$=$};
    \end{tikzpicture}
    \begin{tikzpicture}
        \node[above, rotate=90, scale=0.7] at (-0.5,0.2) {$\bcodif^2$};
        \node[below, rotate=90, scale=0.7] at (0.5,0.2) {$\codif^2$};
        \draw[line width=2,violet] (-0.5,0)--(0.5,0); 
        \draw[violet,thick] (-0.5,0)--(-0.5,3);
        \draw[violet,thick] (0.5,0)--(0.5,3);
        \node[above, rotate=90, scale=0.7] at (-0.5,2.8) {$\codif^2$};
        \draw[decorate, decoration={snake, amplitude=1, segment length=4.5},violet,thick] (-0.5,1.5)--(0.5,1.5);
        \node[below, rotate=90, scale=0.7] at (0.5,2.8) {$\bcodif^2$};
        \node[left, scale=0.6] at (0.5,1.3) {$\square\bcodif^2\codif^2$};
        \node[right] at (1,1.5) {$=$};
    \end{tikzpicture}\\
    &\begin{tikzpicture}
        \node[above, rotate=90, scale=0.7] at (-0.5,0.2) {$\bcodif^2$};
        \node[below, rotate=90, scale=0.7] at (0.5,0.2) {$\codif^2$};
        \draw[line width=2,violet] (-0.5,0)--(0.5,0); 
        \draw[violet,thick] (-0.5,0)--(-0.5,3);
        \draw[violet,thick] (0.5,0)--(0.5,3);
        \node[above, rotate=90, scale=0.7] at (-0.5,2.8) {$\codif^2$};
        \draw[decorate, decoration={snake, amplitude=1, segment length=4.5},violet,thick] (-0.5,1.5)--(0.5,1.5);
        \node[below, rotate=90, scale=0.7] at (0.5,2.8) {$\bcodif^2$};
        \node[left, scale=0.6] at (0.5,1.3) {$\codif^2$};
        \node[right,scale=0.6] at (-0.5,1.3) {$\bcodif^2$};
        \node[scale=0.7] at (0,1.7) {$\square$};
        \node[right] at (1,1.5) {$=$};
    \end{tikzpicture}
    \begin{tikzpicture}
        \node[above, rotate=90, scale=0.7] at (-0.5,0.2) {$\bcodif^2$};
        \node[below, rotate=90, scale=0.7] at (0.5,0.2) {$\codif^2$};
        \draw[line width=2,violet] (-0.5,0)--(0.5,0); 
        \draw[violet,thick] (-0.5,0)--(-0.5,3);
        \draw[violet,thick] (0.5,0)--(0.5,3);
        \node[above, rotate=90, scale=0.7] at (-0.5,2.8) {$\codif^2$};
        \draw[decorate, decoration={snake, amplitude=1, segment length=4.5},violet,thick] (-0.5,1.5)--(0.5,1.5);
        \node[below, rotate=90, scale=0.7] at (0.5,2.8) {$\bcodif^2$};
        \node[below,rotate=90, scale=0.7] at (0.5,1.8) {$\codif^2$};
        \node[above,rotate=90,scale=0.7] at (-0.5,1.8) {$\bcodif^2$};
        \node[scale=0.7] at (0,1.7) {$\square$};
        \node[right] at (1,1.5) {$=$};
    \end{tikzpicture}
    \begin{tikzpicture}
        \node[above, rotate=90, scale=0.7] at (-0.5,1.3) {$\bcodif^2$};
        \node[below, rotate=90, scale=0.7] at (0.5,1.3) {$\codif^2$};
        \draw[line width=2,violet] (-0.5,0)--(0.5,0); 
        \draw[violet,thick] (-0.5,0)--(-0.5,3);
        \draw[violet,thick] (0.5,0)--(0.5,3);
        \node[above, rotate=90, scale=0.7] at (-0.5,2.8) {$\codif^2$};
        \draw[decorate, decoration={snake, amplitude=1, segment length=4.5},violet,thick] (-0.5,1.5)--(0.5,1.5);
        \node[below, rotate=90, scale=0.7] at (0.5,2.8) {$\bcodif^2$};
        \node[below,rotate=90, scale=0.7] at (0.5,1.8) {$\codif^2$};
        \node[above,rotate=90,scale=0.7] at (-0.5,1.8) {$\bcodif^2$};
        \node[scale=0.7] at (0,1.7) {$\square$};
        \node[right] at (1,1.5) {$=$};
    \end{tikzpicture}
    \begin{tikzpicture}
        \node[scale=0.6] at (0,1.3) {$\bcodif^2\codif^2$};
        \draw[line width=2,violet] (-0.5,0)--(0.5,0); 
        \draw[violet,thick] (-0.5,0)--(-0.5,3);
        \draw[violet,thick] (0.5,0)--(0.5,3);
        \node[above, rotate=90, scale=0.7] at (-0.5,2.8) {$\codif^2$};
        \draw[decorate, decoration={snake, amplitude=1, segment length=4.5},violet,thick] (-0.5,1.5)--(0.5,1.5);
        \node[below, rotate=90, scale=0.7] at (0.5,2.8) {$\bcodif^2$};
        \node[below,rotate=90, scale=0.7] at (0.5,1.8) {$\codif^2$};
        \node[above,rotate=90,scale=0.7] at (-0.5,1.8) {$\bcodif^2$};
        \node[scale=0.7] at (0,1.7) {$\square$};
        \node[right] at (1,1.5) {$=$};
    \end{tikzpicture}
    \begin{tikzpicture}
    \begin{scope}
        \fill[violet!20, opacity=0.5]
            (-0.5,0) -- (-0.5,1.5)
            decorate [decoration={snake, amplitude=1, segment length=4.5}]
            { -- (0.5,1.5) }
            -- (0.5,0) -- cycle;
    \end{scope}
    \draw[line width=2,violet] (-0.5,0)--(0.5,0); 
    \draw[violet,thick] (-0.5,0)--(-0.5,3);
    \draw[violet,thick] (0.5,0)--(0.5,3);
    \draw[decorate, decoration={snake, amplitude=1, segment length=4.5},violet,thick] (-0.5,1.5)--(0.5,1.5);
    \node[above, rotate=90, scale=0.7] at (-0.5,2.8) {$\codif^2$};
    \node[below, rotate=90, scale=0.7] at (0.5,2.8) {$\bcodif^2$};
    \node[below,rotate=90, scale=0.7] at (0.5,1.8) {$\codif^2$};
    \node[above,rotate=90,scale=0.7] at (-0.5,1.8) {$\bcodif^2$};
    \node[scale=0.7] at (0,1.7) {$\square$};
    \node[right,scale=0.8] at (1,1.5) {$\to I(\lambda,\mu,\epsilon)$};
\end{tikzpicture}
    \end{split}
\end{equation}

Lastly, for the mixed diagrams with exchange of an $\Ncal=1$ chiral multiplet (\ref{FD:chiralhorizontal1}) and (\ref{FD:chiralhorizontal2}), the D-algebra gives

\begin{equation}
\begin{split}
    &\begin{tikzpicture}
        \node[above, rotate=90, scale=0.7] at (-0.5,0.2) {$\bcodif^2$};
        \node[below, rotate=90, scale=0.7] at (0.5,0.2) {$\codif^2$};
        \draw[line width=2,violet] (-0.5,0)--(0.5,0); 
        \draw[violet,thick] (-0.5,0)--(-0.5,3);
        \draw[violet,thick] (0.5,0)--(0.5,3);
        \node[above, rotate=90, scale=0.7] at (-0.5,2.8) {$\bcodif^2$};
        \node[below, rotate=90, scale=0.7] at (0.5,2.8) {$\codif^2$};
        \node[right, scale=0.7] at (-0.5,1.3) {$\codif^2$};
        \node[left, scale=0.7] at (0.5,1.3) {$\bcodif^2$};
        \draw[violet,thick] (-0.5,1.5)--(0.5,1.5);
        \node[above, rotate=90, scale=0.7] at (-0.5,1.3) {$\codif^2$};
        \node[below, rotate=90, scale=0.7] at (0.5,1.3) {$\bcodif^2$};
        \node[right] at (1,1.5) {$=$};
    \end{tikzpicture}
    \begin{tikzpicture}
        \node[above, rotate=90, scale=0.7] at (-0.5,0.2) {$\bcodif^2$};
        \draw[line width=2,violet] (-0.5,0)--(0.5,0); 
        \draw[violet,thick] (-0.5,0)--(-0.5,3);
        \draw[violet,thick] (0.5,0)--(0.5,3);
        \node[above, rotate=90, scale=0.7] at (-0.5,2.8) {$\bcodif^2$};
        \node[below, rotate=90, scale=0.7] at (0.5,2.8) {$\codif^2$};
        \node[right, scale=0.7] at (-0.5,1.3) {$\codif^2$};
        \node[left, scale=0.7] at (0.5,1.3) {$\bcodif^2$};
        \draw[violet,thick] (-0.5,1.5)--(0.5,1.5);
        \node[above, rotate=90, scale=0.7] at (-0.5,1.3) {$\codif^2$};
        \node[below, rotate=90, scale=0.7] at (0.5,1.3) {$\codif^2\bcodif^2$};
        \node[right] at (1,1.5) {$=$};
    \end{tikzpicture}
    \begin{tikzpicture}
        \node[above, rotate=90, scale=0.7] at (-0.5,0.2) {$\bcodif^2$};
        \draw[line width=2,violet] (-0.5,0)--(0.5,0); 
        \draw[violet,thick] (-0.5,0)--(-0.5,3);
        \draw[violet,thick] (0.5,0)--(0.5,3);
        \node[above, rotate=90, scale=0.7] at (-0.5,2.8) {$\bcodif^2$};
        \node[below, rotate=90, scale=0.7] at (0.5,2.8) {$\codif^2$};
        \node[right, scale=0.6] at (-0.5,1.3) {$\codif^2\bcodif^2\codif^2$};
        \draw[violet,thick] (-0.5,1.5)--(0.5,1.5);
        \node[above, rotate=90, scale=0.7] at (-0.5,1.3) {$\codif^2$};
        \node[below, rotate=90, scale=0.7] at (0.5,1.3) {$\bcodif^2$};
        \node[right] at (1,1.5) {$=$};
    \end{tikzpicture}
    \begin{tikzpicture}
        \node[above, rotate=90, scale=0.7] at (-0.5,0.2) {$\bcodif^2$};
        \draw[line width=2,violet] (-0.5,0)--(0.5,0); 
        \draw[violet,thick] (-0.5,0)--(-0.5,3);
        \draw[violet,thick] (0.5,0)--(0.5,3);
        \node[above, rotate=90, scale=0.7] at (-0.5,2.8) {$\bcodif^2$};
        \node[below, rotate=90, scale=0.7] at (0.5,2.8) {$\codif^2$};
        \node[right, scale=0.7] at (-0.5,1.3) {$\square\codif^2$};
        \draw[violet,thick] (-0.5,1.5)--(0.5,1.5);
        \node[above, rotate=90, scale=0.7] at (-0.5,1.3) {$\codif^2$};
        \node[below, rotate=90, scale=0.7] at (0.5,1.3) {$\bcodif^2$};
        \node[right] at (1,1.5) {$=$};
    \end{tikzpicture}
    \begin{tikzpicture}
        \draw[line width=2,violet] (-0.5,0)--(0.5,0); 
        \draw[violet,thick] (-0.5,0)--(-0.5,3);
        \draw[violet,thick] (0.5,0)--(0.5,3);
        \node[above, rotate=90, scale=0.7] at (-0.5,2.8) {$\bcodif^2$};
        \node[below, rotate=90, scale=0.7] at (0.5,2.8) {$\codif^2$};
        \node[right, scale=0.7] at (-0.5,1.3) {$\square\codif^2$};
        \draw[violet,thick] (-0.5,1.5)--(0.5,1.5);
        \node[above, rotate=90, scale=0.7] at (-0.5,1.3) {$\bcodif^2\codif^2$};
        \node[below, rotate=90, scale=0.7] at (0.5,1.3) {$\bcodif^2$};
        \node[right] at (1,1.5) {$=$};
    \end{tikzpicture}\\
    &\begin{tikzpicture}
        \draw[line width=2,violet] (-0.5,0)--(0.5,0); 
        \draw[violet,thick] (-0.5,0)--(-0.5,3);
        \draw[violet,thick] (0.5,0)--(0.5,3);
        \node[above, rotate=90, scale=0.7] at (-0.5,2.8) {$\bcodif^2$};
        \node[below, rotate=90, scale=0.7] at (0.5,2.8) {$\codif^2$};
        \node[left, scale=0.6] at (0.5,1.3) {$\square\bcodif^2\codif^2$};
        \draw[violet,thick] (-0.5,1.5)--(0.5,1.5);
        \node[above, rotate=90, scale=0.7] at (-0.5,1.3) {$\codif^2$};
        \node[below, rotate=90, scale=0.7] at (0.5,1.3) {$\bcodif^2$};
        \node[right] at (1,1.5) {$=$};
    \end{tikzpicture}
    \begin{tikzpicture}
        \draw[line width=2,color=violet] (-0.5,0)--(0.5,0); 
        \draw[violet,thick] (-0.5,0)--(-0.5,3);
        \draw[violet,thick] (0.5,0)--(0.5,3);
        \node[above, rotate=90, scale=0.7] at (-0.5,2.8) {$\bcodif^2$};
        \node[below, rotate=90, scale=0.7] at (0.5,2.8) {$\codif^2$};
        \node[left, scale=0.6] at (0.5,1.3) {$\square\bcodif^2\codif^2$};
        \draw[violet,thick] (-0.5,1.5)--(0.5,1.5);
        \node[above, rotate=90, scale=0.7] at (-0.5,1.3) {$\codif^2$};
        \node[below, rotate=90, scale=0.7] at (0.5,1.8) {$\bcodif^2$};
        \node[right] at (1,1.5) {$=$};
    \end{tikzpicture}
    \begin{tikzpicture}
        \draw[line width=2,violet] (-0.5,0)--(0.5,0); 
        \draw[violet,thick] (-0.5,0)--(-0.5,3);
        \draw[violet,thick] (0.5,0)--(0.5,3);
        \node[above, rotate=90, scale=0.7] at (-0.5,2.8) {$\bcodif^2$};
        \node[below, rotate=90, scale=0.7] at (0.5,2.8) {$\codif^2$};
        \node[right, scale=0.6] at (-0.5,1.3) {$\square\bcodif^2\codif^2$};
        \draw[violet,thick] (-0.5,1.5)--(0.5,1.5);
        \node[above, rotate=90, scale=0.7] at (-0.5,1.3) {$\codif^2$};
        \node[below, rotate=90, scale=0.7] at (0.5,1.8) {$\bcodif^2$};
        \node[right] at (1,1.5) {$=$};
    \end{tikzpicture}
    \begin{tikzpicture}
        \draw[line width=2,violet] (-0.5,0)--(0.5,0); 
        \draw[violet,thick] (-0.5,0)--(-0.5,3);
        \draw[violet,thick] (0.5,0)--(0.5,3);
        \node[above, rotate=90, scale=0.7] at (-0.5,2.8) {$\bcodif^2$};
        \node[below, rotate=90, scale=0.7] at (0.5,2.8) {$\codif^2$};
        \node[right, scale=0.6] at (-0.5,1.3) {$\square\codif^2$};
        \draw[violet,thick] (-0.5,1.5)--(0.5,1.5);
        \node[above, rotate=90, scale=0.7] at (-0.5,1.3) {$\bcodif^2\codif^2$};
        \node[below, rotate=90, scale=0.7] at (0.5,1.8) {$\bcodif^2$};
        \node[right] at (1,1.5) {$=$};
    \end{tikzpicture}
    \begin{tikzpicture}
        \node[above, rotate=90, scale=0.7] at (-0.5,0.2) {$\bcodif^2$};
        \draw[line width=2,violet] (-0.5,0)--(0.5,0); 
        \draw[violet,thick] (-0.5,0)--(-0.5,3);
        \draw[violet,thick] (0.5,0)--(0.5,3);
        \node[above, rotate=90, scale=0.7] at (-0.5,2.8) {$\bcodif^2$};
        \node[below, rotate=90, scale=0.7] at (0.5,2.8) {$\codif^2$};
        \node[right, scale=0.6] at (-0.5,1.3) {$\square\codif^2$};
        \draw[violet,thick] (-0.5,1.5)--(0.5,1.5);
        \node[above, rotate=90, scale=0.7] at (-0.5,1.3) {$\codif^2$};
        \node[below, rotate=90, scale=0.7] at (0.5,1.8) {$\bcodif^2$};
        \node[right] at (1,1.5) {$=$};
    \end{tikzpicture}\\
    &\begin{tikzpicture}
        \node[above, rotate=90, scale=0.7] at (-0.5,0.2) {$\bcodif^2$};
        \draw[line width=2,violet] (-0.5,0)--(0.5,0); 
        \draw[violet,thick] (-0.5,0)--(-0.5,3);
        \draw[violet,thick] (0.5,0)--(0.5,3);
        \node[above, rotate=90, scale=0.7] at (-0.5,2.8) {$\bcodif^2$};
        \node[below, rotate=90, scale=0.7] at (0.5,2.8) {$\codif^2$};
        \node[scale=0.7] at (0,1.7) {$\square$};
        \draw[violet,thick] (-0.5,1.5)--(0.5,1.5);
        \node[above, rotate=90, scale=0.7] at (-0.5,1.3) {$\codif^2$};
        \node[below, rotate=90, scale=0.7] at (0.5,1.8) {$\bcodif^2$};
        \node[above, rotate=90, scale=0.7] at (-0.5,1.8) {$\codif^2$};
        \node[right] at (1,1.5) {$=$};
    \end{tikzpicture}
        \begin{tikzpicture}
    \fill[violet!20, opacity=0.5] (-0.5,0) rectangle (0.5,1.5);
    \draw[line width=2,violet] (-0.5,0)--(0.5,0); 
    \draw[violet,thick] (-0.5,0)--(-0.5,3);
    \draw[violet,thick] (0.5,0)--(0.5,3);
    \draw[violet,thick] (-0.5,1.5)--(0.5,1.5);
    \node[above, rotate=90, scale=0.7] at (-0.5,2.8) {$\bcodif^2$};
    \node[below, rotate=90, scale=0.7] at (0.5,2.8) {$\codif^2$};
    \node[below,rotate=90, scale=0.7] at (0.5,1.8) {$\bcodif^2$};
    \node[above,rotate=90,scale=0.7] at (-0.5,1.8) {$\codif^2$};
    \node[scale=0.7] at (0,1.7) {$\square$};
    \node[right,scale=0.8] at (1,1.5) {$\to-I(\lambda,\mu,\epsilon)$};
\end{tikzpicture}
\end{split}
\end{equation}

Having expressed all the required Feynman diagrams in terms of $I(\lambda,\mu,\epsilon)$, we can now extract the divergent parts using (\ref{Iresidue}). They will be simply $\pm 1$ times the standard $\lambda/(16\pi^2)$ prefactor. So, up to this factor, the one-loop dilatation operator is simply equal to the group-theoretic coefficients of the diagrams in Section \ref{sec:Reduction}. We will write out the resulting Hamiltonian in Section \ref{sec:ADEHamiltonian}. Before that, however, we need to consider a special class of non-planar contributions which need to be considered for operators of length $2$.

\subsection{Non-planar contributions at $L=2$} \label{sec:Nonplanar}

\begin{figure}[ht]
\begin{center}
    \begin{tikzpicture}[scale=0.8]

  \draw[red,thick](0.2,0)--(1.8,0);\draw[blue,thick](0,0)--(-0.2,0)--(-0.2,-0.2)--(2.2,-0.2)--(2.2,0)--(2,0);
  \draw[red,thick](1.8,0) arc (0:180:0.8);
  \draw[blue,thick](2,0) arc (0:85:1);
  \draw[blue,thick](0,0) arc (180:95:1);
  \draw[blue,thick](0.91,0.99)--(0.9,2);
  \draw[blue,thick](1.09,0.99)--(1.09,2);
  \draw[red,thick](1.8,3) arc (0:-180:0.8);
  \draw[blue,thick](2,3) arc (0:-85:1);
  \draw[blue,thick](0,3) arc (-180:-95:1);
  \draw[red,thick](0.2,3)--(1.8,3);\draw[blue,thick](0,3)--(-0.2,3)--(-0.2,3.2)--(2.2,3.2)--(2.2,3)--(2,3);
  \node at (0.2,-0.6) {$Q_{ij}$};\node at (2,-0.6) {$Q_{ji}$};
\end{tikzpicture}
\begin{tikzpicture}[scale=0.8]
\node at (-1,1.5) {$-\frac{1}{N}$};
  \draw[red,thick](0.2,0)--(1.8,0);\draw[blue,thick](0,0)--(-0.2,0)--(-0.2,-0.2)--(2.2,-0.2)--(2.2,0)--(2,0);
  \draw[red,thick](1.8,0) arc (0:180:0.8);
  \draw[blue,thick](2,0) arc (0:85:1);
  \draw[blue,thick](0,0) arc (180:95:1);
  \draw[blue,thick](0.91,0.99)--(0.9,1.3);
  \draw[blue,thick](1.09,0.99)--(1.09,1.3);
  \draw[blue,thick](1.09,1.3) arc (0:180:0.095);
    \draw[blue,thick](0.91,1.7)--(0.9,2);
\draw[blue,thick](1.09,1.7)--(1.09,2);
  \draw[blue,thick](1.09,1.7) arc (0:-180:0.09);
  \draw[red,thick](1.8,3) arc (0:-180:0.8);
  \draw[blue,thick](2,3) arc (0:-85:1);
  \draw[blue,thick](0,3) arc (-180:-95:1);
  \draw[red,thick](0.2,3)--(1.8,3);\draw[blue,thick](0,3)--(-0.2,3)--(-0.2,3.2)--(2.2,3.2)--(2.2,3)--(2,3);
  \node at (0.2,-0.6) {$Q_{ij}$};\node at (2,-0.6) {$Q_{ji}$};
\node at (3,1.5) {\large ,};
\end{tikzpicture}\hspace{0.5cm}
  \begin{tikzpicture}[scale=0.8]
  \draw[red,thick](0.2,0)--(1.8,0);  \draw[blue,thick](0,0)--(-0.2,0)--(-0.2,-0.2)--(2.2,-0.2)--(2.2,0)--(2,0);
  \draw[red,thick](1.8,0) arc (0:180:0.8);
  \draw[blue,thick](2,0) arc (0:85:1);
  \draw[blue,thick](0,0) arc (180:95.5:1);
  \draw[blue,thick,decorate, decoration={snake, amplitude=-1.1, segment length=8.7}](0.9,0.99)--(0.9,2);
  \draw[blue,thick,decorate, decoration={snake, amplitude=-1.1, segment length=8.7}](1.09,0.99)--(1.09,2);
  \draw[green,thick](1.8,3) arc (0:-180:0.8);
  \draw[blue,thick](2,3) arc (0:-85.5:1);
  \draw[blue,thick](0,3) arc (-180:-95:1);
  \draw[green,thick](0.2,3)--(1.8,3);\draw[blue,thick](0,3)--(-0.2,3)--(-0.2,3.2)--(2.2,3.2)--(2.2,3)--(2,3);
  \node at (0.2,-0.6) {$Q_{ij}$};\node at (2,-0.6) {$\Qb_{ji}$};
    
  \end{tikzpicture}  \begin{tikzpicture}[scale=0.8]
    \node at (-1,1.5) {$-\frac{1}{N}$};
  \draw[red,thick](0.2,0)--(1.8,0);  \draw[blue,thick](0,0)--(-0.2,0)--(-0.2,-0.2)--(2.2,-0.2)-- (2.2,0)--(2,0);
  \draw[red,thick](1.8,0) arc (0:180:0.8);
  \draw[blue,thick](2,0) arc (0:85:1);
  \draw[blue,thick](0,0) arc (180:95.5:1);
  \draw[blue,thick,decorate, decoration={snake, amplitude=-1.1, segment length=8.7}](0.9,0.99)--(0.9,1.3);
  \draw[blue,thick,decorate, decoration={snake, amplitude=-1.1, segment length=8.7}](1.09,0.99)--(1.09,1.3);
    \draw[blue,thick,decorate, decoration={snake, amplitude=-1.1, segment length=8.7}](0.9,1.7)--(0.9,2);
    \draw[blue,thick,decorate, decoration={snake, amplitude=-1.1, segment length=8.7}](1.09,1.7)--(1.09,2);
    \draw[blue,thick](1.09,1.3) arc (0:180:0.095);
  \draw[blue,thick](1.09,1.7) arc (0:-180:0.09);
  \draw[red,thick](1.8,3) arc (0:-180:0.8);
  \draw[green,thick](1.8,3) arc (0:-180:0.8);
  \draw[blue,thick](2,3) arc (0:-85.5:1);
  \draw[blue,thick](0,3) arc (-180:-95:1);
  \draw[green,thick](0.2,3)--(1.8,3);\draw[blue,thick](0,3)--(-0.2,3)--(-0.2,3.2)--(2.2,3.2)--(2.2,3)--(2,3);
  \node at (0.2,-0.6) {$Q_{ij}$};\node at (2,-0.6) {$\Qb_{ji}$};
   
\end{tikzpicture}

\caption{For $L=2$, the $1/N$ part of the adjoint propagators gives a leading contribution in the large $N$ limit. There are two more diagrams, for $\bar{Q}Q$ and $\bar{Q}\bar{Q}$.}\label{L2enhancement}
\end{center}
\end{figure}
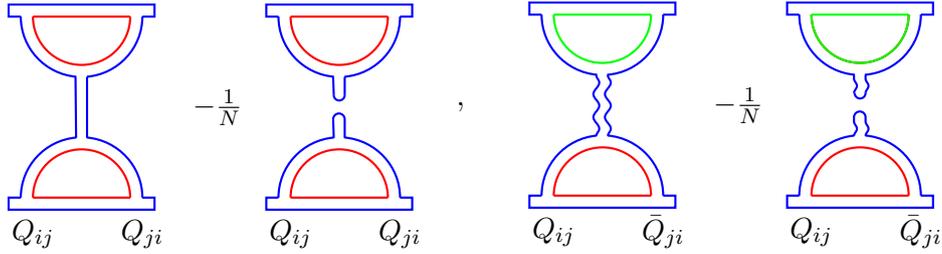

As we are working in the strict planar limit, in the above computations we have ignored non-planar effects. In particular, we have treated the propagators of the adjoint fields as if they were in $\Urm(N)$ rather than $\SU(N)$. The difference is subleading in all cases, apart from the case of length-2 operators, where the diagrams shown in Fig. \ref{L2enhancement} are both of order $N^4$. Therefore, these effects need to be taken into account. In the component formalism, these contributions lead to double-trace terms in the Lagrangian (see e.g. \cite{Gadde:2010zi} for the $\Zset_2$ orbifold case), arising from integrating out the $D$- terms and the adjoint $F_Z$- terms. As we are working in superspace, we don't have double-trace terms but we can compute these contributions directly. Their effect is to precisely cancel the planar part of the dilatation operator for certain operators, bringing their eigenvalues from their planar values down to $E=0$. 

  It is easy to see that these non-planar diagrams only contribute to twisted sectors, as the two cyclic orderings of the three-point vertex give opposite signs, but result in the same trace operators. Clearly, this non-planar correction affects a whole $\SU(2)_R$ triplet at once:
  \be
  \left(\begin{array}{c} Q_{ij}Q_{ji}\\ \frac{1}{\sqrt2}(Q_{ij}\bar{Q}_{ji}-\bar{Q}_{ij}Q_{ji})\\ \bar{Q}_{ij}\bar{Q}_{ji}\end{array}\right)\;.
  \ee 
Therefore, these non-planar contributions lead to the protection of the twisted-sector $L=2$ $\SU(2)_R$ triplets, on top of the untwisted triplet (which is also present for the cyclic quivers). As explained in \cite{Gadde:2010zi}, these specific triplets are guaranteed to be protected by index arguments, and, as we will see, the same is true in our ADE case. So this non-planar effect is crucial in obtaining agreement with index computations.

\section{The ADE Hamiltonian} \label{sec:ADEHamiltonian}

Having computed all the one-loop diagrams that contribute to the planar dilatation operator, in this section we will present the resulting one-loop Hamiltonian for generic finite subgroups of $\SU(2)$. Recall that, in evaluating the diagrams, we restricted to the lowest scalar components of the superfields, so the Hamiltonian below will be for the scalar components of the $X,Y,Z$ superfields (and their conjugates). Since no more superfields will appear in the remainder of this article, will not specifically indicate this and use $X,Y,Z$ to indicate the scalar components from now on.\footnote{It would be straightforward to include the fermions of the chiral multiplets, as it would simply involve additional covariant derivatives in the Feynman diagrams. However, a full treatment beyond the scalar sector would require the inclusion of gauge field and gaugino insertions, which we leave for future work.}

As already discussed, the Hamiltonian for orbifolds of $\Ncal=4$ SYM is integrable at the orbifold point, where all the gauge couplings are equal \cite{Wang:2003cu,Ideguchi:2004wm,Beisert:2005he,Solovyov:2007pw}. So our main interest will be in the marginally deformed theories, which we will call ADE quiver spin chains. 
For the $\Zset_2$ quiver, the marginally deformed theory was treated in detail in \cite{Gadde:2009dj,Gadde:2010zi}. It was called the \emph{interpolating theory}, since it interpolates from the orbifold point $\gym^{(1)}=\gym^{(2)}$ to superconformal QCD (in the Veneziano limit) when one of the gauge couplings is taken all the way to zero. 

Among the ADE quivers, the $\Zset_2$ theory is special in that it has an additional $\SU(2)_L$ symmetry, under which the $X$ and $Y$ bifundamental fields form a doublet. So in this case the full unbroken symmetry group is $\SU(2)_L\times \SU(2)_R\times \Urm(1)_r$. The unbroken $\SU(2)_L$ played an important role in the study of the $XY$ sector in \cite{Pomoni:2021pbj}, where the spin chain is an alternating Heisenberg model. Here our main focus will be on all the other quiver theories, which do not have this symmetry enhancement.  

The Hamiltonian $\Hcal$ is read off from the one loop dilatation operator as
\be
\mathfrak{D}^{(1)}=-\frac{\lambda}{16\pi^2}\Hcal\;.
\ee
At one loop, the Hamiltonian is nearest-neighbour, so we can express its action on each pair of sites on the spin chain:
\be
\Hcal=\sum_{\ell=1}^L \Hcal_{\ell,\ell+1}\;,
\ee
where for closed chains we have $L+1 \simeq 1$. We will exhibit the one-loop scalar Hamiltonian through its action on the different sectors of the theory. Clearly, the Hamiltonian will depend on the adjacency matrix of a given quiver as well as the ranks of the various gauge groups. These are summarised in the coefficients $d_{ij}$ defined in (\ref{dij}). It is also convenient to define 
the quantity $H_i\equiv \sum_{j=1}^Ma^\mathbf{2}_{ij}$, which is simply the number of hypermultiplets attached to a node.

\subsection{Holomorphic sector}

It is easy to see that 
\be
\Hcal_{\ell,\ell+1}=0\;\text{on} \;\;Z_iZ_i\;,
\ee
and, for $i\ne k$:
\be
\Hcal_{\ell,\ell+1}=0\; \text{on}\;\; Q_{ij}Q_{jk}\;.
\ee
The Hamiltonian on bifundamentals which return to the same node is 
\begin{equation}
\Hcal_{\ell,\ell+1}=\frac{2\kappa_i^2}{n_i^3}\begin{pmatrix}
        \frac{\bar{d}_{ij_{1}}d_{ij_1}}{n_{j_1}}&\frac{\bar{d}_{ij_{1}}d_{ij_{2}}}{n_{j_1}}&\ldots&\frac{\bar{d}_{ij_{1}}d_{ij_{H_i}}}{n_{j_1}}\\\
        \frac{\bar{d}_{ij_{2}}d_{ij_1}}{n_{j_2}}&\frac{\bar{d}_{ij_{2}}d_{ij_{2}}}{n_{j_2}}&\ldots&\frac{\bar{d}_{ij_{2}}d_{ij_{H_i}}}{n_{j_2}}\\
        \vdots&\vdots&\ddots&\vdots\\
        \frac{\bar{d}_{ij_{H_i}}d_{ij_1}}{n_{j_{H_i}}}&\frac{\bar{d}_{ij_{m_i}}d_{ij_{2}}}{n_{j_{H_i}}}&\ldots&\frac{\bar{d}_{ij_{H_i}}d_{ij_{H_i}}}{n_{j_{H_i}}}
\end{pmatrix}\;\text{in the basis}\;
\begin{pmatrix}
        Q_{ij_1}Q_{j_1i}\\
        Q_{ij_2}Q_{j_2i}\\
        \vdots\\
        Q_{ij_{H_i}}Q_{j_{H_i}i}
    \end{pmatrix}
\end{equation}
while on one bifundamental and one adjoint we have

\begin{equation}
 \Hcal_{\ell,\ell+1}=\begin{pmatrix}
        2\kappa_i^2&-2\kappa_i\kappa_j\\
        -2\kappa_i\kappa_j&2\kappa_j^2
    \end{pmatrix} \; \text{in the basis} \; \begin{pmatrix}
        Z_iQ_{ij}\\
        Q_{ij}Z_j\\\end{pmatrix}.
\end{equation}
The action of the Hamiltonian on antiholomorphic fields can be found by conjugation.

\subsection{Mixed sector}

Here we need to distinguish two cases, when the Hamiltonian acts on two sites with the same first and last index, or when the indices are different. For the first case, let us define the column vectors:
\be
\mathbf{Q\Qb}_{i}\equiv\begin{pmatrix}
    Q_{ij_1}\Qb_{j_1i}\\
    \vdots\\
    Q_{ij_{H_i}}\Qb_{j_{H_i}i}
\end{pmatrix},\;\mathbf{\Qb Q}_{i}\equiv\begin{pmatrix}
    \Qb_{ij_1}Q_{j_1i}\\
    \vdots\\
    \Qb_{ij_{H_i}}Q_{j_{H_i}i}
\end{pmatrix}\;.
\ee
We also define the $H_i\times H_i$ matrices
\be
\quad\mathbb{M}_{i}\equiv\frac{\kappa_i^2}{n_i}\begin{pmatrix}
    n_{j_1}&\dots&n_{j_{H_i}}\\
    \vdots&\ddots&\vdots\\
    n_{j_1}&\ldots&n_{j_{H_i}}\\
\end{pmatrix},\quad\mathbb{T}_{i}\equiv\begin{pmatrix}
        2\kappa_{j_1}^2& &\\
         &\ddots&\\
         & &2\kappa_{j_{H_i}}^2\end{pmatrix},
\ee
and the $H_i\times1$ and $1\times H_i$ matrices
\be
\mathbb{L}_i\equiv\begin{pmatrix}
             \frac{\kappa_i^2}{n_i}\\
         \vdots\\
         \frac{\kappa_i^2}{n_i}
         \end{pmatrix}\;,\qquad \mathbb{K}_i\equiv\begin{pmatrix}
    \kappa_i^2n_{j_1}&\ldots&\kappa_i^2n_{j_{H_i}}
\end{pmatrix}\;.
\ee
Using these, we can express the mixed Hamiltonian in compact form as 
\be
\Hcal_{\ell,\ell+1}=\begin{pmatrix}
    3\kappa_i^2&-\kappa_i^2&\mathbb{K}_{i}&\mathbb{K}_{i}\\
    -\kappa_i^2&3\kappa_i^2&\mathbb{K}_{i}&\mathbb{K}_{i}\\
    \mathbb{L}_{i}&\mathbb{L}_{i}&\mathbb{T}_{i}+\mathbb{M}_{i}&\mathbb{T}_{i}-\mathbb{M}_{i}\\
    \mathbb{L}_{i}&\mathbb{L}_{i}&\mathbb{T}_{i}-\mathbb{M}_{i}&\mathbb{T}_{i}+\mathbb{M}_{i}
\end{pmatrix}\; \text{in the basis}\;\; 
\begin{pmatrix}
    Z_i\Zb_i\\
    \Zb_iZ_i\\
    \mathbf{Q\Qb}_{i}\\
    \mathbf{\Qb Q}_{i}
\end{pmatrix}\;.
\ee
We note that, written in this way, the Hamiltonian is not hermitian since $\mathbb{L}$ is not equal to $\mathbb{K}^\dag$. This is due to the non-canonical normalisations used in the Lagrangian (\ref{eq:NN=2generalaction}), and can be easily fixed by rescaling the fields by appropriate factors of $\sqrt{n_i}$. This would, however, introduce such square-root factors in the Hamiltonian. Therefore, and since this does not affect the spectrum (which is of course real) we find it best to keep the non-canonical normalisation.

For the case when $k\ne i$, we have
\be
\Hcal_{\ell,\ell+1}=  \begin{pmatrix}
  2\kappa_j^2&\frac{2\kappa_{j}^2d_{kj}\bar{d}_{ij}}{n_in_kn_j^2}\\
    \frac{2\kappa_{j}^2\bar{d}_{kj}d_{ij}}{n_in_kn_j^2}&2\kappa_j^2
\end{pmatrix}\;\text{in the basis}\; \begin{pmatrix}
    Q_{ij}\Qb_{jk}\\
    \Qb_{ij}Q_{jk}
\end{pmatrix}\;.
\ee
and finally
\be
\Hcal_{\ell,\ell+1}=\begin{pmatrix}
    2\kappa_i^2&-2\kappa_i\kappa_j& &\\
    -2\kappa_i\kappa_j&2\kappa_j^2& &\\
    & &2\kappa_i^2&-2\kappa_i\kappa_j\\
    & &-2\kappa_i\kappa_j&2\kappa_j^2
\end{pmatrix}\;\text{in the basis} \begin{pmatrix}
    Z_i\Qb_{ij}\\
    \Qb_{ij}Z_j\\
    \Zb_iQ_{ij}\\
    Q_{ij}\Zb_j
\end{pmatrix}\;.
\ee
where we note that the Hamiltonian does not mix the first two with the second two rows, as that would violate the $\Urm(1)_r$ symmetry. 

One can straightforwardly check that the ADE Hamiltonian commutes with the $\SU(2)_R\times\Urm(1)_r$ R-symmetry group, whose action is shown in Table \ref{tab:Raction}. Of course, unlike the $\Zset_2$ Hamiltonian computed in \cite{Gadde:2010zi}, it does not commute with the $\SU(2)_L$ symmetry under which the $X$ and $Y$ fields would form a doublet. 

We emphasise that the above Hamiltonian is only valid for $L>2$, as it does not account for the non-planar contributions at $L=2$ discussed in section \ref{sec:Nonplanar}. For $L=2$, the states affected by those contributions (which, as discussed, are twisted sector states in the triplet representation of $\SU(2)_R$) will still appear as eigenstates of the Hamiltonian but with non-zero energy, instead of the correct $E=0$. As these states are easily identifiable, in the following we will use the Hamiltonian also for $L=2$ and set the eigenvalues of these specific states to zero by hand. 

Before proceeding to discuss the spin chains on which this Hamiltonian acts, we need to  develop  a second tool in the study of the ADE theories, namely the superconformal index and its various limits. This will provide crucial information on the protected spectrum of our theories. Readers who are more interested in non-protected quantities can jump ahead to Section \ref{sec:ADEspinchains}.

\section{The Protected Spectrum}\label{sec:Protected}

In the previous sections, we computed the one-loop scalar dilatation operator of the ADE theories and expressed it as a spin-chain Hamiltonian. This provides a wealth of information about the spectrum of the theory, although it is of course limited to scalar operators and will receive higher-loop corrections. As in $\Ncal=4$ SYM, the scalar sector will also not be closed at higher loops. Also, even though (as we will show) it is possible to solve the 2-magnon problem for the ADE spin chains, and thus obtain all-length results, the higher-magnon problem is expected to be significantly more complicated. For these reasons, following in the footsteps of \cite{Gadde:2009dj,Gadde:2010zi} for the $\Zset_2$ orbifold theory, we will also consider another powerful tool to aid us in the study of the ADE theories, namely the superconformal index \cite{Romelsberger:2005eg,Kinney:2005ej}. Together with the Molien series (which has already been applied to the ADE quivers in \cite{Benvenuti:2006qr}), it will provide additional, coupling-independent information on the spectrum of protected states. Besides allowing us to access fermionic parts of the spectrum, it will provide additional checks of the Hamiltonian by considering the overlap of the two approaches.

In the following sections, we will briefly review the multiplets of $\Ncal=2$ supersymmetry and the definition of the superconformal index, before evaluating it for the ADE quiver theories.

\subsection{$\NN=2$ Representations and Shortening Conditions}\label{sec:N=2Shortmultiplets}

In this section, we will briefly introduce some of the $\Ncal=2$ representations which will play a role in the computation of the index, as well as in the study of the spectrum by direct diagonalisation of our Hamiltonian. For more details and proofs, we refer to \cite{Dobrev:1985qv,Ferrara:1999ed,Dolan:2002zh,Cordova:2016emh}, as well as the pedagogical treatments in \cite{Pomoni:2019oib, Eberhardt:2020cxo}.
Our notation will mainly follow \cite{Dolan:2002zh}. In this notation $\ket{R,r}^\text{h.w.}$ satisfies the following under $\SU(2)_R\otimes\U(1)_r$ (see Table \ref{tab:Raction}):
\be
\sigma_R^+\ket{R,r}^\text{h.w.}=0,\quad \sigma_R^3\ket{R,r}^\text{h.w.}=R\ket{R,r}^\text{h.w.},\quad \sigma_r\ket{R,r}^\text{h.w.}=r\ket{R,r}^\text{h.w.}.
\ee
Let us now review some of the representation theory of the $\NN=2$ superconformal algebra. We will begin by giving the necessary commutation and anti-commutation relations. For the full superconformal algebra, see e.g. \cite{Pomoni:2019oib,Eberhardt:2020cxo}. Recall that the algebra includes the bosonic generators $P_\mu,M_\alpha^{\;\beta},\bar{M}_{\dot{\alpha}}^{\;\dot{\beta}},K_\mu, \mathfrak{D}, R^i_{\;j},r$ (where $R^i_{\;j}={\sigma_R^{\pm,3}})$ and fermionic generators $\Qcal^i_\alpha,\bar{\Qcal}_{i\dot{\alpha}},\mathcal{S}^\beta_j,\bar{\mathcal{S}}^{j\dot{\beta}}$, where $i=1,2$ and $\alpha,\dot{\alpha}$ are the spacetime $\SU(2)\times\SU(2)$ indices. As is standard, we will be writing $\alpha=\pm,\dot{\alpha}=\dot{\pm}$. First, we have the following commutation relations 
\be\label{eq:n=2-scaDQS}\begin{split}
  [\mathfrak{D},\Qcal^i_\alpha]=&\,+\frac{1}{2}\Qcal^i_\alpha\,,\quad[\mathfrak{D},\bar{\Qcal}_{i\dot{\alpha}}]=\,+\frac{1}{2}\bar{\Qcal}_{i\dot{\alpha}}\;,\\
    [\mathfrak{D},\Scal_i^\alpha]=&\,-\frac{1}{2}\Scal_i^\alpha\,,\quad[\mathfrak{D},\bar{\Scal}^{i\dot{\alpha}}]=\,-\frac{1}{2}\bar{\Scal}^{i\dot{\alpha}}\,,
\end{split}\ee
which imply that the special conformal supercharges, $\Scal_{i}^\alpha$ and $\bar{\Scal}^{i\dot{\alpha}}$, lower the scaling dimension $\Delta$ by $\half$ and the Poincar\'e supercharges, $\Qcal_{\alpha}^i$ and $\bar{\Qcal}_{i\dot{\alpha}}$, raise the scaling dimension $\Delta$ by $\half$\footnote{Often in the literature $\bar{\Qcal}$ and $\bar{\Scal}$ are denoted $\widetilde{\Qcal}$ and $\widetilde{\Scal}$. The tilde notation is used because, while in Lorentzian signature $\bar{\Qcal}=\Qcal^\dagger$ and $\bar{\Scal}=\Scal^\dagger$, in Euclidean signature the left and right-handed spinors are independent.}. Next, we have the following anti-commutation relations
\be\label{eq:n=2-sca}
    \begin{split}
        \lbrace \Qcal^i_{\alpha},\Scal_j^{\beta} \rbrace &=\tfrac{1}{2}\delta^{i}_j\delta^\beta_\alpha \mathfrak{D} +\delta^i_j M_\alpha^\beta-\delta^\beta_\alpha R_j^i-\tfrac{1}{2}\delta^{i}_j\delta^\beta_\alpha r\;,\\
        \lbrace \bar{\Qcal}_{i\dot{\alpha}},\bar{\Scal}^{j\dot{\beta}} \rbrace &=\tfrac{1}{2}\delta^j_i\delta^{\dot{\beta}}_{\dot{\alpha}}\mathfrak{D} +\delta^j_i \bar{M}^{\dot{\beta}}_{\dot{\alpha}}+\delta^{\dot{\beta}}_{\dot{\alpha}} R_i^j-\tfrac{1}{2}\delta^{j}_i\delta^{\dot{\beta}}_{\dot{\alpha}} r\;.
\end{split}\ee
    Consider a state in the Hilbert space $\Hcal$ on $\mathbb{R}\times S^3$ for a $\NN=2$ SCFT. We let $\ket{\Delta,j_1,j_2;R,r}\in\Hcal$, where $\Delta$ is the scaling dimension (the eigenvalue of the dilatation operator $\mathfrak{D}$), $R$ is the Dynkin label of $\mathfrak{su}(2)_R$, $r$ is the $\mathfrak{u}(1)_r$ charge, and $(j_1,j_2)$ are the Dynkin labels of the Lorentz algebra $\mathfrak{su}(2)_1\times\mathfrak{su}(2)_2$. As mentioned, the $\mathcal{S}$ supercharges lower the scaling dimension by $\half$. Because we want to consider representations with a bounded spectrum of $\mathfrak{D}$, we require that there exist states $\ket{\Delta,j_1,j_2;R,r}\in\Hcal$ satisfying 
\be\label{eq:defSuperconformalPrimary}
K_{\mu}\ket{\Delta,j_1,j_2;R,r}=0\,,\quad\Scal^\alpha_i\ket{\Delta,j_1,j_2;R,r}=0\,,\quad\bar{\Scal}^{i\dot{\alpha}}\ket{\Delta,j_1,j_2;R,r}=0.\ee
These states are called \emph{superconformal primaries} (which we refer to as primaries for brevity). The representations of $\Hcal$ are then spanned by the descendants of the superconformal primaries, obtained by acting with the Poincar\'e supercharges:
\be\label{eq:superconformaldescendants}
\prod_{i,j,\alpha,\dot{\alpha}}(\bar{\Qcal}_{i\dot{\alpha}})^{n_{i\dot{\alpha}}}(\Qcal^{j}_{\alpha})^{n_{j\alpha}}\ket{\Delta,j_1,j_2;R,r}\,\;,\quad n_{i\dot{\alpha}},n_{j\alpha}=0,1\,.
\ee
A superconformal primary \eqref{eq:defSuperconformalPrimary} and its descendants \eqref{eq:superconformaldescendants} form a multiplet. The Hilbert space $\Hcal$ is spanned by the primaries and their descendants.

We consider unitary theories, where all states have non-negative norms,  $||\ket{\Delta,j_1,j_2;R,r}||\ge0$. Conjugation acts as
\be\label{eq:conjPoincaresupercharges}
(\Qcal_{\alpha}^i)^\dagger\equiv\Scal_i^{\alpha},\quad(\bar{\Qcal}_{i\dot{\alpha}})^\dagger\equiv \bar{\Scal}^{i\dot{\alpha}}\;,
 \ee
 so that 
 \be
\bra{\Qcal_{\alpha}^i\Psi}=\bra{\Psi}\Scal_i^{\alpha}\,,\quad\bra{\bar{\Qcal}_{i\dot{\alpha}}\Psi}=\bra{\Psi}\bar{\Scal}^{i\dot{\alpha}}\;.
 \ee
Let us now consider the norms of the first descendants of a superconformal primary $\ket{\Delta,j_1,j_2;R,r}$. Computing the norms of $\Qcal^i_\pm$ and $\bar{\Qcal}_{i\dot{\pm}}$ on this state, applying \eqref{eq:n=2-sca}, and requiring positivity gives
\be\begin{split}\label{eq:unitaritydescdants}
    &(\tfrac{1}{2}\Delta\pm j_1+(-1)^iR-\tfrac{1}{2}r)||\ket{\Delta,j_1,j_2;R,r}||^2\ge0\;,\\
    &(\tfrac{1}{2}\Delta\pm j_2+(-1)^iR+\tfrac{1}{2}r)||\ket{\Delta,j_1,j_2;R,r}||^2\ge0\;.
\end{split}\ee
From unitarity, we have $||\ket{\Delta,j_1,j_2;R,r}||^2\ge0$, hence, from \eqref{eq:unitaritydescdants}, we must have\footnote{There are some short multiplets with stricter unitary bounds such that $f(j_1,j_2,R,r)\ge\Delta\ge1$ \cite{Pomoni:2019oib,Eberhardt:2020cxo}} 
\be\label{eq:unitarybound}
\Delta\ge\max\lbrace2+2j_1+2R+r,2+2j_2+2R-r\rbrace\equiv f(j_1,j_2,R,r)\;.
\ee
In \eqref{eq:unitarybound} we are excluding the trivial representation of $\SU(2,2|2)$ (the vacuum state), with $\Delta=0$ \cite{Eberhardt:2020cxo,Pomoni:2019oib}. The constraint on $\Delta$ in \eqref{eq:unitarybound} is called a \emph{unitary bound}. The unitary bounds \eqref{eq:unitarybound} in SCFTs play an analogous role to the positive semi-definite energy condition in SUSY QM. Zero norm states (or \emph{null states}) decouple from the theory; hence, we only need to consider primaries with positive definite norms $||\ket{\Delta,j_1,j_2;R,r}||>0$ and remove those with zero norms. If a primary state $\ket{\Delta,j_1,j_2;R,r}$ saturates the unitary bound,  then one of its descendants will be null and hence will decouple from the theory, i.e. one of the states will be annihilated by the action of one (or more) Poincar\'e supercharges, $\Qcal_{\alpha}^i$ or $\bar{\Qcal}_{i\dot{\alpha}}$. Therefore, the multiplet generated from this primary will contain fewer states than a multiplet generated by a primary whose dimension is strictly greater than the unitary bound \eqref{eq:unitarybound}. Hence, we call the multiplets that saturate the unitary bound \eqref{eq:unitarybound} \emph{short multiplets} and the multiplets strictly above the unitary bound \eqref{eq:unitarybound} \emph{long multiplets}. For reference, we provide a list of these $\Ncal=2$ SCFT multiplets in Table \ref{tab:Oppsof}, following the classification scheme of \cite{Dolan:2002zh}.  
\begin{table}[h]
    \centering
        \renewcommand{\arraystretch}{1.3}
    \begin{tabular}{|c|c|}\hline
    Multiplet&Operator $(R,\,\ell\ge0,\;n\ge2)$\tabularnewline\hline\hline
    $\hat{\mathcal{B}}_1$&$\Tr\left[\sum\Mtrip_{ij}\right]$\tabularnewline\hline
    $\hat{\mathcal{B}}_{R+1}$&$\Tr\left[\sum(Q_{ij}Q_{ji})^{R+1}\right]$\tabularnewline\hline
    $\bar{\mathcal{E}}_{-(\ell+2)(0,0)}$&$\Tr\left[\sum Z_i^{\ell+2}\right]$\tabularnewline\hline
    $\hat{\mathcal{C}}_{R(0,0)}$&$\Tr\left[\sum T(Q_{ij}Q_{ji})^{R}\right]$\tabularnewline\hline
    $\bar{\mathcal{D}}_{R+1(0,0)}$&$\Tr\left[\sum_{m=0}^1(Q_{ij}Q_{ji})^{R+1}Z_i^mZ_j^{1-m}\right]$\tabularnewline\hline
    $\mathcal{D}_{R+\frac{3}{2}(0,\half)}$&$\Tr\left[\sum_{m=0}^1(Q_{ij}Q_{ji})^{R+1}(\bar{\lambda}_{i})_{Z\dot{+}}^{m}(\bar{\lambda}_{j})_{Z\dot{+}}^{1-m}\right]$\tabularnewline\hline
    $\bar{\mathcal{B}}_{R+1,-(\ell+2)(0,0)}$&$\Tr\left[\sum_{m=0}^{\ell+2}(Q_{ij}Q_{ji})^{R+1}Z_i^{m}Z_j^{\ell+2-m}\right]$\tabularnewline\hline
    $\bar{\mathcal{C}}_{R,-(\ell+1)(0,0)}$&$\Tr\left[\sum_{m=0}^{\ell+1}T(Q_{ij}Q_{ji})^{R}Z_i^{m}Z_j^{\ell+1-m}\right]$\tabularnewline\hline
        $\mathcal{C}_{R,(\ell+1)(0,0)}$&$\Tr\left[\sum_{m=0}^{\ell+1}T(Q_{ij}Q_{ji})^{R}\Zb_i^{m}\Zb_j^{\ell+1-m}\right]$\tabularnewline\hline
    $\mathcal{A}^{2R+\ell+2n}_{R,-\ell(0,0)}$&$\Tr\left[\sum_{m=0}^\ell T^n(Q_{ij}Q_{ji})^{R}Z_i^{m}Z_j^{\ell-m}\right]$\tabularnewline\hline
    \end{tabular}
    \caption{The superconformal primaries of some of the multiplets of the $\NN=2$ superconformal algebra, adapted from \cite{Gadde:2009dj}. In the schematic notation used there, the symbol $\sum$ indicates summation over all symmetric traceless permutations of the component fields as allowed by the gauge index structure, and $T$ stands for an $R$-symmetry-neutral combination of scalar fields. Note that $R$ can be half-integer.}
    \label{tab:Oppsof}
\end{table}\\
A generic long multiplet is denoted by $\mathcal{A}^{2R+\ell+2n}_{R,r(j_1,j_2)}$. For the shortening conditions of each multiplet in Table \ref{tab:Oppsof}, we refer to  \cite{Gadde:2009dj}. In $\NN=2$ theories, $\bar{\mathcal{E}}_{r(0,0)}$ corresponds to Coulomb branch physics and $\hat{\mathcal{B}}_{R}$ corresponds to Higgs branch physics.

In the quantum theory, short multiplets can only acquire anomalous dimensions 
\be
\Delta=f(j_1,j_2,R,r)+\gamma(\lambda)\;,
\ee 
(where $\lambda$ is a coupling) if they recombine with other short multiplets and become long. That is, the descendants that were lost due to the existence of a null vector are returned. The rules that dictate how multiplets can recombine are called \emph{recombination rules} and are listed in \cite{Dolan:2002zh}.

\subsection{Review of the Superconformal Index}

In studying the spectrum of $\NN=2$ theories, it would be of great interest to consider a protected object counting precisely the short multiplets that do not become long. This object, known as the superconformal index, was introduced in \cite{Romelsberger:2005eg,Kinney:2005ej}, see \cite{Rastelli:2016tbz, Gadde:2020yah} for reviews. In our discussion below, we will closely follow the treatment in \cite{Gadde:2011uv}\footnote{However, note that to match the notation in Appendix \ref{sec:ExtendedSusyTransformations}, we use a different convention for the labels for the $\NN=2$ vector-multiplet fermions.}. Let us define the following objects
 \begin{equation}\label{eq:defdeltas}
	\delta_{i\pm}\equiv 2\lbrace \Qcal^i_{\pm},\Scal_i^{\pm}\rbrace\,,\qquad \bar{\delta}_{i\dot{\pm}}\equiv 2\lbrace \bar{\Qcal}_{i\dot{\pm}},\bar{\Scal}^{i\dot{\pm}}\rbrace\,.
\end{equation}
From \eqref{eq:conjPoincaresupercharges}, we have that $\delta_{i\pm}\ge0$ and $\bar{\delta}_{i\dot{\pm}}\ge0$. We can write $\delta_{i\pm}$ and $\bar{\delta}_{i\dot{\pm}}$ in terms of the quantum numbers of the states: 
\begin{align}
	\begin{split}
		\delta_{i\pm}=\Delta\pm 2j_1+(-1)^i\,2R-r\,,\qquad \bar{\delta}_{i\dot{\pm}}=\Delta\pm 2j_2+(-1)^i2R+r\,,
	\end{split}
\end{align}
This leads to the general superconformal index introduced in \cite{Romelsberger:2005eg,Kinney:2005ej},  defined as follows:
\be\label{eq:generalsci}
\mathcal{I}(\mu_i)=\Tr_\Hcal\,(-1)^Fe^{-\mu_iT_i}e^{-\beta\delta}\,,
\ee
where the trace is over the Hilbert space $\Hcal$ of the theory defined on $S^1_\beta\times S^3$. $\beta$ is the circumference of the thermal circle $S^1_\beta$ and $\{T_i\}$ a complete set of generators that commute with the supercharge $\Qcal$ in $\delta$ that we ``define the index with respect to" and with each other, and $\{\mu_i\}$ are the corresponding chemical potentials. The superconformal index \eqref{eq:generalsci} counts the states with $\delta=0$, hence, $\Qcal\ket{\Delta,j_1,j_2;R,r}=0$, and is therefore $\beta$-independent. By the state-operator correspondence, counting (with grading) states on $\mathbb{R}\times S^3$ is equivalent to counting (with grading) local operators on $\mathbb{R}^4$. The states with $\delta=0$ are called \emph{BPS states}, and form short multiplets of the superconformal algebra. The superconformal index is a generalisation of the Witten index \cite{Witten:1982df} for supersymmetric quantum mechanics to the superconformal field theory context. Like the Witten index, it is invariant under marginal deformations of the theory and hence is insensitive to marginal changes of couplings. So it is a protected quantity, and a computation of the index in the free theory will still be valid in the quantum theory.

 For four-dimensional $\NN=2$ SCFTs, which are non-chiral, different choices of $\Qcal$ result in physically equivalent indices. The subalgebra of $\mathfrak{su}(2,2|2)$ commuting with a single supercharge is $\mathfrak{su}(1,1|2)$, which has rank three, so the $\NN=2$ index depends on three superconformal fugacities \cite{Gadde:2011uv}.\footnote{Theories with flavour symmetries have additional fugacities, however we will not be considering such theories in this work.} We choose to define our index with respect to $\bar{\Qcal}_{1\dot{-}}$. A basis of the Cartan generators of the commutant subalgebra
$\mathfrak{su}(1, 1|2)$ is given by
\be
\delta_{1-},\quad\delta_{1+},\quad\bar{\delta}_{2\dot{+}}\;.
\ee
With this choice of generators, \eqref{eq:generalsci} becomes
\begin{align}\label{eq:n2-sci-fuga1}
	\begin{split}
		\mathcal{I}(\rho,\sigma,\tau)&=\mathrm{tr}_\Hcal\, (-1)^F \rho^{\half\delta_{1-}}\,\sigma^{\half\delta_{1+}}\,\tau^{\half\bar{\delta}_{2\dot{+}}}e^{-\beta\bar{\delta}_{1\dot{-}}}\,,
	\end{split}
\end{align}
where the trace runs over the states of the theory on $S^3$, and for convergence we take $|\rho|<1,|\sigma|<1,|\tau|<1$. The charges of the states counted by this index obey the shortening condition
\begin{equation} \label{d1mshortening}
    \bar{\delta}_{1\dot{-}}=\Delta-2j_2-2R+r=0\,.
\end{equation}
As discussed in \cite{Gadde:2011uv}, another common parametrisation of the index is in terms of fugacities $(p, q, t)$, related to $(\sigma, \rho, \tau )$ as
\be\label{eq:fuga2def}
p=\tau\sigma,\quad q=\tau\rho,\quad t=\tau^2\;.
\ee
In terms of these fugacities the index \eqref{eq:n2-sci-fuga1}, reads
\begin{align}\label{eq:n2-sci-fuga2}
	\begin{split}
		\mathcal{I}(p,q,t)&=\mathrm{tr}_\Hcal\, (-1)^F  p^{j_2+j_1-r}\,q^{j_2-j_1-r}\,t^{R+r}e^{-\beta\bar{\delta}_{1\dot{-}}}\,.
	\end{split}
\end{align}
where convergence requires $|p|<1\,,|q|<1\,,|t|<1\,,\left|\frac{pq}{t}\right|<1\;$. We will mostly present our results in this second parametrisation. 

In the following, we will evaluate the superconformal index \eqref{eq:generalsci} for general ADE orbifold quiver theories in the $N\to\infty$ limit\footnote{Note that, in the $N\to\infty$ limit, the superconformal index counts the Kaluza-Klein states in the gravity dual theory \cite{Kinney:2005ej}.}. The $\Zset_k$ index was calculated in \cite{Nakayama:2005mf}; however, in dividing out the $\U(1)$ factors, only the contributions from the $\NN=1$ vector multiplet were considered rather than the $\NN=2$ vector multiplet, leading to an $\Ncal=1$ index. Taking this into account, the $\Ncal=2$ index for the $\Zset_2$ quiver was computed in \cite{Gadde:2009dj, Gadde:2010en}. This theory has an additional $\SU(2)_L$ symmetry compared to the generic ADE case, and therefore has an additional fugacity. For more general $\Ncal=1$ quivers, the general form of the index can be found in \cite{Gadde:2010en,Eager:2012hx}, and would lead to an $\Ncal=2$ index by specialising to $\Ncal=2$ quivers and adding appropriate contributions from adjoint fields. However, to our knowledge, the large-$N$ superconformal index of non-abelian $\Ncal=2$ orbifolds has not been directly computed in the literature. We will present its general form for any $\NN=2$ orbifold theory and then evaluate it in the specific examples that we are covering.

 Following the plethystic programme \cite{Benvenuti:2006qr}, we begin by writing down the generalisation of a single-letter index with the single letters given by the fundamental fields and derivatives. We list the single-letter contributions to the index in Table \ref{tab:n=2-multiplets}.
\begin{table}[h]
    \centering
        \renewcommand{\arraystretch}{1.3}
    \begin{tabular}{|c|c|c|c|c|c|c|c|}
\hline
Letters & $  \Delta$ & $j_1$ & $  j_2$ & $R$ & $r$ &  $\mathcal{I}(\sigma, \rho, \tau)$ &$\mathcal{I}(p, q, t)$\tabularnewline
  \hline\hline
$ Z$ & $1$ & $0$ & $0$ & $0$ & $-1$ &  $\sigma\rho$ &$pqt^{-1}$  \tabularnewline
  \hline
  $  \bar{\lambda}_{Z\dot{+}}$  & $  \frac{3}{2}$ & $0$ & $\frac{1}{2}$ & $  \frac{1}{2}$ & $  \frac{1}{2}$  &  $-\tau^2$ &$-t$ \tabularnewline
  \hline
$  \lambda_{V\pm}$ & $  \frac{3}{2}$ & $\pm \frac{1}{2}$ & $0$ & $  \frac{1}{2}$ & $-\frac{1}{2}$  &$-\sigma\tau,\,-\rho\tau$ &  $-p,\,-q$  \tabularnewline
  \hline

$  \bar{F}_{\dot{+}\dot{+}}$ & $2$ & $0$ & $1$ & $0$ & $0$ &$\sigma\rho\tau^2$ &  $pq$ \tabularnewline
  \hline
  $  \partial_{-\dot{+}}  \lambda_{V+} +  \partial_{+\dot{+}}  \lambda_{V-}=0$ & $  \frac{5}{2}$ & $\frac{1}{2}$ & $ 0$ & $  \frac{1}{2}$ &
 $  -\frac{1}{2}$ & $\sigma\rho\tau^2$& $pq$  \tabularnewline
  \hline\hline
$Q_{ij}$ & $1$ & $0$ & $0$ & $  \frac{1}{2}$ & $0$&   $\tau$  &   $t^{\frac{1}{2}}$ \tabularnewline
  \hline
$  \bar{\psi}_{ji\dot{+}}$ & $  \frac{3}{2}$ & $  0$ & $\half$ & $0$ & $\frac{1}{2}$ & $-\sigma\rho\tau$ & $-pqt^{-\frac{1}{2}}$ \tabularnewline
  \hline\hline
$  \partial_{\pm\dot{+}}$ & $1$ & $   \pm\frac{1}{2}$ & $ \frac{1}{2}$ & $0$ & $0$ & $\sigma\tau,\,\rho\tau$ & $p,\,q$ \tabularnewline
\hline
\end{tabular}
    \caption{ Letters in the $\mathcal{N}=2$ vector multiplet and half-hypermultiplet with $\bar{\delta}_{1\dot{-}}=0$. The other half of the hypermultiplet has contributing letters $(Q_{ji},\bar{\psi}_{ij\dot{+}})$ which have the same charges as $(Q_{ij},\bar{\psi}_{ji\dot{+}})$ but transform in a conjugate representation of the gauge and global symmetry.}
    \label{tab:n=2-multiplets}
\end{table}\\
In addition to the  $(\rho,\sigma,\tau)/(p,q,t)$ fugacities, we need to introduce a fugacity with respect to the gauge symmetries, as the spectrum of the theory is given by gauge invariant operators. The vector multiplet at node $i$ contributes
\be\label{eq:single-letter-vector}
    \begin{split}
        i_{\text{vm}}(U_i;\rho,\sigma,\tau)&=\left(-\frac{\sigma\tau}{1-\sigma\tau}-\frac{\rho\tau}{1-\rho\tau}+\frac{\sigma\rho-\tau^2}{\left(1-\sigma\tau\right)\left(1-\rho\tau\right)}\right)\chi_{\text{adj},i}(U_i)\\
        &=\left(1+\frac{pqt^{-1}+pq-t-1}{(1-p)(1-q)}\right)\chi_{\text{adj}_i}(U_i)\,,
    \end{split}
\ee
where $\chi_{\text{adj}_i}(U_i)$ is the character of the adjoint representation of the $i^{\mathrm{th}}$  gauge group. A hypermultiplet connecting nodes $i$ and $j$ contributes
\begin{align}\label{eq:single-letter-hyper}
    \begin{split}
        i_{\text{hm}}(U_i,U_j;\rho,\sigma,\tau)&=\frac{\tau(1-\rho\sigma)}{(1-\rho\tau)(1-\sigma\tau)}\left(\chi_{{\overline{\square}_i\times\square_j}}(U_i,U_j)+\chi_{{\square_i\times\overline{\square}_j}}(U_i,U_j)\right)\\
        &=\frac{t^{\frac{1}{2}}(1-pqt^{-1})}{(1-p)(1-q)}\left(\chi_{{\square_i\times\overline{\square}_j}}(U_i,U_j)+\chi_{{\overline{\square}_i\times\square_j}}(U_i,U_j)\right)\,,
    \end{split}
\end{align}
where ${\square_i\times\overline{\square}_j}$ denotes the bifundamental representation and ${\overline{\square}_i\times\square_j}$ its conjugate. We will use the following notation for the characters of the adjoint and bifundamental representations:
\begin{subequations}
    \begin{align}
        \chi_{ii}(U_{ii})\equiv&\,\chi_{\text{adj},\,i}(U_i)=\Tr_iU_i\Tr_iU_i^\dagger-1,\\
        \chi_{ij}(U_{ij})\equiv&\,\chi_{\square_i\times\overline{\square}_j}(U_i,U_j)=\Tr_iU_i\Tr_jU_j^\dagger,
    \end{align}
\end{subequations}
where the $U_i$'s are in the fundamental representation. 

For $\Gamma\ne\Zset_2$, there is no flavour group generically. However, through deformations of the couplings, a gauge group can become global and turn into a flavour group, which is not a marginal deformation (as it changes the structure of the Hilbert space). Hence, the following computations will apply to generic couplings where no gauge groups become global. 

From Table \ref{tab:n=2-multiplets}, one finds that the single-letter index associated to node $i$ is 
\be\label{eq:single-letter-sci}
i(U;p,q,t)=\sum_{i,j=1}^M\chi_{ij}(U_{ij})\left(\delta_{ij}f_\text{vm}(p,q,t)+f_\text{hm}(p,q,t)\right).
\ee
where we have defined 
\be\begin{split}
    f_\text{vm}(p,q,t)\equiv&\;1-\frac{(1-pqt^{-1})(1+t)}{(1-p)(1-q)}=1-\frac{(1-\rho\sigma)(1+\tau^2)}{(1-\tau\sigma)(1-\tau\rho)}\;,\\
        f_\text{hm}(p,q,t)\equiv&\;\frac{t^{\frac{1}{2}}(1-pqt^{-1})}{(1-p)(1-q)}=\frac{\tau(1-\rho\sigma)}{(1-\rho\tau)(1-\sigma\tau)}.
\end{split}\ee
For our specific ADE quivers, it is useful to define the quantity 
\be \label{fADE}
f_{ij}(p,q,t)\equiv \delta_{ij}f_\text{vm}(p,q,t)+a^\mathbf{2}_{ij}f_\text{hm}(p,q,t)\,,
\ee
Then, we can write the single-letter index \eqref{eq:single-letter-sci} as 
\be\label{eq:sl-sci}
i(U;p,q,t)=\sum_{i,j=1}^M\chi_{ij}(U_{ij})f_{ij}(p,q,t).
\ee
Notice that this index counts the adjoint fields at each node, but, to avoid over-counting, only one half-hypermultiplet from each arrow linking the node to other nodes. The other half-hypermultiplets will be counted together with the other nodes. 

Now that we have the single-letter index \eqref{eq:sl-sci}, we apply the plethystic programme \cite{Feng:2007ur} to compute the index. Let us first define the plethystic exponential (PE)
\be\label{eq:defPE}
\text{PE}[f(x_i)]:=\exp\left(\sum_{n=1}^\infty\frac{f(x_i^n)}{n}\right)\;.
\ee
Then, the full index can be found via the plethystic exponential of the single letter index \cite{Gadde:2020yah} 
\be\label{eq:FullIndexUPE}
\mathcal{I}(U;p,q,t)=\text{PE}[i(U;p,q,t)]\;.
\ee
However, \eqref{eq:FullIndexUPE} counts all operators, not just gauge invariant operators (which are the physical states in the spectrum). In order to project out the gauge invariant operators, we will need to make use of Schur's orthogonality relation for Lie group characters, which states that 
\be\label{eq:SchurorthonalityLiegroup}
\int_{U\in G}[dU]\chi_R(U)\chi_{R'}(U)=\delta_{RR'}\;,
\ee
where $[dU]$ is the Haar measure of a Lie group $G$. We can rewrite the integral over the matrices in \eqref{eq:SchurorthonalityLiegroup} as an integral over the fugacities (the eigenvalues of $U$) as
\be
\int_{U\in G}[dU]f(U)=\prod_{m=1}^N\oint_{|u_m|=1}\frac{du_m}{2\pi iu_m}\Delta(u)f(u)\;,
\ee
where $N$ is the rank of the group and $\Delta(u)$ is the Vandermonde determinant, which evaluates to
\be
   \Delta(u)=\left(\text{PE}\left[\sum_\alpha u^\alpha\right]\right)^{-1}
   =\text{PE}\left[-\sum_\alpha u^\alpha\right]=\prod_\alpha(1-u^\alpha)\;,
\ee
where $\sum_{\alpha}u^\alpha\equiv\sum_{\alpha}\prod_{m}u_m^{\alpha_i}$, where $\alpha_i$ are the roots of the Lie algebra.
Notice that $\chi_\text{adj}(u)=\text{rank}(G)+\sum_{\alpha}u^\alpha$ so that 
\be \label{vandermonde}
\Delta(u)=\text{PE}[\text{rank}(G)-\chi_\text{adj}(u)]\;.
\ee
We can now write the projection of the index on the trivial representation as 
\be\label{eq:ProjectionWittenindex}
\mathcal{I}|_{\text{trivial}}=\prod_{m=1}^N\oint_{|u_m|=1}\frac{du_m}{2\pi iu_m}\text{PE}[i(x;u_m)-\chi_\text{adj}(u)]\chi_\text{trivial}(u)\;,
\ee
in other words, to obtain the gauge-invariant operators, we just need to integrate over the $u_i$.

Let us introduce some useful functions that will help us express the indices in a compact form. First of all, the $q$-Pochhammer symbol $(z;q)_\infty$ is defined by
\be\label{eq:DefPochhammer}
(z;q)_\infty\equiv\prod_{i=0}^{\infty}(1-zq^i)=\exp\left(-\sum_{n=1}^\infty\frac{1}{n}\frac{z^n}{(1-q^n)}\right)\;,
\ee
The elliptic $\Gamma$ function $\Gamma(t;p,q)$ is defined by \cite{Ruijsenaars:1997aqs}
\be\label{EllipticGamma}
\Gamma(t;p,q)\equiv\prod_{i,j=0}^{\infty}\frac{(1-t^{-1}p^{i+1}q^{j+1})}{(1-tp^iq^j)}=\exp\left(\sum_{n=1}^\infty\frac{1}{n}\frac{t^n-p^nq^nt^{-n}}{(1-p^n)(1-q^n)}\right)\;.
\ee
It satisfies the following identities that will be useful to us later \cite{Ruijsenaars:1997aqs}:
\be\label{eq:GammaIdentities}\begin{split}
\Gamma(pqt^{-1};p,q)&=\,\frac{1}{\Gamma(t;p,q)}\;,\; \Gamma(q;p,q)=\,\frac{(p;p)_\infty}{(q;q)_\infty}\;,\\
\Gamma(t;0,q)&=\,(t;q)_\infty^{-1}\; \text{and}\;
\Gamma(t;0,0)=\,\frac{1}{1-t}\;.\end{split}
\ee
From \eqref{eq:DefPochhammer} and \eqref{EllipticGamma}, we immediately find the following useful expression
\be\begin{split}\label{eq:PEfvm}
    \text{PE}[f_\text{vm}(p,q,t)]=&\exp\left(\sum_{n=1}^{\infty}\frac{f_\text{vm}(p^n,q^n,t^n)}{n}\right)\\
    =&\exp\left(\sum_{n=1}^\infty\frac{1}{n}\biggl[-\frac{p^n}{1-p^n}-\frac{q^n}{1-q^n}+\frac{p^nq^nt^{-n}-t^{n}}{(1-p^n)(1-q^n)}\biggr]\right)\\
    =&\frac{(p;p)_\infty(q;q)_{\infty}}{\Gamma(t;p,q)}\,.
\end{split}\ee
From \eqref{EllipticGamma}, we obtain the expression
\be\begin{split}\label{eq:PEhm}
    \text{PE}[(v+v^{-1})f_\text{hm}(p,q,t)]=&\exp\left(\sum_{n=1}^\infty\frac{(v^{ n}+v^{-n})f_\text{hm}(p^n,q^n,t^n)}{n}\right)\\
    =&\exp\left(\sum_{n=1}^\infty\frac{v^{n}t^{\frac{n}{2}}-p^nq^nv^{-n}t^{-\frac{n}{2}}+v^{-n}t^{\frac{n}{2}}-p^nq^nv^{n}t^{-\frac{n}{2}}}{(1-p^n)(1-q^n)}\right)\\
    =&\Gamma(vt^\half;p,q)\Gamma(v^{-1}t^\half;p,q)\,,
\end{split}\ee
where $v$ is a fugacity associated to the $\SU(2)_L$ subgroup of $\SU(4)_R$.

Note that \eqref{eq:ProjectionWittenindex} counts multi-trace states i.e. products of single trace states. We call \eqref{eq:ProjectionWittenindex} the ``multi-trace index". To compare with the Hamiltonian we would like to find the index of single-trace states (the single-trace index). This can easily be done using the inverse of the plethystic exponential, the ``plethystic logarithm", defined as follows
\be\label{eq:defPL}
\text{PL}[f(x_i)]:=\sum_{k=1}^{\infty}\left[\frac{\mu(k)}{k}\log\left(f(x_i^k)\right)\right]\;,
\ee
where $\mu(k)$ is the M\"obius function
\be
\mu(k):=\begin{cases}
    0\quad&\;\text{if $k$ has repeated prime factors}\\
    1\quad&\;k=1\\
    (-1)^n\quad&\;\text{$k$ is the product of $n$ distinct prime factors.}
\end{cases}
\ee
In evaluating the plethystic logarithm, the following relation between $\mu(k)$ and the Euler totient function $\varphi(n)$ is often useful:\footnote{The Euler totient function $\varphi(r)$ is the number of positive integers less than or equal to $r$ that are coprime with respect to $r$.} 
\be\label{eq:MobiusandEuler}
\sum_{d|n}d\mu\left(\frac{n}{d}\right)=\varphi(n)\;.
\ee
 We will also make use of the following formula 
\be\label{eq:EulerTotieone-loopg}
\sum_{r=1}^\infty\frac{\varphi(r)}{r}\log(1-x^r)=-\frac{x}{1-x}\;.
\ee
Before applying the above to the ADE index, let us take a brief tangent and discuss some limits of the index.

\subsection{Limits of the Index}

We now consider several limits of the superconformal index, such that the index counts states that are annihilated by more than one supercharge. Let us just recall that before taking any limits, we are considering states that are annihilated by $\bar{\Qcal}_{1\dot{-}}$, i.e., satisfy (\ref{d1mshortening}). We will be following the conventions in \cite{Gadde:2011uv}, where the different limits of the index were named by the type of symmetric polynomials relevant for their evaluation. They are summarised in Table \ref{tab:LimitsIndex} and we will discuss some relevant features of each case. 
\begin{table}[h]
    \centering
        \renewcommand{\arraystretch}{1.3}
    \begin{tabular}{|c|c|c|}\hline
    Index Name&Fugacities&Shortening Conditions\tabularnewline\hline\hline
    Macdonald&$\sigma\to0$ with $\rho,\,\tau$ fixed&$\Delta+2j_1-2R-r=0$\tabularnewline\hline
    Schur&$\rho=\tau$ with $\sigma$ arbitrary&$\Delta+2j_1-2R-r=0$\tabularnewline\hline
    Hall-Littlewood&$\sigma,\,\rho\to0$ with $\tau$ fixed&$\Delta\pm2j_1-2R-r=0$\tabularnewline\hline
    Coulomb-branch&$\tau\to0$ with $\rho,\,\sigma$ fixed&$\Delta+2j_2+2R+r=0$
    \tabularnewline\hline \hline
    
    Molien series&$-$&$\Delta-2R=0$\tabularnewline\hline\end{tabular}
    \caption{The four main limits of the superconformal index, the limits taken on the fugacities, and the multiplets that they count. In addition to the limits of the superconformal index, we include the Molien series (which is discussed in Section \ref{sec:MolienSeries}).}
    \label{tab:LimitsIndex}
\end{table}

\subsubsection{Macdonald index}

This limit is defined by taking
\begin{equation}
    \sigma\to0,\quad\rho,\,\tau\,\text{ fixed},
\end{equation}
(or equivalently, $p\to0$ with $q$ and $t$ fixed). The index is given by
\be
\begin{split}
    \mathcal{I}_M=&\tr_M(-1)^F\rho^{\half(\Delta-2j_1-2R-r)}\tau^{\half(\Delta+2R+2j_2+r)} e^{-\beta\bar{\delta}_{1\dot{-}}}\\
    =&\tr_M(-1)^Fq^{\half(\Delta-2j_1-2R-r)}t^{R+r}\,,
\end{split}
\ee
where $\tr_M$ denotes the trace restricted to states with $\delta_{1+}=\Delta+2j_1-2R-r=0$, i.e. the states that are annihilated by $\Qcal_{1+}$. The Macdonald index is a $\frac{1}{4}$-BPS object receiving contributions only
from states annihilated by two supercharges, one chiral ($\Qcal_{1+}$) and one anti-chiral ($\bar{\Qcal}_{1\dot{-}}$). 

The single letter indices are given by 
\be\begin{split}
    f_\text{vm}(\rho,\tau)=&1-\frac{1+\tau^2}{1-\rho\tau}=1-\frac{1+t}{1-q},\\
    f_\text{hm}(\rho,\tau)=&\frac{\tau}{1-\rho\tau}=\frac{t^{\half}}{1-q}\;.
\end{split}\ee
The Macdonald index receives contributions from 
\be\label{eq:Schursector}
\hat{\mathcal{B}}_R\,,\quad\mathcal{D}_{R(0,j_2)}\,,\quad\bar{\mathcal{D}}_{R(j_1,0)}\,\quad\hat{\mathcal{C}}_{R(j_1,j_2)}\,,
\ee
known as the Schur sector. More about the Schur sector can be found in \cite{Beem:2013sza}.

\subsubsection{Schur index}

The Schur index is defined by specialising the fugacities to $\rho=\tau$ with $\sigma$ arbitrary (or equivalently $q=t$ with $p$ arbitrary). It reads
\be\label{eq:Schurdef}
\mathcal{I}_S=\tr(-1)^F\sigma^{\half(\Delta+2j_1-2R-r)}\rho^{\Delta-j_1+j_2}e^{-\beta\bar{\delta}_{1\dot{-}}} \;.
\ee
All the charges in \eqref{eq:Schurdef} commute with $\Qcal_{1+}$. Thus the Schur index receives contributions from states with $\delta_{1+}=\bar{\delta}_{1\dot{-}}=0$ and it is independent of both $\beta$ and $\sigma$. We can then write 
\be
\mathcal{I}_S=\tr_S(-1)^F\rho^{2(\Delta-R)}=\tr_S(-1)^Fq^{\Delta-R},
\ee
where $\tr_S$ denotes the trace restricted to states with $\delta_{1+}=\Delta+2j_1-2R-r=0$. The Schur index can also be obtained as a special case of the Macdonald index by setting $\rho=\tau$ (or equivalently $q=t$), and still counts the Schur sector \eqref{eq:Schursector}.

The single-letter indices are given by 
\be\begin{split}
    f_\text{vm}(\rho)=&1-\frac{1+\rho^2}{1-\rho^2}=1-\frac{q}{1-q},\\
    f_\text{hm}(\rho)=&\frac{\rho}{1-\rho^2}=\frac{q^{\frac{1}{2}}}{1-q}.
\end{split}\ee

\subsubsection{Hall-Littlewood index}\label{sec:Hall-Littlewood}

The Hall-Littlewood index is defined by taking the limit 
\be
\sigma\to0,\quad\rho\to0,\quad\tau\,\text{ fixed}, 
\ee
so it is given by
\be
\mathcal{I}_{HL}=\tr_{HL}(-1)^F\tau^{2(\Delta-R)}=\tr_{HL}(-1)^Ft^{R+r},
\ee
where $\tr_{HL}$ denotes the trace restricted to states with $\delta_{1\pm}=\Delta\pm2j_1-2R-r=0$. The states that contribute to the index obey 
\be
j_1=0,\quad j_2=r,\quad \Delta=2R+r,
\ee
and are annihilated by three supercharges: $\Qcal_{1+}$, $\Qcal_{1-}$ and $\bar{\Qcal}_{1\dot{-}}$. 

The Hall-Littlewood (HL) index only receives single letter contributions $\bar{\lambda}_{Z\dot{+}}$ of the vector multiplet and from the scalars $Q_{ij}$ and $Q_{ji}$ of the hypermultiplet. The single-letter contributions are then given by
\be
   f_\text{vm}(\tau)=-\tau^2=-t \;\;\text{and}\;\;
   f_\text{hm}(\tau)=\tau=t^{\half}.
\ee
Let us take a slight detour and discuss the \emph{chiral ring},which is defined as the cohomology of operators annihilated by a Poincar\'e supercharge of one chirality \cite{Gadde:2009dj}.  For our purpose, we choose to define the chiral ring with respect to a right-handed supercharge, $\bar{\Qcal}$ and hence the operators obey what is known as a $\mathcal{B}$-type shortening condition \cite{Dolan:2002zh}. If the theory has extended supersymmetry we focus on the $\NN=1$ subalgebra. As noted in Appendix \ref{sec:ExtendedSusyTransformations}, for our theories with $\NN=2$ supersymmetry, we define the cohomology with respect to $\bar{\mathcal{Q}}_{1\dot{-}}$. Explicitly, we are considering operators $\Ocal$ that satisfy the following relation \cite{Banerjee:2023ddh}: 
\be\label{eq:defQcohomology}
\bar{\Qcal}_{1\dot{-}}\Ocal=0\;,\quad\text{with }\Ocal\ne\bar{\Qcal}_{1\dot{-}}\Ocal'\,.
\ee 
The chiral cohomology classes can be specified by a set of generators and relations, which are easy to determine at weak coupling\footnote{In fact, the relations that we are considering are from the classical theory. The main difference between determining the chiral ring and the index is that in considering the index, we are setting all gauge couplings equal to zero, while in the chiral ring, we consider the cohomology at non-zero coupling.}. At higher order the relations are expected to be corrected due to quantum effects; however, the basic counting of chiral states is not expected to change \cite{Cachazo:2002ry,Kinney:2005ej}. In an $\NN=2$ theory, we can see from Appendix \ref{sec:ExtendedSusyTransformations} that the fundamental fields that define the chiral ring are given by the scalar fields $A=\lbrace Z_i,\;Q_{ij},\;Q_{ji}\rbrace$. Let us recall from the $\NN=1$ subalgebra that we have the supersymmetric transformations
\be\label{eq:N=1susyacscalar}
\bar{\Qcal}_1^2\bar{A}_i=\bar{\Qcal}_{1\dot{-}}\bar{\psi}_{A\dot{+}}=\bar{F}_{A_i}=-\p_{A_i}\Wcal\,,\quad A_i\in A\,,
\ee
where, in the last equality we have applied the equations of motion of the auxiliary field $\bar{F}_{A_i}$. Thus, $\p_{A_i}\mathcal{W}$ is $\bar{\Qcal}_{1\dot{-}}$ exact, so we need to mod it out. This means that the chiral ring is defined subject to the constraint 
\be\label{eq:superpotentialeom}
\p_{A_i}\Wcal(A)_\text{c.r.}=0\,,
\ee
where the subscript c.r. specifies that the relation \eqref{eq:superpotentialeom} is valid in the chiral ring. Often, in the literature, \eqref{eq:superpotentialeom} is called the ``$F$-term constraints" or the ``superpotential equation of motion". Note that in addition to \eqref{eq:superpotentialeom}, at finite $N$, the trace relations impose additional constraints. In the $N\to\infty$ limit, the trace relations have no effect, and \eqref{eq:superpotentialeom} is the only constraint that the chiral ring is subject to. For the superpotential \eqref{eq:ADEsuperpotential}, the constraints \eqref{eq:superpotentialeom} read 
\begin{subequations}\label{eq:spconds}
    \begin{align}
        \label{seq:spcon1}&\kappa_iZ_iQ_{ij}=\kappa_jQ_{ij}Z_{j}\,,\\
        \label{seq:spcon2}&\sum_{j=1}^Md_{ji}\left[(Q_{ji})^{a}\phan_{c}(Q_{ij})^{c}\phan_{b}-\frac{1}{n_iN}\delta^{a}\phan_{b}(Q_{ji})^{c}\phan_{d}(Q_{ij})^{d}\phan_{c}\right]=0\,.
    \end{align}
\end{subequations}
On the Higgs branch ($\expval{Z_i}=0$ and $\expval{Q_{ij}}=c_{ij}\ne0$) only the second condition applies.

Now, as noted in \cite{Gadde:2011uv}, for quiver theories with spherical topology (i.e., non-circular quivers), the Hall-Littlewood index is equivalent to the partition function of the Higgs branch of the chiral ring. Notice that in computing the Higgs-branch partition function, to account for the $F$-term constraints one needs to subtract $t$ from the single letter partition function to account for the missing degrees of freedom. Note that the contributions of constraints to the partition function is exactly the same as the contribution $\bar{\lambda}_{Z\dot{+}}$ to the superconformal index. The number of constraints on non-circular quivers means that there are $M-1$ length-2 triplets $\Mtrip$.   

\subsubsection{Coulomb-branch index}

Finally, let us consider the limit 
\be
\tau\to0,\quad \rho,\,\sigma\,\text{ fixed}.
\ee
The index then becomes 
\be\label{eq:CoulombIndex}
\mathcal{I}_C=\tr_C(-1)^F\sigma^{\half(\Delta+2j_1-2R-r)}\rho^{\half(\Delta-2j_1-2R-r)}e^{-\beta\bar{\delta}_{1\dot{-}}},
\ee
where $\tr_C$ denotes the trace over the states with $\bar{\delta}_{2\dot{+}}=\Delta+2j_2+2R+r=0$,which are annihilated by $\bar{\Qcal}_{2\dot{+}}$. The index receives contributions from states annihilated by two antichiral supercharges, $\bar{\Qcal}_{1\dot{-}}$ and $\bar{\Qcal}_{2\dot{+}}$. The Coulomb-branch index only receives contributions from the $\bar{\mathcal{E}}_{r(0,0)}$ multiplets.

In the Coulomb branch limit, the contributions to the single-letter partition function are 
\be
    f_\text{v.m.}(\rho,\sigma)=\,\rho\sigma\equiv\,T \; \quad \text{and}\quad 
   f_\text{h.m.}(\rho,\sigma)=\,0,
\ee
that is, only the $Z$ field from the $\Ncal=2$ vector multiplet contributes. Hence, the Coulomb-branch index counts the number of BMN vacua.

\subsection{The ADE Index}\label{sec:ADEindex}

In this section, we will evaluate the large-$N$ index for the ADE quiver theories. As discussed, in order to do this one takes the plethystic exponential \eqref{eq:defPE} of the single-letter index \eqref{eq:sl-sci} and projects onto the singlets of the gauge group by multiplying by the character of the trivial representation ($\chi_{i\,\text{trivial}}(U_{i})=1$) and integrating over the gauge group. Hence, the full index is then given by plethystic exponentiation
\be\label{eq:FullindexintermsU}
\begin{split}
    \mathcal{I}_\Gamma(p,q,t)=&\int\prod_{i=1}^M[dU_i]\exp\left(\sum_{n=1}^\infty\sum_{ij}\frac{\chi_{ij}(U_{ij}^n)}{n}f_{ij}(p^n,q^n,t^n)\right)\;.
\end{split}
\ee
Applying \eqref{eq:ProjectionWittenindex}, we can rewrite \eqref{eq:FullindexintermsU} as an integral over the fugacities $u^{(i)}$:\footnote{Note that we are neglecting the contribution of the rank $n_iN$ to the Vandermonde determinant (\ref{vandermonde}), as we will be employing a saddlepoint approximation, and that term will contribute a trivial factor of 1 in the $N\to\infty$ limit.}:
\be\label{eq:FullindexFuga}
\begin{split}
\mathcal{I}_\Gamma(p,q,t)=&\prod_{i=1}^M\prod_{m=1}^{n_iN}\oint_{|u^{(i)}_m|=1}\frac{du^{(i)}_m}{2\pi i u^{(i)}_m} \exp\left(\sum_{n=1}^\infty\sum_{i,j}\frac{\chi_{ij}(u_{ij}^n)}{n}\bigl(f_{ij}(p^n,q^n,t^n)-\delta_{ij}\bigr)\right).
\end{split}
\ee
Note that, in \eqref{eq:FullindexFuga}, the index $i$ labels the gauge group and the index $m$ labels the eigenvalue.  
Using the large $N$ techniques from \cite{Kinney:2005ej,Gadde:2009dj}, we employ a saddle point expansion.  For now, we treat the gauge groups as $\U(n_iN)$, and at the end we will divide off the $\U(1)$ factors. Note that $\U(1)$ factors only affect the $\NN=2$ vector multiplet contributions to the index. We will start by making a change of variables. Let 
\be
u^{(i)}_m\equiv e^{i\theta^{(i)}_m}.
\ee 
Then \eqref{eq:FullindexFuga} becomes 
\be\label{eq:Fullindextheta}
\begin{split}
\mathcal{I}_\Gamma(p,q,t)=&\prod_{i=1}^M\prod_{m=1}^{n_iN}\int_{-\pi}^\pi\frac{d\theta^{(i)}_m}{2\pi} \exp\left(\sum_{n=1}^\infty\sum_{i,j}\sum_{k,l}\frac{\exp\bigl( in(\theta^{(i)}_k-\theta_l^{(j)})\bigr)}{n}\bigl(f_{ij}(p^n,q^n,t^n)-\delta_{ij}\bigr)\right).
\end{split}
\ee
Now, as we take $N\to\infty$, we have an infinite number of eigenvalues $\theta^{(i)}_m$.
They are described by introducing the eigenvalue density function $\varrho^{(i)}(\theta^{(i)})$, which satisfies:
\be
\int_{-\pi}^{\pi}d\theta^{(i)}\varrho^{(i)}(\theta^{(i)})=1 \;.
\ee
We decompose $\varrho^{(i)}$ into its Fourier modes
\be
\varrho_n^i\equiv \int_{-\pi}^\pi d\theta\varrho^{(i)}(\theta)e^{in\theta^{(i)}}\;, \quad n\in \Zset,
\ee
where we note that $\varrho^{(i)*}_n=\varrho^{(i)}_{-n}$. In the $N\to\infty$ limit, we have 
\be
\sum_{k=1}^{n_iN}\to n_iN\int_0^1 d\theta^{(i)}\,.
\ee
The index \eqref{eq:Fullindextheta} becomes
\be\label{eq:indexintermeffectiveaction}
\mathcal{I}_\Gamma=\int[d\varrho]e^{-S_{\Gamma,\,\text{eff}}[\varrho]}\;,
\ee
with the measure $[d\varrho]$ in \eqref{eq:indexintermeffectiveaction} given by
\be
[d\varrho]=\prod_{i=1}^M[d\varrho^{(i)}]=\prod_{i=1}^M\prod_{n=1}^\infty\frac{ n_i^2N^2d\varrho^{(i)}_nd\varrho^{(i)}_{-n}}{2\pi},
\ee
and, recalling (\ref{fADE})  the effective action given by
\be
\begin{split}\label{eq:effaction}   
S_{\Gamma,\,\text{eff}}[\varrho(\theta)]\equiv&\;\sum_{n=1}^\infty\left[\sum_{i,j=1}^M\frac{1}{n}n_iN\varrho_{n}^{(i)}\bigl(\delta_{ij}(1-f_\text{vm}(p^n,q^n,t^n))-a^\mathbf{2}_{ij}f_\text{hm}(p^n,q^n,t^n)\bigr)n_jN\varrho^{(j)}_{-n}\right].
\end{split}
\ee
Let us now define
\be
\text{v}_n\equiv\begin{pmatrix}
    n_1N\varrho^{(1)}_n\\
    \vdots\\
    n_MN\varrho^{(M)}_n
\end{pmatrix},\quad \text{v}_n^\dagger=\begin{pmatrix}
    n_1N\varrho^{(1)}_{-n}&
    \ldots&
    n_MN\varrho^{(M)}_{-n}
\end{pmatrix}.
\ee
Define the connectivity matrix with the following entries
\be
\left[A_{\Gamma}\right]_{ij}\equiv\;a^\mathbf{2}_{ij}.
\ee
Let us consider the $\U(1)$ factors. From \eqref{eq:PEfvm}, a single $\Urm(1)$ contributes
\be\begin{split} 
    \text{PE}[-f_\text{vm}(p,q,t)]=&\;\frac{\Gamma(t;p,q)}{(p;p)_\infty(q;q)_\infty}\;.
\end{split}\ee
We have $M$ such factors, one for each node. Hence, we need to multiply the saddle-point approximation by
\be \label{U1index}
\text{PE}[-Mf_\text{vm}(p,q,t)]=\;\prod_{n=1}^\infty e^{-\frac{M}{n}f(p^n,q^n,t^n)}=\;\frac{\Gamma(t;p,q)^M}{(p;p)_\infty^M(q;q)_\infty^M}
\;.\ee
Thus, we can write \eqref{eq:indexintermeffectiveaction} as
\be
\begin{split}
    \mathcal{I}_{\Gamma}=\prod_{n=1}^{\infty}e^{-\frac{M}{n}f_\text{vm}(p^n,q^n,t^n)}\int d\text{v}_nd\text{v}_n^\dagger\exp\left[-\text{v}_n\big((1\!-\!f_\text{vm}(p^n,q^n,t^n))I_{M\times M}-f_\text{hm}(p^n,q^n,t^n)A_\Gamma\big)\text{v}_n^\dagger\right]\;.
\end{split}
\ee
Let us notice that $1-f_\text{vm}$ factorises rather nicely:
\be
  1-f_\text{vm}(p,q,t)=\frac{(1-pqt^{-1})(1+t)}{(1-p)(1-q)}
  =\frac{(1-\rho\sigma)(1+\tau^2)}{(1-\tau\sigma)(1-\tau\rho)}\;.
\ee
The action \eqref{eq:effaction} is minimised for $\varrho^{(i)}_0=1$ and $\varrho^{(i)}_{n>0}=0$. We can then perform the Gaussian integral about this minimum to find
\be
\begin{split}\label{eq:MultitraceOrbifoldIndex}
\mathcal{I}_{\Gamma}^\text{m.t.}\simeq&\,\prod_{n=1}^\infty\frac{e^{-\frac{M}{n}f_\text{vm}(p^n,q^n,t^n)}}{\det\left(1-f_\text{vm}(p^n,q^n,t^n)I_{M\times M}-f_\text{hm}(p^n,q^n,t^n)A_{\Gamma}\right)}\\
=&\,\prod_{n=1}^\infty\frac{\left((1-p^n)(1-q^{n})\right)^Me^{-\frac{M}{n}f_\text{vm}(p^n,q^n,t^n)}}{\left(1-(pqt^{-1})^n\right)^M\det\left((1+t^{n})I_{M\times M}-t^{\frac{n}{2}}A_{\Gamma}\right)}\\
=&\,\frac{\Gamma(t;p,q)^M}{(p;p)_\infty^M(q;q)_\infty^M}\prod_{n=1}^\infty\frac{\left((1-p^n)(1-q^{n})\right)^M}{\left(1-(pqt^{-1})^n\right)^M\det\left((1+t^{n})I_{M\times M}-t^{\frac{n}{2}}A_{\Gamma}\right)}\\
=&\,\frac{\Gamma(t;p,q)^M}{(pqt^{-1};pqt^{-1})_\infty^M}\prod_{n=1}^\infty\det\left((1+t^{n})I_{M\times M}-t^{\frac{n}{2}}A_{\Gamma}\right)^{-1}\;.
\end{split}
\ee
where the superscript m.t. stands for multi-trace and $\simeq$ indicates an expression valid in the large-$N$ limit. The matrices $(1+t^{n})I_{M\times M}-t^{\frac{n}{2}}A_\Gamma$ are just the $m_{ij}$ adjacency matrices from Appendix \ref{sec:MattercontentN=2}, with the diagonal entries multiplied by $(1+t^{n})$ and the off-diagonal entries multiplied by $-t^{\frac{n}{2}}$. As a check, this formula can be seen to agree with the $\Ncal=1$ results of \cite{Gadde:2010en,Eager:2012hx}, after adding the required contributions from adjoint fields in our $\Ncal=2$ setting. From \eqref{eq:GammaIdentities}, the various limits of the multi-trace index are: 
\be\begin{split}\label{MultitraceOrbifoldLimits}
    \mathcal{I}_{\Gamma;\;M}^\text{m.t.}\simeq&\,(t;q)_\infty^{-M}\prod_{n=1}^\infty\det\left((1+t^{n})I_{M\times M}-t^{\frac{n}{2}}A_{\Gamma}\right)^{-1}\;,\\
    \mathcal{I}_{\Gamma;\;S}^\text{m.t.}\simeq&\,(q;q)_\infty^{-M}\prod_{n=1}^\infty\det\left((1+q^{n})I_{M\times M}-q^{\frac{n}{2}}A_{\Gamma}\right)^{-1}\;,\\
    \mathcal{I}_{\Gamma;\;HL}^\text{m.t.}\simeq&\,(1-t)^{-M}\prod_{n=1}^\infty\det\left((1+t^{n})I_{M\times M}-t^{\frac{n}{2}}A_{\Gamma}\right)^{-1}\;,\\
    \mathcal{I}_{\Gamma;\;C}^\text{m.t.}\simeq&\,\frac{(1-T)^M}{(T,T)_\infty^{M}}\,.
\end{split}\ee
To extract the contribution from single-traces, we evaluate the plethystic logarithm \eqref{eq:defPL} and obtain 
\be
\begin{split}\label{eq:Singletracegenericindex}
    \mathcal{I}_\Gamma^\text{s.t.}(p,q,t)=&\,\sum_{n=1}^\infty\frac{\mu(n)}{n}\log\left[\mathcal{I}_\Gamma^\text{m.t.}(p^n,q^n,t^n)\right]\\
    =&\,M\biggl[\frac{pqt^{-1}}{1-pqt^{-1}}-\frac{p}{1-p}-\frac{q}{1-q}-f_\text{vm}(p,q,t)\biggr]\\
    &-\sum_{n=1}^{\infty}\frac{\varphi(n)}{n}\log\det\left((1+t^{n})I_{M\times M}-t^{\frac{n}{2}}A_{\Gamma}\right)\\
    =&\,M\biggl[\frac{pqt^{-1}}{1-pqt^{-1}}+\frac{t-pqt^{-1}}{(1-p)(1-q)}\biggr]\\
    &-\sum_{n=1}^{\infty}\frac{\varphi(n)}{n}\log\det\left((1+t^{n})I_{M\times M}-t^{\frac{n}{2}}A_{\Gamma}\right)\,,
\end{split}
\ee
where we have used \eqref{eq:MobiusandEuler} and \eqref{eq:EulerTotieone-loopg}.

Note that in \eqref{eq:Singletracegenericindex}, the factors $M$ and $(1+t^{n})I_{M\times M}-t^{\frac{n}{2}}A_{\Gamma}$ depend on the specifics of the group, however the term $\frac{pqt^{-1}}{1-pqt^{-1}}+\frac{t-pqt^{-1}}{(1-p)(1-q)}$ is universal. From Appendices \ref{sec:Er(0,0)} and \ref{sec:BR}, we see that the $\bar{\mathcal{E}}_{-\ell(0,0)}$ and the $\hat{\mathcal{B}}_1$ multiplets contribute precisely this factor:
\be
\begin{split}
    \sum_{\ell=2}^\infty\mathcal{I}[\bar{\mathcal{E}}_{-\ell(0,0)}]+\mathcal{I}[\hat{\mathcal{B}}_1]=&\,\frac{p^2q^2t^{-2}\left(1-t(p^{-1}+q^{-1})+p^{-1}q^{-1}t^2\right)}{\left(1-pqt^{-1}\right)\left(1-p\right)\left(1-q\right)}+\frac{t-pq}{\left(1-p\right)\left(1-q\right)}\\
    =&\,\frac{pqt^{-1}}{1-pqt^{-1}}+\frac{t-pqt^{-1}}{(1-p)(1-q)}\,.
\end{split}
\ee
Note that the primaries of $\bar{\mathcal{E}}_{-\ell(0,0)}$ correspond to the $\Tr Z^\ell$ BMN vacua and the primaries of $\hat{\mathcal{B}}_1$ correspond to the $\SU(2)_R$ triplets $\Tr\Mtrip$. So we can write 
\be\label{eq:SingleTraceGenericWrittenShort}
\begin{split}
    \mathcal{I}_\Gamma^\text{s.t.}(p,q,t)=&M\left[\sum_{\ell=2}^\infty\mathcal{I}[\bar{\mathcal{E}}_{-\ell(0,0)}]+\mathcal{I}[\hat{\mathcal{B}}_1]\right]-\sum_{n=1}^{\infty}\frac{\varphi(n)}{n}\log\det\left((1+t^n)I_{M\times M}-t^{\frac{n}{2}}A_{\Gamma}\right)\,.
\end{split}
\ee
The final term in \eqref{eq:Singletracegenericindex} involves a determinant that requires us to consider the specific quiver, and we will evaluate it case-by-case in our examples. However, we can evaluate the Coulomb-branch limit of \eqref{eq:SingleTraceGenericWrittenShort} immediately. It is given by 
\be\label{eq:SingletracegenericindexCoulomb}
    \mathcal{I}^\text{s.t.}_{\Gamma;\;C}=\,\frac{MT^2}{1-T^2}
    =\,M\sum_{\ell=2}^\infty\mathcal{I}_C[\bar{\mathcal{E}}_{-\ell(0,0)}]\,.
\ee
This just tells us that, for all orbifold theories, we always have $M$ BMN vacua. 
The other limits of the index \eqref{eq:Singletracegenericindex} all involve evaluating the determinant and are given as follows:
\be\begin{split}\label{eq:limitsgeneriSci}
    \mathcal{I}_{\Gamma;\;M}^\text{s.t.}=&\,\frac{Mt}{(1-q)}-\sum_{n=1}^\infty\frac{\varphi(n)}{n}\det\left((1+t^{n})I_{M\times M}-t^{\frac{n}{2}}A_\Gamma\right)\;,\\
    \mathcal{I}_{\Gamma;\;S}^\text{s.t.}=&\,\frac{Mq}{(1-q)}-\sum_{n=1}^\infty\frac{\varphi(n)}{n}\det\left((1+q^{n})I_{M\times M}-q^{\frac{n}{2}}A_\Gamma\right)\;,\\
    \mathcal{I}_{\Gamma;\;HL}^\text{s.t.}=&\,Mt-\sum_{n=1}^\infty\frac{\varphi(n)}{n}\det\left((1+t^{n})I_{M\times M}-t^{\frac{n}{2}}A_\Gamma\right)\;.
\end{split}\ee

\subsubsection{The untwisted sector}

Protected operators in the untwisted sector are inherited from the $\NN=4$ SYM mother theory. So, to evaluate the contribution to the index from the untwisted sector, we start with the single trace index for $\SU(N)$ $\NN=4$ SYM and project onto a $\Gamma$-invariant subspace. The $\NN=4$ index is found by treating the $\NN=4$ as a $\NN=2$ theory with one adjoint vector multiplet and one adjoint hypermultiplet. One finds \cite{Kinney:2005ej,Gadde:2009dj}
\be\begin{split}\label{eq:N=4mtsci}
     \mathcal{I}_{\NN=4}^\text{m.t.}(p,q,t,v)\simeq&\,\prod_{n=1}^\infty\frac{(1-p^n)(1-q^n)e^{-\frac{1}{n}[f_\text{vm}(p^n,q^n,t^n)+(v^n+v^{-n})f_\text{hm}(p^n,q^n,t^n)]}}{(1-q^np^nt^{-n})(1-v^nt^{\frac{n}{2}})(1-v^{-n}t^{\frac{n}{2}})}\\
     =&\,\frac{\Gamma(t;p,q)}{\Gamma(vt^\half;p,q)\Gamma(v^{-1}t^\half;p,q)(qpt^{-1};qpt^{-1})_\infty(vt^\half;vt^\half)_\infty(v^{-1}t^\half;v^{-1}t^\half)_\infty}\;,
\end{split}\ee
where we have used the definitions (\ref{eq:DefPochhammer}) and (\ref{EllipticGamma}). Note the fugacity $v$, which is related to $\SU(2)_L$, where the index of $X$ is $vt^\half$ and the index of $Y$ is $v^{-1}t^\half$. Applying the identities \eqref{eq:GammaIdentities}, the various limits of \eqref{eq:N=4mtsci} are given by
\be\begin{split}\label{eq:mtN=4limits}
    \mathcal{I}_{M;\;\NN=4}^\text{m.t.}(q,t,v)\simeq&\,\frac{(vt^\half;q)_\infty(v^{-1}t^\half;q)_\infty}{(t;q)_\infty(vt^\half;vt^\half)_\infty(v^{-1}t^\half;v^{-1}t^\half)_\infty}\\
    \mathcal{I}_{S;\;\NN=4}^\text{m.t.}(q,v)\simeq&\,\frac{(vq^\half;q)_\infty(v^{-1}q^\half;q)_\infty}{(q;q)_\infty(vq^\half;vq^\half)_\infty(v^{-1}q^\half;v^{-1}q^\half)_\infty}\\
    \mathcal{I}^{m.t.}_{HL;\;\NN=4}(t,v)\simeq&\,\frac{(1-vt^\half)(1-v^{-1}t^\half)}{(1-t)(vt^\half;vt^\half)_\infty(v^{-1}t^\half;v^{-1}t^\half)_\infty}\\
    \mathcal{I}_{C;\;\NN=4}^\text{m.t.}(T,v)\simeq&\,\frac{(1-T)}{(T;T)_\infty}\;.
\end{split}\ee
The single-trace index is given by\footnote{The map to the fugacities used in \cite{Gadde:2009dj} is
$p= t'^3y\,,\; q=t'^3y^{-1}\,,\; t=t'^4v'^{-1}\,,\; v=w^2\;,$ where primes indicate the fugacities of \cite{Gadde:2009dj}.}\label{fugacitymap}
\be
\begin{split}
    \mathcal{I}_{\NN=4}^\text{s.t.}(p,q,t,v)=&\,\frac{pqt^{-1}}{1-pqt^{-1}}+\frac{t-pqt^{-1}}{(1-p)(1-q)}+\frac{t^\half v}{1-t^\half v}+\frac{t^\half v^{-1}}{1-t^\half v^{-1}}-(v+v^{-1})f_\text{hm}(p,q,t)\\
    =&\,\sum_{\ell=2}^\infty\mathcal{I}[\bar{\mathcal{E}}_{-\ell(0,0)}]+\mathcal{I}[\hat{\mathcal{B}}_1]+\frac{t^\half v}{1-t^\half v}+\frac{t^\half v^{-1}}{1-t^\half v^{-1}}-(v+v^{-1})f_\text{hm}(p,q,t)\,.
\end{split}
\ee
The various limits of the single trace index are given by
\be\begin{split}\label{eq:limitsstN=4}
    \mathcal{I}^\text{s.t.}_{\NN=4;\;M}(q,t,v)=&\,\frac{t-(v+v^{-1})t^\half}{1-q}+\frac{t^\half v}{1-t^\half v}+\frac{t^\half v^{-1}}{1-t^\half v^{-1}}\\
    \mathcal{I}^\text{s.t.}_{\NN=4;\;S}(q,v)=&\,\frac{q-(v+v^{-1})q^\half}{1-q}+\frac{q^\half v}{1-q^\half v}+\frac{q^\half v^{-1}}{1-q^\half v^{-1}}-(v+v^{-1})f_\text{hm}(q)\\
    \mathcal{I}^\text{s.t.}_{\NN=4;\;HL}(t,v)=&\,t+\frac{tv^2}{1-t^\half v}+\frac{tv^{-2}}{1-t^\half v^{-1}}\\
    \mathcal{I}^\text{s.t.}_{\NN=4;\;C}(T)=&\,\frac{T^2}{1-T}\,.
\end{split}\ee
From Appendix \ref{sec:IndexOperators}, we can rewrite the Hall-Littlewood and Coulomb-branch limits in \eqref{eq:limitsstN=4} as
\be\begin{split}\label{eq:limitsstN=4operators}
    \mathcal{I}^\text{s.t.}_{\NN=4;\;HL}=&\,\mathcal{I}[\Mtrip]+\sum_{\ell=2}\left[\mathcal{I}[X^\ell]+\mathcal{I}[Y^\ell]\right]\\
    \mathcal{I}^\text{s.t.}_{\NN=4;\;C}=&\,\sum_{\ell=2}\mathcal{I}[Z^\ell]\,.
\end{split}\ee
Let us define the ``$\SU(2)_L$'' part of the index which depends on the fugacities of the $X$ and $Y$ fields: 
\be \label{IL}
\mathcal{I}^\text{s.t.}_{L}(p,q,t,\text{w}\;)\equiv v_xv_y\mathcal{I}[\hat{\mathcal{B}}_{1}]+\,\frac{t^\half v_x}{1-t^\half v_x}+\frac{t^\half v_y}{1-t^\half v_y}-(v_x+v_y)f_\text{hm}(p,q,t),
\ee
where 
\be
\text{w}=\,\begin{pmatrix}
    v_x\\
    v_y
\end{pmatrix}\,,
\ee
and we have included the $\hat{\mathcal{B}}_1$ term which, having an $XY$ top component, also depends on $v_x,v_y$. 
The index of the untwisted sector can then be written as 
\be
\label{eq:untwistedgeneral}
\mathcal{I}^\text{untwisted}_{\Gamma}(p,q,t)=\,\sum_{\ell=2}^\infty \mathcal{I}[\bar{\mathcal{E}}_{-\ell(0,0)}]+\frac{1}{|\Gamma|}\sum_{g\in\Gamma}\mathcal{I}^\text{s.t.}_{L}(p,q,t,R^{(\mathbf{2})}(g)\text{w})\bigr|_{v_x=v_y=1}\,,
\ee
where we have set $v_x=v_y=1$, since for $\Gamma\ne\Zset_2$, the $\SU(2)_L$ symmetry is broken\footnote{As we will see, when we consider the $\Zset_k$ case, there is restoration of this symmetry for certain operators.}. Then the index of the twisted sector of the theory is given by
\be\label{eq:twistedgeneral}
\mathcal{I}^\text{twisted}_{\Gamma}(p,q,t)=\mathcal{I}^\text{s.t.}_\Gamma(p,q,t)-\mathcal{I}^\text{untwisted}_{\Gamma}(p,q,t)\;.
\ee
From \eqref{eq:limitsstN=4operators}, it follows that 
\be\label{eq:untwistedgeneralcoulomb}
\begin{split}
    \mathcal{I}^\text{untwisted}_{\Gamma;\;C}=&\,\sum_{\ell=2}^\infty\mathcal{I}[Z^\ell]\,.
\end{split}
\ee
Thus, from \eqref{eq:untwistedgeneralcoulomb}, we always have one untwisted BMN vacuum. Hence, from \eqref{eq:untwistedgeneralcoulomb} and \eqref{eq:SingletracegenericindexCoulomb}, we see that 
\be\label{eq:twistedgeneralcoulomb}
    \mathcal{I}^\text{twisted}_{\Gamma;\;C}=\,\frac{(M-1)T^2}{1-T}=\,(M-1)\sum_{\ell=2}^\infty\mathcal{I}[Z^\ell]\,.
\ee
So that we see, from \eqref{eq:twistedgeneralcoulomb}, we will always have $M-1$ twisted BMN vacua.

\subsection{The Molien Series}\label{sec:MolienSeries}

Another means of counting protected states is to count the mesonic BPS gauge invariant operators that appear in the chiral ring (i.e. the Higgs branch operators), taking into account the $F$-term relations. That is
\be
\mathbf{M}(x)\equiv\Tr_{\text{Higgs branch}}x^\Delta\;.
\ee
In our notation, this corresponds to states in the $XY$-sector, modulo the $F$-term relations coming from $F_Z$. We are focusing on the large-$N$ limit. In the finite-$N$ case, one would also have to take into account the trace relations. In the case of an orbifold theory, the problem of counting mesonic BPS operators amounts to counting the number of chiral gauge invariants, which in turn amounts to counting the number of polynomial invariants composed of $(x,y)\in\Cset^2$ under the action of the group $\Gamma$ \cite{Benvenuti:2006qr}. For ADE orbifolds, the object that counts the polynomial invariants is known as the Molien series. It is given by
\be\label{eq:defMolien}
\mathbf{M}(x;\Gamma)=\frac{1}{|\Gamma|}\sum_{g\in G}\frac{1}{\det\left(I_{2\times 2}-x R^{(\mathbf{2})}(g)\right)}\;.
\ee
The Molien series for all the ADE quivers were tabulated in \cite{Benvenuti:2006qr} and are reproduced in Table \ref{tab:SU(2)Molienseries} for convenience.
\begin{table}[h]
    \centering
        \renewcommand{\arraystretch}{1.5}
    \begin{tabular}{|c|c|}
\hline
$\Gamma$&$\mathbf{M}(x;\Gamma)$\tabularnewline\hline\hline
$\Zset_k$&$\frac{1+x^{k}}{(1-x^2)(1-x^{k})}$\tabularnewline\hline
$\hat{\text{D}}_{k}$&$\frac{1+x^{2k-2}}{(1-x^4)(1-x^{2k-4})}$\tabularnewline\hline
$2\text{T}$&$\frac{1-x^4+x^8}{1-x^4-x^6+x^{10}}$\tabularnewline\hline
$2\text{O}$&$\frac{1-x^6+x^{12}}{1-x^6-x^8+x^{14}}$\tabularnewline\hline
$2\text{I}$&$\frac{1+x^2-x^6-x^8-x^{10}+x^{14}+x^{16}}{1+x^2-x^6-x^8-x^{10}-x^{12}+x^{16}+x^{18}}$\tabularnewline
\hline\end{tabular}
\caption{The Molien series for the finite subgroups of $\SU(2)$, reproduced from \cite{Benvenuti:2006qr}.}
    \label{tab:SU(2)Molienseries}
\end{table}\\
By its definition, the Molien series counts the untwisted superconformal primaries of $\hat{\mathcal{B}}_{R\geq 1}$, corresponding to states in the $XY$ sector. It does not count the twisted $R=1$ triplets, but fortunately these states are counted by the superconformal index. 

As explained in \cite{Benvenuti:2006qr,Hanany:2006uc}, plethystic exponentiation of the Molien series produces the corresponding finite-$N$ Higgs-branch Hilbert series. In \cite{Hayling:2017cva}, this was applied to the $\Zset_k$ Molien series in Table \ref{tab:SU(2)Molienseries} to show agreement between the large-$k$, finite-$N$ $\Zset_k$-quiver Hilbert series with the $\half$-BPS index in the 6d $(2,0)$ theory, as expected by dimensional-deconstruction arguments \cite{Arkani-Hamed:2001wsh}.

\subsection{Example: The $\Zset_k$ theory}\label{sec:Zkindex}

The evaluation of the indices for the non-abelian orbifold cases needs to be done on a case-by-case basis. For the $\Dfour$ and $\hat{E}_6$ quivers, this is done in Sections  \ref{sec:D4} and \ref{sec:E6}. However, for $\Zset_k$ the circulant matrix structure allows us to evaluate the index for general $k$, so we will present it in this section.\footnote{Consistency with the ADE notation would require us to use $\hat{A}_{k-1}$ instead of $\Zset_k$, however we opt for this label as $\Zset_k$ is the most popular naming convention for the cyclic quivers.} 
  
Before we consider the $k\ge3$ case, let us check that our formalism can reproduce the $\Zset_2$ index as computed in \cite{Gadde:2009dj}: If we replace the 2's on the off-diagonal of the adjacency matrix of $\Zset_2$ by $v+v^{-1}$ to account for the unbroken $\SU(2)_L$ symmetry,
we have 
\be\begin{split}\label{eq:I-AZ_2}
    (1+t)I_{2\times 2}-t^\half(v+v^{-1})A_{\Zset_2}=&\begin{pmatrix}
        1+t&-(v+v^{-1})t^\half\\
        -(v+v^{-1})t^\half&1+t
    \end{pmatrix}\;,
\end{split}\ee
where 
\be
A_{\Zset_2}=\begin{pmatrix}
    0&1\\
    1&0
\end{pmatrix}\,.
\ee
Taking the determinant of \eqref{eq:I-AZ_2} (with $t\ra t^n$), we find 
\be
    \det\left((1+t^{n})I_{2\times 2}-t^{\frac{n}{2}}(v+v^{-1})A_{\Zset_2}\right)
    =(1-v^2t^n)(1-v^{-2}t^n)\,.
\ee
Then, from \eqref{eq:MultitraceOrbifoldIndex}, the multi-trace index for $\Zset_2$ is given by 
\be\begin{split}\label{eq:MultitracesciZ2}
    \mathcal{I}^\text{m.t.}_{\Zset_2}\simeq&\,\prod_{n=1}^\infty\frac{\left((1-p^n)(1-q^{n})\right)^2e^{-\frac{2}{n}f_\text{vm}(p^n,q^n,t^n)}}{\left(1-(pqt^{-1})^n\right)^2\left(1-v^{2n}t^{n}\right)\left(1-v^{-2n}t^{n}\right)}\\
    =&\,\frac{\Gamma(t;p,q)^2}{(pqt^{-1};pqt^{-1})_\infty^2(v^{2}t;v^{2}t)_\infty(v^{-2}t;v^{-2}t)_\infty}\,.
\end{split}\ee
which, after a relabelling of fugacities (see footnote \ref{fugacitymap}), reproduces the result found in \cite{Gadde:2009dj}. 

Now for $k\ge3$, we have 
\be\begin{split}\label{eq:I-AZ_k}
    (1+t)I_{k\times k}-t^\half A_{\Zset_k}=\begin{pmatrix}
    1+t&-t^\half&0&0&\ldots&0&-t^\half\\
    -t^\half&1+t&-t^\half&0&\ldots&0&0\\
    0&-t^\half&1+t&-t^\half&\ldots&0&0\\
    \vdots&\vdots&\vdots&\vdots&\ddots&\vdots&\vdots\\
    -t^\half&0&0&0&\ldots&-t^\half&1+t
\end{pmatrix}.
\end{split}\ee
Since \eqref{eq:I-AZ_k} is a circulant matrix we find (with $\omega_k=e^{2\pi i/k}$) 
\be
\begin{split}
    \det\left((1+t^{n})I_{k\times k}-t^{\frac{n}{2}}A_{\Zset_k}\right)=&\prod_{j=0}^{k-1}\left(1+t^{n}-\left(\omega_k^j+\omega_k^{-j}\right)t^{\frac{n}{2}}\right)\\
    =&\prod_{j=0}^{k-1}\left(1-\omega_k^jt^{\frac{n}{2}}\right)\left(1-\omega_k^{-j}t^{\frac{n}{2}}\right)\\
    =&(1-t^{\frac{kn}{2}})^2\;.
\end{split}
\ee
Thus, from \eqref{eq:MultitraceOrbifoldIndex}, the multi-trace index for $\Zset_k$ is given by
\be\begin{split}\label{eq:MultitracesciZk}
\mathcal{I}^\text{m.t.}_{\Zset_k}\simeq&\,\prod_{n=1}^\infty\frac{\left((1-p^n)(1-q^{n})\right)^ke^{-\frac{k}{n}f_\text{vm}(p^n,q^n,t^n)}}{\left(1-(pqt^{-1})^n\right)^k\left(1-t^{\frac{kn}{2}}\right)^2}\\
&=\,\frac{\Gamma(t;p,q)^k}{(pqt^{-1};pqt^{-1})_\infty^k(t^{\frac{k}{2}};t^{\frac{k}{2}})_\infty^2}\,.
\end{split}\ee
where we used (\ref{U1index}).
As already mentioned, the large-$N$ index for the $\Zset_k$ case was computed in \cite{Nakayama:2005mf}. As discussed in \cite{Gadde:2009dj}, that work does not compute $\Ncal=2$ indices as only part of the $\Ncal=2$ vector multiplets is considered when subtracting the $\Urm(1)$'s. More recently, finite-$N$ corrections to the index (taking a similar approach to the $\Urm(1)$'s) were studied in \cite{Arai:2019aou,Fujiwara:2023bdc}.  

From \eqref{MultitraceOrbifoldLimits}, the limits of \eqref{eq:MultitracesciZk} are
\be\begin{split}\label{eq:MultitraceZklimits}
    \mathcal{I}_{\Zset_k;\;M}^\text{m.t.}\simeq&\,(t;q)_\infty^{-k}(t^{\frac{k}{2}};t^{\frac{k}{2}})_\infty^{-2}\;,\\
    \mathcal{I}_{\Zset_k;\;S}^\text{m.t.}\simeq&\,(q;q)_\infty^{-k}(q^{\frac{k}{2}};q^{\frac{k}{2}})_\infty^{-2}\;,\\
    \mathcal{I}_{\Zset_k;\;HL}^\text{m.t.}\simeq&\,(1-t)^{-k}(t^{\frac{k}{2}};t^{\frac{k}{2}})_\infty^{-2}\;,\\
    \mathcal{I}_{\Zset_k;\;C}^\text{m.t.}\simeq&\,\frac{(1-T)}{(T,T)_\infty^{k}}\,.
\end{split}\ee
The Schur index for the $\Zset_k$ quiver theories was evaluated in \cite{Bourdier:2015sga} using the free-fermion techniques introduced in \cite{Bourdier:2015wda}, at large $N$ for any $k$ and beyond large-$N$ for $k=2$. Our result can be seen to precisely agree with the large-$N$ result of \cite{Bourdier:2015sga}. The asymptotics of the finite-$N$ $\Zset_k$ Schur index in a Cardy-like limit, relevant for comparison with black hole microstates in the dual theory, were considered in \cite{Eleftheriou:2022kkv}.

From \eqref{eq:Singletracegenericindex} and \eqref{eq:SingleTraceGenericWrittenShort}, the single trace index is given by
\be\label{eq:SingleTraceSciZk}
    \mathcal{I}^\text{s.t.}_{\Zset_k}=\,k\biggl[\frac{pqt^{-1}}{1-pqt^{-1}}+\frac{t-pqt^{-1}}{(1-p)(1-q)}\biggr]+\frac{2t^{\frac{k}{2}}}{1-t^{\frac{k}{2}}}\,.
\ee
From Appendix \ref{sec:IndexOperators}, we can rewrite this as
\be\label{eq:SingleTraceSciZkoperators}
\mathcal{I}^\text{s.t.}_{\Zset_k}=\,k\left[\sum_{\ell=2}^\infty\mathcal{I}[\bar{\mathcal{E}}_{-\ell(0,0)}]+\mathcal{I}[\hat{\mathcal{B}}_1]\right]+\sum_{\ell=1}^\infty\left[\mathcal{I}[X^{\ell k}]+\mathcal{I}[Y^{\ell k}]\right]\,.
\ee
Explicitly, the operators $X^{k\ell}$ and $Y^{k\ell}$ are of the form
\be
\Tr\left((X_{12}X_{23}\ldots X_{k1})^\ell\right)\;\text{ and }\;\Tr\left((Y_{k1}\ldots Y_{21}Y_{1k})^\ell\right),
\ee
which are the $\Zset_k$ generalisations of the alternating vacua that were studied in \cite{Pomoni:2021pbj}. They belong to the multiplet $\hat{\mathcal{B}}_{\frac{\ell k}{2}}$. From Appendix \ref{sec:IndexOperators}, we can see that the Hall-Littlewood and Coulomb-branch limits of \eqref{eq:SingleTraceSciZkoperators} take the form 
\be\begin{split}\label{eq:SingleTraceSciZkoperatorslimits}
    \mathcal{I}^\text{s.t.}_{\Zset_k;\;HL}=&\,k\mathcal{I}[\Mtrip]+\sum_{\ell=1}^\infty\left[\mathcal{I}[X^{\ell k}]+\mathcal{I}[Y^{\ell k}]\right]\;,\\
    \mathcal{I}^\text{s.t.}_{\Zset_k;\;C}=&\,k\sum_{\ell=2}^\infty\mathcal{I}[Z^\ell]\,.
\end{split}\ee
So we find that the index counts $k$ protected triplets $\Mtrip$, two protected bifundamental vacua $X^{k\ell}$ and $Y^{k\ell}$, and $k$ protected BMN vacua $Z^\ell$. As we will see in Sections \ref{sec:D4index} and \ref{sec:E6index}, the bifundamental vacua  are unique to the $\Zset_k$ case, which follows from the path algebra of circular quivers. There are further protected states which combine $X$ and $Y$ fields, which as we will see are counted by the Molien series, however due to the high symmetry of the $\Zset_k$ quivers their contributions appear to precisely cancel with fermionic states of the type $\bar{\lambda}_{Z\dot{+}}(XY)^\ell$, and are thus not counted by the index.\footnote{Confirming this would require extending our dilatation operator to the fermionic sector and is therefore beyond the scope of the current work.}

Let us now consider the untwisted sector. The $\Zset_k$ action takes $v\to\omega_k^mv$ for $m=0,1,\ldots,k-1$. Then
\be\label{eq:projectILZk}
\frac{1}{k}\sum_{m=0}^{k-1}\mathcal{I}^\text{s.t.}_L(p,q,t,\omega_k^mv)=\mathcal{I}[\hat{\mathcal{B}}_{1}]+\,\frac{v^kt^{\frac{k}{2}}}{1-v^kt^{\frac{k}{2}}}+\frac{v^{-k}t^{\frac{k}{2}}}{1-v^{-k}t^{\frac{k}{2}}}\,.
\ee
From \eqref{eq:projectILZk}, there is a clear separation of states with fugacities $v$ and $v^{-1}$. This separation hints to a remainder of the $\SU(2)_L$ symmetry remaining unbroken. This is manifest if we consider the operators $(X_{12}X_{23}\ldots X_{k1})^\ell$ and $(Y_{1k}Y_{k1}\ldots Y_{21})^\ell$ which are both in the adjoint representation. Then we have a symmetry that takes 
\be
(X_{12}X_{23}\ldots X_{k1})^\ell\,\leftrightarrow\,(Y_{1k}Y_{k1}\ldots Y_{21})^\ell\,.
\ee
Setting $v=1$ in \eqref{eq:projectILZk}, we find, from \eqref{eq:untwistedgeneral}
\be\begin{split}\label{eq:untwistedZksci}
    \mathcal{I}^\text{untwisted}_{\Zset_k}=&\,\frac{pqt^{-1}}{1-pqt^{-1}}+\frac{t-pqt^{-1}}{(1-p)(1-q)}+\frac{2t^{\frac{k}{2}}}{1-t^{\frac{k}{2}}}\\
    =&\,\sum_{\ell=2}^\infty\mathcal{I}[\bar{\mathcal{E}}_{-\ell(0,0)}]+\mathcal{I}[\hat{\mathcal{B}}_1]+\sum_{\ell=1}\left[\mathcal{I}[X^{k\ell}]+\mathcal{I}[Y^{k\ell}]\right]\,.
\end{split}\ee
The Hall-Littlewood and Coulomb-branch limits of \eqref{eq:untwistedZksci} are
\be\begin{split}\label{eq:limitsuntwistedZksci}
    \mathcal{I}^\text{untwisted}_{\Zset_k;\;HL}=&\,\mathcal{I}[\Mtrip]+\sum_{\ell=1}\left[\mathcal{I}[X^{k\ell}]+\mathcal{I}[Y^{k\ell}]\right]\;\text{and}\;\\
    \mathcal{I}^\text{untwisted}_{\Zset_k;\;C}=&\,\sum_{\ell=2}^\infty\mathcal{I}[Z^\ell]\,,
\end{split}\ee
from which we see that in the untwisted sector of the $\Zset_k$ theory, there is one protected $L=2$ triplet $\Mtrip$, two protected bifundamental vacua $X^{k\ell}$ and $Y^{k\ell}$, and one protected BMN vacuum $Z^\ell$. Hence, from \eqref{eq:twistedgeneral} and \eqref{eq:untwistedZksci}, we can find the twisted index
\be\begin{split}\label{eq:twistedZksci}
    \mathcal{I}^\text{twisted}_{\Zset_k}=&\,(k-1)\left[\sum_{\ell=2}^\infty\mathcal{I}[\bar{\mathcal{E}}_{-\ell(0,0)}]+\mathcal{I}[\hat{\mathcal{B}}_1]\right]\,.
\end{split}\ee
The Hall-Littlewood and Coulomb-branch limits of \eqref{eq:twistedZksci} are
\be\begin{split}\label{eq:limitstwistedZksci}
   \mathcal{I}^\text{twisted}_{\Zset_k;\;HL}=&\,(k-1)\mathcal{I}[\Mtrip]\\
   \mathcal{I}^\text{twisted}_{\Zset_k;\;C}=&\,(k-1)\sum_{\ell=2}^\infty\mathcal{I}[Z^\ell]\,.
\end{split}\ee
Therefore, we expect to find $k-1$ protected triplets $\Mtrip$ and $k-1$ protected BMN vacua $\Tr(Z^\ell)$ in the twisted sector.
 
We summarise the protected spectrum of the $\Zset_k$ theory from the index in Table \ref{tab:ZkProtectedStatesIndex} 
\begin{table}[h]
    \centering
        \renewcommand{\arraystretch}{1.5}
    \begin{tabular}{|c|c|c|c|c|}
\hline
Protected Operator&Multiplet&Untwisted Sector&Twisted Sector&Total\tabularnewline\hline\hline
$\Tr Z^\ell$ $(\ell\ge2)$&$\bar{\mathcal{E}}_{-\ell(0,0)}$&1&$k-1$&$k$\tabularnewline\hline
$\Tr\Mtrip$&$\hat{\mathcal{B}}_1$&1&$k-1$&$k$\tabularnewline\hline
$\Tr X^{k\ell}$ $(\ell\ge1)$&$\hat{\mathcal{B}}_{\frac{\ell k}{2}}$ &1&0&1\tabularnewline\hline
$\Tr Y^{k\ell}$ $(\ell\ge1)$&$\hat{\mathcal{B}}_{\frac{\ell k}{2}}$ &1&0&1
\tabularnewline\hline
\end{tabular}
\caption{The scalar protected states in the $\Zset_k$ theory from the superconformal index.}
    \label{tab:ZkProtectedStatesIndex}
\end{table}\\

Of course, not all protected scalar states listed in Table \ref{tab:Oppsof} are captured by the index. For instance, one always has the $\mathcal{C}_{R,(\ell+1)(0,0)}$ states and their conjugates, which contain the neutral combination $\mathcal{T}_\Gamma$ (see (\ref{TGamma})) times hypermultiplets and $Z$ or $\bar{Z}$ fields. 

It is important to note that even though the states discussed in this section are protected, and therefore do not acquire anomalous dimensions, their correlation functions are far from trivial. Although the correlation functions of untwisted states are the same as those of the $\Ncal=4$ SYM mother theory by the inheritance principle \cite{Bershadsky:1998mb, Bershadsky:1998cb}, those of twisted states acquire 't Hooft-coupling dependence which can be studied both perturbatively and using supersymmetric localisation \cite{Pestun:2016zxk}.  For the orbifold-point $\Zset_2$ and $\Zset_k$ Coulomb-branch operators, this was done in \cite{Pini:2017ouj,Galvagno:2020cgq,Preti:2022inu}. Correlators between these and the Higgs-branch ones were studied in \cite{Pini:2024uia}. The matching of conformal anomalies for these Coulomb-branch operators on the Higgs branch, extracted both using localisation and multi-loop perturbation theory, was shown in \cite{Niarchos:2019onf,Niarchos:2020nxk}.

 Making use of the effective 6d supergravity theory describing the dual of this sector \cite{Gukov:1998kk}, it is possible to show large-$N$ agreement between the localisation results and supergravity \cite{Billo:2021rdb,Billo:2022gmq,Billo:2022fnb,Billo:2022lrv} at strong coupling (at the orbifold point). For the $\Zset_2$ quiver, the subleading-in-$\lambda$ terms (which require going beyond the supergravity limit and applying a string-theoretic treatment) were also successfully matched in \cite{Skrzypek:2023fkr}. However, as explained in \cite{Martinez:2025jjq}, the corresponding computation for $\Zset_k$ appears to be more challenging.

\section{The ADE Spin chains} \label{sec:ADEspinchains}

In the previous sections we constructed the one-loop Hamiltonian as well as the superconformal index for any ADE theory. Of course, many features are highly dependent on the specific theory, so in the following sections we will focus on specific examples. To set the stage, however, it is useful to discuss some generic aspects of the ADE spin chains. 

The main distinctive feature of the orbifold spin chains is that the Hilbert space is constrained, in the sense that not all compositions of fields are possible. This is due to gauge invariance, which requires the indices of the fields to be contracted. Consider a 1-dimensional chain of length $L$, with the sites labelled by $\ell=1,\ldots,L$. As one goes along the chain, the second matrix index of the field at site $\ell$ must be the same as the first matrix index of the field at site $\ell+1$. Since our fields are represented as arrows between nodes of the quiver, this condition simply tells us that the source of the arrow at site $\ell+1$ must be the same as the endpoint of the arrow at site $\ell$. In other words, only sequences of fields which follow the \emph{path algebra} of the quiver can be composed to form a spin-chain state. This is illustrated below with a sample path on the $\hat{D}_4$ quiver starting from node 2:
  \begin{center}
    \begin{tikzpicture}[scale=0.7]
\draw[-,thick,blue] (0,0)--(2,2);
\draw[-,thick,blue] (0,2)--(2,0);
\draw[fill=black] (0,0) circle (.5ex);\draw[fill=black] (2,2) circle (.5ex);
\draw[fill=black] (1,1) circle (.5ex);\draw[fill=black] (2,0) circle (.5ex);
\draw[fill=black] (0,2) circle (.5ex);
\draw[->,red,thick] (-0.1,-0.2) arc (-20:-310:.4);
\draw[->,red,thick](0,0.2)--(0.7,0.9);
\draw[->,red,thick](0.75,1)--(0.15,1.6);
\draw[->,red,thick] (-0.1,1.8) arc (300:0:.4);
\draw[->,red,thick] (-0.18,1.6) arc (280:10:.58);
\draw[->,red,thick](0.2,2)--(0.8,1.4);
\node at (-0.2,2) {\scriptsize$1$};
\node at (-0.2,0) {\scriptsize$2$};
\node at (2.3,2) {\scriptsize$4$};
\node at (2.3,0) {\scriptsize$3$};
\node at (1.4,1) {\scriptsize$5$};
\draw[->,thick,green] (3,1)--(4,1);
\node at (8,1) {{\small $\cdots Z_2 Q_{25} Q_{51} Z_1 Z_1 Y_{15} \cdots$}};
    
    \end{tikzpicture}
    \end{center}

  Of course, spin chains with such Hilbert space restrictions have been considered in various other contexts. A well-known example is the Rydberg atom chain, see \cite{Corcoran:2024ofo} for a recent discussion and references. A powerful way to represent the restrictions is in terms of a fusion category, see e.g. \cite{Trebst_2008, Aasen:2020jwb} for reviews and \cite{Blakeney:2025ext} for recent work. In this context, one can for instance describe interesting (and integrable) chains based on the path algebra of dihedral groups \cite{Braylovskaya:2016btd}. An alternative way to express the path algebra restrictions is to reformulate the vertex model associated to the spin chain as a (restricted)-solid-on-solid model (RSOS) \cite{Andrews:1984af}, using the vertex/face map. Recall that RSOS models are statistical models where heights are placed at the corners of each lattice square, with the allowed heights defined by an adjacency diagram. In \cite{Pomoni:2021pbj} this link to RSOS models was developed for the $\Ncal=2$ $\Zset_k$ quivers, which were argued to correspond to dilute, cyclic SOS models.
  
The restricted state space defined in this way is the same for the orbifold point as well as deformations away from it (at least away from limits where gauge groups become global). However, in the deformed case where $g_i=\kappa_i g_{YM}$, the action of the Hamiltonian depends on the deformation parameters $\kappa_i$. A way to keep track of the action of the Hamiltonian (and associated $R$-matrix, when an Algebraic Bethe Ansatz approach is applicable) is to introduce a dynamical parameter (playing a similar role to that appearing in the construction of elliptic quantum groups \cite{Felder:1994pb,Felder:1994be}) whose values along the chain are determined by the path algebra. This was the approach followed in \cite{Pomoni:2021pbj} for the $\Zset_2$ case where the spin chain was expressed as a dynamical 6-vertex or 15-vertex model (in the $\SU(2)$ or $\SU(3)$ sectors respectively). Such restricted dynamical models have recently been studied from the quantum group perspective in \cite{Felder:2020tct}. 

As indicated above, it is natural to describe the path algebra of a given quiver in categorical language, with the objects being the nodes and the morphisms being the arrows between them. In \cite{Bertle:2024djm} this path category was used to define the coproduct of a further Lie algebroid structure describing the action of those global symmetry generators which were broken during the orbifolding process and thus effectively restoring them as symmetries of the theory.  Although the construction of the Lie algebroid had its roots in the spin chain description, which only arises in the planar limit, it was shown in \cite{Bertle:2024djm} that it can be used to act directly on the $\Zset_2$ quiver Lagrangian and show invariance under the (twisted algebroid version of) the full $R$-symmetry group of $\Ncal=4$ SYM.

We expect all of the above structures, originally defined for the $\Zset_2$ quiver, to be present in the ADE quivers as well. In particular, the RSOS models related to the ADE adjacency graphs are well known \cite{Pasquier:1986jc, Pearce:1990ila}, and we expect the dilute versions of those models to be of relevance to our spin chains. This would provide guidance in introducing appropriate dynamical parameters as well as defining an ADE version of the Lie algebroid of \cite{Bertle:2024djm}. Leaving these constructions for future work, it is important to remark that for the abelian quivers, a given mother theory state and initial index uniquely specifies the path algebra. For instance, choosing the first index to be 1, one has:
\be
XYXZZYY\cdots \ra X_{12}Y_{21}X_{12}Z_2Z_2Y_{21}Y_{1k}\cdots
\ee
However, this is not true for the non-abelian quivers, where the same mother theory state can map to multiple paths on the quiver, as can be seen by considering the reflection of the $\Dfour$ path above, which would turn towards node 3 rather than 1. Therefore, additional information will be needed to specify a given path uniquely. 

To conclude this section, we recall that physical operators, which are gauge-theory traces, correspond to closed paths on the quiver. However, they do not necessarily form linear combinations which are invariant under the finite  group action, but can instead transform non-trivially. From the perspective of the mother theory, these are known as twisted sectors and, as already mentioned, can be constructed by inserting the quiver-basis matrix corresponding to any of the group elements in the gauge theory trace, as represented in the figure below.  

\begin{center}
\begin{tikzpicture}[scale=0.8,baseline=0cm]
  \draw[thick, blue, -] (0,0) [partial ellipse=0:90:3cm and 1.5cm];
  \draw[thick, blue, -] (0,0) [partial ellipse=0:-90:3cm and 1.5cm];
    \draw[thick, blue, -] (0,0.5) [partial ellipse=90:270:2cm and 1cm];
    \draw[-,dashed,purple,thick] (0,-0.5)--(0,-1.5);\node at (-0.7,-1) {{\small $\gamma(g)$}};
\draw[->,thick,green] (3.5,0)--(4.5,0);

\node at (9,0) {$\Tr(\gamma(g) ZZXYZYX\cdots) \;,\;\quad g\in \Gamma$};
\end{tikzpicture}
\end{center}

As discussed, the number of twisted states (including the untwisted state) is the number of conjugacy classes of $\Gamma$, which is the same as the number of nodes of the quiver. However, not all nontrivial mother-theory states give nonzero states in the orbifold theory. In particular, for the non-abelian cases, we will find examples of untwisted projections (obtained by setting $\gamma(e)=I$ above) of BPS states that trivially vanish. Conversely, states which would vanish in the mother theory (for instance, states of the form $\Tr(XY-YX)$) can produce nontrivial twisted-sector states as $\Tr(\gamma(g)(XY-YX))$ does not necessarily vanish for $g\neq e$.

\subsection{States of length 2}

As in the following sections we will be studying the spectrum of short chains, it is convenient to introduce some notation which will apply to all of our examples. In particular, it will be useful to organise operators into multiplets of the $\SU(2)_R$ symmetry, which acts on the bifundamental fields. Denoting the $\SU(2)_R$ indices by  $\mathcal{I},\;\mathcal{J}=1,2$ we have
\be
(Q_{ij})_\mathcal{I}\equiv\begin{pmatrix}
    Q_{ij}\\
    \Qb_{ij}
\end{pmatrix}_\mathcal{I}\;,\quad (Q_{ji})^\mathcal{J}\equiv\begin{pmatrix}
    \Qb_{ji}&Q_{ji}
\end{pmatrix}^\mathcal{J}.
\ee
When we pass to the $X$, $Y$ notation we will typically choose 
\be
(Q_{ij})_\mathcal{I}=\begin{pmatrix}
    X_{ij}\\
    \Yb_{ij}
\end{pmatrix}_\mathcal{I},\quad (Q_{ji})^\mathcal{J}=\begin{pmatrix}
    \Xb_{ji}&Y_{ji}
\end{pmatrix}^\mathcal{J}.
\ee
Notice that our convention for the bifundamentals is that they are in the $\Box_i\otimes \bar{\Box}_j$ representation, which implies that for each arrow on the quiver diagram, the base of the arrow is the fundamental while the tip of the arrow is the conjugate fundamental representation of the corresponding gauge groups. In the $\Zset_2$ quiver case, this does not uniquely specify the fields as they also form multiplets of the additional $\SU(2)_L$ symmetry present there. Therefore, in that case one often introduces an additional $\tilde{Q}$ notation to distinguish the fields. In the cases we consider the choice of $i$, $j$ nodes plus the $\SU(2)_R$ index is enough to distinguish the fields, and in particular if $(Q_{ij})_1$ denotes an $X_{ij}$ field then $(Q_{ij})^2$ will denote a $Y_{ji}$ field. 
Following \cite{Gadde:2010zi}, we can then define the ``mesonic combinations''
\be
\left(\Mcal_{ij}\right)_\mathcal{I}\phan^\mathcal{J}\equiv 
(Q_{ij})_\mathcal{I}(Q_{ji})^\mathcal{J}\;,
\ee
or, more explicitly: 
\be
\Mcal_{ij}=\begin{pmatrix}
    Q_{ij}\Qb_{ji}&Q_{ij}Q_{ji}\\
    \Qb_{ij}\Qb_{ji}&\Qb_{ij}Q_{ji}
\end{pmatrix}\;.
\ee
Note that
\be
\Tr_i(\Mcal_{ij})=\Tr_j(\Mcal_{ji}).
\ee
We can then decompose the mesonic field $\Mcal_{ij}$ into the singlet and the traceless triplet
\be
\Mcal_{ij}^{(\mathbf{1})}\equiv \left(\Mcal_{ij}\right)_\mathcal{I}\phan^\mathcal{I}\;,\quad\left(\Mcal_{ij}^{(\mathbf{3})}\right)_\mathcal{I}\phan^\mathcal{J}\equiv\left(\Mcal_{ij}\right)_\mathcal{I}\phan^{\mathcal{J}}-\half\left(\Mcal_{ij}\right)_\mathcal{K}\phan^\mathcal{K}\delta_\mathcal{I}\phan^\mathcal{J}.
\ee
Explicitly
\begin{subequations}
    \begin{align}
        \Mcal_{ij}^{(\mathbf{1})}=&Q_{ij}\Qb_{ji}+\Qb_{ij}Q_{ji},\\
        \Mcal_{ij}^{(\mathbf{3})}=&\begin{pmatrix}
            \half(Q_{ij}\Qb_{ji}-\Qb_{ij}Q_{ji})&Q_{ij}Q_{ji}\\
            \Qb_{ij}\Qb_{ji}&\half(\Qb_{ij}Q_{ji}-Q_{ij}\Qb_{ji})
        \end{pmatrix}.
    \end{align}
\end{subequations}
The triplet is always protected,
\be
\Hcal\left(\Tr_i\Mtrip_{ij}\right)=0 \;,
\ee
which, as discussed in Section \ref{sec:Nonplanar}, for the twisted sector triplets is due to the enhancement of non-planar diagrams at length-2. 

On the other hand, acting on the singlet with the ADE Hamiltonian we find  
\be
\Hcal\left(\Tr_i\Mcal^{(\mathbf{1})}_{ij}\right)=4(\kappa_i^2+\kappa_j^2)\Tr_i\Mcal^{(\mathbf{1})}_{ij}+\frac{4\kappa_i^2a^\mathbf{2}_{ij}}{n_i}\Tr_i Z_i\bar{Z}_i+\frac{4\kappa_j^2a^\mathbf{2}_{ji}}{n_j}\Tr_j\Phi_j\bar{\Phi}_j.
\ee
Combining this with the action on the adjoint fields, 
\be
\Hcal\left(\Tr_iZ_i\bar{Z}_i\right)=4\kappa_i^2\Tr_iZ_i\bar{Z}_{i}+2\kappa_i^2\sum_{j=1}^Ma^\mathbf{2}_{ij}n_j\Tr_i\Mcal_{ij}^{(\mathbf{1})}\;,
\ee
it is straightforward to check that the following $\kappa_i$-independent combination
\be \label{TGamma}
\Tr(\mathcal{T}_\Gamma)\equiv\sum_{i,j=1}^{M}\biggl[\delta_{ij}n_i\Tr_i\Zb_iZ_i-\frac{a^\mathbf{2}_{ij}n_in_j}{4}\Tr_i\Msing_{ij}\biggr]\;,
\ee
has a zero eigenvalue, $\Hcal\left(\Tr\mathcal{T}_\Gamma\right)=0$, for all values of $\kappa_i$. This state is the conformal primary of
$\hat{\mathcal{C}}_{0(0,0)}$, which descends from the $\Ncal=4$ SYM state $\Tr(X\Xb+Y \Yb-2Z\Zb)$.

At the orbifold point, $\kappa_i=1$, the  other eigenstate of the Hamiltonian in this sector is the classical Konishi operator:
\be \label{Konishiclassical}
\mathcal{K}_\Gamma=\sum_{i,j=1}^M\biggl[\delta_{ij}n_i\Tr_i\Zb_iZ_i+\frac{a^\mathbf{2}_{ij}n_in_j}{2}\Tr_i\Msing_{ij}\biggr]\;.
\ee
As is well known, this operator receives corrections and develops a one-loop anomalous dimension of 12:  
\be
\Hcal^{\text{o.p.}}\left(\mathcal{K}_\Gamma\right)=12\mathcal{K}_\Gamma,
\ee
which corresponds to the $\NN=4$ value \cite{Bianchi:2001cm}, as required by the inheritance principle for untwisted-sector states \cite{Bershadsky:1998mb, Bershadsky:1998cb}. In the deformed theory, $\kappa_i\neq 1$,  the coefficients of (\ref{Konishiclassical}) will become $\kappa_i$-dependent and attempting to write their generic form is not very illuminating. We will indicate the coefficients for some simple deformations in the specific examples we will consider. 

As shown in detail in Appendix \ref{sec:Konishi}, by acting with the unbroken supercharges on the Konishi operator, one obtains the $L=3$ ``superpotential'' operator in the $XYZ$ sector as well as a $L=4$ descendant operator in the $XY$ sector. Clearly, these operators will have the same anomalous dimensions as the Konishi operator, and their coefficients can also be written in terms of the $\kappa_i$-dependent coefficients of Konishi.

\subsection{Magnons on the ADE chains}

As we will be interested in applying Bethe ansatz techniques to our theories, it is important to see what are the possible pseudovacua around which we can consider excitations. The $\Zset_k$ chains are special in this regard, as they can have pseudovacua made up of the bifundamental fields:
\be
\Tr(X_{12}X_{23}\ldots X_{k-1k}X_{k1})^m,\quad \Tr(Y_{1k}Y_{kk-1}\ldots Y_{32}Y_{21})^m\;.
\ee
where $L=m k$. These have $E=0$ and are called the $X$-- and $Y$-- vacuum respectively. In the $\Zset_2$ orbifold context, they played an important role in \cite{Pomoni:2021pbj}, as in that case the $X$ and $Y$ fields have the same gauge indices and it is possible to define an $XY$ sector, with $Y$ excitations on top of the $X$ vacuum or vice versa. This is not possible for the other $\Zset_k$ cases, but one can still consider adjoint $Z$ excitations over the $X$ or $Y$ vacuum. As discussed in \cite{Pomoni:2021pbj}, the dispersion relation for magnon excitations in these vacua appears to be hyperelliptic for $k>2$. Understanding the 2-magnon problem in these vacua is an important question, which we will however not consider in this work as our focus will be on the generic ADE case.

It is easy to see that the non-abelian orbifold $\hat{D}$ and $\hat{E}$ chains cannot have $X$ or $Y$ vacua, as bifundamentals of the same type cannot be composed an arbitrary number of times. Therefore, one only has $Z$ vacua, on top of which one can consider bifundamental or $\Zb$
excitations. The bifundamentals act as domain walls separating different $Z$ vacua, e.g. for an asymptotic chain:
\be \label{QZvac}
\cdots Z_iZ_i Q_{ij} Z_j Z_j \cdots
\ee
Therefore, in the holomorphic sector closeability requires two magnons of different flavour, for instance
\be
\Tr(\cdots Z_i Z_i Q_{ij} Z_j\cdots Z_j Q_{ji} Z_iZ_i \cdots)\;,
\ee
where, as discussed,  if $Q_{ij}$ is an $X$ field $Q_{ji}$ has to be a $Y$ field and vice versa. This means that two magnons of the same type ($X$ or $Y$) do not lead to closeable chains. 

Of course, one can have pairs of a bifundamental and its conjugate,
\be
\Tr(\cdots Z_i Z_i Q_{ij} Z_j\cdots Z_j \bar{Q}_{ji} Z_iZ_i \cdots)\;,
\ee
however this choice takes us outside of the holomorphic sector and we will not consider it here. 

Now consider a single $Q$ magnon in the $Z$ vacuum, which we parametrise in the usual way as a plane wave,
\be
\ket{\psi}=\sum_{\ell=1}^L e^{ip\ell} \ket{\cdots Z_iZ_i Q_{ij} Z_j Z_j \cdots}\;.
\ee
It is easy to find the dispersion relation
\be \label{DispersionGeneral}
E_{1-\text{mag}}(p;\kappa_i,\kappa_j)=2\left(\kappa_i^2+\kappa_j^2-\kappa_i\kappa_j\bigl(e^{ip}+e^{-ip}\bigr)\right),
\ee
which of course reduces to the XXX dispersion relation at the orbifold point $\kappa_i=1$:
\be
E_{1-\text{mag}}(p;1,1)=4-4\cos(p)=4\sin^2\left(\frac{p}{2}\right)\;.
\ee
Note that the dispersion relation has a gap, which means that $E=0$ magnons will have complex momenta. 

The two-magnon problem is highly dependent on the specific ADE case, so although it should be possible to write a generic solution in terms of the general ADE Hamiltonian and adjacency matrices $a^{\bf 3}_{ij}$, it will likely not be very illuminating. Therefore, in the following we will discuss the solution on a case-by-case basis, which should be sufficient to illustrate the general features.

\section{Example: The $\Zset_3$ theory} \label{sec:Z3}

\begin{figure}[ht]
 \begin{center}
   \begin{tikzpicture}[scale=0.5]
     
  \draw[->,violet,thick] (-0.1,0.1)--(2.9,5.3);  \draw[->,violet,thick] (3.1,5.1)-- (0.1,-0.1);
  \draw[->,yellow,thick] (3.1,5.3)--(6.1,0.1);   \draw[->,yellow,thick] (5.9,-0.1)--(2.9,5.1);
  \draw[->,cyan,thick] (6,-0.12)--(0,-0.12);   \draw[->,cyan,thick] (0,0.12)--(6,0.12);
 \draw[->,violet,thick] (-0.1,0.1)--(1.4,2.7); \draw[->,violet,thick] (3.1,5.1)-- (1.6,2.5);
 \draw[->,yellow,thick] (3.1,5.29)--(4.6,2.7); \draw[->,yellow,thick] (5.9,-0.1)--(4.4,2.5);
 \draw[->,cyan,thick] (6,-0.12)--(3,-0.12);  \draw[->,cyan,thick] (0,0.12)--(3,0.12);
 
  \draw[fill=blue] (0,0) circle (2ex);
\draw[fill=red] (3,5.19) circle (2ex);
\draw[fill=green] (6,0) circle (2ex);

\draw[->,blue,thick] (-0.1,-0.7) arc (-20:-310:1);
\draw[->,red,thick] (2.4,5.3) arc (-130:-410:1);
\draw[->,green,thick] (6.4,0.5) arc (100:-170:1);

  \node at (-0.5,-0.5) {$1$};\node at (3,5.9) {$2$};\node at (6.5,-0.5) {$3$};
\node at (0.6,3) {$X_{12}$};\node at (2.3,2) {$Y_{21}$};
\node at (5.3,3) {$X_{23}$};\node at (4,2) {$Y_{32}$};
\node at (3,0.8) {$Y_{13}$};\node at (3,-0.7) {$X_{31}$};

\node at (-1.8,-1.8) {$Z_1$};\node at (3,7.6) {$Z_2$};\node at (7.8,-1.8) {$Z_3$};

\end{tikzpicture}\vspace{-0.5cm}
 \end{center}
 \caption{The $Z_3$ quiver. All the gauge groups are $\SU(N)$. For clarity, we only indicate the holomorphic fields. The arrows for the corresponding conjugate fields are reversed.} \label{Fig:Z3}
 \end{figure}
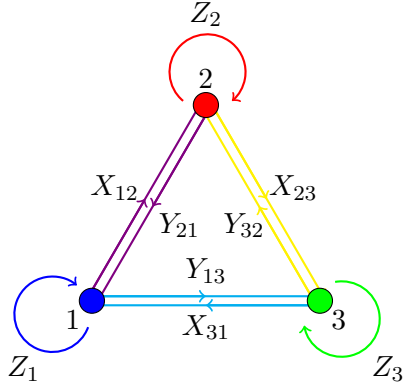

As a first application of our general formalism, we consider the spin chain for the marginally-deformed $\Zset_3$ orbifold theory, which corresponds to the $\hat{A}_2$ quiver. This is the next-simplest example after the $\Zset_2$ case which has already been extensively studied \cite{Gadde:2009dj,Gadde:2010zi,Gadde:2010ku,Pomoni:2021pbj}. However, given that it does not have the additional $\SU(2)_L$ symmetry, it is more indicative of the generic behaviour of the $\Zset_k$ theories. The $\Zset_3$ group is of order 3 and is defined by a single generator $a$:
\be
 \{a|a^3=1\}.
 \ee
\begin{table}[ht]
    \centering
    \begin{tabular}{c|ccc}
          & $e$& $a^2$& $a$\tabularnewline\hline
        $e$ & $e$& $a^2$ &$a$\tabularnewline
        $a$& $a$&$e$&$a^2$\tabularnewline
        $a^2$&$a^2$&$a$&$e$
    \end{tabular}
    \caption{The Cayley table of $\Zset_3$}
    \label{tab:Z3Cayleytable}
\end{table}
 The Cayley table is given in Table \ref{tab:Z3Cayleytable}. From this one can read off the orbit basis representation matrices (\ref{orbitmatrices}), which are
\be
\tau(e)=\begin{pmatrix}
    1&&\\
    &1&\\
    &&1
\end{pmatrix},\quad\tau(a)=\begin{pmatrix}
    &&1\\
    1&&\\
    &1&
\end{pmatrix} \;\text{and}\; \quad\tau(a^2)=\begin{pmatrix}
    &1&\\
    &&1\\
    1&&
\end{pmatrix},
\ee
while, in terms of $\omega_3=e^{\frac{2\pi i}{3}}$, the quiver basis matrices (\ref{gammareg}) are
\be
\gamma(e)=\begin{pmatrix}
    1&&\\
    &1&\\
    &&1
\end{pmatrix},\quad\gamma(a)=\begin{pmatrix}
    1&&\\
    &\omega_3&\\
    &&\omega_3^{2}
\end{pmatrix},\quad\gamma(a^2)=\begin{pmatrix}
    1&&\\
    &\omega_3^2&\\
    &&\omega_3
\end{pmatrix}.
\ee
There are three $\SU(N)$ gauge nodes in the orbifolded theory. Like all the cyclic quivers, it is convenient to choose $Q_{i,i+1}=X_{i,i+1}$ and $Q_{i,i-1}=Y_{i,i-1}$, i.e. the $X$ fields are the arrows from a node to the next one, while the $Y$ fields point in the opposite direction. Then we have
\be
Z=\threebythree{Z_1}{0}{0}{0}{Z_2}{0}{0}{0}{Z_3}\;,\;X=\threebythree{0}{X_{12}}{0}{0}{0}{X_{23}}{X_{31}}{0}{0}\;,\;Y=\threebythree{0}{0}{Y_{13}}{Y_{21}}{0}{0}{0}{Y_{32}}{0}
\ee
where each entry is an $N\times N$ block. The conjugate fields are given by hermitian conjugation. The field content is summarised in Fig. \ref{Fig:Z3}.
The action of $\Zset_3$ on the node indices is given by conjugation by the orbit matrices, and is simply $\tau(a): i\ra i+1$.  

Writing the couplings as $g_i=\kappa_i \gym$, the superpotential is
\be
\begin{split}
    \Wcal_{\Zset_3}=&i\gym\bigl[\kappa_1\Tr_2(Y_{21}Z_1X_{12})-\kappa_2\Tr_1(X_{12}Z_2Y_{21})+\kappa_2\Tr_{3}(Y_{32}Z_2X_{23})\\&\quad\quad -\kappa_3\Tr_2(X_{23}Z_3Y_{32})+\kappa_3\Tr_1(Y_{13}Z_3X_{31})-\kappa_1\Tr_3(X_{31}Z_1Y_{13})\bigr].
\end{split}
\ee

 \subsection{The $\Zset_3$ Hamiltonian}

To write out the Hamiltonian, first recall that 
\be
\Hcal_{\ell,\ell+1} (Z_iZ_i)=0\;\;,\;\;\Hcal_{\ell,\ell+1}(Q_{ij}Q_{jk})=0 \;\;,\;\;\text{for} \;\; {i\neq k}\;.
\ee
Given the $\Zset_3$ symmetry of the problem, it is simplest to write the action of the Hamiltonian on node 1, as the action on the other fields is just the $\Zset_3$ conjugate given by $i\ra i+1\ra i+2$. 

In the holomorphic sector we simply have 
\be
\Hcal_{\ell,\ell+1}=\begin{pmatrix}
2\kappa_1^2&-2\kappa_1^2\\
-2\kappa_1^2&2\kappa_1^2
\end{pmatrix} \;\; \text{in the basis}\;\;
\begin{pmatrix}
  X_{12}Y_{21}\\
  Y_{13}X_{31}
\end{pmatrix}\;,
\ee
and
\be
\Hcal_{\ell,\ell+1}=\begin{pmatrix}2\kappa_1^2&-2\kappa_1\kappa_2& & \\ -2\kappa_1\kappa_2&2\kappa_2^2 & & \\
& &  2\kappa_1^2&-2\kappa_1\kappa_3\\
& & -2\kappa_1\kappa_3&2\kappa_3^2\end{pmatrix} \;\;\text{in the basis} \;\;
\begin{pmatrix} Z_1X_{12}\\
  X_{12}Z_2\\ 
  Z_1Y_{13}\\ Y_{13} Z_3
\end{pmatrix}
\ee
For the mixed sector, it is convenient to define 
\be
\mathbf{Q\bar{Q}}_1=\begin{pmatrix}
  X_{12}\Xb_{21}\\
  Y_{13}\Yb_{31}
\end{pmatrix}\;,\quad\mathbf{\bar{Q}Q}_1=\begin{pmatrix}
\Yb_{12}Y_{21}\\
\Xb_{13}X_{31}
\end{pmatrix}
\ee
and the matrices 
\be
\mathbb{K}_1=\begin{pmatrix}
  \kappa_1^2&\kappa_1^2
\end{pmatrix},\quad\mathbb{T}_1=\begin{pmatrix}
2\kappa_2^2&\\
&2\kappa_3^2
\end{pmatrix},\quad\mathbb{M}_1=\begin{pmatrix}
\kappa_1^2&\kappa_1^2\\
\kappa_1^2&\kappa_1^2
\end{pmatrix},\quad\mathbb{L}_1=\begin{pmatrix}
\kappa_1^2\\
\kappa_1^2
\end{pmatrix}\;,
\ee
in terms of which the Hamiltonian on fields starting at node 1 is 
\be
\Hcal_{\ell,\ell+1}=\begin{pmatrix}
            3\kappa_1^2&-\kappa_1^2&\mathbb{K}_1&\mathbb{K}_1\\
            -\kappa_1^2&3\kappa_1^2&\mathbb{K}_1&\mathbb{K}_1\\
            \mathbb{L}_1&\mathbb{L}_1&\mathbb{T}_1+\mathbb{M}_1&\mathbb{T}_1-\mathbb{M}_1\\
            \mathbb{L}_1&\mathbb{L}_1&\mathbb{T}_1-\mathbb{M}_1&\mathbb{T}_1+\mathbb{M}_1
        \end{pmatrix} \;\;\text{on}\;\; \quad\begin{pmatrix}
            Z_1\Zb_1\\
            \Zb_1Z_1\\
            \mathbf{Q\Qb}_1\\
            \mathbf{\Qb Q}_1
        \end{pmatrix}\;.
\ee
Finally, on mixed fields with different first and last indices, we have 
\be
\Hcal_{\ell,\ell+1}=
\begin{pmatrix}
            2\kappa_1^2&-2\kappa_1\kappa_2& &\\
            -2\kappa_1\kappa_2&2\kappa_2^2& &\\
            & &2\kappa_1^2&-2\kappa_1\kappa_2\\
            & &-2\kappa_1\kappa_2&2\kappa_2^2 
        \end{pmatrix} \;\;\text{on}\;\;\begin{pmatrix}
            Z_1\Yb_{12}\\
            \Yb_{12}Z_2\\
            \Zb_1X_{12}\\
            X_{12}\Zb_2
\end{pmatrix}
\ee
and
\be
\Hcal_{\ell,\ell+1}=\begin{pmatrix}
            2\kappa_1^2&-2\kappa_1\kappa_3& & \\
            -2\kappa_1\kappa_3&2\kappa_3^2& & \\
            & &2\kappa_1^2&-2\kappa_1\kappa_3\\
            & &-2\kappa_1\kappa_3&2\kappa_3^2
        \end{pmatrix}\;\;\text{on}\;\;\begin{pmatrix}
            Z_1\Xb_{13}\\
            \Xb_{13}Z_3\\
            \Zb_1Y_{13}\\
            Y_{13}\Zb_{3}
\end{pmatrix}\;.
\ee
Of course, all the above actions are supplemented by their $\Zset_3$ conjugates.

\subsection{Protected spectrum}

As we already discussed the $\Zset_k$ index in the previous section, here we will simply write out the specific forms that the protected states take for $\Zset_3$. This will be helpful in comparing with the direct diagonalisation of the Hamiltonian.

As usual, we define the ``meson" operators
\be
    \Mcal_{12}=\begin{pmatrix}
        X_{12}\Xb_{21}&X_{12}Y_{21}\\
        \Yb_{12}\Xb_{21}&\Yb_{12}Y_{21}
    \end{pmatrix}\;,
    \ee
and similarly for their $\Zset_3$ conjugates $\Mcal_{23}$ and $\Mcal_{31}$. We can then form the $\SU(2)_R$ singlets    
\be
        \Mcal_{12}^{(\mathbf{1})}=X_{12}\Xb_{21}+\Yb_{12}Y_{21} \;,
\ee
and triplets 
\be
      \Mcal_{12}^{(\mathbf{3})}=\begin{pmatrix}
            \half(X_{12}\Xb_{21}-\Yb_{12}Y_{21})&X_{12}Y_{21}\\
            \Yb_{12}\Xb_{21}&\half(\Yb_{12}Y_{21}-X_{12}\Xb_{21})
        \end{pmatrix} \;.
      \ee
In terms of these we can write the (untwisted) superconformal primary of $\hat{\mathcal{C}}_{0(0,0)}$ as
\be \label{TstateZ3}
\Tr(\mathcal{T}_{\Zset_3})=\Tr_1\Zb_1Z_1+\Tr_2\Zb_2Z_2+\Tr_3\Zb_3Z_3-\frac{1}{2}\left[\Tr_1\Mcal^{(\mathbf{1})}_{12}+\Tr_2\Mcal^{(\mathbf{1})}_{23}+\Tr_3\Mcal^{(\mathbf{1})}_{31}\right]\;.
\ee
At any length $L>1$ we have the $Z$-vacuum states 
\be
    \begin{split}
        \Tr(\gamma(e) Z^{L})\equiv&\Tr_1Z_1^{L}+\Tr_2Z_2^{L}+\Tr_3Z_3^{L},\\
        \Tr(\gamma(a) Z^{L})\equiv&\Tr_1Z_1^{L}+\omega_3\Tr_2Z_2^{L}+\omega_3^2\Tr_3Z_3^{L},\\
        \Tr(\gamma(a^2) Z^{L})\equiv&\Tr_1Z_1^{L}+\omega_3^2\Tr_2Z_2^{L}+\omega_3\Tr_3Z_3^{L},
    \end{split}
\ee
and at $L=2$ we also have the protected triplets:
\be \label{Z3L2triplets}
    \begin{split}
      \Tr(\gamma(e)~ \Mtrip )\equiv&\Tr_1\Mtrip_{12}+\Tr_2\Mtrip_{23}+\Tr_3\Mtrip_{31},\\
      \Tr(\gamma(a)~ \Mtrip)\equiv&\Tr_1\Mtrip_{12}+\omega_3\Tr_2\Mtrip_{23}+\omega_3^2\Tr_3\Mtrip_{31},\\
      \Tr(\gamma(a^2)~ \Mtrip)\equiv&\Tr_1\Mtrip_{12}+\omega_3^2\Tr_2\Mtrip_{23}+\omega_3\Tr_3\Mtrip_{31}.
    \end{split}
\ee
We note that, as for all cyclic quivers, the twisted states have definite eigenvalues under conjugation by the $\tau(g)$ matrices. For example, we see that
\be\begin{split}
\tau(a) \Tr(\gamma(e) Z^{L}) \tau(a)^{-1}&= \Tr(\gamma(e) Z^{L})\;,\\
\tau(a) \Tr(\gamma(a) Z^{L}) \tau(a)^{-1}&=\omega_3 \Tr(\gamma(a) Z^{L})\;,\\
\tau(a) \Tr(\gamma(a^2) Z^{L}) \tau(a)^{-1} &=\omega_3^2 \Tr(\gamma(a^2) Z^{L})\;,
\end{split}
\ee
and similarly for the triplet states. 

To obtain all-length results in the $XY$ sector, we turn to the Molien series. From Table \ref{tab:SU(2)Molienseries}, the Molien series of $\Zset_3$ is given by
\be\label{eq:Z3Molien}
\begin{split}
\mathbf{M}(x;\Zset_3)&=\,\frac{1+x^3}{(1-x^2)(1-x^3)}\\&=\,1\!+\!x^2\!+\!2x^3\!+\!x^4\!+\!2x^5\!+\!3x^6\!+\!2x^7\!+\!3x^8\!+\!4x^9\!+\!3x^{10}\!+\!4x^{11}\!+\!5x^{12}\!+\!\Ocal(x^{13})\,.
\end{split}
\ee
The powers of $x$ correspond to the length, so from the expansion we can read off the protected $\hat{\mathcal{B}}_{R}$ multiplets, where $R>1$ since the Molien series only correctly counts states from length 3 and above. Hence, we find 2 $\hat{\mathcal{B}}_{\frac{3}{2}}$ states at $L=3$, 1 $\hat{\mathcal{B}}_{2}$ state at $L=4$ etc. All the above multiplicities agree with explicit diagonalisation of the $\Zset_3$ Hamiltonian in the $XY$ sector.  

Since the Molien series counts only the states of the form $(XY)^{\ell}$ while the Hall-Littlewood
index counts the number of states of the form $(XY)^{\ell}$ minus the number of the states of the form $\bar{\lambda}_{Z\dot{+}}(XY)^{\ell-1}$, if we consider the following quantity
\be\label{eq:NumberFermionsMolien}
\mathbf{M}(x;\Zset_3)-\mathcal{I}^\text{s.t.}_{\Zset_3;\,HL}(x=t^\half)\,,
\ee
we can count the number of fermionic protected states (we must just subtract off the $\Mtrip$ states that are not counted by the Molien series). We cannot do an explicit check of these numbers as we only have the scalar dilatation operator, so they can be thought of as checks on a future extension to include fermions. 

One can wonder whether one can also count the protected states in the $XZ$ or $YZ$ sectors. As discussed, these sectors only exist for a general length for the cyclic quivers. Although we are not aware of a series that would perform this counting, empirically it is easy to see which are the protected states. Focusing on the $XZ$ sector, for $L\leq 3$ one only has the 3 $Z^L$ vacua, while at $L=3$ one also has the $\Tr(X_{12}X_{23}X_{31})$ state.\footnote{Of course, such states of type $\Tr(X^{3\ell})$ are also included in (\ref{eq:Z3Molien}).}  At $L=4$ there is an $E=0$ state made up of a single $Z$ magnon on this $X$-vacuum state, suitably $\kappa$-symmetrised (generalising the states discussed for $\Zset_2$ in \cite{Gadde:2010zi}). The $\kappa$-symmetrisation works for any number of $Z$'s on the $L=3$ $X$-vacuum, giving one state per length. To illustrate it, we give the explicit forms of these BPS states for $L=4$ and $L=5$ (belonging to $\bar{\mathcal{D}}_{\frac{3}{2}(0,0)}$ and $\bar{\mathcal{B}}_{\frac{3}{2},-2(0,0)}$, respectively):
\be
\begin{split}
\Ocal^{L=4}_{XZ}&=\kappa_1\kappa_3\Tr(X_{12}Z_2X_{23}X_{31})+\kappa_1\kappa_2\Tr(X_{12}X_{23}Z_3X_{31})+\kappa_2\kappa_3\Tr(X_{12}X_{23}X_{31}Z_1)\;,\\
\Ocal^{L=5}_{XZ}&=\kappa_1^2\kappa_3^2\Tr(X_{12}Z_2Z_2X_{23}X_{31})\!+\!\kappa_1^2\kappa_2\kappa_3\Tr(X_{12}Z_2X_{23}Z_3X_{31})
\!+\!\kappa_1^2\kappa_2^2\Tr(X_{12}X_{23}Z_3^2X_{31})\\\!+&\kappa_1\kappa_2\kappa_3^2\Tr(X_{12}Z_2 X_{23}X_{31}Z_1)\!+\!
\kappa_1\kappa_2^2\kappa_3\Tr(X_{12}X_{23}Z_3 X_{31}Z_1)\!+\!\kappa_2^2\kappa_3^2\Tr(X_{12}X_{23}X_{31}Z_1 Z_1)
\end{split}\ee
At $L=6$ one can also have the doubly-wrapped $X$-vacuum state $\Tr((X_{12}X_{23}X_{31})^2)$, and for $L>6$ there is always one $\kappa$-symmetrised state with $Z$ magnons on this state. We see that a new state is added every time the length increases by 3. From this we conclude that the $XZ$-sector protected states are simply given by the step-like pattern:
\be
\mathbf{M}(x;\Zset_3)^{XZ}=1+3x^2+4x^3+4x^4+4x^5+5x^6+5x^7+5x^8+6x^9+6x^{10}+\cdots
\ee
Focusing on the non-trivial states as above (i.e. removing the 3 $Z$-vacuum states) one has
\be \label{XZMolien}
\mathbf{M}(x;\Zset_3)^{XZ}-\frac{3x^2}{1-x}=1+x^3+x^4+x^5+2x^6+2x^7+2x^8+3x^9+3x^{10}+\cdots
\ee

\subsection{Short chains}

Let us now move beyond the protected states and discuss the main features of the $\Zset_3$-chain spectrum. 
We will consider chains of length 2, where we will consider all states, and length 3  and 4, where we will only look at the holomorphic sector.

\subsubsection{Length 2}

At $L=2$ the (cyclically identified) state basis is 21-dimensional. There are 16 $E=0$ states, corresponding to 
\be
\{\Tr(\gamma(g) Z^2) ,\Tr(\gamma(g) \Zb^2) ,\Tr(\mathcal{T}_{\Zset_3}) \;,\Tr(\gamma(g)\mathcal{M}^{(3)})\}\;,
\ee
which contribute $3,3,1$ and $9$ states, respectively. As discussed, the two twisted triplet states (i.e. for $g=a,a^2$) cannot be obtained by directly applying the naive Hamiltonian, as there they would appear as $E=6$ states. As shown in Section \ref{sec:Nonplanar}, non-planar corrections at $L=2$ have the effect of bringing them down to $E=0$, as of course required by index considerations. Their explicit form is
given in (\ref{Z3L2triplets}).

The remaining (non-protected) states are the (untwisted) Konishi state, which has $E=12$ at the orbifold point  
\be\begin{split}
    \mathcal{K}_{\Zset_3}=&
     \Tr_1(\Xb_{13}X_{31})\!+\!\Tr_2(\Xb_{21}X_{12})\!+\!\Tr_3(\Xb_{32}X_{23})\!+\!\Tr_1(\Yb_{12}Y_{21})\!+\!\Tr_2(\Yb_{23}Y_{32})
    \!+\!\Tr_3(\Yb_{31}Y_{13})\\&+\Tr_1(\Zb_1Z_1)+\Tr_2(\Zb_2Z_2)+\Tr_3(\Zb_3Z_3)\;,
\end{split}\ee
and two ``twisted Konishi'' states each at  $E=2(3\pm \sqrt{3})$, so-called because, like the Konishi operator, they are $\SU(2)_R\times\Urm(1)$ singlets. In the mother theory context, these states derive from the $\SU(4)$ singlet state as
\be\begin{split}
\mathcal{K}_{\Zset_3}&=\Tr(\gamma(e)(X\Xb+\Xb X+Y\Yb+\Yb Y+Z\Zb+\Zb Z))\;,\\
\mathcal{K}_{\Zset_3}^{(a)}&=\Tr(\gamma(a)(c~(X\Xb+\Xb X+Y\Yb+\Yb Y)+Z\Zb+\Zb Z))\;, \\
\mathcal{K}_{\Zset_3}^{(a^2)}&=\Tr(\gamma(a^2)(c~(X\Xb+\Xb X+Y\Yb+\Yb Y)+Z\Zb+\Zb Z))\;,
\end{split}
\ee
where the coefficient $c=1\mp\sqrt{3}$ for the states with energy $E=2(3\pm\sqrt{3})$. Such twisted Konishi operators (technically, their descendants in the $\mathrm{SL}(2)$ sector) were studied in \cite{deLeeuw:2011rw} from the perspective of higher-loop integrability, though the orbifolds considered there were non-supersymmetric. See \cite{Skrzypek:2022cgg} for a preliminary discussion of such operators in the $\Ncal=2$ SYM context. 

Away from the orbifold point, the untwisted Konishi state mixes with one each of the $E=2(3\pm\sqrt{3})$ states, and all states acquire $\kappa_i$-dependence. The resulting characteristic polynomial is
\be\begin{split} \label{Z3charpoly}
P(E)=&E^{16}\left(E^3-8E^2(\kappa_1^2+\kappa_2^2+\kappa_3^2)+56E(\kappa_1^2\kappa_2^2+\kappa_2^2\kappa_3^2+\kappa_3^2\kappa_1^2)-288\kappa_1^2\kappa_2^2\kappa_3^2\right)  \\
&\quad\times\left(E^2-4E(\kappa_1^2+\kappa_2^2+\kappa_3^2)+8(\kappa_1^2\kappa_2^2+\kappa_2^2\kappa_3^2+\kappa_3^2\kappa_1^2)\right)\;.
\end{split}
\ee
The largest eigenvalue of the cubic polynomial is the Konishi anomalous dimension. We see that the degeneracy between the two twisted sectors is split in the deformed theory.

\begin{figure}
  \begin{center}
    \begin{tikzpicture}
      \node at(0,0){\includegraphics[width=12cm]{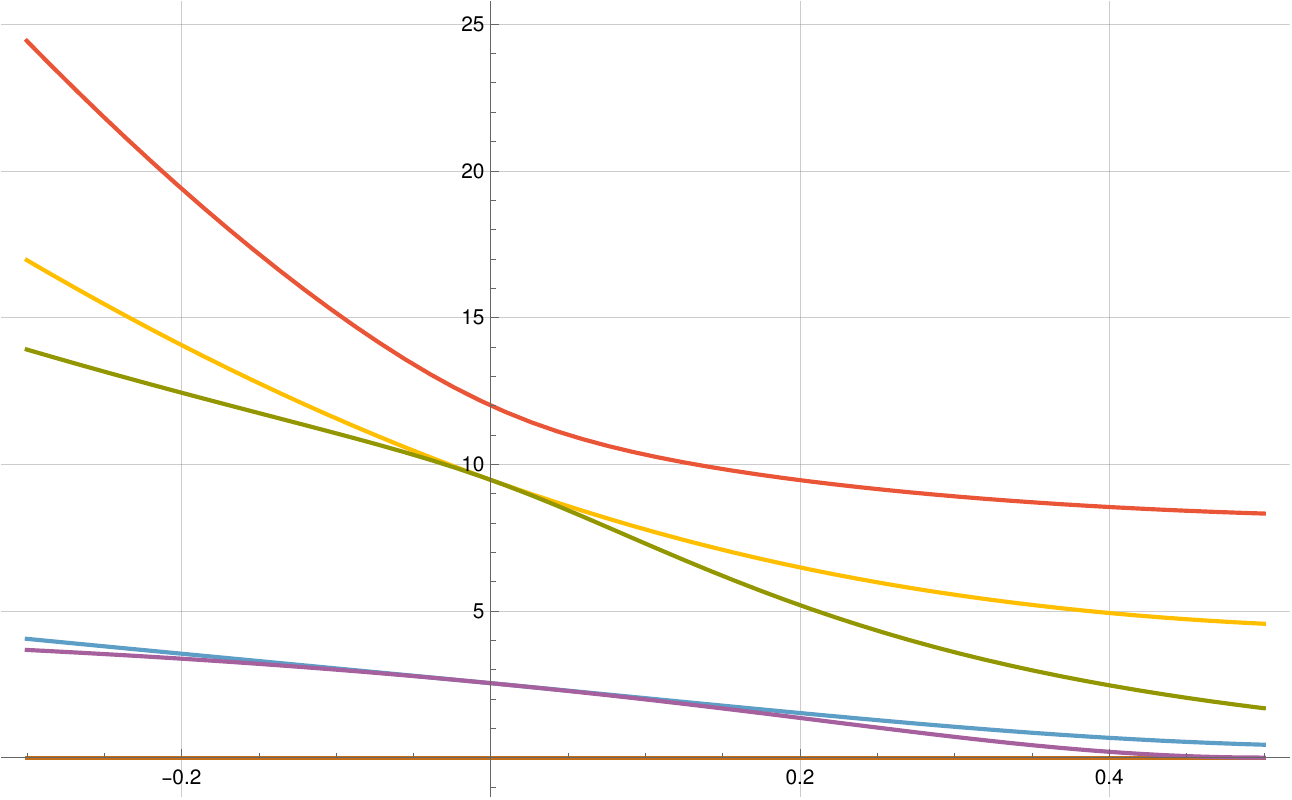}};
\node at (6.3,-3.3){$k$};\node at (-1.4,4){$E$};\end{tikzpicture}
  \caption{The $L=2$ neutral/$L=3$ holomorphic $\Zset_3$ spin chain spectrum for the case $\kappa_1=1-2k,\kappa_2=1-k,\kappa_3=1$. Notice that all orbifold-point degenerate twisted-sector states split under this deformation. We do not extend the plot beyond $k=0.5$ as our Hamiltonian is not applicable to the case of gauge groups becoming global.} \label{GraphZ3L2}
  \end{center}
  \end{figure}

As one approaches a ``SCQCD-like'' limit where one of the $\kappa_i\ra 0$ and the corresponding gauge group becomes global, one obtains further $E=0$ states, due to long multiplets reaching unitarity bounds and breaking up into short ones, as discussed in some detail in \cite{Gadde:2009dj} for the $\Zset_2$ case. In Figs. \ref{GraphZ3L2} and \ref{GraphZ3L3hola} we plot the spectrum for two possible deformations away from the orbifold point, with the latter one limiting to actual SCQCD with $\SU(N)$ gauge group and $N_f=2N$. As expected, in both cases we see enhancement of the protected spectrum. 

\subsubsection{Length 3: Holomorphic sector} \label{Section:Z3L3}

For $L=3$ we will focus on the holomorphic sector, where there are 11 cyclically identified states, 6 of which are 
BPS. As always, we have the $\Tr(\gamma(g) Z^3)$ states in $\bar{\mathcal{E}}_{-3(0,0)}$ and for this length we also have the two untwisted states (in $\hat{\mathcal{B}}_{\frac{3}{2}}$):
\be
\Ocal=\Tr(X_{12}X_{23}X_{31}) \;\;,\;\;\Ocal=\Tr(Y_{13}Y_{32}Y_{21})\;.
\ee
Clearly, these states have no twisted versions. The final $E=0$ state is in the XYZ sector and resembles a symmetrised version of the superpotential:
\be \label{Z3L3holBPS}
  \Ocal =\frac{1}{\kappa_1}\Tr(Z_{1}(X_{12}Y_{21}\!+\!Y_{13}X_{31}))+\frac{1}{\kappa_2}\Tr(Z_{2}(X_{23}Y_{32}\!+\!Y_{21}X_{12}))
  +\frac{1}{\kappa_3}\Tr(Z_{3}(X_{31}Y_{13}\!+\!Y_{32}X_{23}))\;.
  \ee
It is an $\SU(2)_R$ triplet and can be identified as belonging to $\bar{\mathcal{D}}_{1(0,0)}$. 
We note that in the $\Ncal=4$ SYM context, the above six states would all belong to the symmetric ${\mathbf{10}}$ of the unbroken $\SU(4)$. The other four states are projected out by the orbifold projection. We expect that the surviving states can still be connected by a deformed twisted coproduct by following an opening-up procedure similar to that of \cite{Bertle:2024djm} for the $\Zset_2$ case. 

\begin{figure}[ht]
  \begin{center}
    \begin{tikzpicture}
      \node at (0,0){\includegraphics[width=12cm]{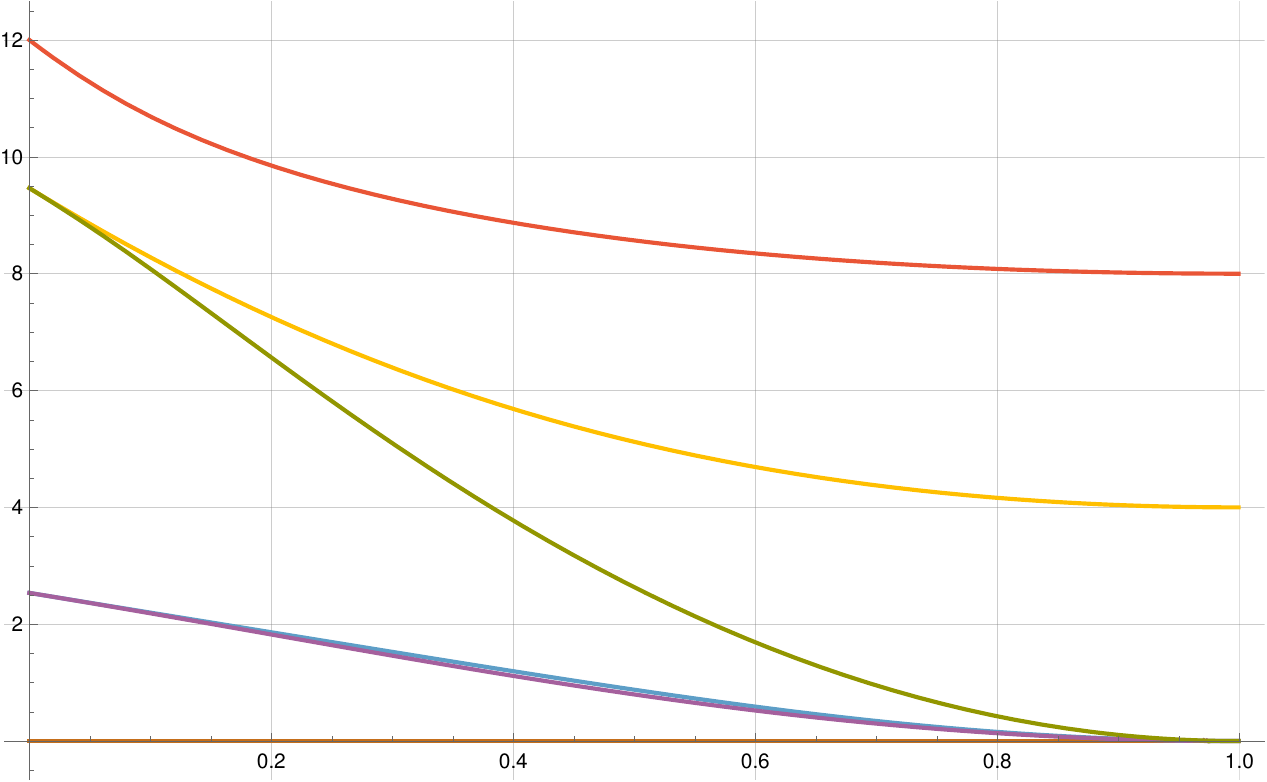}};
\node at (6.3,-3.3){$k$};\node at (-6.1,2.8){$E$};
    \end{tikzpicture}
    \caption{The $L=2$ neutral/$L=3$ holomorphic $\Zset_3$ holomorphic spin chain spectrum for the deformation $\kappa_1=\kappa_2=1-k,\kappa_3=1$. The theory approaches SCQCD as $k\ra 1$, where we see three initially non-protected states approaching $E=0$.} \label{GraphZ3L3hola}
  \end{center}
  \end{figure}

The non-protected states are those that reduce to the superpotential operator at the orbifold point (with $E=12$) as well as two twisted-sector states each at $E=2(3\pm \sqrt{3})$. All these states are superconformal descendants of the corresponding $L=2$ states, and therefore their eigenvalues are still given by (\ref{Z3charpoly}). So, apart from the protected states, the energy spectrum in the $L=3$ holomorphic sector is indistinguishable from that of $L=2$. Therefore the plots in Fig. \ref{GraphZ3L2} and Fig. \ref{GraphZ3L3hola} are relevant for $L=3$ as well. 
 In Section \ref{Z3BetheAnsatz} we will show how the energy spectrum (and states) in this sector can be reproduced in a coordinate Bethe ansatz approach. 

\subsubsection{Length 4: Holomorphic sector} \label{Section:Z3L4}

The state space in this case is 24-dimensional, with 7 of these states having $E=0$. They are the 3 $\Tr(\gamma(g)Z^4)$ states, as well as 1 $XY$-sector state as required by the Molien series (\ref{eq:Z3Molien}), and the 2 states counted in (\ref{XZMolien}) (one in the $XZ$ and one in the $YZ$ sector). The final and perhaps more interesting state is a two-magnon state in the $Z$-vacuum  
\be \label{Z3L4holBPS}
\Ocal=\frac{1}{\kappa_1^2}\Tr(Z_1Z_1X_{12}Y_{21})+\frac{1}{\kappa_1\kappa_2}\Tr(Z_1X_{12}Z_2Y_{21})+\frac{1}{\kappa_1^2}\Tr(Z_1Z_1Y_{13}X_{31}) +\cdots 
\ee
where the $\cdots$ are the $\Zset_3$ conjugates starting at nodes 2 and 3. It belongs to $\bar{\mathcal{B}}_{1,-2(0,0)}$. This is the type of state that one can access through the Bethe ansatz in Section \ref{Z3BetheAnsatz}, which indeed shows an $E=0$ state at every length and correctly predicts its coefficients.  

We will not list the full non-protected spectrum, apart from noting that it contains the $XY$-sector descendant of the Konishi operator with orbifold-point anomalous dimension $E=12$, see Appendix \ref{Z3Konishi}. As a check of the Hamiltonian, one can confirm that the eigenvalue of this $L=4$ state is given by the largest root of the cubic polynomial in (\ref{Z3charpoly}) and its coefficients agree with those expected from (\ref{Z3L4Konishi}).

\begin{figure}[h!]
  \begin{center}\begin{tikzpicture}
      \node at (0,0) {\includegraphics[width=14cm]{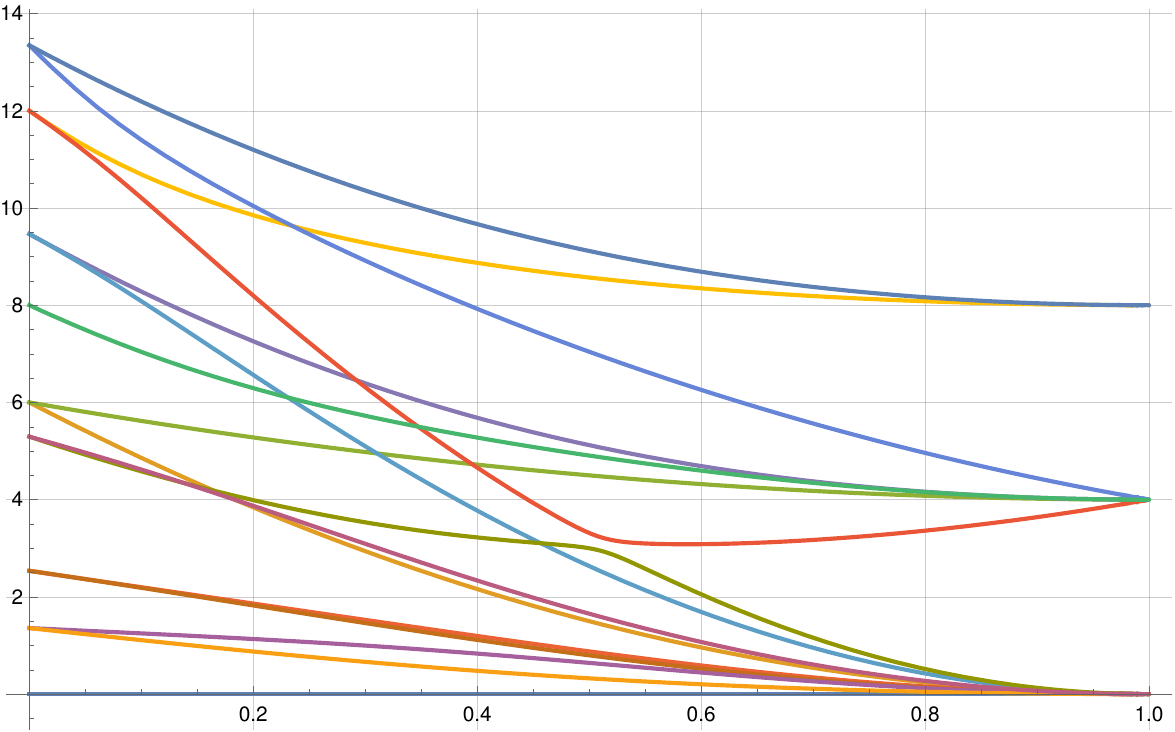}};
\node at (7.2,-3.9){$k$};\node at (-7.1,3.7){$E$};
\end{tikzpicture}
  \caption{The $L=4$ $\Zset_3$ holomorphic spin chain spectrum for the case $\kappa_1=\kappa_2=1-k,\kappa_3=1$. Notice that two states emanating from $E=12$ and $E=5.29$ experience an avoided crossing  at $k\simeq 0.51$.} \label{GraphZ3L4h}
  \end{center}
  \end{figure}

 It is interesting to consider the evolution of the spectrum as one deforms in the direction of SCQCD. One way to approach the limit is to take $\kappa_1=\kappa_2=\kappa=1-k,\kappa_3=1$. In the $k\ra 1$ limit, only the $\SU(N)$ group at node 3 remains while the other two become global. By $\Ncal=2$ supersymmetry, this gives the $\SU(N)$ theory with $2N$ flavours, plus additional decoupled vector multiplets. The evolution of the spectrum is given in Fig. \ref{GraphZ3L4h}. As discussed in \cite{Gadde:2010zi}, in the SCQCD limit several long multiplets of the interpolating theory are expected to reach unitarity bounds and fragment into short multiplets, something which is clearly seen in the plot. 

A noteworthy feature of Fig. \ref{GraphZ3L4h} is the presence of an avoided crossing as the deformation parameter increases. In general the energies involved are roots of an 8${}^{\mathrm{th}}$-order polynomial, but for this very symmetric deformation it factorises into two quartic ones. The polynomial relevant for the avoided crossing is
{\small\be
P(E)=E ^4-8 \left(3 \kappa ^2\!+\!1\right) E ^3\!+\!16 \left(11 \kappa ^4\!+\!9 \kappa ^2\!+\!1\right) E ^2-32 \left(14 \kappa ^4\!+\!17 \kappa ^2\!+\!8\right) \kappa ^2 E \!+\!128 \left(2 \kappa ^4\!+\!2 \kappa ^2\!+\!5\right) \kappa ^4
\ee}
which has roots at $(13.3456,12,5.29612,1.35823)$ at the orbifold point $\kappa=1$. The two middle roots are the ones which experience the avoided crossing.\footnote{Of course, the other $E=12$ eigenvalue at the orbifold point is the Konishi descendant in the $XY$ sector, and is therefore a root of the same cubic polynomial in (\ref{Z3charpoly}). It follows the exact same path as in Fig. \ref{GraphZ3L3hola}.}  By the von Neumann-Wigner eigenvalue-repulsion theorem \cite{vonNeumannWigner29}, such avoided crossings are what one expects to see in generic (i.e. non-integrable) quantum systems (see e.g. the book \cite{HaakeChaos} for a pedagogical introduction to eigenvalue repulsion). What is interesting in our case is that the two states which experience the avoided crossing are two-magnon states whose momenta can be tracked by the Bethe ansatz, so we will come back to discuss this feature at the end of the next section.  

Of course, Fig. \ref{GraphZ3L4h} also shows multiple level crossings. These are explainable by the large number of conserved quantities in this system, i.e. $\SU(2)_R$ spin, magnon number, and type of twisted sector (in general, twisted sectors mix away from the orbifold point, but in the symmetric deformation considered here some degeneracies are preserved). A full study of the eigenvalue distribution to confirm or not the presence of quantum chaos (as was done in \cite{McLoughlin:2022jyt} for the $\Ncal=1$ marginal deformation of $\Ncal=4$ SYM) would require focusing on sectors with identical quantum numbers, and thus going to higher lengths in order to acquire enough statistics. We leave this very interesting problem for future study.

\subsection{Two-magnon Bethe Ansatz} \label{Z3BetheAnsatz}

In this section we consider the $Z$-vacuum two-magnon problem in the holomorphic sector, for the $\Zset_3$ orbifold case. Unlike the $\Zset_2$ case studied in \cite{Gadde:2010zi}, two $X$ magnons on the $Z$ vacuum do not lead to a closeable state. However one $X$ and one $Y$ magnon do produce a closeable state, so we will focus on this case. Of course this implies that we need to be in an $\SU(3)$ sector.  

\subsubsection{Open chain}

As always, we will first consider magnon scattering on the open (asymptotic) chain and impose cyclicity later. As discussed, a single $X$ or $Y$ excitation can be seen as a defect separating two $Z$ vacua of different node index:
\be
\ket{\ell}_{i,i+1}^X= \cdots Z_i Z_i \underset{{\color{red} \ell}\phantom{AB}}{X_{i,i+1}} Z_{i+1} Z_{i+1}\cdots \;\;,\;\;
\ket{\ell}_{i,i-1}^Y= \cdots Z_i Z_i \underset{{\color{red} \ell}\phantom{AB}}{Y_{i,i-1}} Z_{i-1} Z_{i-1}\cdots \;\;,
\ee
where in this section $i=1,2,3$. We express the single-magnon state as
\be
\ket{\psi}^{(1)}_{i,j}=\sum_{\ell} e^{ip\ell} \ket{\ell}_{i,j}\;,
\ee
where $j=i\pm 1$ depending on the type of magnon. Acting with the Hamiltonian, the dispersion relation is easily seen to be
\be
E_{i,j}= 2(\kappa_i^2+\kappa_{j}^2)-2\kappa_{i}\kappa_{j}(e^{ip}+e^{-ip})\;.
\ee
Proceeding to two-magnon states, for a given external vacuum labelled by node index $i$, we have two distinct configurations:
\be \label{BetheZ3XY}
\ket{\ell_1,\ell_2}_{i}^{XY}=\cdots Z_i Z_i \underset{{\color{red}\ell_1}\phantom{AB}}{X_{i,i+1}} Z_{i+1} \cdots Z_{i+1} \underset{{\color{red}\ell_2}\phantom{AB}}{Y_{i+1,i}} Z_{i} Z_{i}\cdots
\ee
and 
\be \label{BetheZ3YX}
\ket{\ell_1,\ell_2}_{i}^{YX}=\cdots Z_i Z_i \underset{{\color{red}\ell_1}\phantom{AB}}{Y_{i,i-1}} Z_{i-1} \cdots Z_{i-1} \underset{{\color{red}\ell_2}\phantom{AB}}{X_{i-1,i}} Z_{i} Z_{i}\cdots
\ee
Let us, for concreteness, fix the exterior vacuum to be the $Z_1$ vacuum. Considering the non-interacting case where the two magnons are separated by more than one $Z$ field, we find the corresponding dispersion relations:
\be \label{EXY1}
E^{XY}_1=4(\kappa_{1}^2+\kappa_{2}^2)-2 \kappa_1 \kappa_{2}\left(e^{ip_1}+e^{-ip_1}+e^{ip_2}+e^{-ip_2}\right)\;,
\ee
and 
\be \label{EYX1}
E^{YX}_1=4(\kappa_{1}^2+\kappa_{3}^2)-2 \kappa_1 \kappa_{3}\left(e^{iq_1}+e^{-iq_1}+e^{iq_2}+e^{-iq_2}\right)\;.
\ee
The ``1'' subscript labels the exterior vacuum. We note that the dispersion relations are different if $\kappa_2\neq \kappa_3$. Therefore, for both of these configurations to live on the same chain (as they have to, as they will mix under the action of the Hamiltonian) we need to take the magnon momenta of each configuration to be different, here denoted by $p$ and $q$. We can of course always solve for $(q_1,q_2)$ in terms of $(p_1,p_2)$ by equating the total momentum $K=p_1+p_2=q_1+q_2$ and total energy $E^{XY}_1=E^{YX}_1$. We can now write the Bethe ansatz
combining (\ref{BetheZ3XY}) and (\ref{BetheZ3YX}):
\be \label{BetheZ32mag}
\begin{split}
  \ket{\psi}^{\text{2-mag}}_{1}&=\sum_{1\leq\ell_1<\ell_2\leq L}\left[\left(A_1 e^{i p_1\ell_1+i p_2\ell_2}+B_1 e^{ip_2 \ell_1+i p_1 \ell_2}\right)\ket{\ell_1,\ell_2}^{XY}_1\right.\\
    &\qquad\qquad\quad+
\left.\left(C_1 e^{i q_2\ell_1+i q_1\ell_2}+D_1 e^{iq_1 \ell_1+i q_2 \ell_2}\right)\ket{\ell_1,\ell_2}^{YX}_1\right]\;.
\end{split}
\ee
To find the interacting equations, consider all the terms in (\ref{BetheZ32mag}) which under the action of the Hamiltonian lead to $\ket{\ell_1,\ell_1+1}_1^{XY}$, i.e. the case where the $X$ and $Y$ magnons are next to each other, as well as all the terms which lead to $\ket{\ell_1,\ell_1+1}_1^{YX}$, where a $Y$ and $X$ magnon are next to each other. The resulting equations are 
\be\begin{split} \label{int1}
6\kappa_1^2&\left(A_1 e^{i p_2}+B_1 e^{i p_1}\right)-2\kappa_1\kappa_2\left(A_1 e^{-i p_1+i p_2}+B_1 e^{-ip_2+i p_1}\right)\\&-2\kappa_1\kappa_2\left(A_1 e^{2i p_2}+B_1 e^{2i p_1}\right)-2\kappa_1^2
\left(C_1 e^{+i q_1}+D_1 e^{i q_2}\right)\\&=E^{XY}_1 \left(A_1 e^{i p_2}+B_1 e^{i p_1}\right)\;,
\end{split}
\ee
and 
\be\begin{split} \label{int2}
6\kappa_1^2&\left(C_1 e^{i q_1}+D_1 e^{+i q_2 }\right)-2\kappa_1\kappa_3\left(C_1 e^{-i q_2+i q_1}+D_1 e^{-iq_1+i q_2 }\right)\\
&-2\kappa_1\kappa_3\left(C_1 e^{2i q_1}+D_1 e^{i q_2}\right)-2\kappa_1^2\left(A_1 e^{i p_2}+B_1 e^{i p_1}\right)\\
&=E^{YX}_1\left(C_1 e^{i q_1}+D_1 e^{i q_2 }\right)\;.
\end{split}
\ee
As always, it is convenient to express the solution of these equations in terms of an $S$-matrix, relating the incoming waves (which we take to be $A,D$, i.e. the modes with momentum $p_1,q_1$ at site $\ell_1$) to the outgoing ones ($B,C$, i.e. those with momentum $p_2,q_2$ at $\ell_1$). Hence, we can write (with the 1's still indicating the $Z_1$ exterior vacuum)
\be
\doublet{A_1}{D_1}=\twobytwo{S_{AB}^{(1)}}{S_{AC}^{(1)}}{S_{DB}^{(1)}}{S_{DC}^{(1)}} \;\; \doublet{B_1}{C_1}\;,
\ee
and solving (\ref{int1},\ref{int2}) we obtain
\be\begin{split} \label{SmatZ3}
S_{AB}^{(1)}&=1+(e^{ip_1}-e^{ip_2})(\kappa_1\kappa_3(\kappa_1^2-2\kappa_2^2)(1+e^{i(q_1+q_2)})-2e^{iq_2}(\kappa_1^2(\kappa_2^2+\kappa_3^2)-2\kappa_2^2\kappa_3^2))/\Dcal_1\;,\\
S_{AC}^{(1)}&=\kappa_1^3\kappa_3e^{i(q_1+q_2)} \left(e^{iq_1}-e^{-iq_1}-e^{iq_2}+e^{-iq_2}\right)/\Dcal_1\;,\\
S_{DB}^{(1)}&=\kappa_1^3\kappa_2e^{i(p_1+p_2)} \left(e^{ip_1}-e^{-ip_1}-e^{ip_2}+e^{-ip_2}\right)/\Dcal_1\;,\\
S_{DC}^{(1)}&=1+(e^{iq_1}-e^{iq_2})(\kappa_1\kappa_2(\kappa_1^2-\kappa_3^2)(1+e^{i(p_1+p_2)})+e^{ip_2}(4\kappa_2^2\kappa_3^2-2\kappa_1^2(\kappa_2^2+\kappa_3^2)))/\Dcal_1\;,
\end{split}\ee
where
\be\begin{split}
\mathcal{D}_1&=(1+e^{i(p_1+p_2)})(1+e^{i(q_1+q_2)})\kappa_1^2\kappa_2\kappa_3+e^{ip_2}(1+e^{iq_1+iq_2})\kappa_1(\kappa_1^2-2\kappa_2^2)\kappa_3\\
&\quad +e^{iq_2}(1+e^{ip_1+ip_2})\kappa_1\kappa_2(\kappa_1^2-2\kappa_3^2)-2e^{ip_2+iq_2}(\kappa_1^2(\kappa_2^2+\kappa_3^2-2\kappa_2^2\kappa_3^2))\;.
\end{split}\ee
We note that, despite appearances, the $S$-matrix only depends on one pair of momenta, as $(q_1,q_2)$ are uniquely determined by $(p_1,p_2)$ through the total energy and momentum conditions. Defined in this way, the $S$-matrix can be seen to satisfy\footnote{An alternative way of writing this is  
$
|C_1|^2-|D_1|^2=\frac{\kappa_2(\sin p_2-\sin p_1)}{\kappa_3(\sin q_2-\sin q_1)}\left(|A_1|^2-|B_1|^2\right)
$.}
\be
(S^{(1)})^*(p_1,p_2) S^{(1)}(p_1,p_2)=I_{2\times 2}\;.
\ee
This non-unitary condition is due to the momentum mismatch between the states before and after scattering, due to the different interior vacua.\footnote{The situation is similar to quantum-mechanical scattering off of a step potential, where one needs to account for the different velocities of the incident and transmitted particles.} A more careful analysis should lead to an improved $S$-matrix, but this already achieves our goal of relating the Bethe coefficients, and is sufficient for the closed-chain scattering we are ultimately interested in. 
At the orbifold point $\kappa_i=1$, the dispersion relations become the same, the $q_{1,2}$ momenta become equal to the $p_{1,2}$ momenta and the $S$-matrix reduces to
\be \label{SmZ3op}
S_{\text{o.p.}}^{(1)}=\frac{1}{1-2e^{ip_2}+e^{i(p_1+p_2)}}\twobytwo{-(1-e^{ip_1})(1-e^{ip_2})}{e^{ip_1}-e^{ip_2}}{e^{ip_1}-e^{ip_2}}{-(1-e^{ip_1})(1-e^{ip_2})}\;,
\ee
which, being symmetric, does satisfy $(S^{(1)}_\text{o.p.})^\dag S^{(1)}_{\text{o.p.}}=I_{2\times 2}$. 

The $S$-matrices for the other two external vacua are easily obtained by applying the $\Zset_3$ symmetry, as discussed in more detail in the next section. 

\subsubsection{Closed chain}

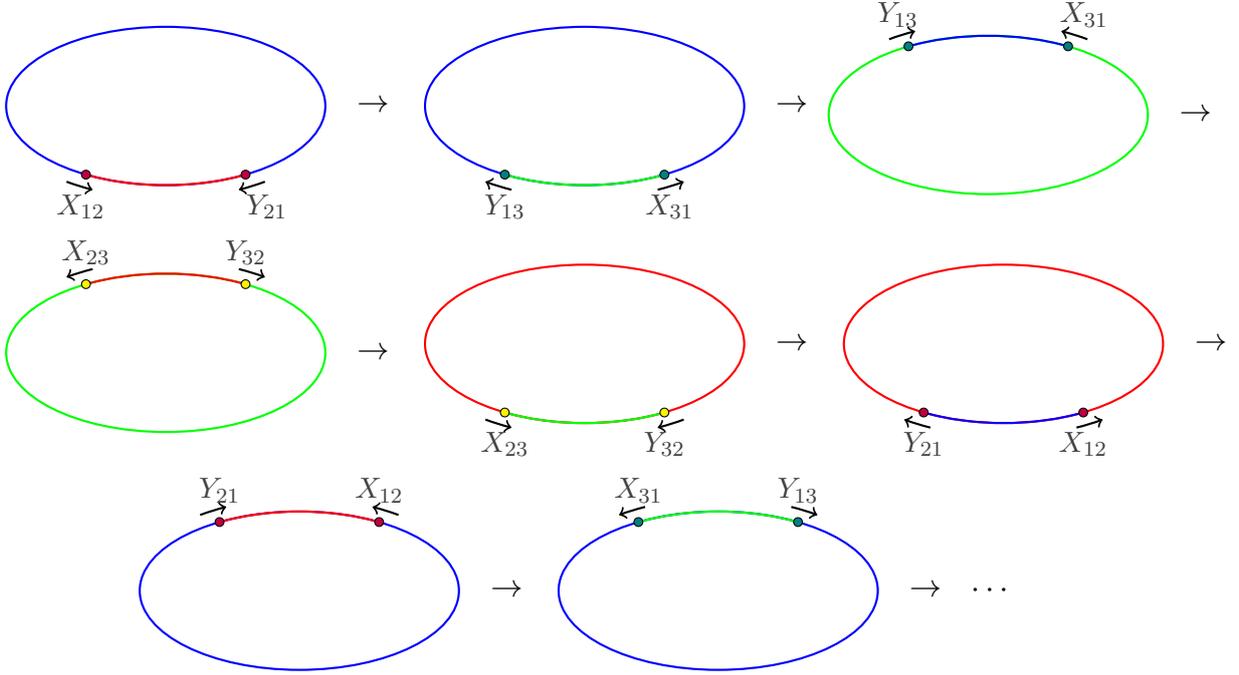
\begin{figure}[ht]
\begin{center}
  \begin{tikzpicture}[scale=0.65]
  \draw[thick, blue, -] (0,0) [partial ellipse=0:360:3cm and 1.5cm];
  \draw[thick, red, -] (0,0) [partial ellipse=240:300:3cm and 1.5cm];
  \draw[thick,->] (0,0) [partial ellipse=235:245:3.25cm and 1.75cm];
  \draw[thick,->] (0,0) [partial ellipse=305:295:3.25cm and 1.75cm];
\draw[fill=purple] (-1.5,-1.3) circle (.5ex);\draw[fill=purple] (1.5,-1.3) circle (.5ex);
  \node at (-1.6,-1.9) {\color{darkgray}$X_{12}$};\node at (1.9,-1.9) {\color{darkgray}$Y_{21}$};
\node at (3.9,0) {\large $\rightarrow$};
\end{tikzpicture}
\begin{tikzpicture}[scale=0.65]
  \draw[thick, blue, -] (0,0) [partial ellipse=0:360:3cm and 1.5cm];
  \draw[thick, green, -] (0,0) [partial ellipse=240:300:3cm and 1.5cm];
  \draw[thick,->] (0,0) [partial ellipse=245:235:3.25cm and 1.75cm];
  \draw[thick,->] (0,0) [partial ellipse=295:305:3.25cm and 1.75cm];
  \draw[fill=teal] (-1.5,-1.3) circle (.5ex);\draw[fill=teal] (1.5,-1.3) circle (.5ex);
  \node at (-1.5,-1.9) {\color{darkgray}$Y_{13}$};\node at (1.6,-1.9) {\color{darkgray}$X_{31}$};
\node at (3.9,0) {\large $\rightarrow$};
\end{tikzpicture}
\begin{tikzpicture}[scale=0.65,baseline=-1.5cm]
  \draw[thick, green, -] (0,0) [partial ellipse=0:360:3cm and 1.5cm];
  \draw[thick, blue, -] (0,0) [partial ellipse=60:120:3cm and 1.5cm];
  \draw[thick,->] (0,0) [partial ellipse=55:65:3.25cm and 1.75cm];
  \draw[thick,->] (0,0) [partial ellipse=125:115:3.25cm and 1.75cm];
    \draw[fill=teal] (-1.5,1.3) circle (.5ex);\draw[fill=teal] (1.5,1.3) circle (.5ex);
  \node at (-1.7,1.9) {\color{darkgray}$Y_{13}$};\node at (1.8,1.9) {\color{darkgray}$X_{31}$};
\node at (3.9,0) {\large $\rightarrow$};
\end{tikzpicture}

\begin{tikzpicture}[scale=0.65,baseline=-1.5cm]
  \draw[thick, green, -] (0,0) [partial ellipse=0:360:3cm and 1.5cm];
  \draw[thick, red, -] (0,0) [partial ellipse=60:120:3cm and 1.5cm];
  \draw[thick,->] (0,0) [partial ellipse=65:55:3.25cm and 1.75cm];
  \draw[thick,->] (0,0) [partial ellipse=115:125:3.25cm and 1.75cm];
      \draw[fill=yellow] (-1.5,1.3) circle (.5ex);\draw[fill=yellow] (1.5,1.3) circle (.5ex);
      \node at (-1.5,1.9) {\color{darkgray}$X_{23}$};\node at (1.5,1.9) {\color{darkgray}$Y_{32}$};
      \node at (3.9,0) {\large $\rightarrow$};
\end{tikzpicture}
\begin{tikzpicture}[scale=0.65]
  \draw[thick, red, -] (0,0) [partial ellipse=0:360:3cm and 1.5cm];
  \draw[thick, green, -] (0,0) [partial ellipse=240:300:3cm and 1.5cm];
  \draw[thick,->] (0,0) [partial ellipse=235:245:3.25cm and 1.75cm];
  \draw[thick,->] (0,0) [partial ellipse=305:295:3.25cm and 1.75cm];
 \draw[fill=yellow] (-1.5,-1.3) circle (.5ex);\draw[fill=yellow] (1.5,-1.3) circle (.5ex);
  \node at (-1.5,-1.9) {\color{darkgray}$X_{23}$};\node at (1.5,-1.9) {\color{darkgray}$Y_{32}$};
\node at (3.9,0) {\large $\rightarrow$};
\end{tikzpicture}
\begin{tikzpicture}[scale=0.65]
  \draw[thick, red, -] (0,0) [partial ellipse=0:360:3cm and 1.5cm];
  \draw[thick, blue, -] (0,0) [partial ellipse=240:300:3cm and 1.5cm];
  \draw[thick,->] (0,0) [partial ellipse=245:235:3.25cm and 1.75cm];
  \draw[thick,->] (0,0) [partial ellipse=295:305:3.25cm and 1.75cm];
   \draw[fill=purple] (-1.5,-1.3) circle (.5ex);\draw[fill=purple] (1.5,-1.3) circle (.5ex);
   \node at (-1.5,-1.9) {\color{darkgray}$Y_{21}$};\node at (1.5,-1.9) {\color{darkgray}$X_{12}$};
   \node at (3.9,0) {\large $\rightarrow$};
\end{tikzpicture}

\begin{tikzpicture}[scale=0.65,baseline=-1.5cm]
  \draw[thick, blue, -] (0,0) [partial ellipse=0:360:3cm and 1.5cm];
  \draw[thick, red, -] (0,0) [partial ellipse=60:120:3cm and 1.5cm];
  \draw[thick,->] (0,0) [partial ellipse=55:65:3.25cm and 1.75cm];
  \draw[thick,->] (0,0) [partial ellipse=125:115:3.25cm and 1.75cm];
   \draw[fill=purple] (-1.5,1.3) circle (.5ex);\draw[fill=purple] (1.5,1.3) circle (.5ex);
  \node at (-1.5,1.9) {\color{darkgray}$Y_{21}$};\node at (1.5,1.9) {\color{darkgray}$X_{12}$};
\node at (3.9,0) {\large $\rightarrow$};
\end{tikzpicture}
\begin{tikzpicture}[scale=0.65,baseline=-1.5cm]
  \draw[thick, blue, -] (0,0) [partial ellipse=0:360:3cm and 1.5cm];
  \draw[thick, green, -] (0,0) [partial ellipse=60:120:3cm and 1.5cm];
  \draw[thick,->] (0,0) [partial ellipse=65:55:3.25cm and 1.75cm];
  \draw[thick,->] (0,0) [partial ellipse=115:125:3.25cm and 1.75cm];
  \draw[fill=teal] (-1.5,1.3) circle (.5ex);\draw[fill=teal] (1.5,1.3) circle (.5ex);
  \node at (-1.5,1.9) {\color{darkgray}$X_{31}$};\node at (1.5,1.9) {\color{darkgray}$Y_{13}$};
\node at (3.9,0) {\large $\rightarrow$};
\end{tikzpicture}
\begin{tikzpicture}[scale=0.65,baseline=-1.5cm]
  \node at (2,0) {\large $\cdots$};
  \node at (3.9,0) {\large\mbox{}};
\end{tikzpicture}\hspace{2cm}
\end{center}
\caption{A schematic depiction of 2-magnon scattering on the $\Zset_3$ chain. The domains coloured blue, red and green correspond to the $Z_1,Z_2$ and $Z_3$-vacuum, respectively. For clarity, only the transmitted modes are depicted. One also needs to consider reflection where the same fields move away from each interaction point (i.e. only the momenta are exchanged). } \label{Fig:Z3scattering}
\end{figure}

In order to consider the closed chain, we note that as two magnons with an initial $Z_1$ exterior vacuum scatter, when they meet again at the back of the chain the exterior vacuum will be $Z_3$, and on scattering again the interior vacuum will be $Z_2$. Further scatterings will produce all possible permutations of exterior and interior vacua, as depicted schematically in Figure \ref{Fig:Z3scattering}.  We therefore need to include those configurations in the Bethe ansatz. Still labelling the 2-magnon energies by the exterior vacua, it is easy to check that $E^{XY}_1=E^{YX}_3$, $E^{XY}_2=E^{YX}_1$ and $E^{XY}_3=E^{YX}_2$, so on top of (\ref{EXY1}) and (\ref{EYX1}) we only need one additional dispersion relation, with its associated set of additional momenta. We take it to be
\be
E^{XY}_2=4(\kappa_{2}^2+\kappa_{3}^2)-2 \kappa_2 \kappa_{3}\left(e^{ir_1}+e^{-ir_1}+e^{ir_2}+e^{-ir_2}\right)\;,
\ee
where the additional momenta are called $(r_1,r_2)$. They are of course not independent of $(p_1,p_2)$ as they are uniquely given by solving $r_1+r_2=p_1+p_2$ and $E^{XY}_2=E^{XY}_1$.

To find the Bethe Ansatz and corresponding $S$-matrices for the sectors with exterior $Z_2$ and $Z_3$ vacua, one takes advantage of the $\Zset_3$ symmetry of the problem, which acts as $(A_i,B_i,C_i,D_i,\kappa_i)\ra (A_{i+1},B_{i+1},C_{i+1},D_{i+1},\kappa_{i+1})$ as well as
\be
(p_1,p_2)\overset{\Zset_3}\longrightarrow (r_1,r_2)\overset{\Zset_3}\longrightarrow (q_1,q_2)\;.
\ee
So the solution in each sector is just given by the appropriate $\Zset_3$ conjugate of (\ref{BetheZ32mag}) and (\ref{SmatZ3}). Now recall that for gauge theory applications we need to impose not just periodicity but also cyclicity, as these spin chain states correspond to trace operators. Specialising to $L=3$ for simplicity, this implies that the following three states need to be equal:
\be
\begin{split} \label{Z3cyclicity1}
X_{12}Y_{21} Z_1 &\ra A_1 e^{ip_1+2ip_2}+B_1 e^{ip_2+2ip_1}\;,\\
Z_1 X_{12}Y_{21} &\ra A_1 e^{2ip_1+3ip_2}+B_1 e^{2ip_2+3ip_1} \;,\\
Y_{21} Z_1 X_{12} &\ra C_2 e^{ip_2+3ip_1}+D_2 e^{ip_1+3ip_2}\;,
\end{split}
\ee
as well as their $\Zset_3$ conjugates as described above, and similarly
\be
\begin{split} \label{Z3cyclicity2}
Z_2 Y_{21} X_{12}&\ra C_2 e^{2i p_2+3ip_1}+D_2e^{2ip_1+3ip_2}\;,\\
Y_{21}X_{12}Z_2&\ra C_2e^{ip_2+2ip_1}+D_2e^{ip_1+2ip_2}\;,\\
X_{12}Z_2 Y_{21}&\ra A_1 e^{ip_1+3ip_2}+B_1 e^{ip_2+3ip_1}\;,
\end{split}
\ee
and their $\Zset_3$ conjugates. Equating the first two equations of each set imposes the condition that the centre-of-mass momentum vanishes,
\be \label{BAcom}
K=p_1+p_2=r_1+r_2=q_1+q_2=0\;,
\ee
while the first and third equations impose the Bethe ansatz relations
\be \label{Z3cyclicityfinal}
A_1=C_2 e^{3ip_1}\;,\;\;B_1=D_2 e^{-3ip_1}\;,
\ee
and their $\Zset_3$ conjugates. Combining these equations with the solution of (\ref{int1}) and (\ref{int2}), and relating the $r$ and $q$ momenta to the $p$ momenta through (\ref{BAcom}) and the energy relations $E_1^{XY}=E_2^{XY}=E_3^{XY}$, finally leads to the momenta $p_1$ characterising the solutions of the two-magnon problem.

At the orbifold point the solutions can be found analytically and are listed in Table \ref{tab:Z3BAop}. In this case it is important to keep all solutions for the momenta, namely
\be
r_1=\pm p_1 \;,\; q_1=\pm p_1\;,
\ee
as the different twisted sectors are distinguished by different choices of sign.

\begin{table}[ht]
  \begin{center}
  \begin{tabular}{|c|c|c|}\hline  $p_1$ & E & Sector \\\hline 
    ``0'' & 0 & Untwisted\\
    $\frac{2\pi}{3}$ & 12 & Untwisted\\
    $\arctan(\sqrt{15-8\sqrt{3}})$ & $2(3-\sqrt{3})$ & $a$-twisted\\ 
    $\arctan(\sqrt{15-8\sqrt{3}})$ & $2(3-\sqrt{3})$ &$a^2$-twisted\\ 
    $\pi-\arctan(\sqrt{15+8\sqrt{3}})$ & $2(3+\sqrt{3})$ & $a$-twisted\\ 
    $\pi-\arctan(\sqrt{15+8\sqrt{3}})$ & $2(3+\sqrt{3})$ &$a^2$-twisted\\\hline
  \end{tabular}\caption{The $\Zset_3$ $L=3$ $XYZ$-sector momenta and energies at the orbifold point. We write ``0'' for the E=0 state as for that value $p_2=p_1$ and the state should be thought of as a descendant of the 0-magnon state. The twisted sector states have the same $p_1$ but differ by signs in the $q$ and $r$ momenta.}  \label{tab:Z3BAop}
\end{center}
\end{table}

The first state, with $E=0$, is the BPS state
\be
\Ocal=\Tr_1(X_{12}Y_{21}Z_1+X_{12}Z_2Y_{21})+\Tr_2(X_{23}Y_{32}Z_2+X_{23}Z_3Y_{32})+\Tr_3(X_{31}Y_{13}Z_3+X_{31}Z_1Y_{13})\;,
\ee
while the second, with $E=12$, is the superpotential 
\be
\Ocal=\Tr_1(X_{12}Y_{21}Z_1-X_{12}Z_2Y_{21})+\Tr_2(X_{23}Y_{32}Z_2-X_{23}Z_3Y_{32})+\Tr_3(X_{31}Y_{13}Z_3-X_{31}Z_1Y_{13})\;.
\ee
The states with $E=2(3\pm \sqrt{3})$ belong to the twisted sector. We note that at the orbifold point the two twisted sectors corresponding to the $\Zset_3$ elements $a$ and $a^2$ are degenerate. To obtain all the states once, we need to restrict to momenta $0\leq p_1 <\pi$. The value $p_1=\pi$ is a solution of the BA equations, with energy $E=16$, however it is spurious: One can check that for this value the $S$-matrix (\ref{SmZ3op}) is equal to minus the identity matrix, and substituting in the Bethe ansatz wavefunction one finds that the state vanishes. Besides this value, all other solutions are physical at the orbifold point and are listed in Table \ref{tab:Z3BAop}.

\begin{figure}[t]
  \begin{center}
\begin{tikzpicture}
  \node at (0,0){\includegraphics[width=6.5cm]{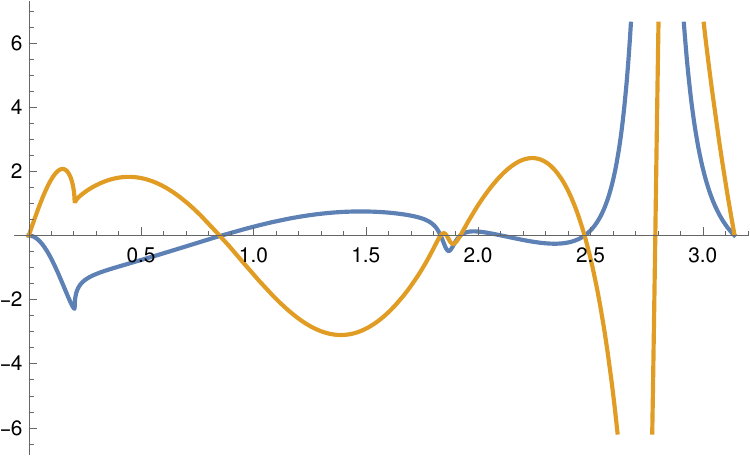}};
  \node[draw] at (-0.3,1.8){\includegraphics[width=2cm]{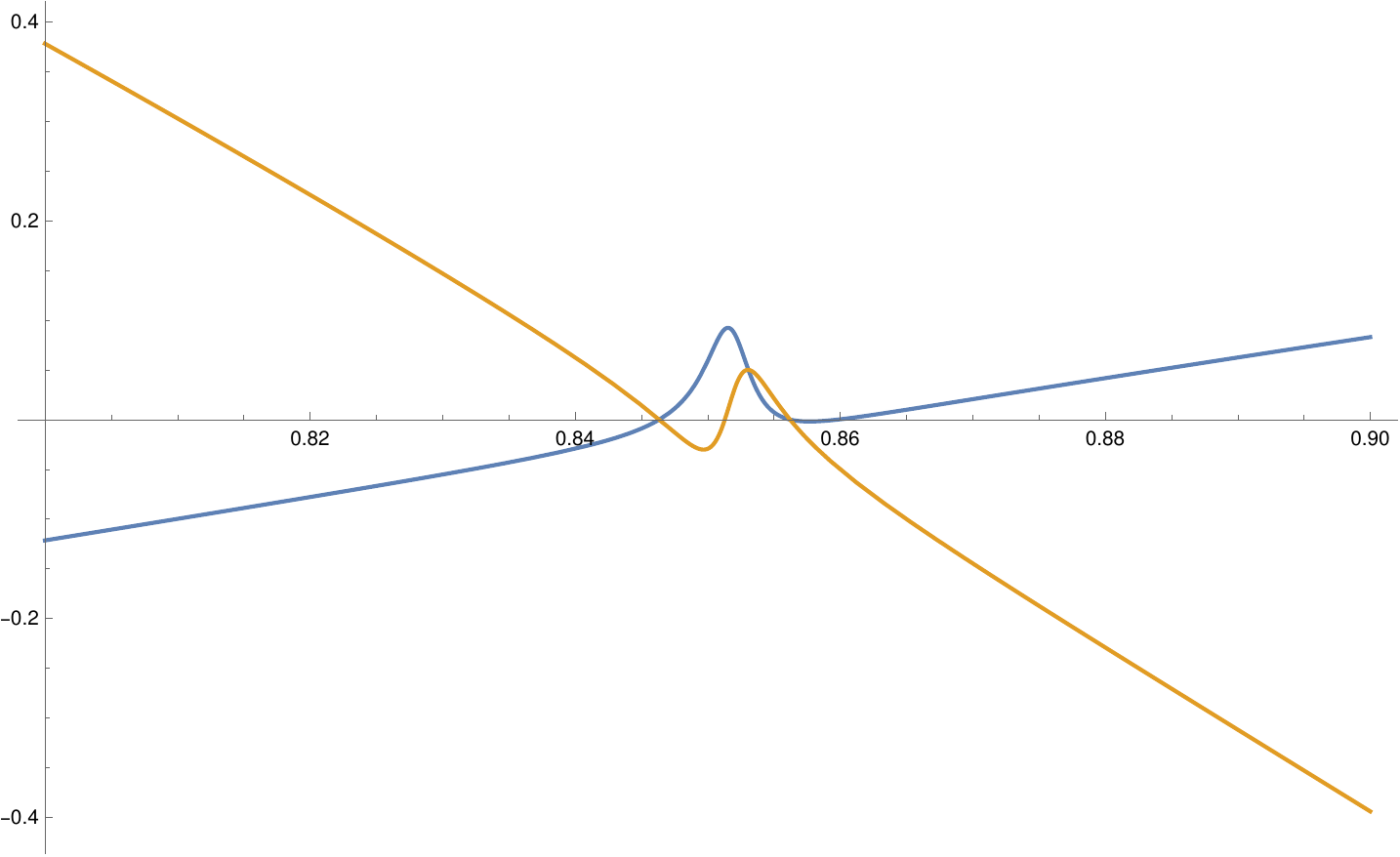}};
    \node at (3.2,-0.4){$p_1$};\node at (-3.3,1.9) {$f$};\node at (0,-2.2){$(a)$};\draw[->,thick,dotted,gray] (-1.2,0.2)--(-0.4,0.9);
    \end{tikzpicture}\hspace{0.4cm}\begin{tikzpicture}
    \node at (0,0){\includegraphics[width=6.5cm]{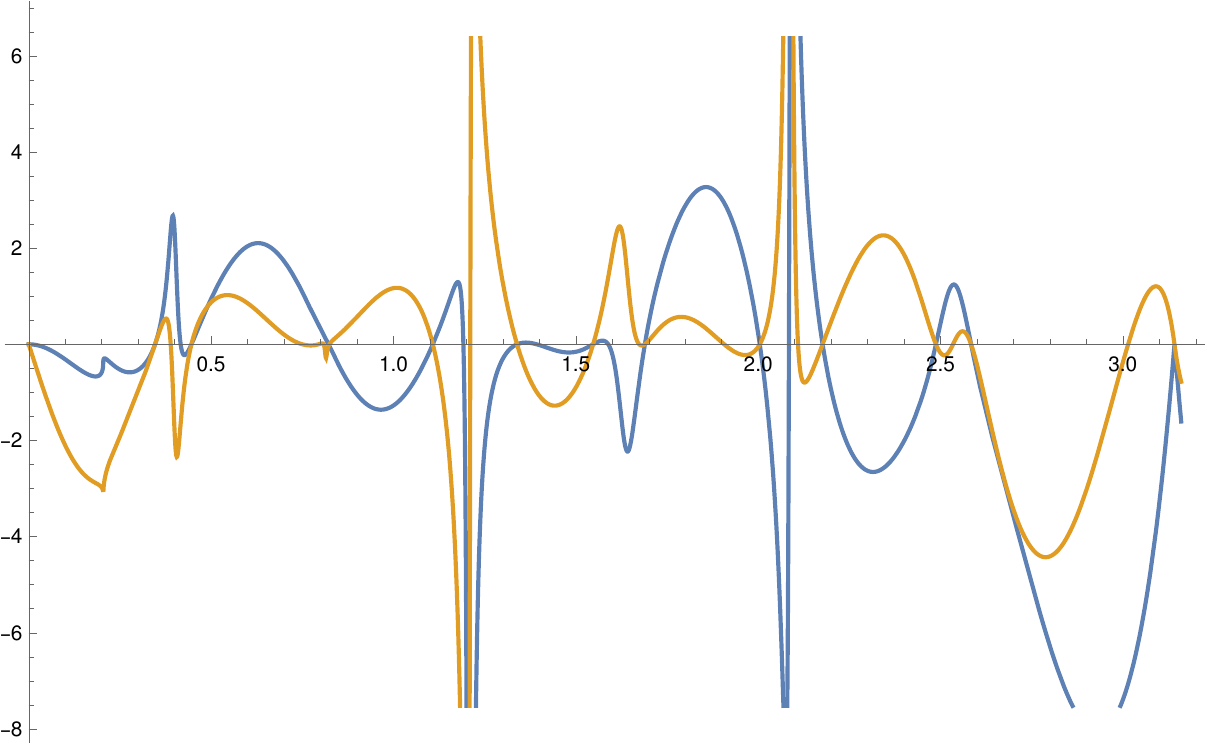}};
    \node at (3.1,-0.4){$p_1$};\node at (-3.3,1.9) {$f$};\node at (0,-2.2){$(b)$};
    \end{tikzpicture}
  \end{center}
  \caption{Plot of the real (blue) and imaginary (orange) part of the function $f=B_3-D_1e^{-Li q_1}$, for $\Zset_3$ with $(\kappa_1,\kappa_2,\kappa_3)=(0.8,0.9,1)$,  after substituting all the coefficients and expressing the momenta in terms of $p_1$. (a) is for $L=3$ while (b) is for $L=6$. The inset in (a) resolves the two roots close to $p_1=0.85$.} \label{Fig:Bethezeroes}
  \end{figure}

It is straightforward to numerically solve the corresponding equations for any marginally deformed case. To better visualise the solution, it is convenient to use the interacting equations ((\ref{int1},\ref{int2}) and their $\Zset_3$ conjugates) to solve for six of the coefficients (e.g. $C_i,D_i$ in terms of $A_i,B_i$), then use five of the cyclicity conditions (\ref{Z3cyclicityfinal}) to solve for the five remaining coefficients (as one of them, e.g. $A_1$, can always be set to 1), and substitute in the sixth condition (in this case, $D_1=B_3e^{3i q_1}$). Expressing the $q$ and $r$ momenta in the centre-of-mass frame in terms of $p_1$, one finally obtains a function of just $p_1$, whose zeroes are the Bethe momenta. For the choices above, this function is shown in (a) of Fig. \ref{Fig:Bethezeroes}. Noting where the real and imaginary parts  simultaneously vanish, one can then provide appropriate input to a routine such as \texttt{FindRoot} and extract the precise roots.

Table \ref{tab:Z3BAdef} provides a comparison of the values one obtains at the orbifold point and a sample deformed case for $L=3$. All the energy values agree with explicit diagonalisation of the Hamiltonian. Note the splitting of the twisted-sector momenta and energies as one deforms away from the orbifold point. The approach described above now leads to two spurious roots in the deformed case, at $p_1=0$ and $p_1=\pi$ (as can be seen in Fig. \ref{Fig:Bethezeroes}). It is straightforward to check that for these momenta the coefficients $A_i,B_i$ are either zero or minus each other, and the same for the $C_i,D_i$, and therefore the Bethe wavefunction vanishes. All other roots in the region $0<p_1<\pi$ are physical. 

\begin{table}[h]
\begin{center}
  \begin{tabular}{|cc|cc|}\hline \multicolumn{2}{|c|}{$(\kappa_1,\kappa_2,\kappa_3)=(1,1,1)$} & \multicolumn{2}{c|}{$(\kappa_1,\kappa_2,\kappa_3)=(0.8,0.9,1)$}\\  \hline $p_1$ & E & $p_1$ & E \\\hline 
    ``0'' & 0 & $0.1179 i$ & 0\\
    2.0944 & 12 & 2.4737& 10.3225\\
    0.8189 & 2.5359 & 0.8463 & 1.9827\\
    0.8189 & 2.5359 & 0.8562 &2.02549\\
    1.7538 & 9.4641 & 1.9207 & 7.7745\\
    1.7538 & 9.4641 & 1.8333& 7.2948 \\ \hline
    \end{tabular}\caption{A comparison of the $\Zset_3$ $L=3$ $XYZ$-sector momenta and energies at the orbifold point and a sample deformation, corresponding to $k=0.1$ in Fig. \ref{GraphZ3L2}.}\label{tab:Z3BAdef}   
\end{center}
\end{table}
The XYZ BPS state which was at $p_1=0$ at the orbifold point now corresponds to an imaginary value of $p_1$, which cannot be obtained directly via the above approach, as the $S$-matrix vanishes in that case. But of course the imaginary momentum of this state can be found by directly solving the condition $E^{XY}_1=0$
\be \label{pBPS}
p_1=i\ln\frac{\kappa_1}{\kappa_2}\;,
\ee
and is also compatible with the other momenta leading to the same energy:
\be \label{qrp}
q_1=i\ln\frac{\kappa_3}{\kappa_1}\;\;,\;\;r_1=i\ln\frac{\kappa_2}{\kappa_3}\;.
\ee
Such momenta are familiar from the study of the XXZ model (see e.g. \cite{Gomez96}), which is not surprising given the similarity of the $Z$-vacuum single-magnon dispersion relations to those of XXZ (as already noticed in \cite{Gadde:2010zi} for the $\Zset_2$ case). Of course, the details of magnon scattering in this model are very different from XXZ.

Substituting the above values in the Bethe ansatz, it is easy to reproduce the BPS states (\ref{Z3L3holBPS}) and (\ref{Z3L4holBPS}), by first noticing that for the momenta (\ref{pBPS}) corresponding to these states the $B_i$ coefficients vanish, so only the plane wave parts are relevant. For instance, in (\ref{Z3L4holBPS}), one correctly finds that the ratio of the first to the second term should be
\be
\frac{e^{ip_13+ip_24}}{e^{ip_12+ip_24}}=e^{ip_1}=\frac{\kappa_2}{\kappa_1}\;.
\ee
The ratios to the remaining terms (depending on  $q$ or $r$ momenta) can also be found, by using (\ref{qrp}). 

In the above discussion we specified $L=3$ in order to explicitly compare with Section \ref{Section:Z3L3}. Of course, the solution can be extended for any $L$ by suitably modifying the conditions (\ref{Z3cyclicity1}) and (\ref{Z3cyclicity2}), which simply gives $3\ra L$ in (\ref{Z3cyclicityfinal}). For $L$ large, these states correspond to near-BMN operators \cite{Berenstein:2002jq}, which have been studied in the ADE quiver context in \cite{Oz:2002wy}.

It is important to verify whether the Bethe ansatz can correctly account for all the 2-magnon states in a higher-length example. For concreteness, let us consider $L=6$, where by directly diagonalising the Hamiltonian one finds 15 2-magnon states. Solving in the same way as above for a specific deformation, we obtain the function shown in (b) of Fig. \ref{Fig:Bethezeroes}. The roots at $p_1=0$ and $p_1=\pi$ are again spurious and are excluded. Apart from the usual imaginary momentum of the BPS state, one finds 12 real roots in the region $0<p_1<\pi$, however the final two roots are now complex, with real value $\pi$. They are not visible from the plot, but are easily found using \texttt{FindRoot}. The resulting energies are listed in Table \ref{Table:L6Z3Bethe} and are of course in precise agreement with explicit diagonalisation. 

\begin{table}[h]
\begin{center}
  \begin{tabular}{|cc|cc|cc|}\hline 
    $p_1$ & $E$ &$p_1$ &$E$ &$p_1$ &$E$ \\ \hline
    $0.1179i$ & 0       & 1.10800 & 3.22823 & 2.17675 & 9.08058 \\
    0.34635 & 0.38204 & 1.3389 & 4.47621  & 2.48748 & 10.3711 \\
    0.44614 & 0.60379 & 1.54754 & 5.6660  & 2.58395 & 10.6874 \\
    0.80859 & 1.82244 & 1.68941 & 6.4816  & $\pi+0.37927i$ & 11.9793 \\
    0.82452 & 1.88949 & 2.00446 & 8.22034 & $\pi+0.76133i$ & 13.3115 \\ \hline
  \end{tabular}
\end{center}\caption{The $\Zset_3$-quiver two-magnon Bethe momenta and energies for $L=6$ and $(\kappa_1, \kappa_2, \kappa_3)=(0.8,0.9,1)$. For the complex roots, we have chosen the imaginary part to be positive.} \label{Table:L6Z3Bethe}
\end{table}

It should be noted that the complex values appearing in this case seem to be an artifact of having solved for $p_1$. One could equally well have solved for $q_1$ or $r_1$, and it turns out that in this case the $r_1$ values for all the states (apart from the $E=0$ state) are real. Specifically, for $(\kappa_1,\kappa_2, \kappa_3)=(0.8,0.9,1)$, one finds
\be
p_1=\pi+0.76133i \;\Rightarrow \; q_1=\pi+0.32994i \;,\;r_1=2.57413  \;\; 
\ee
Considering several other lengths and deformations, this appears to be a general feature: When some of the highest two-magnon energies are higher that what can be obtained from some of the dispersion relations with real momenta, the corresponding momentum will take a complex value, but one can always switch to one of the other dispersion relations, where all energies can be accessed with real momenta.\footnote{It should be noted that at the orbifold point, where the $p$, $q$, $r$ momenta are equal up to signs, all the momenta are real, and similarly for small deviations from the orbifold point, where all three dispersion relations can reproduce all the energies with real momenta.}

The above analysis (which we have repeated for other lengths and choices of $\kappa_i$) indicates that, as usual, the Bethe ansatz includes some spurious solutions, which however are easily identified as they simply correspond to $p_1=0,\pi$. Otherwise, all the two-magnon states are correctly accounted for and appear only once if we restrict to the region $0<p_1<\pi$ for the real roots, and positive imaginary part for the complex roots. As discussed above, the BPS state (where the $S$-matrix vanishes) can be explicitly seen analytically. We conclude that the Bethe ansatz is complete at the two-magnon level for generic deformations of the $\Zset_3$ orbifold.

\begin{figure}[h]
  \begin{center}
\begin{tikzpicture}
    \node at (0,0){\includegraphics[width=8cm]{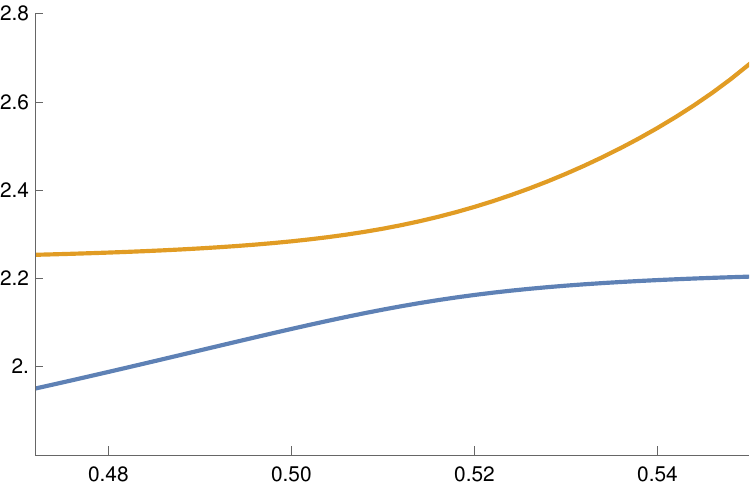}};
\node at (4.3,-2.2){$k$};
    \node at (3,2){$p_1^{(12)}$};
    \node at (3.2,0.1){$p_1^{(5.29)}$};
    \end{tikzpicture}
  \caption{The $p_1$ momenta of the two holomorphic $L=4$ spin chain states which experience the avoided crossing in Fig. \ref{GraphZ3L3hola}, as a function of $k$, where $\kappa_1=\kappa_2=1-k,\kappa_3=1$. We focus on a small region surrounding the closest approach. The higher momentum describes the state starting at $E=12$ at the orbifold point, while the lower one the state starting at $E=5.29$ at the orbifold point.} \label{fig:Avoided}
  \end{center}
  \end{figure}

As mentioned in the previous section, the states which experience an avoided crossing at $L=4$ (see Fig. \ref{GraphZ3L4h}) are two-magnon states, and we can therefore find their associated Bethe momenta. These are plotted in Fig. \ref{fig:Avoided}. Effectively, in this sector one can rephrase the usual energy eigenvalue repulsion as repulsion between the Bethe momenta. It is rather remarkable that a feature typically taken as a signature of non-integrability can be observed in this explicitly solvable Bethe ansatz setting. 

Extending the above Bethe ansatz approach to more magnons should be possible by following the approach of \cite{Bozkurt:2024tpz,Bozkurt:2025exl} for the $\Zset_2$ case, suitably adapted to $XY$-magnon scattering. However, one advantage of the $\Zset_3$ quiver compared to $\Zset_2$ is that three $X$ magnons lead to a closeable state, while for the $\Zset_2$ case, to compare with the physical spectrum one has to consider four $X$ magnons \cite{Bozkurt:2025exl}. We did not consider purely $X$ magnon scattering here, as we are focusing on the generic closeable case, but it should be a straightforward and relevant construction.

\section{Example: The $\Dfour$ theory} \label{sec:D4}

For our second example, we will consider the orbifold by $\hat{\mathrm{D}}_4$, the binary dihedral group of order $8$. This will be our first example of a spin chain based on a non-abelian orbifold. We refer to \cite{Feng:2000af} for more details on such orbifolds. It has the presentation
\be
\{a,b | a^{4}=e,b^2=a^2=z, bab^{-1}=a^{-1}\}\;.
\ee
where $z$ is a central element. The Cayley table is shown in Table \ref{tab:DfourCayleytable}.
\begin{table}[ht]
    \centering
    \begin{tabular}{c|cccccccc}
         &$e$&$a$&$a^2b$&$ab$&$a^2$&$a^3$&$b$&$a^3b$\\
         \hline
         $e$&$e$&$a$&$a^2b$&$ab$&$a^2$&$a^3$&$b$&$a^3b$\\
         $a^3$&$a^3$&$e$&$ab$&$b$&$a$&$a^2$&$a^3b$&$a^2b$\\
         $b$&$b$&$a^3b$&$e$&$a$&$a^2b$&$ab$&$a^2$&$a^3$\\
         $a^3b$&$a^3b$&$a^2b$&$a^3$&$e$&$ab$&$b$&$a$&$a^2$\\
         $a^2$&$a^2$&$a^3$&$b$&$a^3b$&$e$&$a$&$a^2b$&$ab$\\
         $a$&$a$&$a^2$&$a^3b$&$a^2b$&$a^3$&$e$&$ab$&$b$\\
         $a^2b$&$a^2b$&$ab$&$a^2$&$a^3$&$b$&$a^3b$&$e$&$a$\\
         $ab$&$ab$&$b$&$a$&$a^2$&$a^3b$&$a^2b$&$a^3$&$e$
    \end{tabular}
    \caption{The Cayley table of $\hat{\text{D}}_4$, from which one can read off the orbit-basis matrices $\tau$.}
    \label{tab:DfourCayleytable}
\end{table}

The eight elements organise themselves into five conjugacy classes:
\be
r_0\equiv\{e\},\quad r_1\equiv\{z\}\,,\quad r_2\equiv\{a,a^3\},\quad r_3\equiv\{b,a^2b\},\quad r_4\equiv\{ab,a^3b\}\;.
\ee
This matches the number of nodes of the affine $\Dfour$ Dynkin diagram and implies that there will be one untwisted and four twisted sectors. We will label the sectors by the first element of each conjugacy class. The character table of $\hat{\mathrm{D}}_4$ is given in Table \ref{tab:D4charactertable}.

\begin{table}[ht]
    \centering
    \begin{tabular}{|c|c|c|c|c|c|}\hline
         &$r_0$&$r_1$&$r_2$&$r_3$&$r_4$\tabularnewline\hline
         $\chi_1$&$1$&$1$&$1$&$1$&$1$\tabularnewline
         $\chi_2$&$1$&$1$&$-1$&$1$&$-1$\tabularnewline
         $\chi_3$&$1$&$1$&$1$&$-1$&$-1$\tabularnewline
         $\chi_4$&$1$&$1$&$-1$&$-1$&$1$\tabularnewline
         $\chi_5$&$2$&$-2$&$0$&$0$&$0$\tabularnewline\hline
    \end{tabular}
    \caption{The character table of $\hat{\mathrm{D}}_4$.}
    \label{tab:D4charactertable}
\end{table}

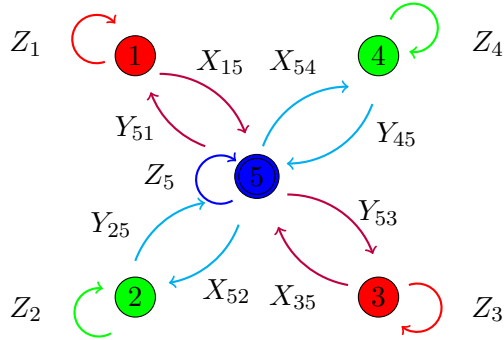
\begin{figure}[ht]
    \centering
    \begin{tikzpicture}[scale=0.8]
      \draw[fill=blue] (0,0) circle (2.2ex);
        \draw[fill=blue] (0,0) circle (1.8ex);
  \draw[fill=green] (2,2) circle (2ex);
  \draw[fill=red] (-2,2) circle (2ex);
  \draw[fill=red] (2,-2) circle (2ex);
  \draw[fill=green] (-2,-2) circle (2ex);

\draw[->,blue,thick] (-0.4,-0.4) arc (-60:-310:0.4);
\draw[->,green,thick] (2.2,2.6) arc (160:-100:0.4);
\draw[->,green,thick] (-2.4,-2.6) arc (120:-100:-0.4);
\draw[->,red,thick] (-2.5,1.9) arc (-70:-320:0.4);
\draw[->,red,thick] (2.5,-1.8) arc (110:-130:0.4);

\draw[->,purple,thick] (-1.6,1.7) arc (90:20:1.5);
\draw[->,purple,thick] (0.5,-0.3) arc (90:20:1.5);
\draw[->,cyan,thick] (-0.3,-0.8) arc (-20:-80:1.5);
\draw[->,cyan,thick] (1.9,1.2) arc (-20:-90:1.5);
\draw[->,purple,thick] (-0.85,0.5) arc (250:200:1.5);
\draw[->,cyan,thick] (-2,-1.4) arc (-200:-260:1.5);
\draw[->,purple,thick] (1.5,-1.8) arc (260:200:1.5);
\draw[->,cyan,thick] (0.1,0.5) arc (-200:-270:1.5);

\node at (0,0) {$5$};\node at (-2,2) {$1$};\node at (2,2) {$4$};\node at (2,-2) {$3$};\node at (-2,-2) {$2$};
\node at (-1.6,0) {$Z_{5}$};\node at (-3.6,2.4) {$Z_{1}$};\node at (3.5,2.4){$Z_4$};
\node at (-3.6,-2.2) {$Z_{2}$}; \node at (3.5,-2.2) {$Z_{3}$};
\node at (-2,0.8) {$Y_{51}$};
\node at (-0.6,1.9) {$X_{15}$};
\node at (0.6,1.9) {$X_{54}$};
\node at (2.1,0.5) {$Y_{45}$};
\node at (0.6,-2) {$X_{35}$};
\node at (2,-0.5) {$Y_{53}$};
\node at (-0.5,-1.9) {$X_{52}$};
\node at (-2.4,-0.8) {$Y_{25}$};

\end{tikzpicture}
    \caption{The $\Dfour$ quiver diagram. The exterior nodes are $\SU(N)$ gauge groups while the interior node is $\SU(2N)$ (symbolised by the double circle). For clarity, only the holomorphic fields are shown. The corresponding antiholomorphic ones can be found by reversing the arrows.}
    \label{fig:D4Quiver}
\end{figure}

$\hat{\mathrm{D}}_4$ has 5 irreducible representations, four of them being one-dimensional and one two-dimensional. The matrix elements of the one-dimensional representations can be directly read off from Table \ref{tab:D4charactertable}, and are given as follows:
\begin{center}
\begin{tabular}{|c|c|c|c|c|} 
 \hline
     & $a^{2m}$ & $a^{2m+1}$ & $a^{2m}b$& $a^{2m+1}b$ \\
\hline
 $(\rho^1)(g)^1\phan_1$ & 1 & 1 & 1 & 1 \\ 
 $(\rho^2)(g)^1\phan_1$ & 1 &$ -1$ & 1 & $-1$ \\ 
 $(\rho^3)(g)^1\phan_1$ & 1 & 1 & $-1$ & $-1$ \\ 
 $(\rho^4)(g)^1\phan_1$ & 1 & $-1$ & $-1$ & 1 \\
 \hline
\end{tabular}
\end{center}
with $m=0,1$. The matrix elements of the two dimensional representation are:
\begin{equation} \label{D42drep}
(\rho_5)(a^p)=\begin{pmatrix}
        i^p&0\\
        0&(-i)^p
        \end{pmatrix}\;\;\;\;\text{and}\;\;\;\;\\
        (\rho_5)(ba^p)=\begin{pmatrix}
            0&(-i)^{p}\\
            -i^p&0
        \end{pmatrix},
\end{equation}
with $p=0,1,2,3$ and $i,j=1,2$. Therefore, the regular representation matrices (\ref{gammareg}), or in other words the   
quiver basis generators, are
\be
\gamma(a)=\begin{pmatrix}1 &0&0&0&0&0\\
0&-1 &0&0&0&0\\
0&0&1 &0&0&0\\
0&0&0&-1 &0&0\\
0&0&0&0&i I_{2\times2} &0\\
0&0&0&0&0&-iI_{2\times2} \end{pmatrix}\;\;\text{and}\;\;
\gamma(b)=\begin{pmatrix}1 &0&0&0&0&0\\
0&1 &0&0&0&0\\
0&0&-1 &0&0&0\\
0&0&0&-1 &0&0\\
0&0&0&0&0 & I_{2\times2}\\
0&0&0&0&-I_{2\times2}& 0\end{pmatrix}\;,
\ee
while the representations of the other matrices can be found by applying the algebra relations. For the induced representation we simply use the two-dimensional representation (\ref{D42drep}):
\be
R(a)^I_{\;J}=\twobytwo{i}{0}{0}{-i}\;\;\text{and}\;\;R(b)^I_{\;J}=\twobytwo{0}{1}{-1}{0}\;.
\ee
To apply the orbifolding process, we start with $\Ncal=4$ SYM with $\SU(8N)$ gauge group and impose the conditions of invariance under the group action (\ref{invarianceVZ}), (\ref{invarianceXY}), with the $\gamma(g)$ matrices acting on the $8\times 8$ gauge fields where each element is an $N\times N$ block. In this way (and after removing the $\Urm(1)$'s as discussed) we obtain a quiver theory with $\SU(N)^4\times \SU(2N)$ gauge group. The field content is
\be \label{ZD4}
Z=\begin{pmatrix}Z_{1} &0&0&0&0&0\\
0&Z_{2} &0&0&0&0\\
0&0&Z_{3} &0&0&0\\
0&0&0&Z_{4} &0&0\\
0&0&0&0&Z_{5} &0\\
0&0&0&0&0&Z_{5} \end{pmatrix}\;,
\ee
\be
X=\begin{pmatrix}0 &0&0&0&X_{15}&0\\
0&0 &0&0&0&-Y_{25}\\
0&0&0 &0&X_{35}&0\\
0&0&0&0 &0&Y_{45}\\
0&X_{52}&0&X_{54}&0 &0\\
-Y_{51}&0&Y_{53}&0&0&0 \end{pmatrix}\;,
\ee
and
\be
Y=\begin{pmatrix}0 &0&0&0&0&X_{15}\\
0&0 &0&0&Y_{25}&0\\
0&0&0 &0&0&-X_{35}\\
0&0&0&0 &Y_{45}&0\\
Y_{51}&0&Y_{53}&0 &0&0\\
0&X_{52}&0&-X_{54}&0&0 \end{pmatrix}\;
\ee
Note that these are $8N\times 8N$ matrices, as the $Z_i$ are $N\times N$ blocks, the $Z_5$ is a $2N\times 2N$ block, the $Q_{i5}$ $N\times 2N$ and the $Q_{5i}$ $2N\times N$ blocks. The quiver diagram of $\Dfour$ is given in Fig. \ref{fig:D4Quiver}. There are 5 nodes (one for each gauge group), with the external nodes 1$\ldots$4 being $\SU(N)$ gauge groups, while the central node 5 corresponds to an $\SU(2N)$ gauge group. This quiver can of course also be directly obtained by applying the $\Dfour$ adjacency matrix (\ref{D4a}), however note that we have relabelled the nodes so that the central node is 5.

Rescaling the gauge couplings as $g_i=\gym \kappa_i$, the marginally deformed superpotential is
\be
\begin{split}
  \mathcal{W}_{\hat{\text{D}}_4}&=2ig_{\YM}\biggl(\Tr_5\bigl(\kappa_1Y_{51}Z_1X_{15}-\kappa_2X_{52}Z_2Y_{25}+\kappa_3Y_{53}Z_3X_{35}-\kappa_4X_{54}Z_4Y_{45}\bigr)\\
  &\quad +\!\kappa_5\bigl(\Tr_2\bigl(Y_{25}Z_5X_{52}\bigr)\!-\!\Tr_1\bigl(X_{15}Z_5Y_{51}\bigr)\!+\!\Tr_4\bigl(Y_{45}Z_5X_{54}\bigr)\!-
  \!\Tr_3\bigl(X_{35}Z_5Y_{53}\bigr)\bigr)\biggr)\;.
\end{split}\ee

The $\Dfour$ quiver has an $S_4$ permutation symmetry, given by exchanging any two of the outer nodes. This is clearest in the $Q_{ij}$ notation, since given our labelling one needs to be careful when expressing it in $X,Y$ notation. In the following we will mainly make use of the $\Zset_4$ subgroup of $S_4$ (given by $i\ra i+1$, with $i=1,2,3,4$ and $i+4\sim i$) in writing more compact expressions for the Hamiltonian and operators. Of course, we could have chosen to label the fields such that the $X$'s point inwards and the $Y$'s point outwards: we can achieve this via the following relabelling
\be
\begin{split}
    Y_{25}\to-X_{25}, \quad X_{52}\to Y_{25}, \quad Y_{45}\to -X_{45},\quad X_{54}\to Y_{54}\;.
\end{split}
\ee
Although this choice would be natural given the symmetries of the $\Dfour$ quiver (and is indeed the convention we follow in the $\hat{E}_6$ case of Section \ref{sec:E6}), for the $\hat{D}_k$ quivers with $k>4$ it might be best to allow for a string of $X$ fields starting at node $1$ and ending at node $k+1$ (which here is node 4 given the relabelling mentioned above). We will therefore keep the labels of the fields as in Fig. \ref{fig:D4Quiver}.

\subsection{The $\Dfour$ Hamiltonian}

Let us now spell out the Hamiltonian of Section \ref{sec:ADEHamiltonian} for the $\Dfour$ case. 
In this section $i=1,\ldots 4$ will denote the four exterior nodes. We will also find it convenient to use the notation

\be
Q_{i5}=\{X_{15},Y_{25},X_{35},Y_{45}\}\;,\;Q_{5i}=\{Y_{51},X_{52},Y_{53},X_{54}\}\;.
\ee
In the holomorphic sector, the Hamiltonian vanishes on two $Z$'s and two $Q$'s with different first and last index. For the same first and last index, we have 
\be \label{HD4QQ}
\Hcal_{\ell,\ell+1}=4\kappa_i^2 \;\;\text{on}\;\; Q_{i5}Q_{5i}\;,
\ee
and
\be \label{HD4holom}
\Hcal_{\ell,\ell+1}=\begin{pmatrix}
\kappa_5^2&-\kappa_5^2&\kappa_5^2&-\kappa_5^2\\
-\kappa_5^2&\kappa_5^2&-\kappa_5^2&\kappa_5^2\\
\kappa_5^2&-\kappa_5^2&\kappa_5^2&-\kappa_5^2\\
-\kappa_5^2&\kappa_5^2&-\kappa_5^2&\kappa_5^2
\end{pmatrix} \;\; \text{on} \;\; \begin{pmatrix}
  Y_{51}X_{15}\\
  X_{52}Y_{25}\\
  Y_{53}X_{35}\\
  X_{54}Y_{45}
\end{pmatrix}\;.
\ee
For one adjoint and one bifundamental, we have
\be
\Hcal_{\ell,\ell+1}=\begin{pmatrix}
           2\kappa_i^2&-2\kappa_i\kappa_5\\
            -2\kappa_i\kappa_5&2\kappa_5^2
        \end{pmatrix}\;\;\text{on}\;\; \begin{pmatrix}
            Z_iQ_{i5}\\
            Q_{i5}Z_5
\end{pmatrix}\;,
\ee
as well as 
\be
\Hcal_{\ell,\ell+1}=\begin{pmatrix}
           2\kappa_5^2&-2\kappa_i\kappa_5&\\
            -2\kappa_i\kappa_5&2\kappa_i^2
        \end{pmatrix} \;\;\text{on}\;\;\begin{pmatrix}
            Z_5Q_{5i}\\
            Q_{5i}Z_i
\end{pmatrix}\;.
\ee
Turning to the mixed sector, to write the action of the Hamiltonian in compact form, it is convenient to define the two-site combinations
\be\begin{split}
\mathbf{Q\Qb}_1&=\begin{pmatrix}
            X_{15}\Xb_{51}
\end{pmatrix}\;,\mathbf{Q\Qb}_2=\begin{pmatrix}
            Y_{25}\Yb_{52}
        \end{pmatrix}\;,\mathbf{Q\Qb}_3=\begin{pmatrix}
            X_{35}\Xb_{53}
        \end{pmatrix}\;,\mathbf{Q\Qb}_4=\begin{pmatrix}
            Y_{45}\Yb_{54}
\end{pmatrix}\;,\\
\mathbf{\Qb Q}_1&=\begin{pmatrix}
            \Yb_{15}Y_{51}
        \end{pmatrix}\;,\mathbf{\Qb Q}_2=\begin{pmatrix}
            \Xb_{25}X_{52}
        \end{pmatrix}\;,\mathbf{\Qb Q}_3=\begin{pmatrix}
            \Yb_{35}Y_{53}
        \end{pmatrix}\;,\mathbf{\Qb Q}_4=\begin{pmatrix}
            \Xb_{45}X_{54}
\end{pmatrix},
\end{split}\ee
and
\be
        \mathbf{Q\Qb}_5=\begin{pmatrix}
            Y_{51}\Yb_{15}\\
            X_{52}\Xb_{25}\\
            Y_{53}\Yb_{35}\\
            X_{54}\Xb_{45}
        \end{pmatrix}\;,\quad\mathbf{\Qb Q}_5=\begin{pmatrix}
            \Xb_{51}X_{15}\\
            \Yb_{52}Y_{25}\\
            \Xb_{53}Y_{35}\\
            \Yb_{54}Y_{45}
        \end{pmatrix}.
        \ee
as well as the matrices 

\be
\mathbb{K}_i=\begin{pmatrix}
           2\kappa_i^2
        \end{pmatrix}\;,\;\mathbb{L}_i=\begin{pmatrix}
            \kappa_i^2
        \end{pmatrix}\;,\;\mathbb{T}_{i}=\begin{pmatrix}
            2\kappa_5^2
        \end{pmatrix}\;,\;\quad\mathbb{M}_i=\begin{pmatrix}
            2\kappa_i^2
\end{pmatrix}\;,
\ee
\be
\mathbb{K}_5= \begin{pmatrix}
  \kappa_5^2&\kappa_5^2&\kappa_5^2&\kappa_5^2
\end{pmatrix},\quad\mathbb{L}_5=\frac{\kappa_5^2}{2}\begin{pmatrix}
1\\
1\\
1\\
1
\end{pmatrix}\;,\;\mathbb{T}_{5}=\begin{pmatrix}
            2\kappa_1^2& & & \\
            &2\kappa_2^2& & \\
            & &2\kappa_3^2& \\
            & & &2\kappa_4^2
\end{pmatrix}\;
\ee
and
\be
\mathbb{M}_5= \frac{\kappa_5^2}{2}\begin{pmatrix}
  1 &1& 1& 1 \\
  1 &1& 1& 1 \\
  1 &1& 1& 1 \\
  1 &1& 1& 1 
\end{pmatrix}.
\ee

Then we can write 
\be
\Hcal_{\ell,\ell+1}=
\begin{pmatrix}
  3\kappa_i^2&-\kappa_i^2&\mathbb{K}_i&\mathbb{K}_i\\
  -\kappa_i^2&3\kappa_i^2&\mathbb{K}_i&\mathbb{K}_i\\
  \mathbb{L}_i&\mathbb{L}_i&\mathbb{T}_i+\mathbb{M}_i&\mathbb{T}_i-\mathbb{M}_i\\
  \mathbb{L}_i&\mathbb{L}_i&\mathbb{T}_i-\mathbb{M}_i&\mathbb{T}_i+\mathbb{M}_i
\end{pmatrix}\;\text{on}\;\begin{pmatrix}
Z_i\Zb_i\\
\Zb_iZ_i\\
\mathbf{Q\Qb}_i\\
\mathbf{\Qb Q}_i
\end{pmatrix}\;,
\ee
and 
\be
\Hcal_{\ell,\ell+1}=
\begin{pmatrix}
  3\kappa_5^2&-\kappa_5^2&\mathbb{K}_5&\mathbb{K}_5\\
  -\kappa_5^2&3\kappa_5^2&\mathbb{K}_5&\mathbb{K}_5\\
  \mathbb{L}_5&\mathbb{L}_5&\mathbb{T}_5+\mathbb{M}_5&\mathbb{T}_5-\mathbb{M}_5\\
  \mathbb{L}_5&\mathbb{L}_5&\mathbb{T}_5-\mathbb{M}_5&\mathbb{T}_5+\mathbb{M}_5
\end{pmatrix}\;\text{on}\;
\begin{pmatrix}
Z_5\Zb_5\\
\Zb_5Z_5\\
\mathbf{Q\Qb}_5\\
\mathbf{\Qb Q}_5
\end{pmatrix},
\ee
Note the different coefficients of the $\mathbb{K}$ and $\mathbb{L}$ matrices, which make the Hamiltonian non-Hermitian. As discussed in Section \ref{sec:ADEHamiltonian}, this is just an artifact of the non-canonical normalisation of the Lagrangian due to the $n_i$ factors. It can be easily fixed by a rescaling to canonical normalisation, at the cost of introducing factors of $\sqrt{2}$ in the Hamiltonian. Next we need to consider the cases where the first and last node indices are not equal. We have:
\be
\Hcal_{\ell,\ell+1}=
\begin{pmatrix}
            2\kappa_5^2&2\kappa_5^2\\
            2\kappa_5^2&2\kappa_5^2
        \end{pmatrix}\;\text{on}\;
\begin{pmatrix}
    X_{15}\Xb_{53}\\
    \Yb_{15}Y_{53}
\end{pmatrix},\;\begin{pmatrix}
    X_{35}\Xb_{51}\\
    \Yb_{35}Y_{51}
\end{pmatrix},\;\begin{pmatrix}
    Y_{25}\Yb_{54}\\
    \Xb_{25}X_{54}
\end{pmatrix}\;\;\text{and}\;\;\begin{pmatrix}
    Y_{45}\Yb_{52}\\
    \Xb_{45}X_{52}
\end{pmatrix},\;
\ee
and
\be
\Hcal_{\ell,\ell+1}=\begin{pmatrix}
    2\kappa_5^2&-2\kappa_5^2\\
    -2\kappa_5^2&2\kappa_5^2
\end{pmatrix}\;\text{on}\;
\begin{pmatrix}
    X_{15}\Yb_{52}\\
    \Yb_{15}X_{52}
\end{pmatrix}\;,\begin{pmatrix}
    X_{15}\Yb_{54}\\
    \Yb_{15}X_{54}
\end{pmatrix}\;,\begin{pmatrix}
    X_{35}\Yb_{52}\\
    \Yb_{35}X_{52}
\end{pmatrix}\;\;\text{and}\;\;\begin{pmatrix}
    X_{35}\Yb_{54}\\
    \Yb_{35}X_{54}
\end{pmatrix},
\ee
and similarly
\be
\Hcal_{\ell,\ell+1}=\begin{pmatrix}
    2\kappa_5^2&-2\kappa_5^2\\
    -2\kappa_5^2&2\kappa_5^2
\end{pmatrix}\;\text{on}\;
\begin{pmatrix}
    Y_{25}\Xb_{51}\\
    \Xb_{25}Y_{51}
\end{pmatrix}\;,\begin{pmatrix}
    Y_{25}\Xb_{53}\\
    \Xb_{25}Y_{53}
\end{pmatrix}\;,
\begin{pmatrix}
    Y_{45}\Xb_{51}\\
    \Xb_{45}Y_{51}
\end{pmatrix}\;\;\text{and}\;\;
\begin{pmatrix}
    Y_{45}\Xb_{53}\\
    \Xb_{45}Y_{53}
\end{pmatrix}.
\ee
Finally we have 
\be
\Hcal_{\ell,\ell+1}=
     \begin{pmatrix}
         2\kappa_i^2&-2\kappa_i\kappa_5& &\\
         -2\kappa_i\kappa_5&2\kappa_5^2& &\\
         & &2\kappa_i^2&-2\kappa_i\kappa_5\\
         & &-2\kappa_i\kappa_5&2\kappa_5^2
     \end{pmatrix}\;\text{on}\;\begin{pmatrix}
        Z_i\Qb_{i5}\\
        \Qb_{i5}Z_5\\
        \Zb_iQ_{i5}\\
        Q_{i5}\Zb_5
     \end{pmatrix},\\
     \ee
     and
     \be
 \Hcal_{\ell,\ell+1}=    
    \begin{pmatrix}
        2\kappa_5^2&-2\kappa_i\kappa_5&&\\
        -2\kappa_i\kappa_5&2\kappa_i^2&&\\
        &&2\kappa_5^2&-2\kappa_i\kappa_5\\
        &&-2\kappa_i\kappa_5&2\kappa_5^2
    \end{pmatrix}\;\text{on}\;\begin{pmatrix}
        Z_5\Qb_{5i}\\
        \Qb_{5i}Z_i\\
        \Zb_5Q_{5i}\\
        Q_{5i}\Zb_i
    \end{pmatrix}\;.
\ee
The Hamiltonian can be seen to commute with the $\SU(2)_R\times\Urm(1)$ symmetry.

Unlike the abelian orbifold case, here we cannot have $X$ or $Y$ vacua, as the maximum number of holomorphic $X$ or $Y$ chains is 2. So the only BPS states which appear at any length are made up of $Z$ fields:
\be\label{ZD4twisted}\begin{split}
\Tr(\gamma(e) Z^L)&=\Tr_1 Z_1^L+\Tr_2 Z_2^L+\Tr_3 Z_3^L+\Tr_4 Z_4^L+2\Tr_5 Z_5^L\;,\\
\Tr(\gamma(z) Z^L)&=\Tr_1 Z_1^L+\Tr_2 Z_2^L+\Tr_3 Z_3^L+\Tr_4 Z_4^L-2\Tr_5 Z_5^L\;,\\
\Tr(\gamma(a) Z^L)&=\Tr_1 Z_1^L-\Tr_2 Z_2^L+\Tr_3 Z_3^L-\Tr_4 Z_4^L\;,\\
\Tr(\gamma(b) Z^L)&=\Tr_1 Z_1^L+\Tr_2 Z_2^L-\Tr_3 Z_3^L-\Tr_4 Z_4^L\;,\\
\Tr(\gamma(ab) Z^L)&=\Tr_1 Z_1^L-\Tr_2 Z_2^L-\Tr_3 Z_3^L+\Tr_4 Z_4^L\;.
\end{split}
\ee
These 5 states can be labelled by their eigenvalues under the $\Zset_2^{(1)}\times \Zset_2^{(2)}$ subset of the $S_4$ symmetry group of the Dynkin diagram, which we can take to be
\be
\Zset_2^{(1)}: (1\leftrightarrow 2,3\leftrightarrow 4)\;\;\; \text{and} \;\;\; \Zset^{(2)}_2: (1\leftrightarrow 4,2\leftrightarrow 3)
\ee
under which their eigenvalues are: $(1,1),(1,1),(-1,-1),(1,-1),(-1,1)$, respectively. Note that the first two states cannot be distinguished by this abelian group. They can however be distinguished by conjugation by $\tau(a^2)$, which from the Cayley table (Table \ref{tab:DfourCayleytable}) takes the form
\be
\tau(a^2)=\twobytwo{0_{4\times 4}}{I_{4\times 4}}{I_{4\times 4}}{0_{4\times 4}}
\ee
and effectively takes $Z_i\leftrightarrow Z_5$ (note that there are four copies of $Z_5$, which is slightly obscured by the block structure used in (\ref{ZD4})). Under this conjugation, the first two states in (\ref{ZD4twisted}) have eigenvalues $+1$ and $-1$, respectively, while the remaining states map to zero.

In the next section we will see how these, as well as other protected states arising at specific lengths, are captured by the index and Molien series. 

\subsection{Protected spectrum} \label{sec:D4index}

The matrix entering the multitrace ADE index (\ref{eq:MultitraceOrbifoldIndex}) is 
\be
(1+t)I_{5\times 5}-t^\half A_{\hat{\text{D}}_4}=\begin{pmatrix}
    1+t&0&0&0&-t^\half\\
    0&1+t&0&0&-t^\half\\
    0&0&1+t&0&-t^\half\\
    0&0&0&1+t&-t^\half\\
    -t^\half&-t^\half&-t^\half&-t^\half&1+t
\end{pmatrix}.
\ee
where the difference to (\ref{D4a}) is due to the relabelling leading to 5 being the central node. We find
\be
    \det\left((1+t)I_{5\times 5}-t^\half A_{\hat{\text{D}}_4}\right)=\,(1-t^2)^2(1+t)
    =\frac{(1-t^2)^3}{(1-t)}.
\ee
Hence, from \eqref{eq:MultitraceOrbifoldIndex}, the multi-trace index for $\Dfour$ is given by
\be\begin{split}\label{eq:MultiTraceSciD4}
\mathcal{I}^\text{m.t.}_{\hat{\text{D}}_4}\simeq&\prod_{n=1}^\infty\frac{\left((1-p^n)(1-q^{n})\right)^5(1-t^{n})e^{-\frac{5}{n}f_\text{vm}(p^n,q^n,t^n)}}{\left(1-(pqt^{-1})^n\right)^5\left(1-t^{2n}\right)^3}\\
=&\frac{\Gamma(t;p,q)^5(t;t)_\infty}{(pqt^{-1};pqt^{-1})_\infty^5(t^2;t^2)^3}\,.
\end{split}\ee
From \eqref{MultitraceOrbifoldLimits}, the limits of \eqref{eq:MultiTraceSciD4} are
\be\begin{split}\label{eq:MultiTraceD4limits}
    \mathcal{I}^\text{m.t.}_{\hat{\text{D}}_4;\;M}\simeq&\,\frac{(t;t)_\infty}{(t;q)_{\infty}^{5}(t^2;t^2)_\infty^{3}}\;,\\
    \mathcal{I}^\text{m.t.}_{\hat{\text{D}}_4;\;S}\simeq&\,(q;q)_{\infty}^{-4}(q^2;q^2)^{-3}_\infty\;,\\
    \mathcal{I}^\text{m.t.}_{\hat{\text{D}}_4;\;HL}\simeq&\,\frac{(t;t)_\infty}{(1-t)^{5}(t^2;t^2)^{3}_\infty}\;,\\
    \mathcal{I}^\text{m.t.}_{\hat{\text{D}}_4;\;C}\simeq&\,\frac{(1-T)}{(T;T)_{\infty}^{5}}\,.
\end{split}\ee
It would be interesting to see whether applying the free-fermion methods used in \cite{Bourdier:2015sga} could reproduce the Schur index in the $\Dfour$ case and provide a check of our result.

From \eqref{eq:Singletracegenericindex}, the single trace index is given by
\be\begin{split}\label{eq:SingleTraceSciD4}
\mathcal{I}^\text{s.t.}_{\hat{\text{D}}_4}=&\,5\biggl[\frac{pqt^{-1}}{1-pqt^{-1}}+\frac{t-pqt^{-1}}{(1-p)(1-q)}\biggr]+\frac{3t^2}{1-t^2}-\frac{t}{1-t}\;\\
=&\,5\biggl[\frac{pqt^{-1}}{1-pqt^{-1}}+\frac{t-pqt^{-1}}{(1-p)(1-q)}\biggr]-t+\frac{3t^2}{1-t^2}-\frac{t^2}{1-t}\;\\
=&\,5\left[\sum_{\ell=2}^\infty\mathcal{I}[\bar{\mathcal{E}}_{-\ell(0,0)}]+\mathcal{I}[\hat{\mathcal{B}}_1]\right]-\mathcal{I}[\Mtrip]+\frac{3t^2}{1-t^2}-\frac{t^2}{1-t}\,.
\end{split}\ee
Note that we extracted a factor of $-t$, which corresponds to the $F$-term constraint in a quiver with spherical topology as noted in Section \ref{sec:Hall-Littlewood}. The Hall-Littlewood and Coulomb limits are 
\be\begin{split}\label{eq:limitsSingleTraceSciD4}
    \mathcal{I}^\text{s.t.}_{\Dfour;\;HL}=&\,4\mathcal{I}[\Mtrip]+\frac{3t^2}{1-t^2}-\frac{t^2}{1-t}\;,\;\\
    \mathcal{I}^\text{s.t.}_{\Dfour;\;C}=&\,5\sum_{\ell=2}^\infty\mathcal{I}[Z^\ell]\;.
\end{split}\ee
We note that, unlike the $\Zset_k$ case where the final terms could be identified as $X$ and $Y$ vacua, here their interpretation is not as clear. In particular, we see that both bosonic and fermionic states have survived in the index. In the following we will take advantage of our knowledge of the Molien series to clarify their origin. 

But first let us briefly consider what information we can extract about the untwisted and twisted sectors from the $\Ncal=4$ SYM index. Let us recall the $\SU(2)_L$-averaged index (\ref{IL}):
\be
\mathcal{I}^\text{s.t.}_{L}(p,q,t,\text{w}\;)\equiv v_xv_y\mathcal{I}[\hat{\mathcal{B}}_{1}]+\,\sum_{\ell=1}^\infty t^\frac{\ell}{2}(v_x^\ell+v_y^\ell) -(v_x+v_y)f_\text{hm}(p,q,t),
\ee
where we have expanded the denominators to simplify the averaging process. The linear and quadratic terms in $v_x,v_y$ cancel under the sum over the $\hat{\mathrm{D}}_4$ elements, as there are no $\hat{\mathrm{D}}_4$ invariants at those degrees. It is easy to check that the only $\hat{\mathrm{D}}_4$ invariants that we can obtain by averaging are of the type $\half(x^{4n}+y^{4n})$, with $n=1,2,\ldots$. We are left with 
\be\label{eq:projectionILD4}
\frac{1}{8}\sum_{g\in\hat{\text{D}}_4}\mathcal{I}^\text{s.t.}_{L}(p,q,t,R^{(\mathbf{2})}(g)\text{w})\biggr|_{v_x=v_y=1}
=\sum_{n=1}^\infty t^{2n}(v_x^{4n}+v_y^{4n})|_{v_x=v_y=1}
=\,\frac{2t^2}{1-t^2}\;.
\ee
However, this is still overcounting the untwisted states, as it is easy to check that mother-theory states of type $\Tr(X^{4n})$ and $\Tr(Y^{4n})$ project to the same state (see e.g. (\ref{D4L4XXXX})). So we need to further divide by a factor of two, to finally obtain (referring also to Appendix \ref{sec:IndexOperators})
\be\label{eq:untwistedD4}
    \mathcal{I}^\text{untwisted}_{\hat{\text{D}}_4}=\,\sum_{\ell=2}^\infty\mathcal{I}[\bar{\mathcal{E}}_{-\ell(0,0)}]+\sum_{\ell=1}^\infty\mathcal{I}[(XY)^{2\ell}]\,.
\ee
where $(XY)^{2\ell}$ is a schematic form of a state of length $4\ell$ which projects from the $X$ or $Y$ vacuum state of $\Ncal=4$ SYM. These states contain both $X$ and $Y$ daughter-theory fields, as there are no $X$ or $Y$ vacua (given that the path algebra of the $\Dfour$ quiver does not permit closed loops consisting of only $X$'s or $Y$'s). To conclude, we find that the index counts one BMN vacuum $\Tr(Z^\ell)$ and one protected primary of $\hat{\mathcal{B}}_{2\ell}$ in the untwisted sector. Comparing with (\ref{eq:limitsSingleTraceSciD4}), we see that we expect 4 twisted BMN vacua and 4 twisted $\hat{\mathcal{B}}_{2\ell}$ states, which we will confirm by explicit diagonalisation.

Of course, there are also other untwisted $XY$-sector primaries which do not descend from the $X$ or $Y$ vacua but from non-trivial $XY$-sector states in the mother theory. These are not counted by the untwisted index, but do contribute to the Molien series, which we now turn to. 
From Table \ref{tab:SU(2)Molienseries}, the Molien series of $\Dfour$ is given by
\be\label{eq:D4Molien}
\mathbf{M}(x;\hat{\text{D}}_4)=\,\frac{1+x^{6}}{(1-x^4)^2}=\,1+2x^4+x^6+3x^8+2x^{10}+4x^{12}+\Ocal(x^{14})\,.
\ee
where the powers correspond to the length. The invariant polynomials contributing to the series at low orders are listed in \cite{Benvenuti:2006qr}. We have checked all the listed multiplicities against the explicit diagonalisation of the Hamiltonian, finding full agreement.\footnote{For $\Dfour$, the dimension of the $XY$-sector cyclically identified basis is $10,24,70,208,352$ for $L=4,6,8,10,12$ respectively, so the diagonalisation for $L=12$ and beyond is already a rather time-consuming exercise.}

As all the states entering the Molien series are the highest-weight components of the $\hat{\mathcal{B}}_R$ multiplets (see (\ref{eq:qtildeqR})), with $R\geq2$, it is interesting to rewrite the series to explicitly see which ones contribute at any length. It is straightforward to obtain 
\be\begin{split}
    \mathbf{M}(t^\half;\hat{\text{D}}_4)&=1+\,\sum_{\ell=1}^\infty(\ell+1)t^{2\ell}+\sum_{\ell=1}^\infty\ell ~ t^{2\ell+1}\\
    &=\,\sum_{\ell=1}^\infty(\ell+1)\mathcal{I}[(XY)^{2\ell}]+\sum_{\ell=1}^{\infty}\ell~\mathcal{I}[(XY)^{2\ell+1}]\;.\\
\end{split}\ee
where in the schematic notation we are using, $(XY)^{2\ell}$ is a length-$4\ell$ state. So we see that $(\ell+1)$ states contribute at length $4\ell$ and $\ell$ states contribute at length $4\ell+2$, for $\ell\geq 1$. 

Now consider the Hall-Littlewood limit of the $\hat{\text{D}}_4$ index (\ref{eq:SingleTraceSciD4}):
\be\mathcal{I}^\text{s.t.}_{\hat{\text{D}}_4;\; HL}(x=t^\half)=4t+\frac{3t^2}{1-t^2}-\frac{t^2}{1-t}\,.
\ee
Since this only receives bosonic contributions from the $\Mtrip$ states, which are precisely those that are counted by the Molien series, we can count the number of single-fermion states that cancel with bosonic states in the HL index. We write 
\be\begin{split}\label{eq:difD4MolienHL}
\mathcal{I}^F_{\hat{\text{D}}_4;\; HL}(x=t^\half)&=\mathbf{M}(x;\hat{\text{D}}_4)-\mathcal{I}_{\hat{\text{D}}_4;\; HL}(x=t^\half)+4t\\&=1+2x^6+2x^8+3x^{10}+3x^{12}+\cdots
\end{split}
\ee
Note that we have subtracted from the index the four triplet states, which are twisted and are therefore not counted by the Molien series. 
The terms in (\ref{eq:difD4MolienHL}) should correspond to fermionic states of the schematic form $\bar{\lambda}_{Z\dot{+}}(XY)^{R-1}$. A verification of the above multiplicities will need to wait for the construction of the full dilatation operator including fermions.

\subsection{Short chains} \label{D4short}

Let us now consider the eigenstates of the $\Dfour$ Hamiltonian for some short chains.

\subsubsection{Length 2}

There are 31 cyclically-identified states at $L=2$, 23 of which are at $E=0$. These are the 5 $\Tr(\gamma(g) Z^2)$ states and their conjugates, and 4 twisted-sector triplets, which are not $E=0$ eigenstates of the naive Hamiltonian but drop to zero when including the non-planar corrections discussed in Section \ref{sec:Nonplanar}. Their top components are
\be\begin{split}
\Tr(\gamma(a)(XY-YX))&=2\Tr(X_{15}Y_{51}+X_{52}Y_{25}+X_{35}Y_{53}+X_{54}Y_{45})\;,\\
\Tr(\gamma(b)(XY-YX))&=2\Tr(X_{15}Y_{51}-X_{52}Y_{25}-X_{35}Y_{53}+X_{54}Y_{45})\;,\\
\Tr(\gamma(ab)(XY-YX))&=2\Tr(X_{15}Y_{51}+X_{52}Y_{25}-X_{35}Y_{53}-X_{54}Y_{45})\;,\\
\Tr(\gamma(z)(XY-YX))&=4\Tr(X_{15}Y_{51}-X_{52}Y_{25}+X_{35}Y_{53}-X_{54}Y_{45})\;.
\end{split}
\ee
The final BPS state is of course $\Tr(\mathcal{T}_{\Dfour})$ :
{\small
\be\begin{split}
\Tr(\mathcal{T}_{\Dfour})&=\Tr(X_{15}\Xb_{51}+X_{52}\Xb_{25}+X_{35}\Xb_{53}+X_{54}\Xb_{45}+Y_{51}\Yb_{15}+Y_{25}\Yb_{52}+Y_{53}\Yb_{35}+Y_{45}\Yb_{54}\\
&\qquad- Z_1\Zb_1- Z_2\Zb_2- Z_3\Zb_3-Z_4\Zb_4-2Z_5\Zb_5)\;.
\end{split}\ee
}
At the orbifold point we find the Konishi operator with $E=12$
{\small
  \be\begin{split}
  \mathcal{K}_{\Dfour}&=\Tr(X_{15}\Xb_{51}+X_{52}\Xb_{25}+X_{35}\Xb_{53}+X_{54}\Xb_{45}+Y_{51}\Yb_{15}+Y_{25}\Yb_{52}+Y_{53}\Yb_{35}+Y_{45}\Yb_{54}\\
&\qquad+\half Z_1\Zb_1+\half Z_2\Zb_2+\half Z_3\Zb_3+\half Z_4\Zb_4+Z_5\Zb_5)\;,
\end{split}
\ee
}
as well as an $E=4$ $z$-twisted state
\be
\Tr(\gamma(z)(Z\Zb+\Zb Z))=2\left(\Tr(Z_1\Zb_1+Z_2\Zb_2+Z_3\Zb_3+Z_4\Zb_4-2Z_5\Zb_5)\right)\;,
\ee
and $a,b,ab$-twisted $\SU(2)_R$-singlet states at $E=2(3\pm \sqrt{5})$. Their mother-theory form is
\be
\Tr\left(\gamma(g)(c(X\Xb+\Xb X+Y\Yb+\Yb Y)+Z\Zb+\Zb Z)\right)\;\;\text{with}\; g=a,b,ab\;,
\ee
with $c=\half(1\mp\sqrt{5})$ for $E=2(3\pm\sqrt{5})$. 

All of these non-protected states acquire $\kappa_i$-dependence when deforming away from the orbifold point.

\subsubsection{Length 3: Holomorphic spectrum}

In the holomorphic sector we have 11 states, five of which are BPS of type $\Tr(\gamma(g)ZZZ)$. It is notable that the equivalent of (\ref{Z3L3holBPS}) is absent in this case as the trace of the corresponding mother theory state vanishes,
\be
\Tr(\gamma(e)(XYZ+XZY))=0\;.
\ee
However, the $z$-twisted version of this state is an eigenstate with $E=4$:
\be\begin{split}
\Ocal^{(z)}&=\half\Tr(\gamma(z)(XYZ+XZY))\\
&=\Tr(X_{15}Y_{51}Z_1+X_{15}Z_5Y_{51}+X_{35}Y_{53}Z_3+X_{35}Z_5Y_{53}\\
&\quad-(X_{52}Y_{25}Z_5+X_{52}Z_{2}Y_{25}+X_{54}Y_{45}Z_5+X_{54}Z_4Y_{45}))\;.
\end{split}
\ee
The other three twisted sector states are degenerate at the orbifold point, with energies $E=2(3\pm\sqrt{5})$. They are:
\be\begin{split}
\Ocal^{(b)}&=\Tr(\gamma(b)(XYZ+\half(1\mp\sqrt{5})XZY))\;,\\
\Ocal^{(ab)}&=\Tr(\gamma(ab)(XYZ+\half(1\mp\sqrt{5})XZY))\;,\\
\Ocal^{(a)}&=\Tr(\gamma(a)(XYZ\mp\frac{2+i}{\sqrt{5}}XZY))\;.
\end{split}\ee
Although the last state appears complex, it can be suitably rescaled to have real coefficients. All these states are descendants of the corresponding neutral $L=2$ states, and therefore have the same energies both at the orbifold point and in the marginally deformed theory. 

\begin{figure}[h!]
  \begin{center}
    \begin{tikzpicture}
      \node at (0,0) {\includegraphics[width=12cm]{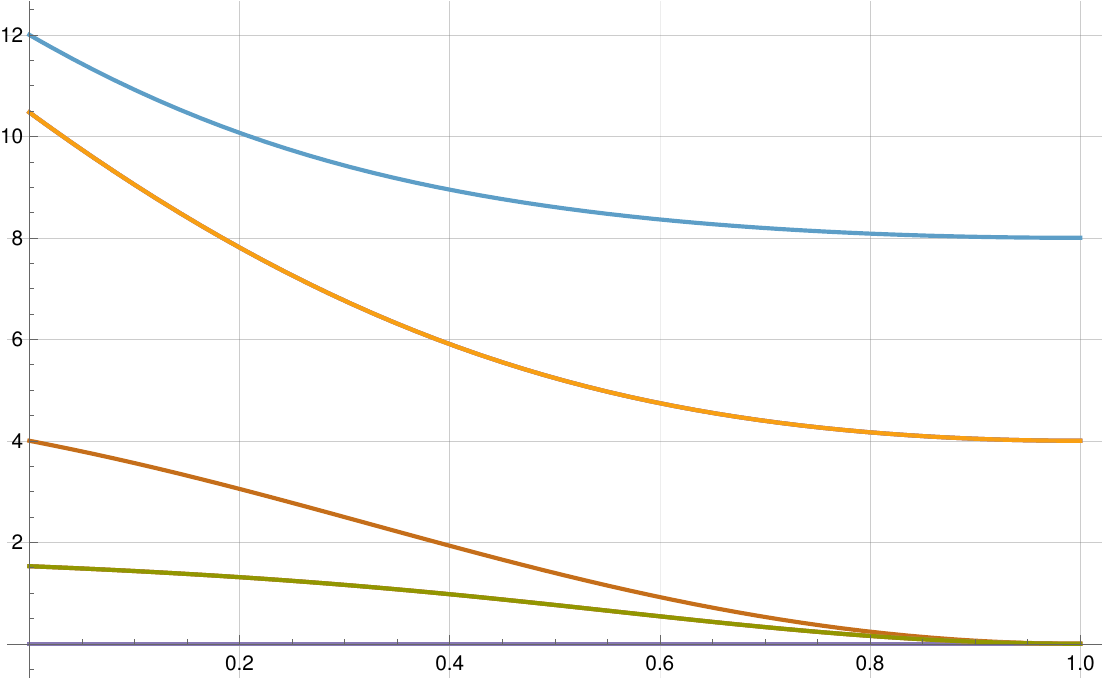}};
\node at (6.2,-3.3){$k$};\node at (-6.2,2.7){$E$};
\end{tikzpicture}
  \caption{The spectrum of neutral $L=2$/holomorphic $L=3$ states for the case $\{\kappa_i=1-k,\kappa_5=1\}$, where $k=0$ corresponds to the orbifold point and $k=1$ to SCQCD (plus decoupled vector multiplets). There are 3 states each at $E=2(3\pm\sqrt{5})$ at the orbifold point, which stay degenerate under the deformation.}\label{GraphD4L3ha}
  \end{center}
  \end{figure}

\begin{figure}[h!]
  \begin{center}
 \begin{tikzpicture}
      \node at (0,0) {\includegraphics[width=12cm]{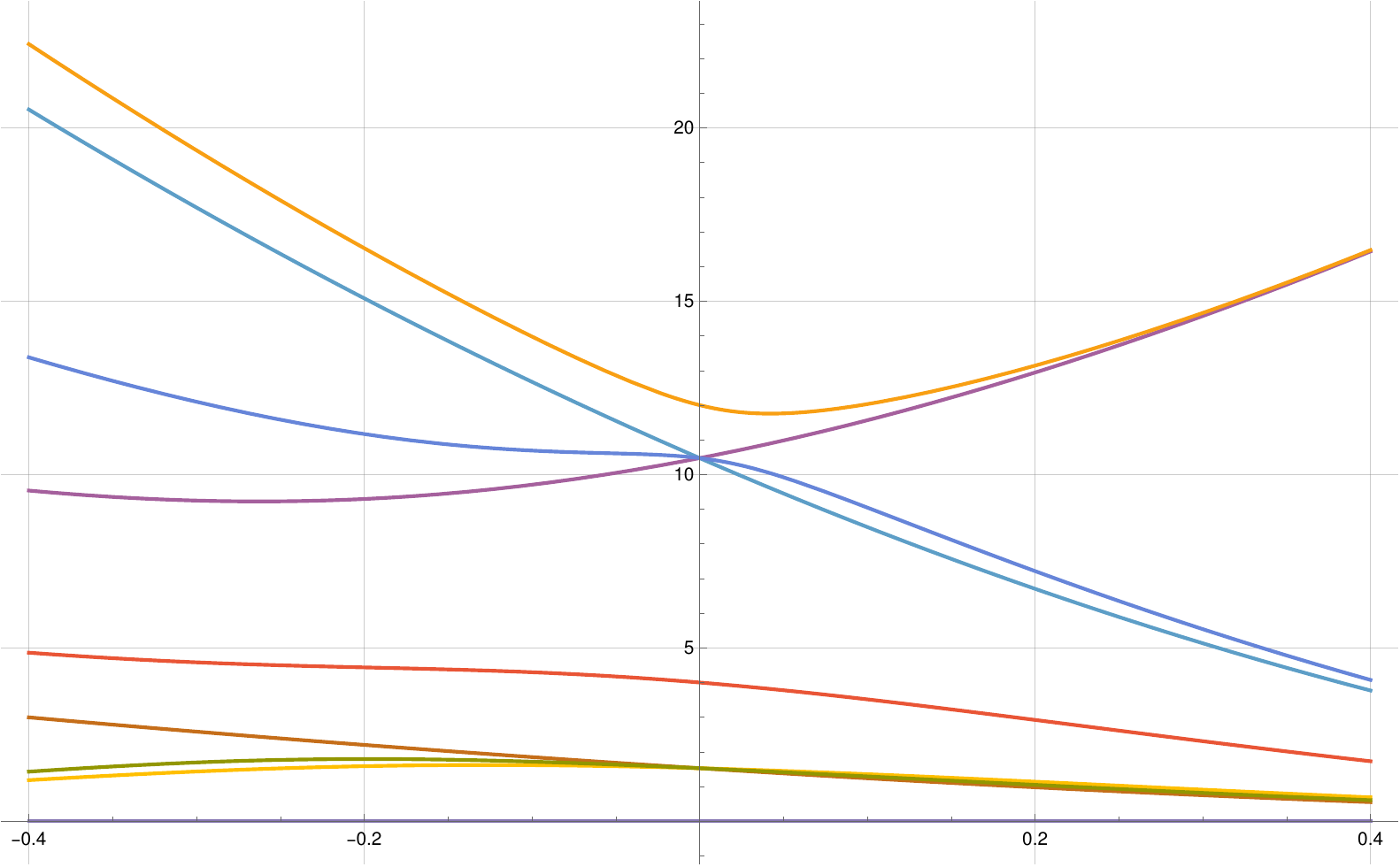}};
\node at (6.4,-3.3){$k$};\node at (-0.4,3.4){$E$};
\end{tikzpicture}
  \caption{The spectrum of neutral $L=2$/holomorphic $L=3$ states for the case $\{\kappa_1=\kappa_3=1-k,\kappa_2=\kappa_4=1+k,\kappa_5=1-k\}$, where $k=0$ corresponds to the orbifold point. All the degenerate twisted states with $E=2(3\pm\sqrt{5})$ at the orbifold point split under this deformation.}\label{GraphD4L3hb}
  \end{center}
\end{figure}

Clearly there are many inequivalent ways to marginally deform away from the orbifold point. The generic characteristic polynomial is not particularly illuminating so we just present a few of its terms: 
\be \label{charpolyD4}
P(E)=E^5\left(E^8-(8(\kappa_1^2+\kappa_2^2+\kappa_3^2+\kappa_4^2)+20\kappa_5^2)E^7+\cdots+196608\kappa_1^2\kappa_2^2\kappa_3^2\kappa_4^2\kappa_5^8\right)\;.
\ee
As expected, it is symmetric under exchanges of the $\kappa_i$, $i=1,2,3,4$, in line with the symmetries of the $\Dfour$ theory. A special case is when all the exterior couplings are equal, $\kappa_i=\kappa$, while $\kappa_5=1$, where it simplifies to just
\be
P(E)=E^5(E^2-8(1+\kappa^2)E+48\kappa^2)(E^2-(4+8\kappa^2)+16\kappa^2)^3\;.
\ee
The first term factors as $(E-12)(E-4)$ at the orbifold point, corresponding to the untwisted $E=12$ and twisted $E=4$ states. These mix under the marginal deformation. However, under this very symmetric deformation, the multiplicity-3 twisted states with $E=2(3\pm \sqrt5)$ at the orbifold point stay degenerate along the deformation. Taking $\kappa\ra0$, the outer gauge groups become global, and the theory approaches SCQCD with $\SU(2N)$ gauge group and $4N$ flavours. The spectrum along this deformation is plotted in Fig.\ref{GraphD4L3ha}. A less symmetric deformation, where all the twisted-sector degeneracies are lifted, is illustrated in Fig. \ref{GraphD4L3hb}.

\subsubsection{Length 4: Holomorphic spectrum}

At $L=4$ there are 27 cyclically identified holomorphic states, with 7 of those being at $E=0$. They are the 5 $\Tr(\gamma(g) Z^4)$ states, plus the two $XY$-sector states counted by the Molien series (\ref{eq:D4Molien}). These are both untwisted and descend from the corresponding $\Ncal=4$ SYM symmetrised states:
\be \label{D4L4XXXX}
\frac{1}{4}\Tr(XXXX)=\frac{1}4\Tr(YYYY)=\Tr\left(X_{52}Y_{25}(Y_{51}X_{15}\!-\!Y_{53}X_{35})\!+\!X_{54}Y_{45}(Y_{51}X_{15}\!-\!Y_{53}X_{35})\right)
\ee
and
\be\begin{split}
\half\Tr(XXYY)+\frac{1}{4}\Tr(XYXY)&=\Tr\left(X_{52}Y_{25}(Y_{51}X_{15}+Y_{53}X_{35}+2X_{54}Y_{45})\right)\\
&\quad+\Tr\left(X_{54}Y_{45}(Y_{51}X_{15}+Y_{53}X_{35}+2X_{52}Y_{25})\right)\;.
\end{split}
\ee
These states are clearly related to the two degree-4 invariant polynomials of $\hat{\mathrm{D}}_4$, which are $\half(x^4+y^4)$ and $x^2y^2$ \cite{Benvenuti:2006qr}.

\begin{figure}[h!]
  \begin{center}
    \begin{tikzpicture}
      \node at (0,0) {\includegraphics[width=12cm]{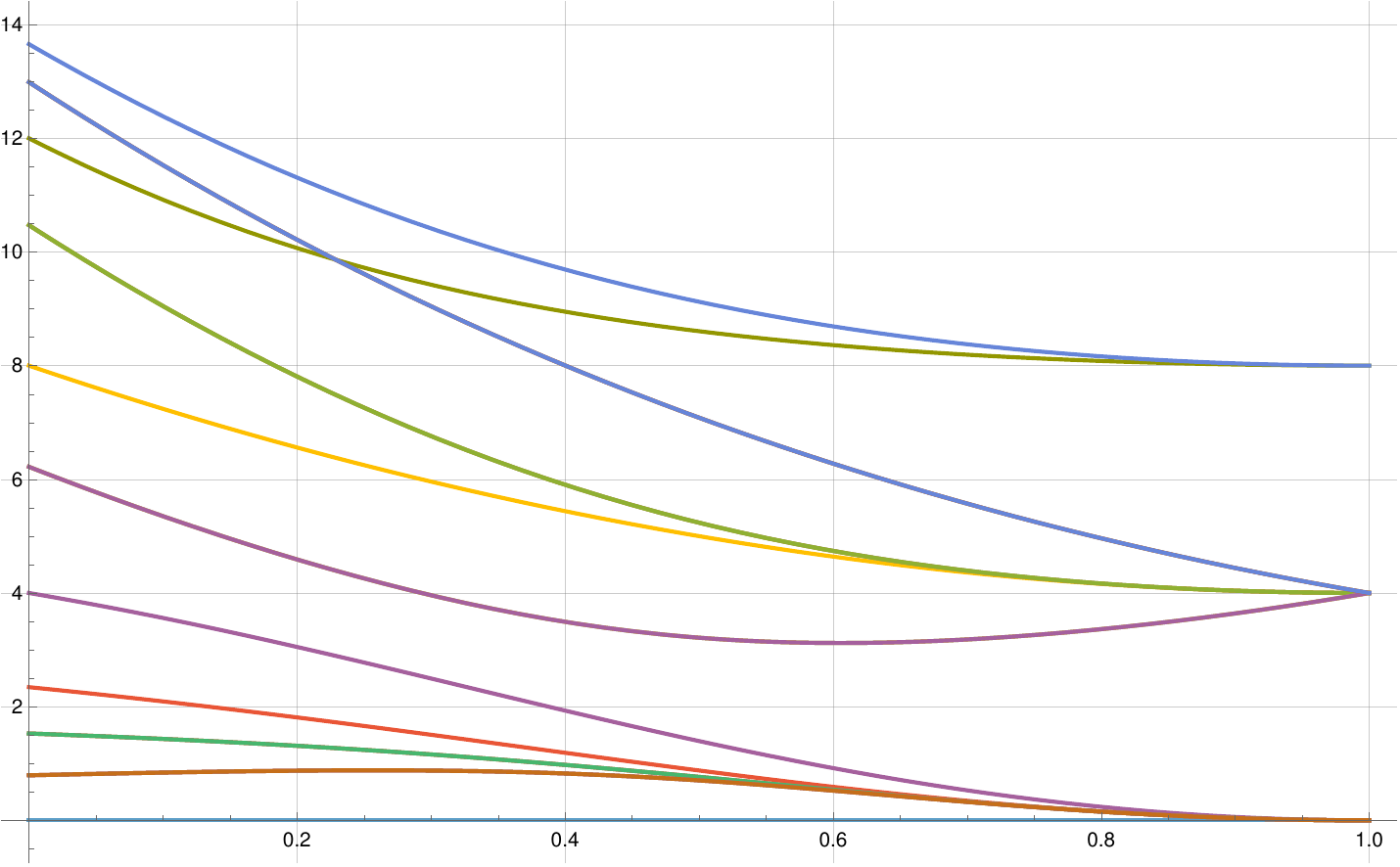}};
\node at (6.2,-3.3){$k$};\node at (-6.2,3.1){$E$};
\end{tikzpicture}
  \caption{The spectrum of $\Dfour$ $L=4$ holomorphic states for the case $\{\kappa_i=1-k,\kappa_5=1\}$, where $k=0$ corresponds to the orbifold point and $k=1$ to the SCQCD limit (plus decoupled vector multiplets). }\label{GraphD4L4ha}
  \end{center}
\end{figure}

\begin{figure}[h!]
  \begin{center}
        \begin{tikzpicture}
      \node at (0,0) {\includegraphics[width=12cm]{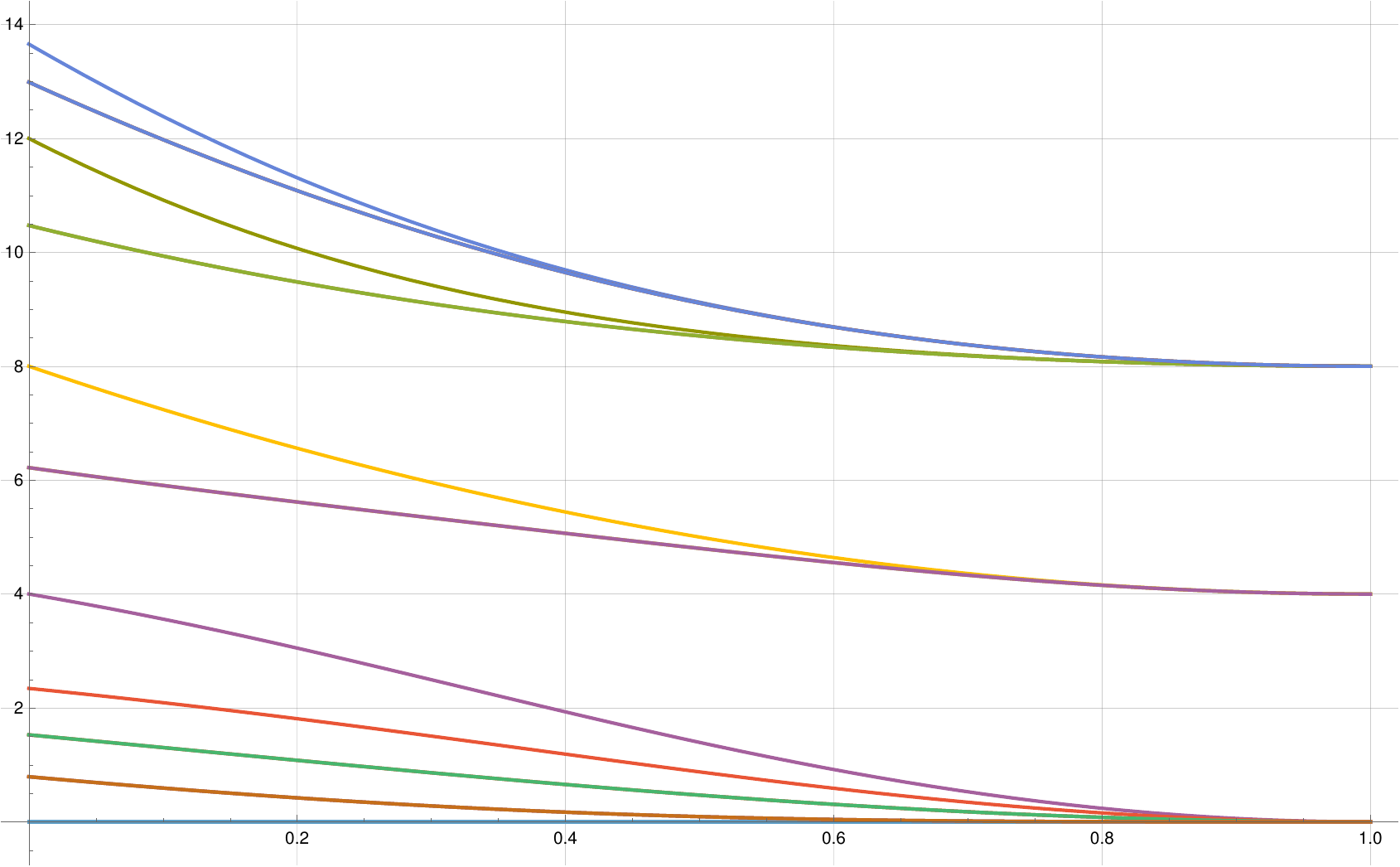}};
\node at (6.2,-3.3){$k$};\node at (-6.2,3.1){$E$};
\end{tikzpicture}
  \caption{The spectrum of $\Dfour$ $L=4$ holomorphic states for the case $\{\kappa_i=1,\kappa_5=1-k\}$, where $k=0$ corresponds to the orbifold point and $k=1$ to four decoupled copies of SCQCD (plus decoupled vector multiplets). Notice that more states reach $E=8$ as compared to Fig. \ref{GraphD4L4ha}. }\label{GraphD4L4hb}
  \end{center}
\end{figure}

\begin{figure}[h!]
  \begin{center}
        \begin{tikzpicture}
      \node at (0,0) {\includegraphics[width=12cm]{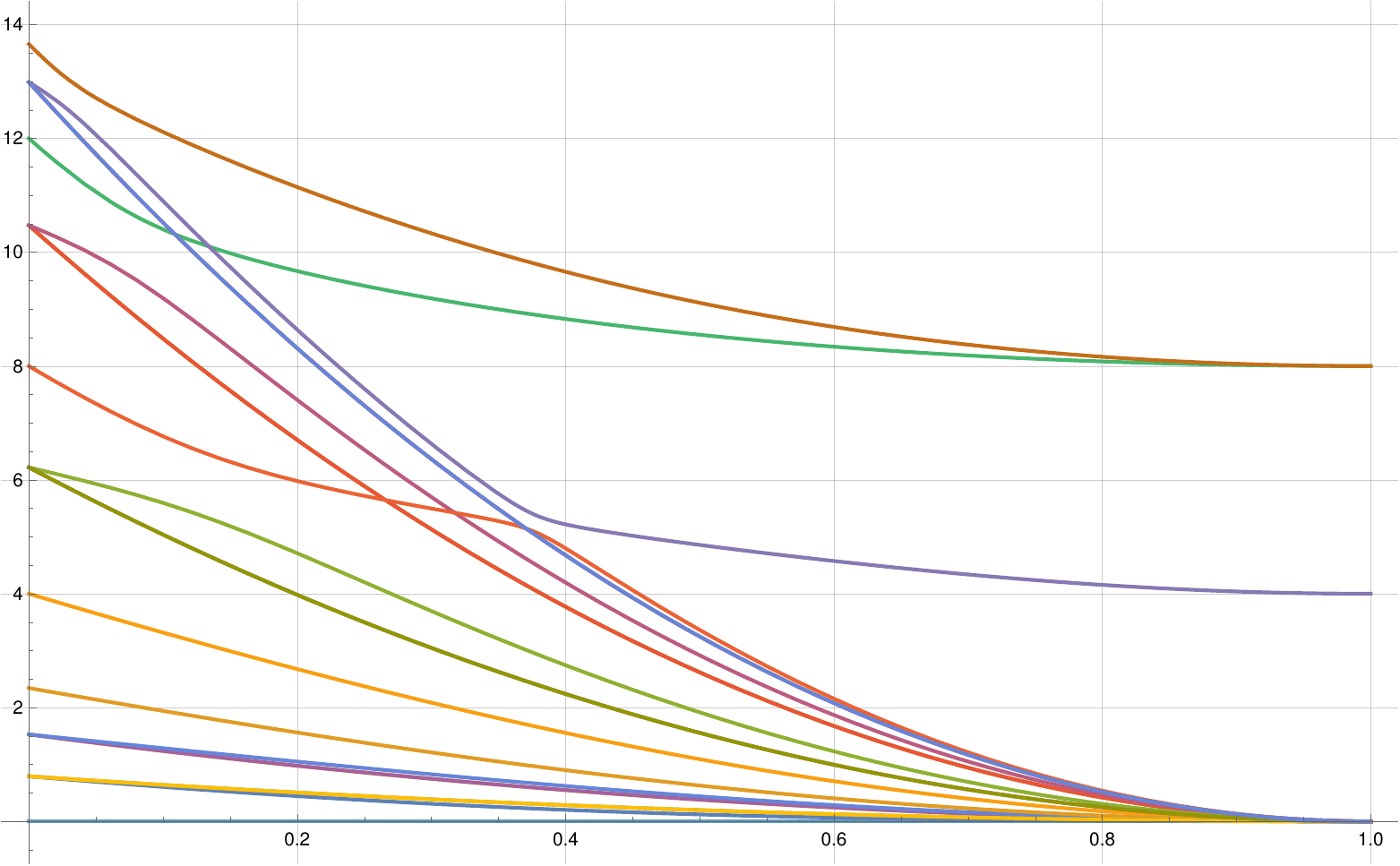}};
\node at (6.2,-3.3){$k$};\node at (-6.2,3.1){$E$};
\end{tikzpicture}
  \caption{The spectrum of $\Dfour$ $L=4$ holomorphic states for the case $\{\kappa_1=1,\kappa_{2\ldots5}=1-k\}$, where $k=0$ corresponds to the orbifold point and $k=1$ to SCQCD (plus additional decoupled vector multiplets). Notice an avoided crossing at $k\simeq 0.39$.}\label{GraphD4L4hc}
  \end{center}
\end{figure}

We will not discuss the rest of the holomorphic $L=4$ spectrum in detail, but instead consider the approach to SCQCD, which, as discussed, can be done in several ways. Firstly, one can take all the outer node couplings to zero ($\kappa_i\ra 0$ for $i=1,2,3,4$), in which case one ends up with a single node with $\SU(2N)$ and $4N$ flavours. Alternatively, one can take the central node coupling to zero, $\kappa_5\ra 0$. In that case one ends up with four decoupled $\SU(N)$ nodes, with $2N$ flavours each, so one has four decoupled copies of SCQCD. Finally, one can take all the couplings to zero apart from an outer one. (Of course for each case there are also decoupled vector multiplets at the global nodes). In the case where $\kappa_5\ra 0$ one expects to find the same spectrum as for $\kappa_i\ra 0$, but with four times the multiplicity. This is precisely what one sees by explicit diagonalisation.\footnote{Specifically, the multiplicity at $E=8$ is four times higher, while there are fewer states arriving at $E=4$, from which we conclude that $E=4$ is not a true SCQCD energy but due to the decoupled multiplets.} We plot these three cases in Fig. \ref{GraphD4L4ha}, Fig. \ref{GraphD4L4hb} and Fig. \ref{GraphD4L4hc}, respectively.

\begin{figure}[h!]
  \begin{center}
        \begin{tikzpicture}
      \node at (0,0) {\includegraphics[width=12cm]{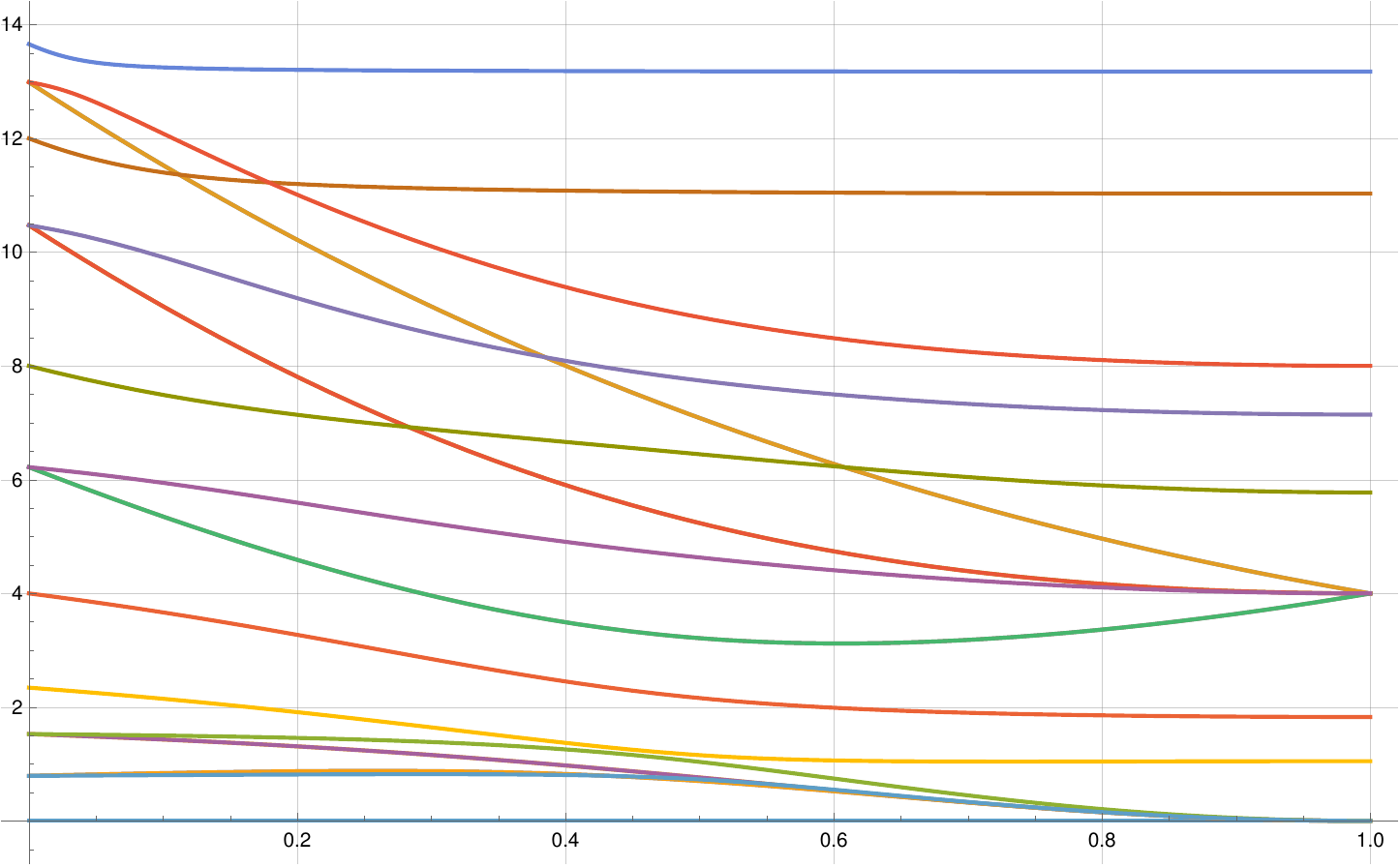}};
\node at (6.2,-3.3){$k$};\node at (-6.2,3.1){$E$};
\end{tikzpicture}
  \caption{The spectrum of $\Dfour$ $L=4$ holomorphic states for the case $\{\kappa_i=\kappa_5=1,\kappa_{2,3,4}=1-k\}$, where $k=0$ corresponds to the orbifold point and $k=1$ to the $\SU(N)\times\SU(2N)$ theory with $N_f=3N$ attached to the $\SU(2N)$ node. }\label{GraphD4L4hd}
  \end{center}
\end{figure}

Of course, there are multiple other ways to obtain theories with flavour nodes from the $\Dfour$ quiver. As a final example, one can take $\kappa_{2,3,4}\ra 0$ while $\kappa_1=\kappa_5=1$. This produces the 
\begin{tikzpicture}[baseline=-0.05cm]
  \draw[-] (0,0) circle (1.5ex);
  \draw[-] (0.75,0) circle (1.5ex);
  \draw[-] (1.25,-0.25)--(1.25,0.25)--(1.75,0.25)--(1.75,-0.25)--(1.25,-0.25);
  \draw[-] (0.25,0)--(0.5,0);\draw[-] (1,0)--(1.25,0);
  \node at (0,0) {\footnotesize $N$};
  \node at (0.75,0) {\footnotesize $2N$};
  \node at (1.5,0) {\footnotesize $3N$};
\end{tikzpicture}
balanced linear quiver with $\SU(N)\times\SU(2N)$ gauge group and an $N_f=3N$ flavour node, see \cite{Nunez:2023loo} for a recent discussion and background. The spectrum of this deformation is plotted in Fig. \ref{GraphD4L4hd}, with the energies at $k=1$ being the roots of
\be
P(E)=E^{13}(E-4)^7(E-8)(E^3-20E^2+122E-144)(E^3-20E^2+96E-80)\;.
\ee
Of course, as also emphasised in \cite{Gadde:2010zi} in the $\Zset_2$ interpolating theory context, taking the limit where some gauge groups become global does not give the full information on the spectrum of the limiting theory, as it will not capture states which are not singlets of the original gauge groups. However, one expects that the above energies form part of the spectrum of the limiting theory, and it would be interesting to reproduce them directly. 


\subsubsection{Length 6: $XY$ sector}

Although we leave a detailed study of higher-length chains for the future, it is worth writing out the single (as predicted by the Molien series (\ref{eq:D4Molien})), untwisted $XY$-sector protected state at $L=6$:
\be\begin{split}
\frac14\Tr\left(XY(XXXX-YYYY)\right)&=\Tr\left(X_{54}Y_{45}Y_{53}X_{35}(Y_{51}X_{15}+X_{52}Y_{25})\right)\\
&\;+\Tr\left(X_{52}Y_{25}Y_{51}X_{15}(X_{54}Y_{45}+Y_{53}X_{35})\right)\\
&\;-\Tr\left(X_{52}Y_{25}Y_{53}X_{35}(Y_{51}X_{15}+X_{54}Y_{45})\right)\\
&\;-\Tr\left(X_{54}Y_{45}Y_{51}X_{15}(Y_{53}X_{35}+X_{52}Y_{25})\right)
\end{split}\ee
The descent of this state from the degree-6 $\hat{\mathrm{D}}_4$ invariant polynomial $\half xy(x^4-y^4)$ \cite{Benvenuti:2006qr} is clear. One expects to be able to  similarly understand the structure of the higher-length BPS states in this sector as derived from the invariants of $\hat{\mathrm{D}}_4$.

\subsection{Two-magnon Bethe Ansatz}

\begin{figure}[h]
  \begin{center}
    \begin{tikzpicture}[scale=0.7]
      \draw[thick, red, -] (0,0) [partial ellipse=0:360:3cm and 1.5cm];
      \draw[thick, blue, -] (0,0) [partial ellipse=240:300:3cm and 1.5cm];
      \draw[thick,->] (0,0) [partial ellipse=235:245:3.25cm and 1.75cm];
      \draw[thick,->] (0,0) [partial ellipse=305:295:3.25cm and 1.75cm];
      \draw[fill=purple] (-1.5,-1.3) circle (.5ex);\draw[fill=purple] (1.5,-1.3) circle (.5ex);
      \node at (-1.6,-1.9) {\color{darkgray}$X_{15}$};\node at (1.9,-1.9) {\color{darkgray}$Y_{51}$};
      \node at (3.9,0) {\large $\rightarrow$};
    \end{tikzpicture}\hspace{0.2cm}
\begin{tikzpicture}[scale=0.7]
  \draw[thick, red, -] (0,0) [partial ellipse=0:360:3cm and 1.5cm];
  \draw[thick, blue, -] (0,0) [partial ellipse=240:300:3cm and 1.5cm];
  \draw[thick,->] (0,0) [partial ellipse=245:235:3.25cm and 1.75cm];
  \draw[thick,->] (0,0) [partial ellipse=295:305:3.25cm and 1.75cm];
        \draw[fill=purple] (-1.5,-1.3) circle (.5ex);\draw[fill=purple] (1.5,-1.3) circle (.5ex);
        \node at (-1.5,-1.9) {\color{darkgray}$X_{15}$};\node at (1.6,-1.9) {\color{darkgray}$Y_{51}$};
         \node at (3.9,0) {\large $\rightarrow$};
\end{tikzpicture}

\begin{tikzpicture}[scale=0.7]
  \draw[thick, blue, -] (0,0) [partial ellipse=0:360:3cm and 1.5cm];
  \draw[thick, red, -] (0,0) [partial ellipse=60:120:3cm and 1.5cm];
  \draw[thick,->] (0,0) [partial ellipse=55:65:3.25cm and 1.75cm];
  \draw[thick,->] (0,0) [partial ellipse=125:115:3.25cm and 1.75cm];
        \draw[fill=purple] (-1.5,1.3) circle (.5ex);\draw[fill=purple] (1.5,1.3) circle (.5ex);
  \node at (-1.5,1.9) {\color{darkgray}$X_{15}$};\node at (1.7,1.9) {\color{darkgray}$Y_{51}$};
 \node at (3.9,0) {\Large$=$};
\end{tikzpicture}\hspace{0.2cm}
\begin{tikzpicture}[scale=0.7,baseline=-1.2cm]
  \draw[thick, blue, -] (0,0) [partial ellipse=0:360:3cm and 1.5cm];
  \draw[thick, red, -] (0,0) [partial ellipse=240:300:3cm and 1.5cm];
  \draw[thick,->] (0,0) [partial ellipse=235:245:3.25cm and 1.75cm];
  \draw[thick,->] (0,0) [partial ellipse=305:295:3.25cm and 1.75cm];
        \draw[fill=purple] (-1.5,-1.3) circle (.5ex);\draw[fill=purple] (1.5,-1.3) circle (.5ex);
  \node at (-1.6,-1.9) {\color{darkgray}$Y_{51}$};\node at (1.9,-1.9) {\color{darkgray}$X_{15}$};
 \node at (3.9,0) {\large $\rightarrow$};
\end{tikzpicture}

\vspace{0.3cm}
\begin{tikzpicture}[scale=0.7]
  \draw[thick, blue, -] (0,0) [partial ellipse=0:360:3cm and 1.5cm];
  \draw[thick, red, -] (0,0) [partial ellipse=240:300:3cm and 1.5cm];
  \draw[thick,->] (0,0) [partial ellipse=245:235:3.2cm and 1.7cm];
  \draw[thick,->] (0,0) [partial ellipse=295:305:3.2cm and 1.7cm];
   \draw[fill=purple] (-1.5,-1.3) circle (.5ex);\draw[fill=purple] (1.5,-1.3) circle (.5ex);
  \node at (-1.5,-1.9) {\color{darkgray}$Y_{51}$};\node at (1.5,-1.9) {\color{darkgray}$X_{15}$};
 \node at (3.9,0) {\large $+$};
\end{tikzpicture}
\begin{tikzpicture}[scale=0.7]
  \draw[thick, blue, -] (0,0) [partial ellipse=0:360:3cm and 1.5cm];
  \draw[thick, red, -] (0,0) [partial ellipse=240:300:3cm and 1.5cm];
  \draw[thick,->] (0,0) [partial ellipse=245:235:3.2cm and 1.7cm];
  \draw[thick,->] (0,0) [partial ellipse=295:305:3.2cm and 1.7cm];
  \draw[fill=purple] (-1.5,-1.3) circle (.5ex);\draw[fill=purple] (1.5,-1.3) circle (.5ex);
  \node at (-1.5,-1.9) {\color{darkgray}$Y_{53}$};\node at (1.5,-1.9) {\color{darkgray}$X_{35}$};
 \node at (3.9,0) {\large $+$};
\end{tikzpicture}

\vspace{0.3cm}

\begin{tikzpicture}[scale=0.7]
  \draw[thick, blue, -] (0,0) [partial ellipse=0:360:3cm and 1.5cm];
  \draw[thick, green, -] (0,0) [partial ellipse=240:300:3cm and 1.5cm];
  \draw[thick,->] (0,0) [partial ellipse=245:235:3.2cm and 1.7cm];
  \draw[thick,->] (0,0) [partial ellipse=295:305:3.2cm and 1.7cm];
   \draw[fill=teal] (-1.5,-1.3) circle (.5ex);\draw[fill=teal] (1.5,-1.3) circle (.5ex);
  \node at (-1.5,-1.9) {\color{darkgray}$X_{52}$};\node at (1.5,-1.9) {\color{darkgray}$Y_{25}$};
  \node at (3.9,0) {\large $+$};
  \end{tikzpicture}
\begin{tikzpicture}[scale=0.7]
  \draw[thick, blue, -] (0,0) [partial ellipse=0:360:3cm and 1.5cm];
  \draw[thick, green, -] (0,0) [partial ellipse=240:300:3cm and 1.5cm];
  \draw[thick,->] (0,0) [partial ellipse=245:235:3.2cm and 1.7cm];
  \draw[thick,->] (0,0) [partial ellipse=295:305:3.2cm and 1.7cm];
  \draw[fill=teal] (-1.5,-1.3) circle (.5ex);\draw[fill=teal] (1.5,-1.3) circle (.5ex);
  \node at (-1.5,-1.9) {\color{darkgray}$X_{54}$};\node at (1.5,-1.9) {\color{darkgray}$Y_{45}$};
\end{tikzpicture}\hspace{0.6cm}\mbox{}
  \end{center}\caption{Scattering on the $\Dfour$ chain. Two magnons in an exterior $Z_1$ vacuum can only reflect (first row). As they meet again at the back of the chain the exterior vacuum will be $Z_5$ (second row). Then they can both reflect with either $Z_1$ or $Z_3$ interior vacuum (third row) as well as transmit, with either $Z_2$ or $Z_4$ interior vacuum (fourth row). For clarity, in the second row the chain has simply been rotated so that the scattering happens at the front.} \label{Fig:XYmagD4}
    
\end{figure}
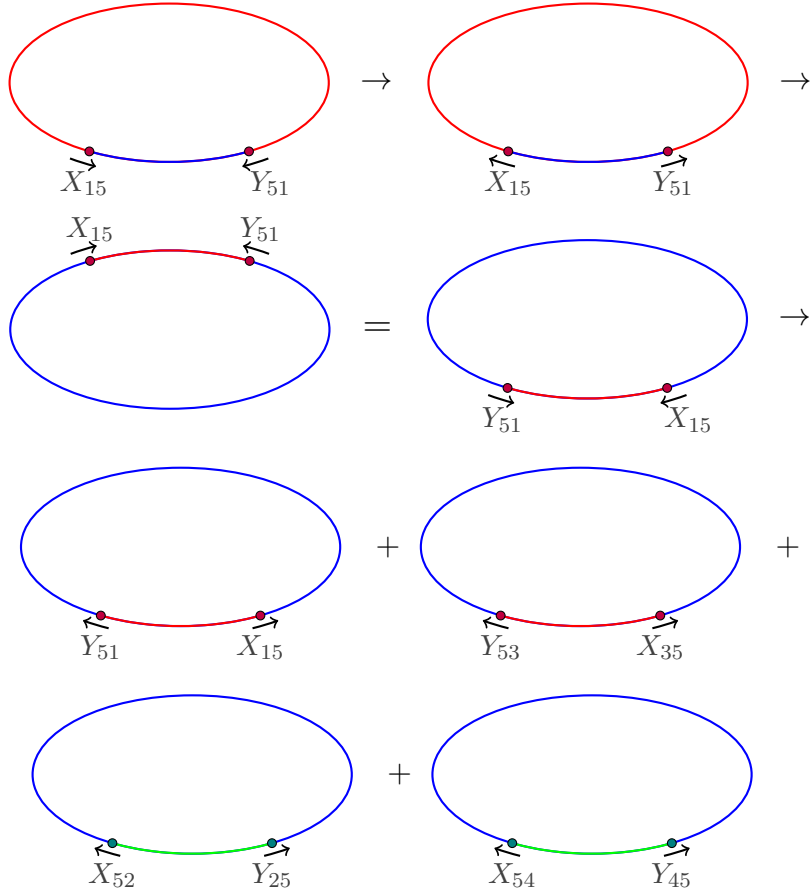

Let us now turn to the the study of magnons on the $\Dfour$ chain. As discussed, for the non-abelian orbifolds the only vacua available at any length are the $Z$ vacua. As we are interested in describing physical (closed chain) states, the simplest case to consider is that of one $X$ and one $Y$ magnon of the same type (i.e. separating the same two vacua). Given the $\Dfour$ quiver structure, one of the vacua always has to be the $Z_5$ vacuum. The main novelty compared to the $\Zset_k$ case is that two magnons scattering with an exterior $Z_i$ vacuum, where $i\neq 5$,  and an interior $Z_5$ vacuum, can only scatter reflectively. On the other hand, two magnons approaching each other with an exterior $Z_5$ vacuum and an interior $Z_i$ vacuum can reflect as well as transmit, creating all possible interior vacua in the process. An example of this behaviour is illustrated in Fig. \ref{Fig:XYmagD4}.

We therefore need to construct a Bethe ansatz that captures the above behaviour. Given the symmetries of the problem, we distinguish between nodes $i=1,2,3,4$ and node 5. We write
\be
Q_{i5}=\{X_{15},Y_{25},X_{35},Y_{45}\} \;\;\text{and}\;\; Q_{5i}=\{Y_{51},X_{52},Y_{53},X_{54}\}\;.
\ee

\subsubsection{Open chain}

We start by considering an asymptotic chain with a single magnon, which  can be thought of as a domain wall between the $Z_5$ vacuum and one of the $Z_i$ vacua:
\be
\ket{\ell}_{i5}= \cdots Z_i Z_i \underset{{\color{red} \ell}\phantom{A}}{Q_{i5}} Z_{5} Z_{5}\cdots \;\; \text{and}\quad
\ket{\ell}_{5i}= \cdots Z_5 Z_5 \underset{{\color{red} \ell}\phantom{A}}{Q_{5i}} Z_{i} Z_{i}\cdots \;\;,
\ee
The dispersion relation (\ref{DispersionGeneral}) applied to this case is
\be
E_{i5}=E_{5i}= 2(\kappa_i^2+\kappa_{5}^2)-2\kappa_{i}\kappa_{5}(e^{ip^{(i)}}+e^{-ip^{(i)}})\;,
\ee
where we have labelled the momenta according to the $i$ vacuum. We see that zero-energy states have imaginary momenta $p^{(i)}=\pm i\ln\frac{\kappa_5}{\kappa_i}$. At the orbifold point, the dispersion relation reduces to the usual $E=4(1-\cos(p))$ where the zero-energy condition is $p=0$.

We are interested in solving the 2-magnon problem with one magnon of $Q_{i5}$ and one magnon of $Q_{5i}$ type. This implies that the exterior indices to the left and right of the magnons will be the same, and thus we can form closed states, which will be trace operators in the gauge theory.
We therefore distinguish two cases: either the exterior vacuum is made up of $Z_{i}$ fields or $Z_{5}$ fields. For the first case we write
\be
\ket{\ell_1,\ell_2}_{i}=\cdots Z_i Z_i \underset{{\color{red}\ell_1}\phantom{A}}{Q_{i5}} Z_{5} \cdots Z_{5} \underset{{\color{red}\ell_2}\phantom{A}}{Q_{5i}} Z_{i} Z_{i}\cdots
\ee
It is easy to see that there is no transmission possible in this case, as for neighbouring magnons the $\Dfour$ Hamiltonian (\ref{HD4QQ}) acts simply as $\Hcal(Q_{i5}Q_{5i})=4\kappa_i^2 Q_{i5}Q_{5i}$. Of course, the momenta of the two magnons can still be exchanged. 
We therefore write a Bethe ansatz of the form
\be
\ket{\psi}_i=\sum_{\ell_1<\ell_2} \left(A_i e^{i(p^{(i)}_1\ell_1+p^{(i)}_2\ell_2)}+B_i e^{i(p^{(i)}_2\ell_1+p^{(i)}_1\ell_2)}\right)\ket{\ell_1,\ell_2}_i\;.
\ee
As above, the non-interacting equations simply give us the 2-magnon energies
\be \label{Energyi}
E_{i}^{(2)}=4(\kappa_i^2+\kappa_5^2)-2\kappa_i\kappa_5\left(e^{i p^{(i)}_1}+e^{-i p^{(i)}_1}+e^{i p^{(i)}_2}+e^{-i p^{(i)}_2}\right)\;.
\ee
The interacting equations, arising by asking for the state $\ket{\ell,\ell+1}$ to be an eigenstate, are
\be\begin{split}
&8\kappa_i^2(A_ie^{ip^{(i)}_2}+B_i e^{ip^{(i)}_1})-2\kappa_i\kappa_5(A_ie^{-ip^{(i)}_1+ip^{(i)}_2}+B_i e^{-ip^{(i)}_2+ip^{(i)}_1}
+A_ie^{2ip^{(i)}_2}+B_i e^{2ip^{(i)}_1})\\
&\quad=E_i(A_ie^{ip^{(i)}_2}+B_i e^{ip^{(i)}_1})\;.
\end{split}
\ee
We can then write the $S$-matrix for the exterior $Z_i$-vacuum chain as
\be
S_i=\frac{B_i}{A_i}=-\frac{\kappa_i\kappa_5(1+e^{i(p^{(i)}_1+p^{(i)}_2)})+2e^{ip^{(i)}_2}(\kappa_i^2-\kappa_5^2)}{\kappa_i\kappa_5(1+e^{i(p^{(i)}_1+p^{(i)}_2)})+2e^{ip^{(i)}_1}(\kappa_i^2-\kappa_5^2)}\;,
\ee
which satisfies $S_i^*S_i=1$. Notice that it reduces to $S_i=-1$ at the orbifold point $\kappa_i=\kappa_5=1$.

The case with exterior $Z_5$ vacuum is more complicated as, for example, a $Y_{51}X_{15}$ pair of magnons can scatter to $X_{52}Y_{25}$, $Y_{53}X_{35}$ and $Y_{54}X_{45}$. Therefore, defining the generic two-magnon state as
\be
\ket{\ell_1,\ell_2}^{(5)}_{i}=\cdots Z_5 Z_5 \underset{{\color{red}\ell_1}\phantom{A}}{Q_{5i}} Z_{i} \cdots Z_{i} \underset{{\color{red}\ell_2}\phantom{A}}{Q_{i5}} Z_{5} Z_{5}\cdots \;,
\ee
we need to write a Bethe ansatz of the form
\be
\ket{\psi}^{(5)}_i=\sum_{\ell_1<\ell_2}\sum_{i=1}^4 (-1)^{i+1}(C_i e^{i p^{(i)}_2\ell_1+i p^{(i)}_1\ell_2}+D_i e^{i p^{(i)}_1\ell_1+i p^{(i)}_2\ell_2}) \ket{\ell_1,\ell_2}^{(5)}_{i}\;.
\ee
Now, $i$ labels the \emph{interior} vacuum. Note the sign for even $i$, which takes into account that the states $X_{52}Y_{25}$ and $X_{54}Y_{45}$ appear with opposite sign in the action of the Hamiltonian (\ref{HD4holom}) compared to $Y_{51}X_{15}$ and $Y_{53}X_{35}$. Note also that we have exchanged the meaning of the incoming and outgoing momenta, a convention which will come out useful when making the chain periodic. 

The energies are still given by (\ref{Energyi}). For magnons of type $i$ to coexist with magnons of type $j$ on the same chain, their corresponding momenta need to be related by solving
\be \label{D4energymomentum}
p^{(i)}_1+p^{(i)}_2=p^{(j)}_1+p^{(j)}_2 \;\; \text{and}\;\; E_i^{(2)}(p_1^{(i)},p_2^{(i)})=E_j^{(2)}(p_1^{(j)},p_2^{(j)})\;.
\ee
Of course, at the orbifold point all the momenta are equal (up to sign), $p^{(1)}_k =\pm p^{(2)}_k$ $=\pm p^{(3)}_k=\pm p^{(4)}_k$ for $k=1,2$. 

Let us now consider the interacting equations for the exterior $Z_5$ vacuum. Given the mixing of all the states with different interior vacua, we obtain the four equations
\be\begin{split}
&5\kappa_5^2(C_ie^{ip^{(i)}_1}\!+\!D_ie^{ip^{(i)}_2})-2\kappa_i\kappa_5(C_ie^{-i p^{(i)}_2+i p^{(i)}_1}\!+\!D_ie^{-i p^{(i)}_1+i p^{(i)}_2}\!+\!
C_ie^{2i p^{(i)}_1}\!+\!D_ie^{2i p^{(i)}_2})\\&+\kappa_5^2\sum_{j\neq i}(C_je^{ip^{(j)}_1}\!+\!D_je^{i p^{(j)}_2})=E_i
(C_ie^{ip^{(i)}_1}+D_ie^{ip^{(i)}_2})\;.
\end{split}\ee
Note that there is a $\Zset_4$ symmetry under ($\kappa_i,C_{i},D_{i}) \ra (\kappa_{i+1},C_{i+1},D_{i+1}$).

To solve the system, we treat the $C_i$ as incoming states and the $D_i$ as outgoing. So our $S$-matrix for this case will be defined through
\be
D_i=\sum_{j} S^{(5)}_{ij} C_j\;.
\ee
The solution can be expressed in a compact way by defining the combinations
\be
n^{(i)}_k=\kappa_5-2\kappa_ie^{ip^{(i)}_k}+\kappa_5 e^{i(p^{(i)}_1+p^{(i)}_2)}\;,
\ee
as well as
\be
m^{(i)}_k=8 e^{ip^{(i)}_k}\kappa_{i+1}\kappa_{i+2}\kappa_{i+3}\kappa_5^6 n^{(i+1)}_kn^{(i+2)}_k n^{(i+3)}_k\;,
\ee
and the $\Zset_4$-invariant combination
\be
\Ncal(k^1,k^2,k^3,k^4)=m^{(1)}_{k^1}+m^{(2)}_{k^2}+m^{(3)}_{k^3}+m^{(4)}_{k^4}+16\kappa_1\kappa_2\kappa_3\kappa_4\kappa_5^4 n^{(1)}_{k^2}n^{(2)}_{k^2}n^{(3)}_{k^3}n^{(4)}_{k^4}\;.
\ee
Above, $k=1,2$ labels the first or second momentum of each pair of magnons. Then the diagonal components of the $S$-matrix are 
\be
S^{(5)}_{11}=-\frac{\Ncal(1,2,2,2)}{\Ncal(2,2,2,2)}\;,S^{(5)}_{22}=-\frac{\Ncal(2,1,2,2)}{\Ncal(2,2,2,2)}\;,S^{(5)}_{33}=-\frac{\Ncal(2,2,1,2)}{\Ncal(2,2,2,2)}\;,S^{(5)}_{44}=-\frac{\Ncal(2,2,2,1)}{\Ncal(2,2,2,2)}\;.
\ee
Effectively, in the numerator of $S^{(5)}_{ii}$  we need to exchange the momenta $p^{(i)}_1 \leftrightarrow p^{(i)}_2$, leaving the other momenta unchanged. For the off-diagonal components we have
\be
S^{(5)}_{ij}=-\frac{8}{\Ncal(2,2,2,2)}(\prod_{j\neq i}\kappa_i)\kappa_5^7 \left(e^{ip^{(j)}_1}-e^{ip^{(j)}_2}+e^{i(2p^{(j)}_1+p^{(j)}_2)}+e^{i(2p^{(j)}_2+p^{(j)}_1)}\right)\prod_{m\neq{i,j}} n^{(m)}_2\;.
\ee
We emphasise that although the $S$-matrix appears to depend on all the momenta, for the scattering to make sense we need to impose the energy and momentum constraints (\ref{D4energymomentum}), which express all the momenta in terms of one set, e.g. $p^{(1)}_{1,2}$. With this choice, we could write it as $S^{(5)}(p^{(1)}_1,p^{(2)}_1)$.

The $S$-matrix satisfies $(S^{(5)})^* S^{(5)}=I_{4\times 4}$, similarly to the $\Zset_3$ case. As discussed there, this is due to the different dispersion relations of each magnon species. At the orbifold point, where all the $p^{(i)}_k=\pm p_k$, we have
\be
S^{(5),\text{o.p}}_{ii}=\frac{-2+3e^{ip_1}+e^{ip_2}-2e^{i(p_1+p_2)}}{2(1-2e^{ip_2}+e^{i(p_1+p_2)})}\;,\; S^{(5),\text{o.p}}_{ij}=S^{(5),\text{o.p}}_{ji}=\frac{e^{ip_2}-e^{ip_1}}{2(1-2e^{ip_2}+e^{i(p_1+p_2)})}\;,
\ee
which is symmetric and therefore the $S$-matrix becomes unitary,  $(S^{(5),\text{o.p}})^\dag S^{(5),\text{o.p}}=I_{4\times 4}$. 

\subsubsection{Closed chain}

We now need to impose cyclicity of the trace, which will relate the $\ket{\psi}^i$ and $\ket{\psi}^{(5)}_i$ wavefunctions. To illustrate this, let us specialise to the case of $L=3$. Then we need to impose that the wavefunctions corresponding to the states
\be
Z_5Y_{5i}X_{i5} \;,Y_{5i}X_{i5}Z_5 \; \text{and} \; X_{i5}Z_5Y_{5i}\;,
\ee
are equal. We therefore have
\be\begin{split}
C_i e^{i(2p^{(i)}_2+3p^{(i)}_1)}+D_i e^{i(2p^{(i)}_1+3p^{(i)}_2)}&=C_i e^{i(p^{(i)}_2+2p^{(i)}_1)}+D_i e^{i(p^{(i)}_1+2p^{(i)}_2)}\\
&=A_i e^{i(p^{(i)}_1+2p^{(i)}_2)}
+B_i e^{i(p^{(i)}_2+2p^{(i)}_1)}\;.
\end{split}
\ee
The first condition imposes the centre-of-mass condition $p^{(i)}_2=-p^{(i)}_1$, as usual, while the second condition relates the Bethe ansatz coefficients as
\be \label{D4L3cyclicity}
C_i=A_i e^{-3i p^{(i)}_1}\;\;,\;\;D_i=B_i e^{3i p^{(i)}_1}\;.
\ee
Using the $S$-matrices to express the $B_i$ in terms of the $A_i$ and the $C_i$ in terms of the $D_i$, and also expressing all the $p^{(i)}_1$ in terms of a single momentum (e.g. $p^{(1)}_1$) through the dispersion relations, we can solve for $p^{(1)}_1$ and find the corresponding energies. The momenta and corresponding energies at the orbifold point are indicated in Table \ref{Table:D4L3op}. We have also indicated which twisted sector each solution corresponds to. Notice that there is no BPS state in this sector in the $D_4$ theory. Instead, we see a twisted sector state with $E=4$, which was not present in the $\Zset_3$ case above. 

\begin{table}[ht]
\begin{center}
  \begin{tabular}{|c|c|c|} \hline $p^{(1)}_1$ & E & Sector \\\hline 
    $\frac{2\pi}{3}$ & 12 & Untwisted\\
    $3\pi/5$ & $2(3+\sqrt{5})$ & $a$-twisted\\
    $3\pi/5$ & $2(3+\sqrt{5})$ & $b$-twisted\\ 
    $3\pi/5$ & $2(3+\sqrt{5})$ & $ab$-twisted\\
    $\pi/3$ & $4$ &$a^2$-twisted\\ 
    $\pi/5$ & $2(3-\sqrt{5})$ & $a$-twisted\\ 
    $\pi/5$ & $2(3-\sqrt{5})$ &$b$-twisted\\
    $\pi/5$ & $2(3-\sqrt{5})$ &$ab$-twisted\\ \hline
  \end{tabular}\caption{The $\hat{D}_4$ $L=3$ $XYZ$-sector momenta and energies at the orbifold point. Note the absence of an $E=0$ state. The degenerate twisted states are distinguished by different signs in the identification of the $p^{(i)}$ momenta.} \label{Table:D4L3op}
\end{center}
\end{table}

We can now consider the momenta and energies for deformations away from the orbifold point. In Table \ref{Table:D4L3def} we list the values for two deformations, one ($\kappa_i=0.9$) which preserves the $\Zset_4$ symmetry, as well as a more general deformation. The values one obtains are in complete agreement with the explicit diagonalisation of the Hamiltonian. 

\begin{table}[h]
\begin{center}
  \begin{tabular}{|cc|cc|cc|}\hline \multicolumn{2}{|c|}{$(1,1,1,1,1)$} & \multicolumn{2}{c|}{$(0.9,0.9,0.9,0.9,1)$} & \multicolumn{2}{c|}{$(0.9,0.8,0.93,0.99,1)$}\\  \hline $p_1^{(1)}$ & E & $p^{(1)}_1$ & E  & $p^{(1)}_1$ & E\\\hline 
    2.0944 & 12 & 2.1072& 10.9193 & 2.1689& 11.2942\\
    1.885 & 10.4721 & 1.8246 &9.0476 &1.9473 & 9.8874\\
    1.885 & 10.4721& 1.8246 & 9.0476&1.8493 &9.2192 \\
    1.885 & 10.4721 & 1.8246 & 9.0476&1.6932 & 8.1193\\
    1.0472 & 4 & 1.0344&3.5607 & 1.0332& 3.5528 \\
    0.6284 & 1.5279 & 0.6324 & 1.4324& 0.6475& 1.4972\\
    0.6284 & 1.5279 & 0.6324& 1.4324& 0.6357& 1.4465\\
    0.6284 & 1.5279 & 0.6324& 1.4324& 0.6111& 1.3433\\\hline 
    \end{tabular}\caption{A comparison of the $\hat{D}_4$ $L=3$ $XYZ$-sector momenta and energies at the orbifold point and two sample deformations, labelled by the values of $(\kappa_1,\kappa_2,\kappa_3,\kappa_4,\kappa_5)$. The middle deformation corresponds to Fig. \ref{GraphD4L3ha} at $k=0.1$.}   \label{Table:D4L3def}
\end{center}
\end{table}

\begin{figure}[t]
  \begin{center}
\begin{tikzpicture}
    \node at (0,0){\includegraphics[width=10cm]{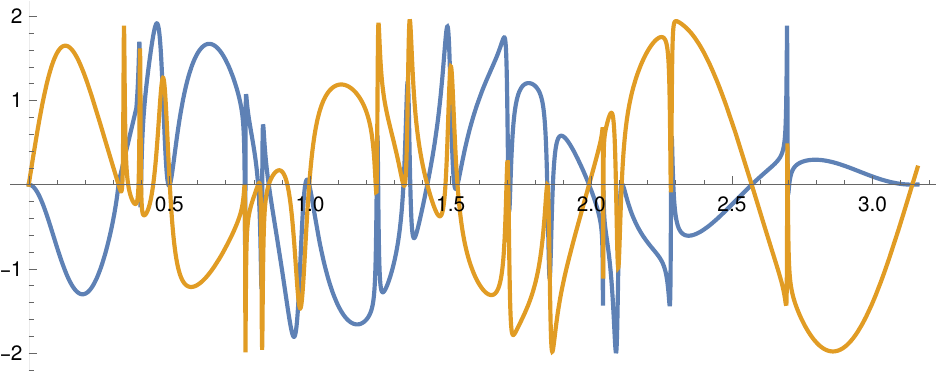}};
    \node at (5.1,-0.4){$p_1^{(1)}$};\node at (-5.1,1.9) {$f$};
    \end{tikzpicture}
  \end{center}
  \caption{Plot of the real (blue) and imaginary (orange) part of the function $f=C_1-A_1e^{6i p_2^{(1)}}$, for the $L=6$ $\hat{D}_4$ deformation $(\kappa_1,\kappa_2,\kappa_3,\kappa_4,\kappa_5)$ $=(0.9,0.8,0.93,0.7,1)$,  after substituting all the coefficients and expressing the $p^{(2,3,4)}$ momenta in terms of $p_1^{(1)}$.} \label{Fig:BethezeroesD4}
\end{figure}

Of course, the above two-magnon Bethe ansatz solution works equally well for any length (replacing $3\ra L$ in (\ref{D4L3cyclicity})) and thus provides an infinite number of one-loop energy values for two-excitation operators. As for the $\Zset_3$ quiver, to illustrate the general features we can consider the $L=6$ case. Using the $S$-matrices to solve for all coefficients, and solving all the cyclicity conditions apart from one, which in this case we take to be $C_1-A_1e^{6ip_2^{(1)}}$, we can substitute the $p^{(2,3,4)}$ momenta in terms of $p_1^{(1)}=-p_2^{(1)}$ and plot the resulting function of $p_1^{(1)}$ to obtain a visual guide to where the roots are located. For a specific deformation this is plotted in Fig.\ref{Fig:BethezeroesD4}. The resulting roots (easily obtained using \texttt{FindRoot}) are listed in Table \ref{Table:L6D4Bethe}. The energies obtained are in exact agreement with the energies of the 20 two-magnon states that one finds by direct diagonalisation.  

\begin{table}[h]
\begin{center}
  \begin{tabular}{|cc|cc|cc|cc|}\hline 
    $p_1^{(1)}$ & $E$ &$p_1^{(1)}$ &$E$ &$p_1^{(1)}$ &$E$ &$p_1^{(1)}$ &$E$ \\ \hline
    0.319216 & 0.403732       & 0.815535 & 2.30455 & 1.41706 & 6.13748 &2.04319& 10.5161\\
    0.353569 & 0.485371 & 0.853333 & 2.50618  & 1.52446 & 6.90647 &2.10822& 10.9258\\
    0.399496 & 0.606947 & 0.997224 & 3.33302  & 1.70514 & 8.20439 & 2.28393& 11.9503\\
    0.501433 & 0.926359 & 1.23728 & 4.88298  & 1.84358 &9.17976  & 2.57056& 13.2977\\
    0.767938 & 2.06072 & 1.33233 & 5.5393 & 1.9922 & 10.1851 & 2.69882& 13.7457\\ \hline
  \end{tabular}
\end{center}\caption{The $\hat{D}_4$-quiver two-magnon Bethe momenta and energies for $L=6$ and $(\kappa_1, \kappa_2, \kappa_3,\kappa_4,\kappa_5)=(0.9,0.8,0.93,0.7,1)$.} \label{Table:L6D4Bethe}
\end{table}

There are again spurious solutions at $p_1^{(1)}=0,\pi$ where the wavefunction vanishes and which we need to omit. All other roots correspond to valid two-magnon energies. Note that, with this choice of deformation, all values of $p_1^{(1)}$ happen to be real. However, some of the other momenta will be complex, for instance the $p_1^{(4)}$ momenta corresponding to the three largest energies are $\pi+0.37121i, \pi+0.76871i$ and $\pi+0.85704i$. As in the $\Zset_3$ case, at the orbifold point and deformations close to it, all four momenta $p^{(i)}$ are real for all two-magnon energies.  

 Having considered further lengths and choices of deformation parameters, we conclude that the two-magnon Bethe ansatz for the $\hat{D}_4$ quiver is complete, and one can choose at least one of the four momenta such that all its required values are real and lie in the range $(0,\pi)$ in order to produce each of the two-magnon energies once. The main difference to the $\Zset_3$ case is the absence of the $E=0$ state, which, as discussed, is projected out in the non-abelian quivers. 

Extensions to different types of excitations (such as one holomorphic and one antiholomorphic) should be straightforward, however extending to three or more magnons is expected to suffer from the same issues as for the $\Zset_2$ case (see \cite{Gadde:2010zi,Pomoni:2021pbj}), whose resolution will likely require a more advanced approach along the lines of \cite{Bozkurt:2024tpz,Bozkurt:2025exl}.

\section{Example: The $\hat{E}_6$ theory} \label{sec:E6}

The $\hat{E}_6$ theory corresponds to the binary tetrahedral group ${2T}$, which is of order 24 and defined as (\ref{2Tdefinition})
\be
\{r,s,t|r^2=s^3=t^3=rts=z\}. 
\ee
The element $z$ is central. As $t=r^{-1}zs^{-1}=r^{-1}s^2=rs^{-1}$, it is sufficient to use $s$ and $r$ to represent all the elements. There are 7 conjugacy classes, represented by $e,z,s,s^2,s^4,s^5,r$. 

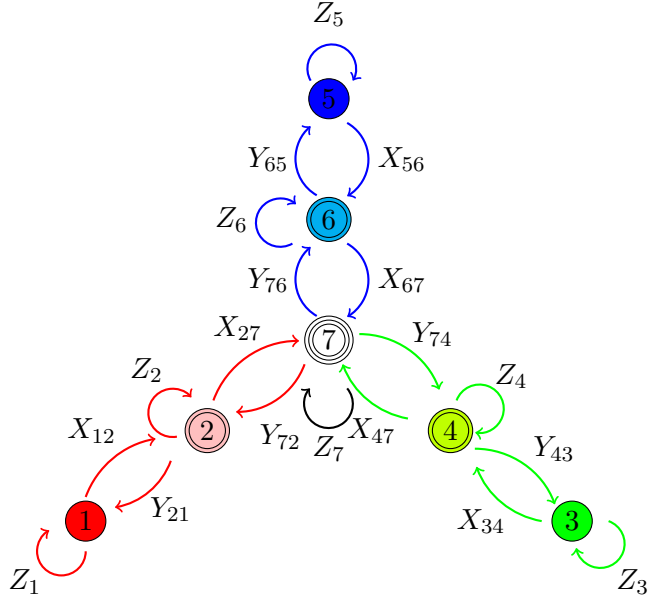
\begin{figure}[h]
    \centering
    \begin{tikzpicture}[scale=0.8]
  \draw[fill=white] (4,4) circle (2.4ex);
  \draw[fill=white] (4,4) circle (2ex);
  \draw[fill=white] (4,4) circle (1.6ex);
  \draw[fill=pink] (2,2.5) circle (2.2ex);
   \draw[fill=pink] (2,2.5) circle (1.8ex);
  \draw[fill=red] (0,1) circle (2ex);
  \draw[fill=lime] (6,2.5) circle (2.2ex);
    \draw[fill=lime] (6,2.5) circle (1.8ex);
  \draw[fill=green] (8,1) circle (2ex);
  \draw[fill=cyan] (4,6) circle (2.2ex);
    \draw[fill=cyan] (4,6) circle (1.8ex);
 \draw[fill=blue] (4,8) circle (2ex);

 \draw[->,blue,thick] (3.4,5.6) arc (-60:-310:0.4);
 \draw[->,blue,thick] (3.7,8.3) arc (210:-30:0.4);
 \draw[->,black,thick] (4.3,3.2) arc (40:-220:0.4);

 \draw[->,green,thick] (6.1,3) arc (160:-100:0.4);
 \draw[->,green,thick] (8.6,1) arc (70:-170:0.4);

 \draw[->,red,thick] (0,0.5) arc (0:-240:0.4);
\draw[->,red,thick] (1.5,2.4) arc (-80:-320:0.4);

\draw[->,green,thick] (4.5,4.1) arc (90:20:1.4);
\draw[->,green,thick] (6.4,2.2) arc (90:20:1.4);
\draw[->,green,thick] (5.3,2.7) arc (260:200:1.4);
\draw[->,green,thick] (7.5,1) arc (260:200:1.4);

\draw[->,red,thick] (1.4,2) arc (-20:-80:1.2);
\draw[->,red,thick] (3.6,3.6) arc (-20:-90:1.2);
\draw[->,red,thick] (0,1.4) arc (-200:-260:1.5);
\draw[->,red,thick] (2.1,3) arc (-200:-270:1.5);

\draw[->,blue,thick] (3.8,4.4) arc (-120:-230:0.7);
\draw[->,blue,thick] (3.8,6.4) arc (-120:-230:0.7);

\draw[->,blue,thick] (4.3,7.6) arc (60:-60:0.7);
\draw[->,blue,thick] (4.3,5.6) arc (60:-60:0.7);

\node at (4,4) {$7$};\node at (2,2.5) {$2$};\node at (0,1) {$1$};
\node at (6,2.5) {$4$};\node at (8,1) {$3$};
\node at (4,6) {$6$};\node at (4,8) {$5$};

\node at (-1,0) {$Z_{1}$};
\node at (1,3.5) {$Z_{2}$};
\node at (9,0) {$Z_{3}$};
\node at (7,3.4) {$Z_{4}$};
\node at (4,9.4) {$Z_{5}$};
\node at (2.4,6) {$Z_{6}$};
\node at (4,2.2) {$Z_{7}$};

\node at (0.1,2.5) {$X_{12}$};
\node at (2.5,4.2) {$X_{27}$};
\node at (6.5,1) {$X_{34}$};
\node at (4.7,2.5) {$X_{47}$};
\node at (5.2,7) {$X_{56}$};
\node at (5.2,5) {$X_{67}$};

\node at (1.4,1.2) {$Y_{21}$};
\node at (3.2,2.4) {$Y_{72}$};
\node at (7.7,2.2) {$Y_{43}$};
\node at (5.7,4.1) {$Y_{74}$};
\node at (3,7) {$Y_{65}$};
\node at (3,5) {$Y_{76}$};

\end{tikzpicture}
    \caption{The $\hat{E}_6$ quiver diagram.}
    \label{fig:E6Quiver}
\end{figure}

For the quiver-basis representation matrices, it is enough to show $\gamma(s)$ and $\gamma(r)$, as all the other matrices can be obtained using the group relations. They are
{\tiny\be
\gamma(s)=\left(
\begin{array}{cccccccccccc}
 1 & 0 & 0 & 0 & 0 & 0 & 0 & 0 & 0 & 0 & 0 & 0 \\
 0 & \frac{1}{2}+\frac{i}{2} & \frac{1}{2}+\frac{i}{2} & 0 & 0 & 0 & 0 & 0 & 0 & 0 & 0 & 0 \\
 0 & -\frac{1}{2}+\frac{i}{2} & \frac{1}{2}-\frac{i}{2} & 0 & 0 & 0 & 0 & 0 & 0 & 0 & 0 & 0 \\
 0 & 0 & 0 & \omega_3^2 & 0 & 0 & 0 & 0 & 0 & 0 & 0 & 0 \\
 0 & 0 & 0 & 0 & \left(\frac{1}{2}+\frac{i}{2}\right) \omega_3^2 & \left(\frac{1}{2}+\frac{i}{2}\right) \omega_3^2 & 0 & 0 & 0 & 0 & 0 & 0 \\
 0 & 0 & 0 & 0 & \left(-\frac{1}{2}+\frac{i}{2}\right)\omega_3^2 & \left(\frac{1}{2}-\frac{i}{2}\right) \omega_3^2 & 0 & 0 & 0 & 0 & 0 & 0 \\
 0 & 0 & 0 & 0 & 0 & 0 & \omega_3  & 0 & 0 & 0 & 0 & 0 \\
 0 & 0 & 0 & 0 & 0 & 0 & 0 & \left(\frac{1}{2}+\frac{i}{2}\right) \omega_3 & \left(\frac{1}{2}+\frac{i}{2}\right) \omega_3 & 0 & 0 & 0 \\
 0 & 0 & 0 & 0 & 0 & 0 & 0 & \left(-\frac{1}{2}+\frac{i}{2}\right) \omega_3 & \left(\frac{1}{2}-\frac{i}{2}\right) \omega_3 & 0 & 0 & 0 \\
 0 & 0 & 0 & 0 & 0 & 0 & 0 & 0 & 0 & 0 & 1 & 0 \\
 0 & 0 & 0 & 0 & 0 & 0 & 0 & 0 & 0 & 0 & 0 & 1 \\
 0 & 0 & 0 & 0 & 0 & 0 & 0 & 0 & 0 & 1 & 0 & 0 \\
\end{array}
\right)\;,
\ee
}
and
{\tiny\be\gamma(r)=\left(
\begin{array}{cccccccccccc}
 1 & 0 & 0 & 0 & 0 & 0 & 0 & 0 & 0 & 0 & 0 & 0 \\
 0 & -i & 0 & 0 & 0 & 0 & 0 & 0 & 0 & 0 & 0 & 0 \\
 0 & 0 & i & 0 & 0 & 0 & 0 & 0 & 0 & 0 & 0 & 0 \\
 0 & 0 & 0 & 1 & 0 & 0 & 0 & 0 & 0 & 0 & 0 & 0 \\
 0 & 0 & 0 & 0 & -i & 0 & 0 & 0 & 0 & 0 & 0 & 0 \\
 0 & 0 & 0 & 0 & 0 & i & 0 & 0 & 0 & 0 & 0 & 0 \\
 0 & 0 & 0 & 0 & 0 & 0 & 1 & 0 & 0 & 0 & 0 & 0 \\
 0 & 0 & 0 & 0 & 0 & 0 & 0 & -i & 0 & 0 & 0 & 0 \\
 0 & 0 & 0 & 0 & 0 & 0 & 0 & 0 & i & 0 & 0 & 0 \\
 0 & 0 & 0 & 0 & 0 & 0 & 0 & 0 & 0 & -1 & 0 & 0 \\
 0 & 0 & 0 & 0 & 0 & 0 & 0 & 0 & 0 & 0 & -1 & 0 \\
 0 & 0 & 0 & 0 & 0 & 0 & 0 & 0 & 0 & 0 & 0 & 1 \\
\end{array}
\right)\;.\ee}
We note that these are $24\times24$ matrices, as the elements in the (2,3), (5,6) and (8,9) blocks are multiplied by $I_2$ so those are really $4\times 4$ blocks, while the elements in the (10,11,12) block are multiplied by $I_3$ so that is actually a $9\times 9$ block. For the induced representation we use
\be
R(s)=\half\left(
\begin{array}{cc}
 1+i & 1+i \\
 -1+i & 1-i \\
\end{array}
\right)\;\text{and}\;\; R(r)=\left(
\begin{array}{cc}
 -i & 0 \\
 0 & i \\
\end{array}
\right)\;.
\ee
Starting from $\Ncal=4$ SYM with $\SU(24N)$ and imposing invariance under 2T, we obtain a gauge theory with 7 gauge groups, with the outer gauge groups being $\SU(N)$, the middle ones $\SU(2N)$ and the central one $\SU(3N)$. There are 6 hypermultiplets connecting the nodes. The quiver diagram is shown in Fig. \ref{fig:E6Quiver}. In this case our conventions are such that all the $X$ fields are inward-pointing arrows, while the $Y$ fields are outward-pointing arrows.\footnote{Notice that we have relabelled the nodes compared to what would be the result of using the adjacency matrix (\ref{E6a}), in order to make $7$ the central node.}

Imposing {2T} invariance as in (\ref{invarianceVZ},\ref{invarianceXY}), the mother theory fields $X,Y$ and $Z$ reduce to 
{\tiny
  \be
  Z=\left(
\begin{array}{cccccccccccc}
 Z_{1} & 0 & 0 & 0 & 0 & 0 & 0 & 0 & 0 & 0 & 0 & 0 \\
 0 & Z_{2} & 0 & 0 & 0 & 0 & 0 & 0 & 0 & 0 & 0 & 0 \\
 0 & 0 & Z_{2} & 0 & 0 & 0 & 0 & 0 & 0 & 0 & 0 & 0 \\
 0 & 0 & 0 & Z_{3} & 0 & 0 & 0 & 0 & 0 & 0 & 0 & 0 \\
 0 & 0 & 0 & 0 & Z_{4} & 0 & 0 & 0 & 0 & 0 & 0 & 0 \\
 0 & 0 & 0 & 0 & 0 & Z_{4} & 0 & 0 & 0 & 0 & 0 & 0 \\
 0 & 0 & 0 & 0 & 0 & 0 & Z_{5} & 0 & 0 & 0 & 0 & 0 \\
 0 & 0 & 0 & 0 & 0 & 0 & 0 & Z_{6} & 0 & 0 & 0 & 0 \\
 0 & 0 & 0 & 0 & 0 & 0 & 0 & 0 & Z_{6} & 0 & 0 & 0 \\
 0 & 0 & 0 & 0 & 0 & 0 & 0 & 0 & 0 & Z_{7} & 0 & 0 \\
 0 & 0 & 0 & 0 & 0 & 0 & 0 & 0 & 0 & 0 & Z_{7} & 0 \\
 0 & 0 & 0 & 0 & 0 & 0 & 0 & 0 & 0 & 0 & 0 & Z_{7} \\
\end{array}
\right)\;,
\ee
}
{\tiny
\be
X=\left( 
\begin{array}{cccccccccccc}
 0 & X_{12} & 0 & 0 & 0 & 0 & 0 & 0 & 0 & 0 & 0 & 0 \\
 0 & 0 & 0 & 0 & 0 & 0 & 0 & 0 & 0 & X_{27} & -i X_{27} & 0 \\
 -Y_{21} & 0 & 0 & 0 & 0 & 0 & 0 & 0 & 0 & 0 & 0 & -X_{27} \\
 0 & 0 & 0 & 0 & X_{34} & 0 & 0 & 0 & 0 & 0 & 0 & 0 \\
 0 & 0 & 0 & 0 & 0 & 0 & 0 & 0 & 0 & i \omega_6 ^2X_{47}  & X_{47} & 0 \\
 0 & 0 & 0 & -Y_{43} & 0 & 0 & 0 & 0 & 0 & 0 & 0 & i\omega_6 X_{47}   \\
 0 & 0 & 0 & 0 & 0 & 0 & 0 & X_{56} & 0 & 0 & 0 & 0 \\
 0 & 0 & 0 & 0 & 0 & 0 & 0 & 0 & 0 & i \omega_6X_{67}   & -X_{67} & 0 \\
 0 & 0 & 0 & 0 & 0 & 0 & -Y_{65} & 0 & 0 & 0 & 0 & i X_{67} \omega_6 ^2 \\
 0 & 0 & Y_{72} & 0 & 0 & i \omega_6  Y_{74} & 0 & 0 & i \omega_6 ^2 Y_{76} & 0 & 0 & 0 \\
 0 & 0 & -i Y_{72} & 0 & 0 & -Y_{74} & 0 & 0 & Y_{76} & 0 & 0 & 0 \\
 0 & Y_{72} & 0 & 0 & -i \omega_6 ^2 Y_{74} & 0 & 0 & -i \omega_6  Y_{76} & 0 & 0 & 0 & 0 \\
\end{array}
\right)\;
\ee}
and
{\tiny
  \be
  Y=\left(
\begin{array}{cccccccccccc}
 0 & 0 & X_{12} & 0 & 0 & 0 & 0 & 0 & 0 & 0 & 0 & 0 \\
 Y_{21} & 0 & 0 & 0 & 0 & 0 & 0 & 0 & 0 & 0 & 0 & -X_{27} \\
 0 & 0 & 0 & 0 & 0 & 0 & 0 & 0 & 0 & -X_{27} & -i X_{27} & 0 \\
 0 & 0 & 0 & 0 & 0 & X_{34} & 0 & 0 & 0 & 0 & 0 & 0 \\
 0 & 0 & 0 & Y_{43} & 0 & 0 & 0 & 0 & 0 & 0 & 0 & i\omega_6 X_{47}   \\
 0 & 0 & 0 & 0 & 0 & 0 & 0 & 0 & 0 & -i\omega_6 ^2 X_{47}  & X_{47} & 0 \\
 0 & 0 & 0 & 0 & 0 & 0 & 0 & 0 & X_{56} & 0 & 0 & 0 \\
 0 & 0 & 0 & 0 & 0 & 0 & Y_{65} & 0 & 0 & 0 & 0 & i \omega_6 ^2X_{67}  \\
 0 & 0 & 0 & 0 & 0 & 0 & 0 & 0 & 0 & -i \omega_6X_{67}   & -X_{67} & 0 \\
 0 & Y_{72} & 0 & 0 & i \omega_6 Y_{74} & 0 & 0 & i \omega_6^2 Y_{76} & 0 & 0 & 0 & 0 \\
 0 & i Y_{72} & 0 & 0 & Y_{74} & 0 & 0 & -Y_{76} & 0 & 0 & 0 & 0 \\
 0 & 0 & -Y_{72} & 0 & 0 & i \omega_6^2 Y_{74} & 0 & 0 & i \omega_6  Y_{76} & 0 & 0 & 0 \\
\end{array}
\right)\;.
\ee
}
where $\omega_6=e^{\pi i/3}$. These are of course $24N\times24N$ matrices, with say $X_{12}$ representing a $N\times 2N$ block, $X_{27}$ a $2N\times 3N$ block etc. 

Rescaling the gauge couplings as $g_i=\kappa_i\gym$, the superpotential of the general marginally deformed theory can be written as:
\be \label{E6superpotential}
\begin{split}
    \Wcal_{\hat{E}_6}=&2i\gym\bigl[\kappa_1\Tr_2(Y_{21}Z_1X_{12})\!-\!\kappa_2\Tr_1(X_{12}Z_2Y_{21})+3\kappa_2\Tr_7(Y_{72}Z_2X_{27})\!-\!3\kappa_7\Tr_2(X_{27}Z_7Y_{72})\\
    &+3\kappa_4\Tr_7(Y_{74}Z_4X_{47})\!-\!3\kappa_7\Tr_4(X_{47}Z_7Y_{74})+\kappa_3\Tr_4(Y_{43}Z_3X_{34})\!-\!\kappa_4\Tr_3(X_{34}Z_4Y_{43})\\
    &+3\kappa_6\Tr_7(Y_{76}Z_6X_{67})\!-\!3\kappa_7\Tr_6(X_{67}Z_7Y_{76})+\kappa_5\Tr_6(Y_{65}Z_{5}X_{56})\!-\!\kappa_6\Tr_5(X_{56}Z_6Y_{65})\bigr].
\end{split}
\ee
The $\hat{E}_6$ quiver has an $S_3$ permutation symmetry given by exchanging any two branches. In the following we will make use of the $\Zset_3$ subgroup (taking nodes $(1,2)\ra(3,4)\ra(5,6)\ra(1,2)$) to write more compact expressions for the Hamiltonian and operators of the theory.

\subsection{The $\hat{E}_6$ Hamiltonian}

In order to present the $\hat{E}_6$ Hamiltonian, let us define in the following the index $i$ to range over the three values $i=1,3,5$, corresponding to the outer nodes with $\SU(N)$ gauge groups. Then $i+1=2,4,6$ corresponds to the middle nodes with $\SU(2N)$ gauge groups. 

In the holomorphic $XY$ sector we have 
\be \label{E6H1}
\Hcal_{\ell,\ell+1}=4\kappa_i^2 \;\;\text{on}\;\; X_{i,i+1}Y_{i+1,i}\;,
\ee
\be \label{E6H2}
\Hcal_{\ell,\ell+1}=\begin{pmatrix}
            \kappa_{i+1}^2&-3\kappa_{i+1}^2\\
            -\kappa_{i+1}^2&3\kappa_{i+1}^2
        \end{pmatrix}\;\;\text{on} \;\;\begin{pmatrix}
            Y_{i+1,i}X_{i,i+1}\\
            X_{i+1,7}Y_{7,i+1}
        \end{pmatrix}\;,
\ee
and
\be \label{E6H3}
\Hcal_{\ell,\ell+1}=\frac{4\kappa_7^2}{3}\begin{pmatrix}1&1&1\\1&1&1\\1&1&1\end{pmatrix}\;\text{on}\;\;
\begin{pmatrix}
            Y_{72}X_{27}\\
            Y_{74}X_{47}\\
            Y_{76}X_{67}
        \end{pmatrix}\;,
  \ee
  while in the holomorphic $XZ$ and $YZ$ sectors we obtain
\be \label{E6H4}
\Hcal_{\ell,\ell+1}=\begin{pmatrix}
            2\kappa_i^2&-2\kappa_i\kappa_{i+1}\\
            -2\kappa_i\kappa_{i+1}&2\kappa_{i+1}^2
        \end{pmatrix}\;\;\text{on}\;\;\begin{pmatrix}
            Z_iX_{i,i+1}\\
            X_{i,i+1}Z_{i+1}
            \end{pmatrix}\;,
\ee
\be \label{E6H5}
\Hcal_{\ell,\ell+1}=\begin{pmatrix}
            2\kappa_{i+1}^2&-2\kappa_i\kappa_{i+1}\\
            -2\kappa_i\kappa_{i+1}&2\kappa_{i+1}^2
        \end{pmatrix}\;\;\text{on}\;\;\begin{pmatrix}
            Z_{i+1}Y_{i+1,i}\\
            Y_{i+1,i}Z_i
            \end{pmatrix}\;,
\ee

\be \label{E6H6}
\Hcal_{\ell,\ell+1}=\begin{pmatrix}
            2\kappa_{i+1}^2&-2\kappa_{i+1}\kappa_7\\
            -2\kappa_{i+1}\kappa_7&2\kappa_7^2
        \end{pmatrix}\;\;\text{on}\;\;\begin{pmatrix}
            Z_{i+1}X_{i+1,7}\\
            X_{i+1,7}Z_7
        \end{pmatrix}\;,
\ee
and 
\be \label{E6H7}
\Hcal_{\ell,\ell+1}=\begin{pmatrix}
            2\kappa_7^2&-2\kappa_{i+1}\kappa_7\\
            -2\kappa_{i+1}\kappa_7&2\kappa_{i+1}^2
        \end{pmatrix}\;\;\text{on}\;\;\begin{pmatrix}
            Z_7Y_{7,i+1}\\
            Y_{7,i+1}Z_{i+1}
\end{pmatrix}\;.
\ee
As above, in the mixed sector it is convenient to define the combinations
\be
\mathbf{Q\Qb}_i=\begin{pmatrix}
            X_{i,i+1}\Xb_{i+1,i}
        \end{pmatrix}\;,\quad\mathbf{\bar{Q}Q}_i=\begin{pmatrix}
            \Yb_{i,i+1}Y_{i+1,i}
\end{pmatrix}\;,
\ee
\be
\mathbf{Q\Qb}_{i+1}=\begin{pmatrix}
            Y_{i+1,i}\Yb_{i,i+1}\\
            X_{i+17}\Xb_{7,i+1}
        \end{pmatrix},\quad\mathbf{\Qb Q}_{i+1}=\begin{pmatrix}
            \Xb_{i+1,i}X_{i,i+1}\\
            \Yb_{i+1,7}Y_{7,i+1}
\end{pmatrix}\;,
\ee
and
\be
        \mathbf{Q\Qb}_7=\begin{pmatrix}
            Y_{72}\Yb_{27}\\
            Y_{74}\Yb_{47}\\
            Y_{76}\Yb_{67}
        \end{pmatrix},\quad\mathbf{\Qb Q}_7=\begin{pmatrix}
            \Xb_{72}X_{27}\\
            \Xb_{74}X_{47}\\
            \Xb_{76}X_{67}
        \end{pmatrix}\;,
   \ee
as well as the matrices
   
\be
\mathbb{K}_i=\begin{pmatrix}
  2\kappa_i^2
\end{pmatrix},\quad\mathbb{L}_i=\begin{pmatrix}
\kappa_i^2
\end{pmatrix},\quad\mathbb{M}_i=\begin{pmatrix}
2\kappa_i^2
\end{pmatrix},\quad\mathbb{T}_i=\begin{pmatrix}
2\kappa_i^2
\end{pmatrix}\;,
\ee
\be
\mathbb{K}_{i+1}=\begin{pmatrix}
            \kappa_{i+1}^2&3\kappa_{i+1}^2
        \end{pmatrix}\;,\;\quad\mathbb{L}_2=\begin{pmatrix}
            \frac{\kappa_{i+1}^2}{2}\\
            \frac{\kappa_{i+1}^2}{2}
        \end{pmatrix}\;,\;\mathbb{M}_2=\begin{pmatrix}
            \frac{\kappa_{i+1}^2}{2}&\frac{3\kappa_{i+1}^2}{2}\\
            \frac{\kappa_{i+1}^2}{2}&\frac{3\kappa_{i+1}^2}{2}
        \end{pmatrix}\;,\;\mathbb{T}_2=\begin{pmatrix}
            2\kappa_i^2&\\
            &2\kappa_7^2
\end{pmatrix}\;,
\ee
and
\be
\mathbb{K}_7=\begin{pmatrix}
        2\kappa_7^2&2\kappa_7^2&2\kappa_7^2
    \end{pmatrix},\;\;\mathbb{L}_7=\begin{pmatrix}
        \frac{\kappa_7^2}{3}\\
        \frac{\kappa_7^2}{3}\\
        \frac{\kappa_7^2}{3}
    \end{pmatrix},\;\mathbb{M}_7=\begin{pmatrix}
        \frac{2\kappa_7^2}{3}&\frac{2\kappa_7^2}{3}&\frac{2\kappa_7^2}{3}\\
        \frac{2\kappa_7^2}{3}&\frac{2\kappa_7^2}{3}&\frac{2\kappa_7^2}{3}\\
        \frac{2\kappa_7^2}{3}&\frac{2\kappa_7^2}{3}&\frac{2\kappa_7^2}{3}
    \end{pmatrix},\;\mathbb{T}_7=\begin{pmatrix}
        2\kappa_2^2&&\\
        &2\kappa_4^2&\\
        &&2\kappa_6^2
    \end{pmatrix}.
\ee
Using this notation, we can write the Hamiltonian on basis elements starting on the outer nodes $i=1,3,5$ as 

\be
\Hcal_{\ell,\ell+1}=\begin{pmatrix}
            3\kappa_i^2&-\kappa_i^2&\mathbb{K}_i&\mathbb{K}_i\\
            -\kappa_i^2&3\kappa_i^2&\mathbb{K}_i&\mathbb{K}_i\\
            \mathbb{L}_i&\mathbb{L}_i&\mathbb{T}_i+\mathbb{M}_i&\mathbb{T}_i-\mathbb{M}_i\\
            \mathbb{L}_i&\mathbb{L}_i&\mathbb{T}_i-\mathbb{M}_i&\mathbb{T}_i+\mathbb{M}_i
        \end{pmatrix},\;\;\text{on}\begin{pmatrix}
            Z_i\Zb_i\\
            \Zb_iZ_i\\
            \mathbf{Q\Qb}_i\\
            \mathbf{\Qb Q}_i
        \end{pmatrix}\;,
\ee
on the middle nodes $i+1=2,4,6$ as 
\be
\Hcal_{\ell,\ell+1}=\begin{pmatrix}
            3\kappa_{i+1}^2&-\kappa_{i+1}^2&\mathbb{K}_{i+1}&\mathbb{K}_{i+1}\\
            -\kappa_{i+1}^2&3\kappa_{i+1}^2&\mathbb{K}_{i+1}&\mathbb{K}_{i+1}\\
            \mathbb{L}_{i+1}&\mathbb{L}_{i+1}&\mathbb{T}_{i+1}+\mathbb{M}_{i+1}&\mathbb{T}_{i+1}-\mathbb{M}_{i+1}\\
            \mathbb{L}_{i+1}&\mathbb{L}_{i+1}&\mathbb{T}_{i+1}-\mathbb{M}_{i+1}&\mathbb{T}_{i+1}+\mathbb{M}_{i+1}
        \end{pmatrix},\;\;\text{on}\begin{pmatrix}
            Z_{i+1}\Zb_{i+1}\\
            \Zb_{i+1}Z_{i+1}\\
            \mathbf{Q\Qb}_{i+1}\\
            \mathbf{\Qb Q}_{i+1}
        \end{pmatrix}\;,
\ee
and on the middle nodes as
\be
\begin{pmatrix}
  3\kappa_7^2&-\kappa_7^2&\mathbb{K}_7&\mathbb{K}_7\\
  -\kappa_7^2&3\kappa_7^2&\mathbb{K}_7&\mathbb{K}_7\\
  \mathbb{L}_7&\mathbb{L}_7&\mathbb{T}_7+\mathbb{M}_7&\mathbb{T}_7-\mathbb{M}_7\\
  \mathbb{L}_7&\mathbb{L}_7&\mathbb{T}_7-\mathbb{M}_7&\mathbb{T}_7+\mathbb{M}_7
        \end{pmatrix},\;\;\text{on}\;\;\begin{pmatrix}
            Z_7\Zb_7\\
            \Zb_7Z_7\\
            \mathbf{Q\Qb}_7\\
            \mathbf{\Qb Q}_7
        \end{pmatrix}\;.
\ee
For $XY$-sector states starting and ending on different nodes, we have 
\be
\Hcal_{\ell,\ell+1}=\begin{pmatrix}
            2\kappa_{i+1}^2&-2\kappa_{i+1}^2\\
            -2\kappa_{i+1}^2&2\kappa_{i+1}^2
        \end{pmatrix}\;\;\text{on}\;\;\begin{pmatrix}
            X_{i,i+1}\Yb_{i+1,7}\\
            \Yb_{i,i+1}X_{i+1,7}
        \end{pmatrix}\;\;\text{and}\;\;\begin{pmatrix}
            Y_{7,i+1}\Xb_{i+1,i}\\
            \Xb_{7,i+1}Y_{i+1,i}
        \end{pmatrix}\;,
\ee
as well as (with $i+7\sim i+1$)
\be\begin{split}
\Hcal_{\ell,\ell+1}=\begin{pmatrix}
            2\kappa_7^2&2\kappa_7^2\\
            2\kappa_7^2&2\kappa_7^2
        \end{pmatrix}\;\;&\text{on}\;\;\begin{pmatrix}
            X_{i+1,7}\Xb_{7,i\pm 3}\\
            \Yb_{i+1,7}Y_{7,i\pm 3}\end{pmatrix}\;.
\end{split}
\ee
Finally, in the mixed $XZ$ and $YZ$ sectors the Hamiltonian acts as:
\be
\Hcal_{\ell,\ell+1}=\begin{pmatrix}
            2\kappa_1^2&-2\kappa_1\kappa_2\\
            -2\kappa_1\kappa_2&2\kappa_2^2
\end{pmatrix}\;\;\text{on}\;\;\begin{pmatrix}
            Z_i\Yb_{i,i+1}\\
            \Yb_{i,i+1}Z_{i+1}
\end{pmatrix}\;\;\text{and}\;\;
\begin{pmatrix}
            \Zb_iX_{i,i+1}\\
            X_{i,i+1}\Zb_{i+1}
\end{pmatrix}\;,
\ee
\be
\Hcal_{\ell,\ell+1}=\begin{pmatrix}
            2\kappa_{i+1}^2&-2\kappa_i\kappa_{+1}\\
            -2\kappa_i\kappa_{i+1}&2\kappa_i^2
        \end{pmatrix}\;\text{on}\;\quad\begin{pmatrix}
            Z_{i+1}\Xb_{i+1,i}\\
            \Xb_{i+1,i}Z_i
            \end{pmatrix}\;\;\text{and}\;\;\begin{pmatrix}
            \Zb_iY_{i+1,i}\\
            Y_{i+1,i}\Zb_i
        \end{pmatrix}\;,
\ee
\be
\Hcal_{\ell,\ell+1}=
\begin{pmatrix}
            2\kappa_{i+1}^2&-2\kappa_{i+1}\kappa_7\\
            -2\kappa_{i+1}\kappa_7&2\kappa_7^2
        \end{pmatrix}\;\text{on}\;\quad\begin{pmatrix}
            Z_{i+1}\Yb_{i+1,7}\\
            \Yb_{i+1,7}Z_7
            \end{pmatrix}\;\;\text{and}\;\;\begin{pmatrix}
            \Zb_{i+1}X_{i+1,7}\\
            X_{i+1,7}\Zb_7
\end{pmatrix}\;,
\ee
and
\be
\Hcal_{\ell,\ell+1}=
        \begin{pmatrix}
            2\kappa_7^2&-2\kappa_{i+1}\kappa_7\\
            -2\kappa_{i+1}\kappa_7&2\kappa_{i+1}^2
        \end{pmatrix}\;\text{on}\;\quad\begin{pmatrix}
            Z_7\Xb_{7,i+1}\\
            \Xb_{7,i+1}Z_{i+1}
            \end{pmatrix}\;\;\text{and}\;\;\begin{pmatrix}
            \Zb_7Y_{7,i+1}\\
            Y_{7,i+1}\Zb_{i+1}
        \end{pmatrix}\;.
\ee

\mbox{}

Having written out the Hamiltonian, let us also write down the ``meson'' operators, which for $\hat{E}_6$ come in six copies:
\be
\Mcal_{i,i+1}=\begin{pmatrix}
X_{i,i+1}\Xb_{i+1,i}&X_{i,i+1}Y_{i+1,i}\\
\Yb_{i,i+1}\Xb_{i+1,i}&\Yb_{i,i+1}Y_{i+1,i}
\end{pmatrix} \;\;\text{and}\;\;
\Mcal_{i+1,7}=\begin{pmatrix}
  X_{i+1,7}\Xb_{7,i+1}&X_{i+1,7}Y_{7,i+1}\\
  \Yb_{i+1,7}\Xb_{7,i+1}&\Yb_{i+1,7}Y_{7,i+1}
\end{pmatrix}\;.
\ee
As discussed, from these operators we can form $\SU(2)_R$ singlets:
\be
        \Msing_{i,i+1}=X_{i,i+1}\Xb_{i+1,i}+\Yb_{i,i+1}Y_{i+1,i} \;\;\text{and}\;\;
        \Msing_{i+1,7}=X_{i+1,7}\Xb_{7,i+1}+\Yb_{i+1,7}Y_{7,i+1}\;,\ee
        and triplets
\be\begin{split}
        \Mtrip_{i,i+1}=&\begin{pmatrix}
         \half(X_{i,i+1}\Xb_{i+1,i}-\Yb_{i,i+1}Y_{i+1,i})&X_{i,i+1}Y_{i+1,i}\\
            \Yb_{i,i+1}\Xb_{i+1,i}&\half(\Yb_{i,i+1}Y_{i+1,i}-X_{i,i+1}\Xb_{i+1,i})
        \end{pmatrix}\\
        \Mtrip_{i+1,7}=&\begin{pmatrix}
            \half(X_{i+1,7}\Xb_{7,i+1}-\Yb_{i+1,7}Y_{7,i+1})&X_{i+1,7}Y_{7,i+1}\\
            \Yb_{i+1,7}\Xb_{7,i+1}&\half(\Yb_{i+1,7}Y_{7,i+1}-X_{i+1,7}\Xb_{7,i+1})
        \end{pmatrix}\;.\end{split}
\ee
In terms of these, the superconformal primary of $\hat{\mathcal{C}}_{0(0,0)}$ can be expressed as:
\be
\begin{split}
    \Tr(\mathcal{T}_{\hat{E}_6})=&\Tr_1\Zb_1Z_1+\Tr_3\Zb_3Z_3+\Tr_5\Zb_5Z_5+2[\Tr_2\Zb_2Z_2+\Tr_4\Zb_4Z_4+\Tr_6\Zb_6Z_6]+3\Tr_7\Zb_7Z_7\\
    &-\Tr_1\Msing_{12}-\Tr_3\Msing_{34}-\Tr_5\Msing_{56}-3\Tr_2\Msing_{27}-3\Tr_4\Msing_{47}-3\Tr_6\Msing_{67}.
\end{split}
\ee
Projecting the mother-theory Coulomb-branch BPS state, we find the following BPS states in the untwisted and six twisted sectors. 
\be \label{BPSE6Z}
    \begin{split}
      \Tr(\gamma(e)Z^L)=&\,\Tr_1Z_1^{L}+\Tr_3Z_3^{L}+\Tr_5Z_5^{L}+2(\Tr_2Z_2^{L}+\Tr_4Z_4^{L}+\Tr_6Z_6^{L})+3\Tr_7Z_7^{L}\;,\\
        \Tr(\gamma(z)Z^L)=&\,\Tr_1Z_1^{L}+\Tr_3Z_3^{L}+\Tr_5Z_5^{L}-2(\Tr_2Z_2^{L}+\Tr_4Z_4^{L}+\Tr_6Z_6^{L})+3\Tr_7Z_7^{L}\;,\\
        \Tr(\gamma(r)Z^L)=&\,\Tr_1Z_1^{L}+\Tr_3Z_3^{L}+\Tr_5Z_5^{L}-\Tr_7Z_7^{L}\;,\\
        \Tr(\gamma(s)Z^L)= &\,\Tr_1Z_1^{L}+\Tr_2Z_2^{L}+\omega_3(\Tr_3Z_3^{L}+\Tr_4Z_4^{L})+\omega_3^2(\Tr_5Z_5^{L}+\Tr_6Z_6^{L})\;,\\
        \Tr(\gamma(s^2)Z^L)=&\,\Tr_1Z_1^{L}-\Tr_2Z_2^{L}+\omega_3(\Tr_3Z_3^{L}-\Tr_4Z_4^{L})+\omega_3^2(\Tr_5Z_5^{L}-\Tr_6Z_6^{L})\;,\\
        \Tr(\gamma(s^4)Z^L)=&\,\Tr_1Z_1^{L}-\Tr_2Z_2^{L}+\omega_3^2(\Tr_3Z_3^{L}-\Tr_4Z_4^{L})+\omega_3(\Tr_5Z_5^{L}-\Tr_6Z_6^{L})\;,\\
        \Tr(\gamma(s^5)Z^L)=&\,\Tr_1Z_1^{L}+\Tr_2Z_2^{L}+\omega_3^2(\Tr_3Z_3^{L}+\Tr_4Z_4^{L})+\omega_3(\Tr_5Z_5^{L}+\Tr_6Z_6^{L})\;.
    \end{split}
\ee
These states can be partially distinguished by their eigenvalues under the $\Zset_3$ taking $(i,i+1)\ra(i+2,i+3)$, under which the $e,z,r$ states have eigenvalue 1, the $s,s^2$ states eigenvalue $\omega_3$ and the $s^4,s^5$ states eigenvalue $\omega_3^2$. Although we have not constructed the $\tau$ matrices for this case, it is clear that one can further distinguish the twisted states by their eigenvalues under further $\Zset_2$ symmetries exchanging $Z_i\leftrightarrow Z_7$ and $Z_i\leftrightarrow Z_{i+1}$. 

Finally, the orbifold-point Konishi operator is given by
\be
\begin{split}
    \mathcal{K}_{\hat{E}_6}=&\Tr_1\Zb_1Z_1+\Tr_3\Zb_3Z_3+\Tr_5\Zb_5Z_5+2(\Tr_2\Zb_2Z_2+\Tr_4\Zb_4Z_4+\Tr_6\Zb_6Z_6)+3\Tr_7\Zb_7Z_7\\
    &+2(\Tr_1\Msing_{12}+\Tr_3\Msing_{34}+\Tr_5\Msing_{56})+6(\Tr_2\Msing_{27}+\Tr_4\Tr\Msing_{47}+\Tr_6\Msing_{67}).
\end{split}
\ee
This operator (which as always has $E=12$ at the orbifold point), will of course receive $\kappa_i$-dependent corrections in the marginally deformed theory. We discuss its descendants in the $L=3$ $XYZ$ and $L=4$ $XY$ sectors in Appendix \ref{sec:E6Konishi}.

\subsection{Protected spectrum} \label{sec:E6index}

In this case the relevant matrix entering the index (\ref{eq:MultitraceOrbifoldIndex}) is
\be
(1+t)I_{7\times 7}-t^\half A_\text{2T}=\begin{pmatrix}
    1+t&-t^\half&0&0&0&0&0\\
    -t^\half&1+t&-t^\half&0&0&0&0\\
    0&-t^\half&1+t&-t^\half&0&-t^\half&0\\
    0&0&-t^\half&1+t&-t^\half&0&0\\
    0&0&0&-t^\half&1+t&0&0\\
    0&0&-t^\half&0&0&1+t&-t^\half\\
    0&0&0&0&0&-t^\half&1+t
\end{pmatrix}\;,
\ee
where we used (\ref{E6a}). Its determinant can be evaluated as 
\be
\det\left((1+t)I_{7\times 7}-t^\half A_\text{2T}\right)=(1-t^3)^2(1+t) =\frac{(1-t^3)^2(1-t^2)}{(1-t)}\;.
\ee
where in the last equation we have brought it in a more appropriate form with which to perform the product in (\ref{eq:MultitraceOrbifoldIndex}).
So we can evaluate the large-$N$ multi-trace index as 
\be\begin{split}\label{MultiTraceSci2T}
\mathcal{I}^\text{m.t.}_\text{2T}\simeq&\prod_{n=1}^\infty\frac{\left((1-p^n)(1-q^{n})\right)^7(1-t^{n})e^{-\frac{7}{n}f_\text{vm}(p^n,q^n,t^n)}}{\left(1-(pqt^{-1})^n\right)^7\left(1-t^{3n}\right)^2\left(1-t^{2n}\right)}\\
=&\frac{\Gamma(t;p,q)^7(t;t)_\infty}{(pqt^{-1};pqt^{-1})_\infty^7(t^2;t^2)_\infty(t^3;t^3)^2_\infty}\,.
\end{split}\ee
Its various limits are 
\be\begin{split}\label{eq:MultiTrace2Tlimits}
    \mathcal{I}^\text{m.t.}_{\text{2T};\;M}\simeq&\,\frac{(t;t)_\infty}{(t;q)_{\infty}^{7}(t^2;t^2)_\infty(t^3;t^3)^2_\infty}\;,\\
    \mathcal{I}^\text{m.t.}_{\text{2T};\;S}\simeq&\,(q;q)_{\infty}^{-5}(q^2;q^2)^{-1}_\infty(q^3;q^3)_\infty^{-2}\;,\\
    \mathcal{I}^\text{m.t.}_{\text{2T};\;HL}\simeq&\,\frac{(t;t)_\infty}{(1-t)^{7}(t^2;t^2)_\infty(t^3;t^3)_\infty^2}\;,\\
    \mathcal{I}^\text{m.t.}_{\text{2T};\;C}\simeq&\,\frac{(1-T)}{(T;T)_{\infty}^{7}}\,.
\end{split}\ee
It would be interesting to see whether similar methods to \cite{Bourdier:2015sga} could reproduce the above Schur index. 
From \eqref{eq:Singletracegenericindex}, the single trace index is given by
\be\begin{split}
\label{eq:SingleTraceSci2T}
\mathcal{I}^\text{s.t.}_{\text{2T}}=&\,7\biggl[\frac{pqt^{-1}}{1-pqt^{-1}}+\frac{t-pqt^{-1}}{(1-p)(1-q)}\biggr]+\frac{2t^3}{1-t^3}+\frac{t^2}{1-t^2}-\frac{t}{1-t}\\
=&\,7\biggl[\frac{pqt^{-1}}{1-pqt^{-1}}+\frac{t-pqt^{-1}}{(1-p)(1-q)}\biggr]-t+\frac{2t^3}{1-t^3}+\frac{t^2}{1-t^2}-\frac{t^2}{1-t}\\
=&\,7\biggl[\sum_{\ell=2}^\infty\mathcal{I}[\bar{\mathcal{E}}_{-\ell(0,0)}]+\mathcal{I}[\hat{\mathcal{B}}_1]\biggr]-\mathcal{I}[\Mtrip]+\frac{2t^3}{1-t^3}+\frac{t^2}{1-t^2}-\frac{t^2}{1-t}\,.
\end{split}
\ee
We have extracted a $-t$ factor corresponding to the $F$-term constraint for quivers with spherical topology, as noted in \ref{sec:Hall-Littlewood}. The Hall-Littlewood and Coulomb-branch limits of the index are
\be\begin{split}\label{eq:limitsSingleTraceSci2T}
    \mathcal{I}^\text{s.t.}_{\text{2T};\;HL}=&\,6~\mathcal{I}[\Mtrip]+\frac{2t^3}{1-t^3}+\frac{t^2}{1-t^2}-\frac{t^2}{1-t}\;,\;\\
    \mathcal{I}^\text{s.t.}_{\text{2T};\;C}=&\,7~\sum_{\ell=2}^\infty\mathcal{I}[Z^\ell]\;.
\end{split}\ee
We see that there are six protected $\Mtrip$ operators and seven protected BMN vacua $Z^\ell$. The remaining states counted in $\mathcal{I}^\text{s.t.}_{\text{2T};\;HL}$ correspond to $\hat{\mathcal{B}}_{R}$ for $R>1$ and to $\mathcal{D}_{\ell+\half(0,\half)}$, which are of the form $\bar{\lambda}_{Z\dot{+}}(XY)^{\ell}$. 

We can now briefly consider the twisted and untwisted sectors. Similarly to the $\Dfour$ case, if we consider the $\SU(2)_L$ part of the $\Ncal=4$ index (\ref{IL}) and average over the group, the linear and quadratic terms in the fugacities will cancel out, as there are no 2T invariants at those degrees. The only terms that survive the averaging are those where an initial term of the form $v_x^\ell+v_y^\ell$ gives an invariant, and it is straightforward to check that this only happens for $\ell=4n$, with $n=2,3,\ldots$. We again need to divide the result by 2 as the states $\Tr X^{(4 n)}$ and $\Tr Y^{(4 n)}$ project to the same state. Via this procedure we arrive at the same answer as for $\Dfour$, i.e. 
\be\begin{split}\label{eq:untwisted2T}
    \mathcal{I}^\text{untwisted}_{\text{2T}}=&\,\sum_{\ell=2}^\infty\mathcal{I}[\bar{\mathcal{E}}_{-\ell(0,0)}]+\sum_{\ell=2}^\infty\mathcal{I}[(XY)^{2\ell}]\,,
\end{split}\ee
where $(XY)^{2\ell}$ is schematic notation for the $XY$-sector state of length $4\ell$ which projects from the vacuum state $X^{4\ell}$ (or equivalently $Y^{(4\ell)}$) in the mother theory. Note that there is no untwisted $\Mtrip$ triplet state, which implies that all the 6 triplets appearing in (\ref{eq:limitsSingleTraceSci2T}) are twisted states. There are of course further untwisted $XY$-sector states, projecting from non-trivial states in the mother theory, which the untwisted index does not count. For these we need to turn to the Molien series.

From Table \ref{tab:SU(2)Molienseries}, the Molien series of 2T is given by
\be\label{eq:2TMolien}
\mathbf{M}(x;\text{2T})=\,\frac{1-x^4+x^8}{1-x^4-x^6+x^{10}}=\,1+x^6+x^8+2x^{12}+\Ocal(x^{13})\,.
\ee
We have verified these numbers (and in particular, the absence of BPS states at $L=4$ and $L=10$) up to order $x^{10}$ by explicit diagonalisation.\footnote{For $\hat{E}_6$, the $XY$-sector cyclically identified basis is $12,26,72,210$-dimensional for $L=4,6,8,10$, respectively. Constructing the basis and diagonalising the Hamiltonian for $L>10$, although straightforward, is time-consuming.}

As in the $\Dfour$ case, one can rewrite the Molien series to better indicate the highest-weight components of the $\hat{\mathcal{B}}_R$ multiplets that it counts. We find
\be
\mathbf{M}(x;\text{2T})=1+\sum_{\ell=1}^\infty\left((\ell+1) t^{6\ell}+\ell (t^{6\ell-3}+ t^{6\ell-2}+t^{6\ell+1}+t^{6\ell+2}+t^{6\ell+5})\right)\;,
\ee
from which we can identify the first term as a contribution of $\mathcal{I}[(XY)^{6\ell}]$ and similarly for the other terms. So we see that $(\ell+1)$ states contribute at length $12\ell$ and $\ell$ states at lengths $12\ell-6,12\ell-4,12\ell+2,12\ell+4,12\ell+10$ for $\ell\geq 1$. 
To count the number of fermionic states that cancel with bosonic states in the index, we can define 
\be\begin{split}\label{eq:difE6MolienHL}
\mathcal{I}^F_{\text{2T};\; HL}(x=t^\half)&=\mathbf{M}(x;\text{2T})-\mathcal{I}^\text{s.t.}_{\text{2T};\; HL}(x=t^\half)+6t\,,\\
&=1+x^8+x^{10}+2 x^{14}+x^{16}+\cdots
\end{split}\ee
where we have again subtracted the 6 twisted states from the index as these are not counted by Molien. As mentioned, the states above should correspond to fermionic states of the form $\bar{\lambda}_{Z\dot{+}}(XY)^{R-1}$, and would be possible to check explicitly with an extension of the ADE dilatation operator to include fermions.

\subsection{Short chains} \label{E6short}

In this section we will present some of the features of the short-chain spectrum of the $\hat{E}_6$ quiver. The overall features are similar to the $\Dfour$ case, so we will be brief. As the expressions are slightly long, we will mostly work in the mother theory language.

\subsubsection{Length 2}

The closed basis for length-2 operators is 45 dimensional. There are 33 $E=0$ states. Firstly, we have the 7 $\Tr(\gamma(g)Z^2)$ states (\ref{BPSE6Z} for $L=2$) plus their conjugates, plus the (untwisted) $\Tr(\mathcal{T})$ state. These states do not acquire $\kappa$ dependence. We also have the 6 twisted-sector triplets with highest components $\Tr(\gamma(g)(XY-YX))$. As an example we write
\be
\Tr(\gamma(r)(XY-YX))=\Tr(X_{12}Y_{21}+X_{27}Y_{72}+X_{34}Y_{43}+X_{47}Y_{74}+X_{56}Y_{65}+X_{67}Y_{76})\;.
\ee
At the orbifold point we can write all the remaining states in mother-theory form as:
\be
\Tr\left(\gamma(g)( c_1(X\Xb+\Xb X+Y\Yb+\Yb Y)+c_2(Z\Zb+\Zb Z))\right) \;,
\ee
with the coefficients and energies given in Table \ref{tab:L2E6}. 

\begin{table}[h]\begin{center}
\begin{tabular}{|c|c|c|c|}\hline
  $g$ & $c_1$ &$c_2$ & $E$\\ \hline
  $e$ &1 &1 & 12\\
  $s$,$s^5$&1 & $-\half(1\pm\sqrt{7})$ &$ 2(3\mp \sqrt{7})$\\
  $s^2$,$s^4$&1 & $-\half(1\pm\sqrt{3})$ & $2(3\mp \sqrt{3})$\\
  $r$ & 1& $-\half(1\pm\sqrt{5})$ & $2(3\mp \sqrt{5})$\\
  $z$ &0 &1 &4\\
  \hline\end{tabular}\caption{The $L=2$ untwisted and twisted non-protected states for the $\hat{E}_6$ quiver.} \label{tab:L2E6}\end{center}
\end{table}
The untwisted state is the Konishi operator, which is discussed in Section \ref{sec:E6Konishi}. For illustration, let us write out the $E=4$ twisted state:
\be
\Tr(\gamma(z)(Z\Zb+\Zb Z))=\sum_{i=1,3,5} \left(\Tr(Z_i\Zb_i)-2\Tr(Z_{i+1}\Zb_{i+1})\right)+3\Tr(Z_7\Zb_7)\;.
\ee
Moving away from the orbifold point, the $E>0$ states will mix and acquire $\kappa$-dependence. The form of the deformed Konishi is discussed in (\ref{sec:E6Konishi}). The characteristic polynomial is too long to write down in the general case, so for illustration we only write it in the very simple case where $\kappa_i=\kappa_{i+1}=1, \kappa_7=\kappa$:
\be\begin{split}\label{charpolyE61}
P(E)&=E^{33}\left(E ^4+\left(-8 \kappa ^2-20\right) E ^3 +\left(144 \kappa ^2+112\right) E ^2+ \left(-688 \kappa ^2-144\right) E+768 \kappa ^2\right)\\
&\times\left(E ^4+\left(-4 \kappa ^2-20\right) E ^3 +\left(64 \kappa ^2+112\right) E ^2+ \left(-240 \kappa ^2-144\right) E+192 \kappa ^2\right)^2\;.
\end{split}
\ee
At the orbifold point $\kappa=1$, the first quartic polynomial factorises as $(E-12)(E-4)(E^2-12E+16)$ with roots $12,4,2(3\pm\sqrt{5})$.  
Looking at Table \ref{tab:L2E6}, we see that away from the orbifold point the untwisted state mixes with the $r$ and $z$ twisted sector states, while the $s$ states deform together. As we approach $\kappa\ra0$ this theory does not reduce to SCQCD, but to three copies of the \begin{tikzpicture}[baseline=-0.05cm]
  \draw[-] (0,0) circle (1.5ex);
  \draw[-] (0.75,0) circle (1.5ex);
  \draw[-] (1.25,-0.25)--(1.25,0.25)--(1.75,0.25)--(1.75,-0.25)--(1.25,-0.25);
  \draw[-] (0.25,0)--(0.5,0);\draw[-] (1,0)--(1.25,0);
  \node at (0,0) {\footnotesize $N$};
  \node at (0.75,0) {\footnotesize $2N$};
  \node at (1.5,0) {\footnotesize $3N$};
\end{tikzpicture} linear quiver theory. Accordingly, the two different factors in (\ref{charpolyE61}) become equal as $\kappa\ra0$ and we get three times the same factor. In section \ref{D4short}, we obtained the same linear quiver from the $\Dfour$ theory by taking three of the exterior couplings to zero and the other two equal. As a consistency check, the eigenvalues all agree, with the multiplicities here being three times higher, as expected.

\subsubsection{Length 3: Holomorphic states}

As always, the holomorphic states at $L=3$ are all descendants of the neutral $L=2$ ones. We will not write them explicitly, apart from noting that, as in the $\Dfour$ case, also here there is no protected state in the $XYZ$ sector as the untwisted $\Tr(\gamma(e)(XYZ+XZY))$ state vanishes.

\begin{figure}[h!]
  \begin{center}
        \begin{tikzpicture}
      \node at (0,0) {\includegraphics[width=12cm]{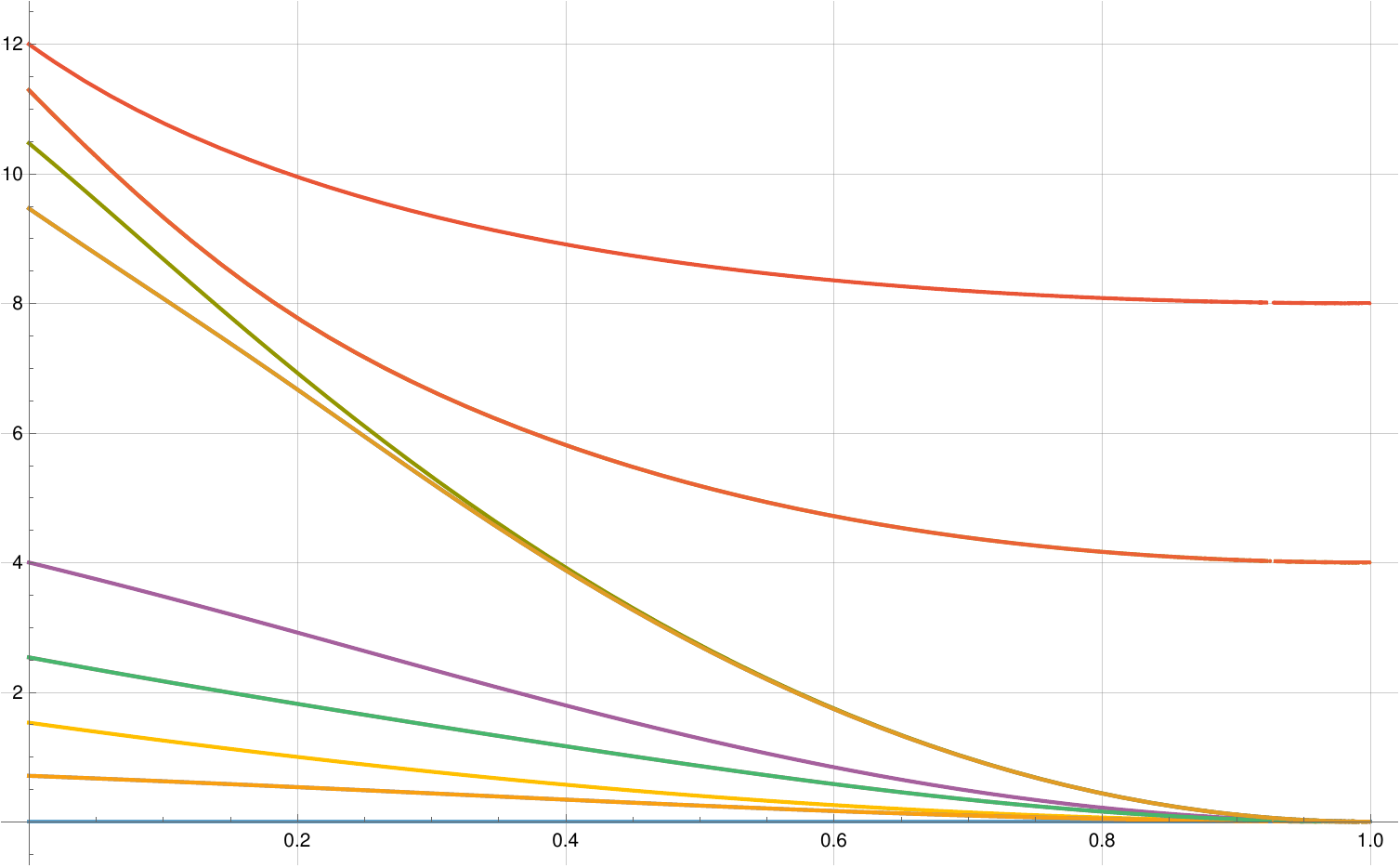}};
\node at (6.2,-3.3){$k$};\node at (-6.2,2.8){$E$};
\end{tikzpicture}
  \caption{The spectrum of $L=3$ holomorphic states of the $\hat{E}_6$ chain for the deformation $\{\kappa_i=1-k,\kappa_7=1\}$, where $k=0$ corresponds to the orbifold point. As $k\ra 1$ all the gauge groups apart from the central $\SU(3N)$ become global and the theory approaches SCQCD + decoupled vector multiplets.}\label{GraphE6L3a}
  \end{center}
\end{figure}

\begin{figure}[h!]
  \begin{center}
            \begin{tikzpicture}
      \node at (0,0) {\includegraphics[width=12cm]{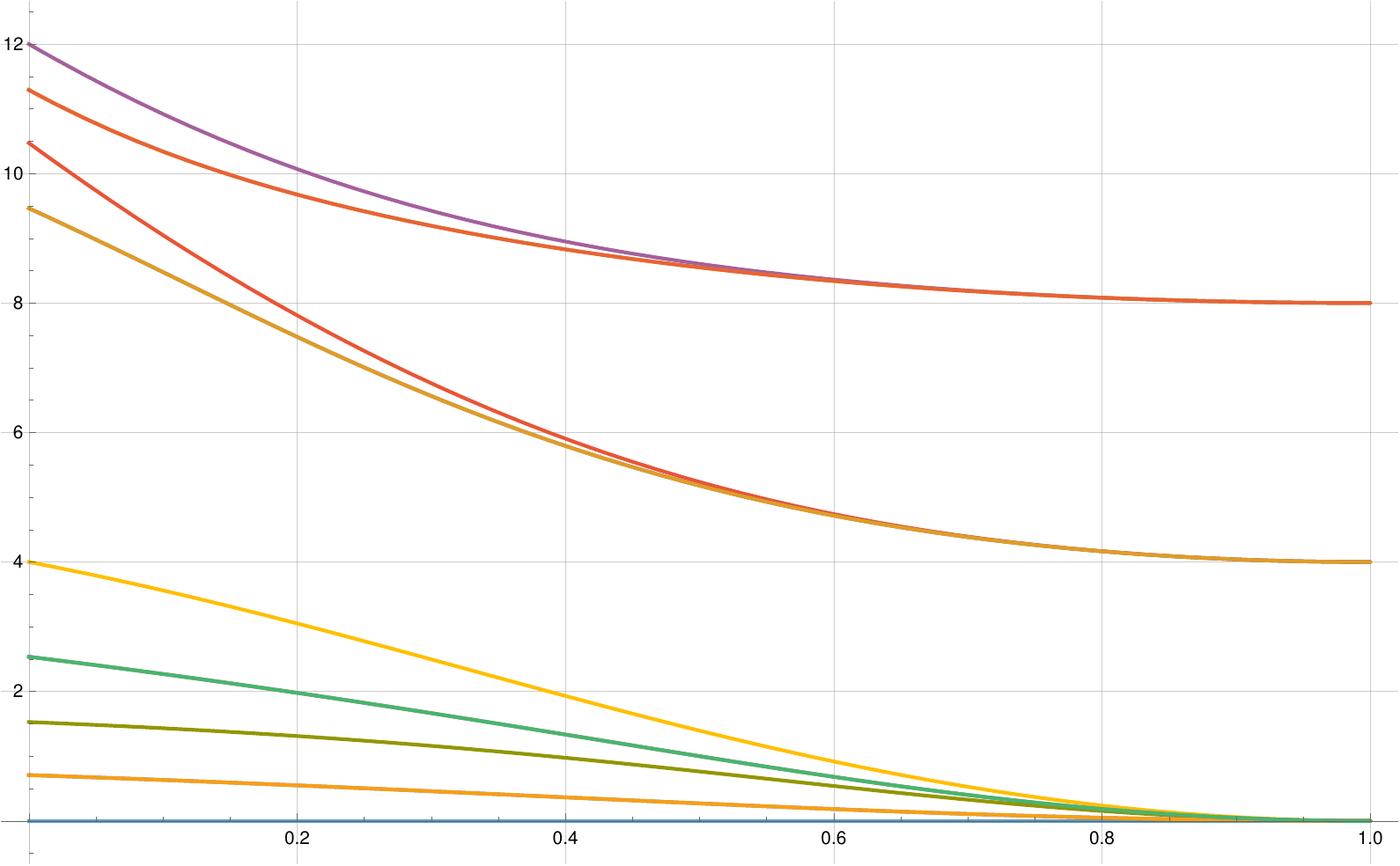}};
\node at (6.2,-3.3){$k$};\node at (-6.2,2.8){$E$};
\end{tikzpicture}
  \caption{The spectrum of $L=3$ holomorphic states of the $\hat{E}_6$ chain for the deformation $\{\kappa_i=\kappa_7=1-k,\kappa_{i+1}=1\}$, where $k=0$ corresponds to the orbifold point. As $k\ra 1$ all groups apart the three middle gauge groups at nodes 2,4, and 6 become global and we are left with three copies of SCQCD plus decoupled vector multiplets. In line with this, two degenerate twisted-sector states join the superpotential state and arrive at $E=8$.}\label{GraphE6L3b}
  \end{center}
\end{figure}

In Fig.\ref{GraphE6L3a} we plot the spectrum of the theory at $L=3$ for a symmetric deformation where all the exterior and middle couplings are taken to be equal. As $\kappa_i\ra0$, $i=1,\ldots, 6$, the theory approaches SCQCD with $\SU(3N)$ gauge group and $6N$ flavours, and one obtains the same eigenvalues as for the $\Zset_3$ and $\Dfour$ cases. Another (of many) ways to approach SCQCD is to take the couplings of the exterior and central nodes to zero, leaving 3 decoupled copies of SCQCD in the middle nodes. The limiting theory exhibits the expected tripling of eigenvalues, and is shown in Fig. \ref{GraphE6L3b}.

\begin{figure}[h!]
  \begin{center}
                \begin{tikzpicture}
      \node at (0,0) {\includegraphics[width=12cm]{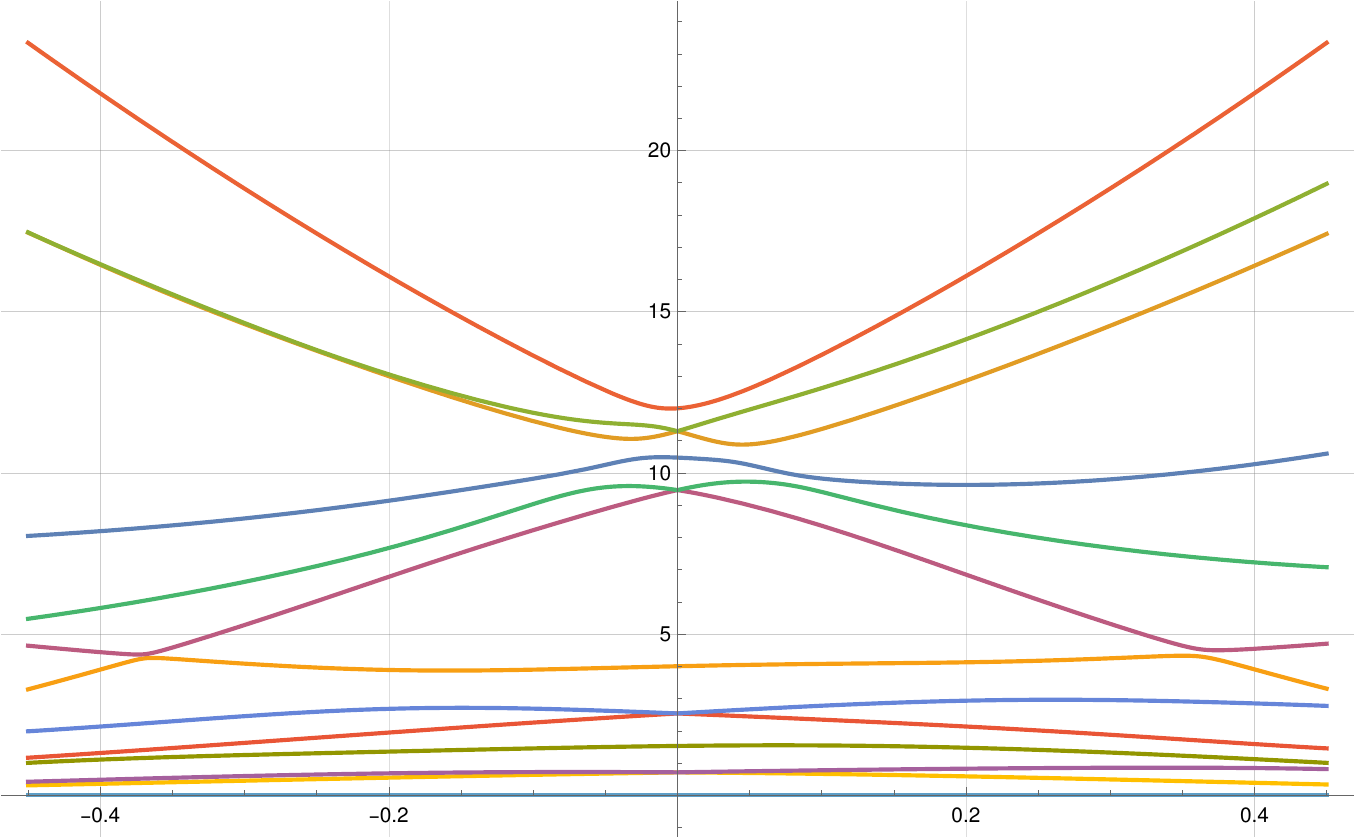}};
\node at (6.2,-3.3){$k$};\node at (-0.3,3.3){$E$};
\end{tikzpicture}
  \caption{The spectrum of $L=3$ states of the $\hat{E}_6$ chain for the less symmetric deformation $\{\kappa_1=\kappa_2=\kappa_3=1-k,\kappa_4=\kappa_5=\kappa_6=1+k,\kappa_7=1\}$, where $k=0$ corresponds to the orbifold point. Note the breaking of twisted-sector degeneracies, and the multiple avoided crossings. }\label{GraphE6L3c}
  \end{center}
\end{figure}

Given the symmetries of the above deformations, the degeneracies of the twisted sectors are unbroken. To illustrate the more generic case, a less symmetric deformation is plotted in Fig. \ref{GraphE6L3c}. 


\subsubsection{Length 4: Holomorphic states}

\begin{figure}[h!]
  \begin{center}
    \begin{tikzpicture}
      \node at (0,0) {\includegraphics[width=12cm]{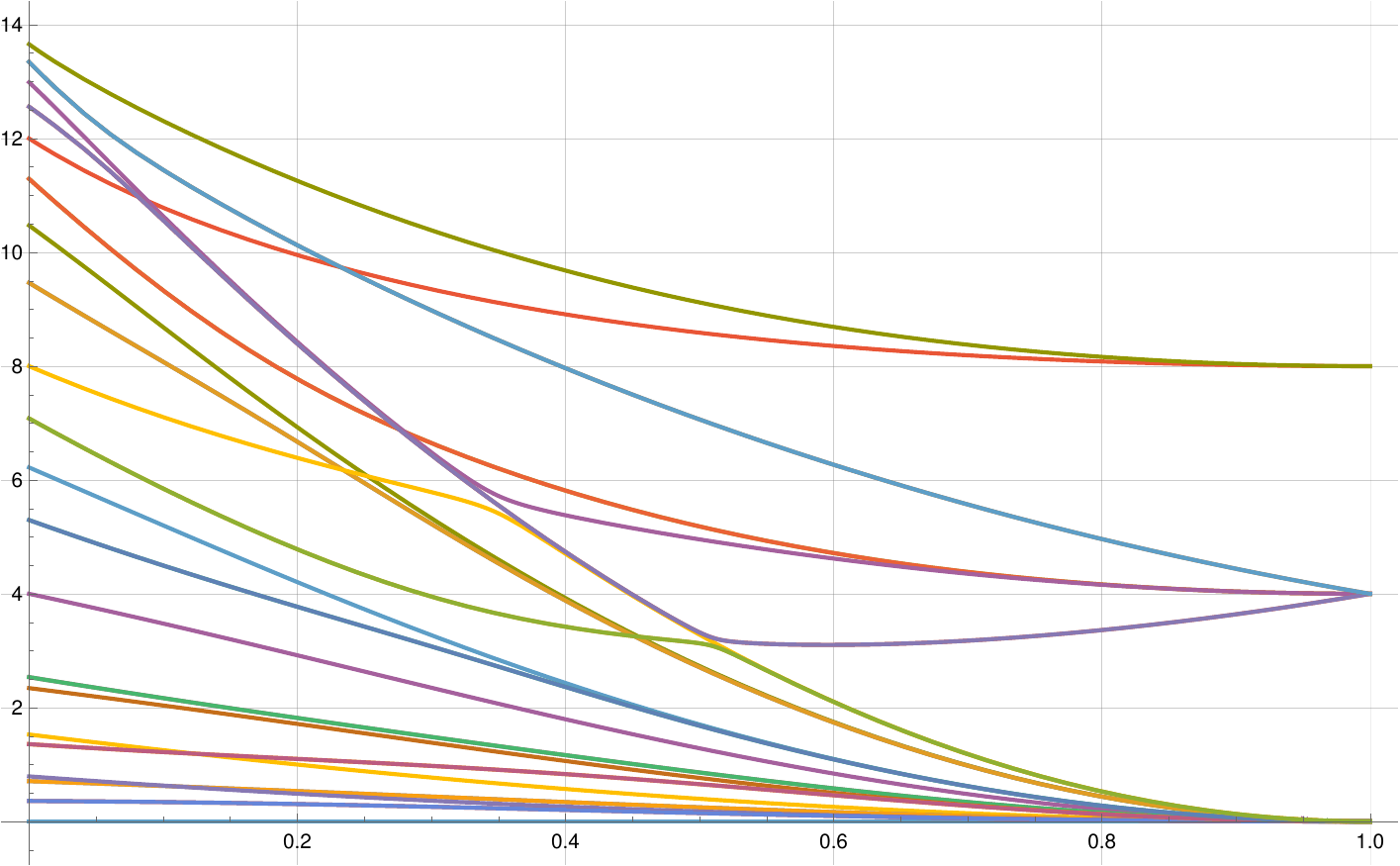}};
\node at (6.2,-3.3){$k$};\node at (-6.2,3){$E$};
\end{tikzpicture}
  \caption{The spectrum of $L=4$ holomorphic states of the $\hat{E}_6$ chain for the deformation $\{\kappa_i=\kappa_{i+1}=1-k,\kappa_7=1\}$, where $k=0$ corresponds to the orbifold point and $k=1$ to SCQCD with additional decoupled vectors.}\label{GraphE6L4ha}
  \end{center}
\end{figure}

\begin{figure}[h!]
  \begin{center}
        \begin{tikzpicture}
      \node at (0,0) {\includegraphics[width=12cm]{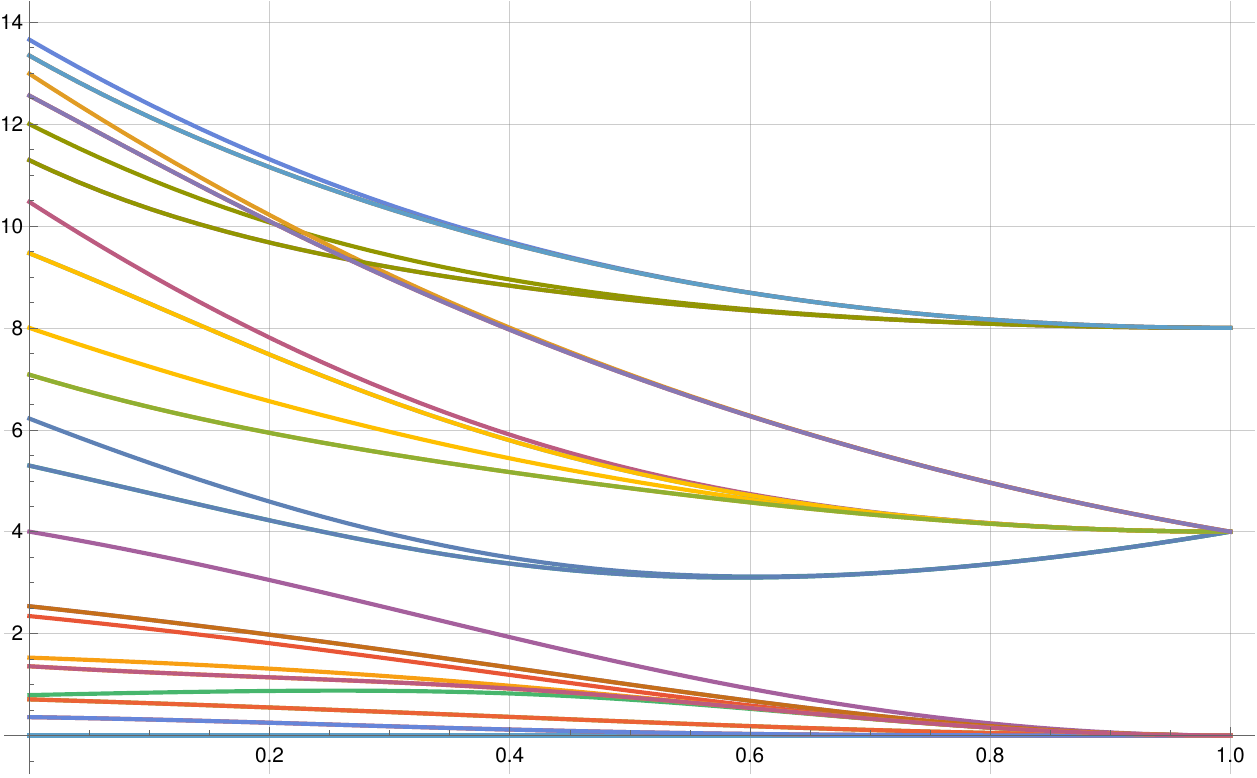}};
\node at (6.2,-3.3){$k$};\node at (-6.2,3){$E$};
\end{tikzpicture}
  \caption{The spectrum of $L=4$ holomorphic states of the $\hat{E}_6$ chain for the deformation $\{\kappa_i=\kappa_{7}=1-k,\kappa_{i+1}=1\}$, where $k=0$ corresponds to the orbifold point and $k=1$ to three copies of SCQCD with additional decoupled vectors. As expected, there are more states reaching $E=4$ and $E=8$ compared to Fig. \ref{GraphE6L4ha}.}\label{GraphE6L4hb}
  \end{center}
\end{figure}

The $L=4$ spectrum for $\hat{E}_6$ exhibits the same general features as our other examples. In accordance with the Molien series, we do not find any $E=0$ states at $L=4$. As before, the marginally deformed spectrum includes several instances of avoided crossings. In Figs \ref{GraphE6L4ha} and \ref{GraphE6L4hb} we plot two deformations which limit to one and three copies of SCQCD, respectively. Although not clearly visible in the plots, we have checked that the number of states arriving at $E=8$ in the second deformation is indeed three times that of the first one. 

\subsubsection{Length 6: $XY$ sector}

As expected from the Molien series (\ref{eq:2TMolien}), length 6 is the first case where we find a protected state in the (untwisted) $XY$ sector. It can be written as 
\be\begin{split}
\Tr(YXXXXX)=-\Tr(XYYYYY)&=6i\sqrt{3}\Tr(X_{12}X_{27}(Y_{76}X_{67}-Y_{74}X_{47})Y_{72}Y_{21})\\
&\;\;+6i\sqrt{3}\Tr(Y_{72}X_{27}Y_{72}X_{27}(Y_{76}X_{67}-Y_{74}X_{47})) + \cdots
\end{split}\ee
where $\cdots$ are the two $\Zset_3$ conjugates of the states shown. The structure of the mother-theory state is in line with the first non-trivial invariant of 2T being $xy(x^4-y^4)$, as can be seen by applying the Reynolds formula (see \cite{Benvenuti:2006qr} for a discussion).

\subsection{Two-magnon Bethe Ansatz}

In this section we construct the two-magnon coordinate Bethe ansatz for the $\hat{E}_6$ quiver theory. The overall features are very similar to the previous cases, with some differences that we will highlight.

\subsubsection{Open Chain}

As usual, we consider one $X$ and one $Y$ excitation on the $Z$ vacuum, with the left and right exterior $Z$ vacua being the same for closeability. Continuing to let the index $i=1,3,5$, the possibilities are:
\begin{itemize}
\item 3 $Z_i$ exterior vacua with $Z_{i+1}$ interior vacua
\item 3 $Z_{i+1}$ exterior vacua with either $Z_i$ or $Z_7$ interior vacua
\item 1 $Z_7$ exterior vacuum with any of the three $Z_{i+1}$ interior vacua.
\end{itemize}
Correspondingly, we define the following 12 two-magnon states:
\be\begin{split}
\ket{\ell_1,\ell_2}_{i}&=\cdots Z_i Z_i {X_{i,i+1}} Z_{i+1} \cdots Z_{i+1} {Y_{i+1,i}} Z_{i} Z_{i}\cdots\;,\\
\ket{\ell_1,\ell_2}^{YX}_{i+1}&=\cdots Z_{i+1} Z_{i+1} {Y_{i+1,i}} Z_{i} \cdots Z_{i} {X_{i,i+1}} Z_{i+1} Z_{i+1}\cdots\;,\\
\ket{\ell_1,\ell_2}^{XY}_{i+1}&=\cdots Z_{i+1} Z_{i+1} {X_{i+1,7}} Z_{7} \cdots Z_{7} {Y_{7,i+1}} Z_{i+1} Z_{i+1}\cdots\;,\\
\ket{\ell_1,\ell_2}^{7}_{i+1}&=\cdots Z_{7} Z_{7} {Y_{7,i+1}} Z_{i+1} \cdots Z_{i+1} {X_{i+1,7}} Z_{7} Z_{7}\cdots\;.
\end{split}
\ee
As always, we will consider the scattering in each exterior vacuum sector separately. 

\mparagraph{Exterior $Z_i$ vacua} As can be seen from the holomorphic Hamiltonian (\ref{E6H1}), or just by the fact that a $Y$ magnon cannot be to the left of a $X$ magnon, the scattering in this sector is purely by reflection. So we define the Bethe state as simply
\be
\ket{\psi}_i=\sum_{\ell_1,\ell_2} \left(A_i e^{p^{(i)}_1\ell_1+p^{(i)}_2\ell_2}+B_i e^{p^{(i)}_2\ell_1+p^{(i)}_1\ell_2}\right)\ket{\ell_1,\ell_2}_i\;,
\ee
with dispersion relation
\be \label{E6idisp}
E_i=4(\kappa_i^2+\kappa_{i+1}^2)-2\kappa_i\kappa_{i+1}(e^{ip^{(i)}_1}+e^{-i p^{(i)}_1}+e^{ip^{(i)}_2}+e^{-i p^{(i)}_2})\;,
\ee
The interacting equations are
\be
(8\kappa_i^2-E_i)(A_i e^{ip^{(i)}_2}+B_i e^{i p^{(i)}_1})-2\kappa_1\kappa_2(A_ie^{-ip^{(i)}_1+ip^{(i)}_2}+B_i e^{-ip^{(i)}_2+ip^{(i)}_1}+A_i e^{2ip^{(i)}_2}+B_i e^{2ip^{(i)}_1})=0\;.
\ee
The $S$-matrix can trivially be found to be
\be
S^{(i)}=B_i/A_i=-\frac{\kappa_i\kappa_{i+1}+e^{ip^{(i)}_1+ip^{(i)}_2}+2 e^{ip^{(i)}_2}(\kappa_i^2-\kappa_{i+1}^2)}{\kappa_i\kappa_{i+1}+e^{ip^{(i)}_1+ip^{(i)}_2}+2 e^{ip^{(i)}_1}(\kappa_i^2-\kappa_{i+1}^2)}\;,
\ee
which reduces to $S^{(i)}=-1$ at the orbifold point. 

\mparagraph{Exterior $Z_{i+1}$ vacua} From (\ref{E6H2}) we see that for $Z_{i+1}$ exterior vacua the states with the two possible interior vacua can mix. We therefore write our Bethe state as
\be\begin{split}
\ket{\psi}_{i+1}&=\sum_{\ell_1,\ell_2} \left(C^{YX}_{i+1} e^{ip^{(i)}_2\ell_1+ip^{(i)}_1\ell_2}+D^{YX}_{i+1} e^{ip^{(i)}_1\ell_1+ip^{(i)}_2\ell_2}\right)\ket{\ell_1,\ell_2}_{i+1}^{YX}\\
&+\sum_{\ell_1,\ell_2} \left(C^{XY}_{i+1} e^{ip^{(i+1)}_1\ell_1+ip^{(i+1)}_2\ell_2}+D^{XY}_{i+1} e^{ip^{(i+1)}_2\ell_1+ip^{(i+1)}_1\ell_2}\right)\ket{\ell_1,\ell_2}_{i+1}^{XY}\;.
\end{split}
\ee
The dispersion relation for the $p^{(i)}$ momenta is (\ref{E6idisp}) while for the $p^{(i+1)}$ momenta we have  
\be \label{E6ip1disp}
E_{i+1}=4(\kappa_{i+1}^2+\kappa_{7}^2)-2\kappa_{i+1}\kappa_7\left(e^{ip^{(i+1)}_1}+e^{-i p^{(i+1)}_1}+e^{ip^{(i+1)}_2}+e^{-i p^{(i+1)}_2}\right)\;,
\ee
There are two types of interacting equations, one when the $YX$ magnons meet:
\be\begin{split}
&(5\kappa_{i+1}^2-E_{i})\left(C^{YX}_{i+1} e^{ip^{(i)}_1}+D^{YX}_{i+1}e^{ip^{(i)}_2}\right)\\
&-2\kappa_i\kappa_{i+1}\left(C^{YX}_{i+1}e^{-ip^{(i)}_2+p^{(i)}_1}+D^{YX}_{i+1} e^{-ip^{(i)}_1+i p^{(i)}_2}
+C^{YX}_{i+1}e^{2i p^{(i)}_1}+D^{YX}_{i+1} e^{2i p^{(i)}_2}\right)\\
&-\kappa_{i+1}^2\left(C^{XY}_{i+1} e^{ip^{(i+1)}_2}+D^{XY}_{i+1}e^{ip^{(i+1)}_1}\right)=0\;,
\end{split}
\ee
and one when the $XY$ magnons meet:
\be\begin{split}
&(7\kappa_{i+1}^2-E_{i+1})(C^{XY}_{i+1} e^{ip^{(i+1)}_2}+D^{XY}_{i+1}e^{ip^{(i+1)}_1})\\
&-2\kappa_{i+1}\kappa_{7}\left(C^{XY}_{i+1}e^{-ip^{(i+1)}_1+p^{(i+1)}_2}+D^{XY}_{i+1} e^{-ip^{(i+1)}_2+i p^{(i+1)}_1}
+C^{XY}_{i+1}e^{2i p^{(i+1)}_2}+D^{XY}_{i+1} e^{2i p^{(i+1)}_1}\right)\\
&-3\kappa_{i+1}^2\left(C^{YX}_{i+1} e^{ip^{(i)}_1}+D^{YX}_{i+1}e^{ip^{(i)}_2}\right)=0\;.
\end{split}
\ee
We can express the solution in terms of an $S$-matrix relating the $D$ to the $C$ coefficients:
\be
\doublet{D^{XY}_{i+1}}{D^{YX}_{i+1}}=S^{(i+1)} \doublet{C^{XY}_{i+1}}{C^{YX}_{i+1}}\;.
\ee
Defining  the combination
\be\begin{split}
&\Ncal(p^{(i)}_1,p^{(i)}_2,p^{(i+1)}_1,p^{(i+1)}_2)=2\kappa_i\kappa_{i+1}^2\kappa_7(1+e^{i(p^{(i)}_1+p^{(i)}_2)})(1+e^{i(p^{(i+1)}_1+p^{(i+1)}_2)})\\
&\qquad+\kappa_{i+1}\kappa_7(\kappa_{i+1}^2-4\kappa_i^2)e^{ip^{(i)}_2}(1+e^{i(p^{(i+1)}_1+p^{(i+1)}_2)})-2\kappa_{i+1}^2\kappa_7^2e^{i(p^{(i)}_2+p^{(i+1)}_1)}\\
&\qquad+e^{ip^{(i+1)}_1}\kappa_i(3\kappa_{i+1}^2-4\kappa_7^2)(\kappa_{i+1}-2e^{ip^{(i)}_2}\kappa_i+\kappa_{i+1} e^{ip^{(i)}_1+ip^{(i)}_2})\;,
\end{split}
\ee
we can express the coefficients of the $S$-matrix as 
\be
S^{(i+1)}_{11}=-\Ncal(p^{(i)}_2,p^{(i)}_1,p^{(i+1)}_2,p^{(i+1)}_1)/\Ncal(p^{(i)}_2,p^{(i)}_1,p^{(i+1)}_1,p^{(i+1)}_2)\;,
\ee
\be
S^{(i+1)}_{12}=3\kappa_i\kappa_{i+1}^3(e^{ip^{(i)}_1}-e^{ip^{(i)}_2}+e^{i(2p^{(i)}_1+p^{(i)}_2)}-e^{i(p^{(i)}_1+2p^{(i)}_2)})/\Ncal(p^{(i)}_1,p^{(i)}_2,p^{(i+1)}_1,p^{(i+1)}_2)\;,
\ee
\be
S^{(i+1)}_{21}=-\kappa_{i+1}^3\kappa_{7}(e^{ip^{(i+1)}_1}\!-\!e^{ip^{(i+1)}_2}\!+\!e^{i(2p^{(i+1)}_1+p^{(i+1)}_2)}\!-\!e^{i(p^{(i+1)}_1+2p^{(i+1)}_2)})/\Ncal(p^{(i)}_1,p^{(i)}_2,p^{(i+1)}_1,p^{(i+1)}_2)\;,
\ee
and
\be
S^{(i+1)}_{22}=-\Ncal(p^{(i)}_2,p^{(i)}_1,p^{(i+1)}_1,p^{(i+1)}_2)/\Ncal(p^{(i)}_1,p^{(i)}_2,p^{(i+1)}_1,p^{(i+1)}_2)
\ee
As before, this $S$-matrix satisfies $(S^{(i+1)})^*(S^{(i+1)})=I_{2\times 2}$. Unlike the previous cases, it still does not reduce to a unitary $S$-matrix at the orbifold point (where $p^{(i+1)}_{1,2}=\pm p^{(i)}_{1,2}$), due to the factor-of-3 mismatch in coefficients between $S^{(i+1)}_{12}$ and $S^{(i+1)}_{21}$. However, as mentioned, this is simply an artifact of our non-canonical normalisation, and switching to canonical form would make the orbifold $S$-matrix unitary.

\mparagraph{Exterior $Z_7$ vacuum} Finally we consider the case where the exterior vacuum is made up of $Z_7$ fields. As can be seen from (\ref{E6H3}), even though the $Y$ and $X$ fields do not transmit, they do all mix under the action of the Hamiltonian. For example, an initial  $Y_{72}X_{27}$ magnon state can scatter to all $Y_{7,i+1}X_{i+1,7}$ states with equal probability. Therefore, the Bethe ansatz is
\be
\ket{\psi}_7=\sum_{i+1} \sum_{\ell_1,\ell_2}\left(F_{i+1} e^{ip^{(i+1)}_2\ell_1+ip^{(i+1)}_1\ell_2}+G_{i+1} e^{ip^{(i+1)}_1\ell_1+ip^{(i+1)}_2\ell_2}\right)\;.
\ee
Note that in this case the momenta are labelled by the interior vacuum, and satisfy the same dispersion relation (\ref{E6ip1disp}). There are three interacting equations:
\be\begin{split}
&(\frac{16}{3}\kappa_7^2-E_{i+1})\left(F_{i+1}e^{ip^{(i+1)}_1}+G_{i+1} e^{ip^{(i+1)}_2}\right)\\
&-2\kappa_{i+1}\kappa_7\left(F_{i+1}e^{-ip^{(i+1)}_2+ip^{(i+1)}_1}+G_{i+1}e^{-ip^{(i+1)}_1+ip^{(i+1)}_2}+F_{i+1}e^{2ip^{(i+1)}_1}+G_{i+1}e^{2ip^{(i+1)}_2}\right)\\
&+\frac{4}{3}\kappa_7^2\left(F_{i+3}e^{ip^{(i+3)}_1}+G_{i+3} e^{ip^{(i+3)}_2}+F_{i+5}e^{ip^{(i+5)}_1}+G_{i+5} e^{ip^{(i+5)}_2}\right)=0\;,
\end{split}
\ee
where of course $i$ is identified modulo 6. We define the $3\times 3$ S-matrix
\be
\triplet{G_{2}}{G_{4}}{G_{6}}= \left(S^{(7)}\right)\triplet{F_{2}}{F_{4}}{F_{6}}\;.
\ee
Let us define the combination 
\be
n^{(i+1)}_k=\kappa_7-2\kappa_{i+1} e^{ip^{(i+1)}_k}+\kappa_7 e^{i(p^{(i+1)}_1+p^{(i+1)}_2)}\;,
\ee
where $k=1,2$, and the  $\Zset_3$-invariant combination
\be
\Dcal=-\frac{64}{9}\sum_{i+1}\left(e^{ip^{(i+1)}_2}\kappa_{i+3}\kappa_{i+5}\kappa_7^4 n^{(i+3)}_2 n^{(i+5)}_2\right)-\frac{32}{3}\kappa_2\kappa_4\kappa_6\kappa_7^2n^{(2)}_2n^{(4)}_2 n^{(6)}_2\;.
\ee
Then the upper-left components of the $S$-matrix are 
\be
S^{(7)}_{11}=-\Dcal(p^{(2)}_2,p^{(2)}_1)/\Dcal(p^{(2)}_1,p^{(2)}_2)\;,
\ee
\be
S^{(7)}_{12}=-\frac{64}{9}(e^{ip^{(4)}_1}-e^{ip^{(4)}_2}+e^{i(2p^{(4)}_1+p^{(4)}_2)}-e^{i(p^{(4)}_1+2p^{(4)}_2)})\kappa_4\kappa_6\kappa_7^5 n^{(6)}_2/\Dcal\;,
\ee
\be
S^{(7)}_{21}=-\frac{64}{9}(e^{ip^{(2)}_1}-e^{ip^{(2)}_2}+e^{i(2p^{(2)}_1+p^{(2)}_2)}-e^{i(p^{(2)}_1+2p^{(2)}_2)})\kappa_2\kappa_6\kappa_7^5 n^{(6)}_2/\Dcal\;,
\ee
with the other components found by $\Zset_3$ conjugation. This $S$-matrix satisfies $(S^{(7)})^*S^{(7)}=I_{3\times 3}$, and becomes symmetric, and therefore unitary, at the orbifold point.

\subsubsection{Closed chain}

Moving to the closed chain, it is clear that the states with exterior $Z_i$ vacua (corresponding to the $A_i,B_i$ coefficients) will live on the same chain as the states with exterior $Z_{i+1}$ vacua, and in particular with those we have denoted by the $C^{YX}_{i+1}$ and $D^{YX}_{i+1}$ coefficients. As we saw, the latter mix with the states we have labelled by $C^{XY}_{i+1}$ and $D^{XY}_{i+1}$, which in turn live on the same chain as the states with exterior $Z_{7}$ vacua, labelled by $F_{i+1},G_{i+1}$. So in the end all the 2-magnon states we have described interact with each other. The cyclicity conditions on the length-$L$ $\hat{E}_6$ chain can be seen to be $p^{(i)}_2=-p^{(i)}_1,p^{(i+1)}_2=-p^{(i+1)}_1$ and:
\be\begin{split}
A_i&=C^{YX}_{i+1} e^{Lip^{(i)}_i}\;,\;B_i=D^{YX}_{i+1} e^{-Lip^{(i)}_1}\;,\\
  C^{XY}_{i+1}&=F_{i+1} e^{Lip^{(i+1)}_1}\;,\; D^{XY}_{i+1}=G_{i+1}e^{-Li p^{(i+1)}_1}\;.
\end{split}\ee
Solving these conditions, taking into account the $S$-matrices of the respective states, one can express the solution of the 2-magnon Bethe ansatz in terms of a single momentum, which we will take to be $p^{(1)}_1$. At the orbifold point, for $L=3$ we find the values shown in Table \ref{Table:E6L3op}. 

\begin{table}[ht]
\begin{center}
  \begin{tabular}{|c|c|c|}\hline  $p^{(1)}_1$ & E & Sector \\\hline 
    $\frac{2\pi}{3}$ & 12 & Untwisted\\
    $\pi/2+\arctan(\frac{4-\sqrt{7}}3)$& $2(3+\sqrt{7})$& $s,s^5$-twisted\\
    $3\pi/5$ & $2(3+\sqrt{5})$ & $r$-twisted\\
    $\pi+\arctan\sqrt{15+8\sqrt{3}}$ & $2(3+\sqrt{3})$ &$s^2,s^4$-twisted\\
    $\pi/3$ & $4$ &$z$-twisted\\
    $\pi/2-\arctan(\frac{4+\sqrt{7}}3)$& $2(3-\sqrt{7})$ & $s,s^5$-twisted\\
    $\pi/5$ & $2(3-\sqrt{5})$ & $r$-twisted\\ 
    $\arctan\sqrt{15-8\sqrt{3}}$ & $2(3-\sqrt{3})$ & $s^2,s^4$-twisted \\ \hline 
  \end{tabular}\caption{The $\hat{E}_6$ $L=3$ $XYZ$-sector momenta and energies at the orbifold point. Note the absence of an $E=0$ state. The degenerate twisted states are distinguished by different signs in the identification of the $p^{(i)}$ momenta.} \label{Table:E6L3op}
\end{center}
\end{table}

Finally, let us consider the momenta and energies for deformations away from the orbifold point. In Table \ref{Table:D4L3def} we compare the orbifold-point values above to the deformation shown in Fig. \ref{GraphE6L3c}. The values one obtains for the two-magnon states are in complete agreement with the explicit diagonalisation of the Hamiltonian. 

\begin{table}[h]
\begin{center}
  \begin{tabular}{|cc|cc|}\hline \multicolumn{2}{|c|}{$\kappa_i=\kappa_{i+1}=\kappa_7=1$} & \multicolumn{2}{c|}{$\kappa_i=0.9,\kappa_{i+1}=1,\kappa_7=0.9$} \\  \hline $p_1^{(1)}$ & E & $p^{(1)}_1$ & E \\\hline 
    2.0944 & 12 & 2.10717& 10.9193 \\
    1.99483 & 11.2915 & 2.01582& 10.3395 \\
    1.88496 & 10.4721& 1.82456& 9.04747  \\
    1.75485 & 9.4641 & 1.74355& 8.47767 \\
    1.0472 &   4   & 1.03442& 3.56065  \\
    0.81892 & 2.5359 & 0.808858 & 2.26966 \\
    0.62832 & 1.5279 & 0.632407 &1.43243\\
    0.42403 & 0.7085 & 0.40877 & 0.63319\\\hline 
    \end{tabular}\caption{A comparison of the $\hat{E}_6$ $L=3$ two-magnon  momenta and energies at the orbifold point and a sample deformation, which corresponds to $k=0.1$ in Figure \ref{GraphE6L3c}.} \label{Table:E6L3def}
\end{center}
\end{table}

Considering a range of lengths and deformation parameters, and similarly to the previous examples of $\Zset_3$ and $\hat{D}_4$, one concludes that (besides the spurious $p_1=0,\pi$ solutions) the two-magnon $\hat{E}_6$ Bethe ansatz correctly accounts for all two-magnon energies.

\section{Discussion} \label{sec:Discussion}

In this work we derived the planar one-loop scalar dilatation operator for all $\Ncal=2$ quiver theories which can be obtained by orbifolding $\Ncal=4$ SYM and marginally deforming by varying the gauge couplings. These ADE theories cover all the perturbatively-finite $\Ncal=2$ superconformal theories with products of $\SU(N)$ gauge groups, which allow for a large-$N$ limit \cite{Katz:1997eq,Lawrence:1998ja,Bhardwaj2013}. 
Interpreting the dilatation operator as a spin-chain Hamiltonian, we discussed some general features of the spectrum of such spin chains, both in terms of explicit diagonalisation (for short chains), the protected spectrum, and the two-magnon Bethe ansatz for general length. The main aim of our analysis was to show agreement between these three approaches, and therefore we focused on relatively simple sectors (mostly that of holomorphic fields for $L>2$). Having succeeded in this, we plan to extend the analysis of the spectrum to more complicated but perhaps physically more interesting sectors in future work.

A major motivation for the study of the $\Zset_2$ quiver theory in \cite{Gadde:2009dj,Gadde:2010zi} was as a way of approaching Superconformal QCD (in the Veneziano limit) as one of the gauge couplings is tuned to zero. As we saw, the ADE theories can also approach SCQCD in multiple ways, but they also have a far larger space of degenerations when only some of the gauge groups become global. Many of these limits, being linear quivers, are of independent interest, see e.g. \cite{Nunez:2023loo}.  It would be relevant to study the spectrum of these theories in their own right and compare with the limit of the ADE theories. In particular, we found qualitative agreement with the arguments in \cite{Gadde:2009dj} that certain long multiplets break up into protected short ones in any degeneration, however it is certainly possible to do a more precise analysis of what is happening at the level of the $\Ncal=2$ recombination rules \cite{Dolan:2002zh}. 

In order to better understand the protected spectrum of the ADE quiver theories, we computed the relevant superconformal indices. Where there is overlap with the results coming from the spin-chain picture, we find exact agreement, providing an important consistency check on our computations. But of course the indices provide additional predictions for quantities beyond the reach of our current Hamiltonian, for instance regarding operators involving fermions. Checking these predictions would require computing the Hamiltonian beyond the scalar sector. We note that, since the computation in Section \ref{sec:ADEDilatationOperator} was performed in superspace, extending to the fermions of the $\Ncal=1$ chiral multiplets requires little new computation beyond acting with superspace covariant derivatives, so only the $\Ncal=1$ vector multiplet contributions would need to be added. 

This work is intended to set the foundations for several future investigations. For the $\Zset_2$ quiver theory, it was recently shown that one can restore the $\SU(4)_R$ generators which are naively broken by the orbifolding process, by moving beyond the Lie-algebraic setting to that of Lie algebroids \cite{Bertle:2024djm}. The construction involved introducing a groupoid coproduct which enables the composition of the fields in a way that respects the $\Zset_2$ path algebra. Using this coproduct, extended away from the orbifold point by appropriate twists, it was possible to show invariance of the 4d Lagrangian under the algebroid version of $\SU(4)_R$. This construction would be expected to generalise to the generic ADE case by defining a suitable coproduct based on the path algebra of each quiver. Constructing such coproducts and finding the corresponding twists is an important avenue of future work. 

In parallel to the above, and as discussed in Section \ref{sec:ADEspinchains}, there have been recent investigations of spin chains with restricted Hilbert spaces \cite{Corcoran:2024ofo,Blakeney:2025ext}, as well as studies of restricted quantum groups \cite{Felder:2020tct}, where the common thread is the categorical description of the path algebra indicating the allowed compositions of spins/fields at each two sites of the chain. In \cite{Pomoni:2021pbj}, it was argued that mapping a quiver spin chain to an equivalent RSOS picture can make these restrictions more apparent, as in that picture the heights are taken from an adjacency diagram related to the quiver. RSOS models based on the ADE Dynkin diagrams are well known (see \cite{Pearce:1990ila} and references therein) and (suitably extended to our ``dilute'' case) would perhaps provide a useful alternative description of our dynamical spin chains. Formalising a restricted quantum group picture for the ADE chains would not only be relevant in defining the above groupoid symmetries but would also inform extensions to include a spectral parameter via Baxterisation (see \cite{Ren:2023mtn} for a recent discussion of Baxterisation in the context of dynamical algebras).

As a side note, even though the ADE spin chains have long been known to be integrable at the orbifold point \cite{Beisert:2005he,Solovyov:2007pw}, to our knowledge the expected Yangian-like structure responsible for this integrability has not yet been explicitly written down. Given the understanding of the symmetries of the $\Zset_2$ quiver achieved in \cite{Bertle:2024djm} in terms of groupoids, one expects the existence of a groupoid version of the $\Ncal=4$ SYM Yangian algebra, which should generalise to the general ADE chains at the orbifold point. 

Another important direction involves a deeper understanding of the exotic spin chains that correspond to the ADE quivers. We have taken a few first steps by computing the magnon dispersion relations and solving the 2-magnon Bethe Ansatz in the $\SU(3)$ sector. Although the question of integrability is still not settled for the marginally deformed $\Zset_2$ case, it has recently been shown that the $Z$-vacuum 3- and 4-magnon problems are tractable by Bethe ansatz techniques \cite{Bozkurt:2024tpz,Bozkurt:2025exl}, suitably extended to include long-range contributions. This indicates that the additional symmetry discussed above is able to constrain the scattering problem. It would be relevant to reach a similar understanding for some of the cases we considered here.

Starting at length-four, we have also started to see avoided crossings in the energy spectrum, which are clearly relevant for the question of integrability in the context of the von Neumann-Wigner theorem \cite{vonNeumannWigner29,HaakeChaos}. As discussed, a full analysis of this and other potential signatures of quantum chaos would require isolating the sectors that can potentially mix among the many sectors which do not due to the large amount of symmetry in the problem, and is left for future work.   

As compared to the $\Zset_2$ case, the ADE theories have several more couplings that can be independently tuned. So one can wonder whether there are specific points in the parameter space which might be ``more integrable'' than the generic case, for instance by asking for the standard Yang-Baxter equation to be satisfied, similarly to the approach taken in \cite{Bundzik:2005zg} for the $\Ncal=1$ marginal deformations of $\Ncal=4$ SYM. 

The $\Zset_2$ and $\Zset_k$ quiver theories have been extensively studied using techniques of supersymmetric localisation, see \cite{Pestun:2016zxk} for a review. These studies include exact effective couplings \cite{Mitev:2014yba,Mitev:2015oty}, correlation functions of chiral operators \cite{Pini:2017ouj,Niarchos:2019onf,Niarchos:2020nxk,Galvagno:2020cgq,Billo:2021rdb,Billo:2022fnb,Billo:2022gmq,Preti:2022inu} as well as of Wilson loops \cite{Rey:2010ry, Fraser:2015xha, Zarembo:2020tpf, Ouyang:2020hwd, Fiol:2020ojn, Beccaria:2021ksw, Galvagno:2021bbj,Pini:2023lyo,Pini:2024zwi,DeSmet:2025mbc,DeLillo:2025eqg}. Since at the orbifold point the integrability structure is well understood, one can also apply integrability-based methods to the study of three-point functions, which were successfully compared with localisation in \cite{Ferrando:2025qkr} and extended to non-protected operators in \cite{lePlat:2025eod}. Given that the majority of the above studies have been at the orbifold point of the cyclic quiver theories, it is clear that one has only just scratched the surface in terms of checks that could potentially be made for the marginally deformed ADE theories using localisation. 

For the $\Zset_k$ quivers, it is relevant to take the long-quiver limit $k\ra \infty$, where the node number becomes a continuous parameter. Correlation functions in this limit have been studied in \cite{Beccaria:2023qnu,Korchemsky:2025eyc}. Taking the couplings $\kappa_i$ to also vary smoothly, and in a suitable double-scaling limit, the localisation equations become tractable \cite{Sobko:2025zci}. It would be interesting to take this limit in our approach as well (potentially also for the $\hat{D}_k$ quivers) and see whether additional structures emerge. 

Extending our results to the non-planar dilatation operator should be straightforward, and could be relevant for questions related to the $1/N^2$ behaviour of observables that can also be approached by localisation \cite{Galvagno:2020cgq,Galvagno:2021bbj,Preti:2022inu,Beccaria:2023kbl}.

It was recently shown that the tensionless limit of string theory correctly captures the spectrum of the free planar $\Zset_k$ theory at the orbifold point \cite{Gaberdiel:2022iot}. It would be important to extend these worldsheet studies to include the marginal deformation parameters, which among other applications could provide another perspective on the fate of the orbifold-point integrable structures when moving away from the orbifold point. 

While our focus has been on the spin chains arising in the planar limit of the ADE-quiver SCFT's, there has been recent work exploring a very different, large $R$-charge, limit related to $M5$-branes probing ADE singularities, which leads to integrable spin chains in the corresponding $6d$ (2,0) SCFT's \cite{Baume:2020ure,Heckman:2020otd,Baume:2022cot}. It would be worth exploring the chains of dualities linking those spin chains to the ones we have considered.

Finally, to obtain a fuller understanding of the ADE theories it will be important to consider the gravity side of the correspondence as well. Even at the orbifold point, where the $\mathrm{AdS}_5\times\mathrm{S}^5/\Gamma$ geometry is well understood \cite{Kachru:1998ys,Oz:1998hr,Gukov:1998kk}, there are many non-trivial questions one can ask especially about the twisted sectors. See \cite{Skrzypek:2022cgg} for a recent detailed discussion of the comparisons that can be made, with a focus on the $\Zset_2$ case. As mentioned, recent work has also gone beyond supergravity to study string theory correlators of twisted sector operators in the $\Zset_2$ \cite{Skrzypek:2023fkr} and $\Zset_k$ orbifold theories \cite{Martinez:2025jjq}. An important element of these latter works was the resolution of the orbifold singularity via a multi-Gibbons-Hawking ALE geometry. Although an implicit description of the corresponding geometries for the non-abelian ALE cases has been known since the work of Kronheimer \cite{Kronheimer:1989zs}, it is only recently that the explicit metrics for the dihedral ALE geometries were constructed \cite{Cherkis:2003wk, Ionas:2016oxr}. It should therefore be possible to perform analogous studies for the $\hat{D}_k$ theories at the orbifold point.

Of course, more relevant to our case would be the (less well-studied) supergravity description for general couplings, which would allow for more detailed comparison with the marginally deformed ADE chains (e.g. through the study of semiclassical strings on the background), but might also provide a useful alternative regularisation of the orbifold limit in which to perform string computations. See \cite{Gadde:2009dj} for a detailed discussion of the expected features of the supergravity/string duals away from the orbifold point in the $\Zset_2$ quiver context, which would be expected to be similar for the other ADE cases.

\mparagraph{Acknowledgements} We are thankful to Robert de Mello Koch, Elli Pomoni and Sam van Leuven for many helpful discussions and critical comments on the manuscript, and to Elli Pomoni and Sam van Leuven for collaboration on related topics. JB wishes to thank Martin Ro\v cek for helpful correspondence. KZ wishes to thank DESY (Hamburg) and the Quantum Universe network, as well as the Niels Bohr Institute, Copenhagen, for their generous hospitality during the later stages of this work. 

\appendix

\section{Character Tables of Finite Subgroups of $\SU(2)$} \label{sec:CharacterTables}

For completeness, we include here the character tables for the finite subgroups of $\SU(2)$. For more details, we refer the reader to e.g. \cite{Lomont1959,Anselmi:1993sm,DiFrancesco,Hanany:1998sd}. In the following $\omega_k\equiv e^{\frac{2\pi i}{k}}$ is the $k$'th primary root of unity and $\varphi\equiv\frac{1+\sqrt{5}}{2}$ is the golden ratio. Note that $n_i=\chi_i(r_0)$.

\begin{table}[h!]
    \centering
    \begin{tabular}{|c|c|c|c|c|c|c|}\hline
     & $r_0$ & $r_1$&$r_2$ &$\ldots$ &$r_{k-2}$& $r_{k-1}$\tabularnewline\hline
    $\chi_1$ & $1$& $1$ &$1$ &$\ldots$ &$1$& $1$\tabularnewline
    $\chi_2$& $1$& $\omega_k$ & $\omega_k^{2}$&$\ldots$&$\omega_k^{k-2}$&$\omega_k^{k-1}$\tabularnewline
    $\chi_3$& $1$& $\omega_k^2$ & $\omega_k^{4}$&$\ldots$&$\omega_k^{2(k-2)}$&$\omega_k^{2(k-1)}$\tabularnewline
    $\vdots$& $\vdots$& $\vdots$ & $\vdots$&$\ddots$&$\vdots$&$\vdots$\tabularnewline
    $\chi_k$& $1$& $\omega_k^{k-1}$ & $\omega_k^{2(k-1)}$&$\ldots$&$\omega_k^{(k-2)(k-1)}$&$\omega_k^{(k-1)^2}$\tabularnewline\hline
\end{tabular}
    \caption{The $\Zset_k$ character table}
    \label{tab:Zncharactertable}
\end{table}
\begin{table}[h!]
    \centering
    \begin{tabular}{|c|c|c|c|c|c|}\hline
          &$r_0$&$r_1$&$r_{l=2,\ldots k-2}$&$r_{k-1}$&$r_{k}$\tabularnewline\hline  
         $\chi_1$&$1$&$1$&$1$&$1$&1\tabularnewline
         $\chi_2$&$1$&$1$&$1$&$-1$&$-1$\tabularnewline
         $\chi_3$&$1$&$-1$&$-1$&$i^k$&$-i^k$\tabularnewline
         $\chi_4$&$1$&$-1$&$-1$&$-i^k$&$i^k$\tabularnewline
         $\chi_{m=5,\ldots,k+1}$&$2$&$(-1)^{m}2$&$2\cos\left(\frac{\pi (m-4)l}{k}\right)$&$0$&$0$\tabularnewline\hline
    \end{tabular}
    \caption{The $\hat{\mathrm{D}}_k$ ($k$ even) character table.}
    \label{tab:Dnevencharactertable}
\end{table}
\begin{table}[h!]
    \centering
    \begin{tabular}{|c|c|c|c|c|c|}\hline
         &$r_0$&$r_1$&$r_{l=2,\ldots k-2}$&$r_{k-1}$&$r_{k}$\tabularnewline\hline  
         $\chi_1$&$1$&$1$&$1$&$1$&1\tabularnewline
         $\chi_2$&$1$&$1$&$1$&$-1$&$-1$\tabularnewline
         $\chi_{m=3,\ldots,k+1}$&$2$&$(-1)^{m}2$&$2\cos\left(\frac{\pi (m-2)l}{k}\right)$&$0$&$0$\tabularnewline\hline
    \end{tabular}
    \caption{The $\hat{\mathrm{D}}_k$ ($k$ odd) character table.}
    \label{tab:Dnoddcharactertable}
\end{table}
\begin{table}[h!]
    \centering
    \begin{tabular}{|c|c|c|c|c|c|c|c|c|}\hline
         &$r_0$&$r_1$&$r_2$&$r_3$&$r_4$&$r_5$&$r_6$\tabularnewline\hline
         $\chi_1$&$1$&$1$&$1$&$1$&$1$&$1$&$1$\tabularnewline
         $\chi_2$&$1$&$1$&$1$&$\omega_3$&$\omega_3$&$\omega_3^2$&$\omega_3^2$\tabularnewline
         $\chi_3$&$1$&$1$&$1$&$\omega_3^2$&$\omega_3^2$&$\omega_3$&$\omega_3$\tabularnewline
         $\chi_4$&$2$&$-2$&$0$&$1$&$-1$&$1$&$-1$\tabularnewline
         $\chi_5$&$2$&$-2$&$0$&$\omega_3$&$-\omega_3$&$\omega_3^2$&$-\omega_3^2$\tabularnewline
         $\chi_6$&$2$&$-2$&$0$&$\omega_3^2$&$-\omega_3^2$&$\omega_3$&$-\omega_3$\tabularnewline
         $\chi_7$&$3$&$3$&$-1$&$0$&$0$&$0$&$0$\tabularnewline\hline
    \end{tabular}
    \caption{The {2T} character table.}
    \label{tab:2Tcharactertable}
\end{table}
\begin{table}[h!]
    \centering
    \begin{tabular}{|c|c|c|c|c|c|c|c|c|c|}\hline
         &$r_0$&$r_1$&$r_2$&$r_3$&$r_4$&$r_5$&$r_6$&$r_7$\tabularnewline\hline
         $\chi_1$&$1$&$1$&$1$&$1$&$1$&$1$&$1$&$1$\tabularnewline
         $\chi_2$&$1$&$1$&$-3$&$-1$&$-1$&$1$&$1$&$0$\tabularnewline
         $\chi_3$&$2$&$-2$&$0$&$\sqrt{2}$&$-\sqrt{2}$&$1$&$-1$&$0$\tabularnewline
         $\chi_4$&$2$&$-2$&$0$&$-\sqrt{2}$&$\sqrt{2}$&$1$&$-1$&$0$\tabularnewline
         $\chi_5$&$2$&$2$&$-2$&$0$&$0$&$-1$&$-1$&$1$\tabularnewline
         $\chi_6$&$3$&$3$&$-1$&$1$&$1$&$0$&$0$&$-1$\tabularnewline
         $\chi_7$&$3$&$3$&$-3$&$-1$&$-1$&$0$&$0$&$0$\tabularnewline
         $\chi_8$&$4$&$-4$&$0$&$0$&$0$&$-1$&$1$&$0$\tabularnewline\hline
    \end{tabular}
    \caption{The {2O} character table.}
    \label{tab:2Ocharactertable}
\end{table}
\begin{table}[h!]
    \centering
    \begin{tabular}{|c|c|c|c|c|c|c|c|c|c|c|}\hline
         &$r_0$&$r_1$&$r_2$&$r_3$&$r_4$&$r_5$&$r_6$&$r_7$&$r_8$\tabularnewline\hline
         $\chi_1$&$1$&$1$&$1$&$1$&$1$&$1$&$1$&$1$&$1$\tabularnewline
         $\chi_2$&$2$&$-2$&$\varphi^{-1}$&$-\varphi$&$\varphi$&$-\varphi^{-1}$&$1$&$-1$&$0$\tabularnewline
         $\chi_3$&$2$&$-2$&$-\varphi$&$\varphi^{-1}$&$-\varphi^{-1}$&$\varphi$&$1$&$-1$&$0$\tabularnewline
         $\chi_4$&$3$&$3$&$-\varphi^{-1}$&$\varphi$&$\varphi$&$-\varphi^{-1}$&$0$&$0$&$-1$\tabularnewline
         $\chi_5$&$3$&$3$&$\varphi$&$-\varphi^{-1}$&$-\varphi^{-1}$&$\varphi$&$0$&$0$&$-1$\tabularnewline
         $\chi_6$&$4$&$-4$&$-1$&$-1$&$1$&$1$&$-1$&$-1$&$0$\tabularnewline
         $\chi_7$&$4$&$4$&$-1$&$-1$&$-1$&$1$&$1$&$1$&$1$\tabularnewline
         $\chi_8$&$5$&$5$&$0$&$0$&$0$&$0$&$-1$&$-1$&$1$\tabularnewline
         $\chi_9$&$6$&$-6$&$1$&$1$&$-1$&$-1$&$0$&$0$&$0$\tabularnewline\hline
    \end{tabular}
    \caption{The {2I} character table.}
    \label{tab:2Icharactertable}
\end{table}

\section{Adjacency matrices for the finite subgroups of $\SU(2)$}\label{sec:MattercontentN=2}

In this appendix we tabulate (again following \cite{Hanany:1998sd}) the adjacency matrices $a^{\bf 3}_{ij}$ for the finite subgroups of $\SU(2)$, which express the connectivity of the corresponding affine Dynkin diagrams (see Fig. \ref{fig:ADEDynkin}). In the $\Ncal=2$ quiver context, a nonzero entry in the $i,j$ position indicates a field (an arrow in the Dynkin diagram) which connects the $i^{\mathrm{th}}$ and $j^{\mathrm{th}}$ node. As the $\Ncal=2$ theories are non-chiral, if there is an arrow from $i$ to $j$ there is also an arrow from $j$ to $i$. If $i=j$ the field returns to the same node and is an adjoint field, while if $i\neq j$ it is bifundamental. Since the ADE quivers are simply-laced, there is at most one arrow from $i$ to $j$. The exception is $\hat{A}_1$, which has two fields (in each direction) between the nodes.  
$a^\mathbf{3}_{ij}$ is equal to McKay’s $m_{ij}$ matrix \cite{McKay1980}.

\mbox{}

\noindent$\hat{A}_1$, $\Gamma=\Zset_2$:
\be
 a^{\bf 3}_{ij}=\begin{pmatrix}
    1&2\\
    2&1
\end{pmatrix}.
\ee
$\hat{A}_{n\ge2}$, $\Gamma=\Zset_{n+1}$:
\be
 a^{\bf 3}_{ij}=\begin{pmatrix}
    1&1&0&0&\ldots&0&1\\
    1&1&1&0&\ldots&0&0\\
    0&1&1&1&\ldots&0&0\\
    \vdots&\vdots&\vdots&\vdots&\ddots&\vdots&\vdots\\
    1&0&0&0&\ldots&1&1
\end{pmatrix}.
\ee
$\hat{D}_{4}$, $\Gamma=\hat{\mathrm{D}}_4$:
\be \label{D4a}
a^{\bf 3}_{ij}=\begin{pmatrix}
    1&0&1&0&0\\
    0&1&1&0&0\\
    1&1&1&1&1\\
    0&0&1&1&0\\
    0&0&1&0&1
\end{pmatrix}\;.
\ee
$\hat{D}_{n\ge5}$, $\Gamma=\text{D}_{n}$:
\be
a^{\bf 3}_{ij}=\begin{pmatrix}
    1&0&1&0&0&\ldots&0&0&0\\
    0&1&1&0&0&\ldots&0&0&0\\
    1&1&1&1&0&\ldots&0&0&0\\
    0&0&1&1&1&\ldots&0&0&0\\
    \vdots&\vdots&\vdots&\vdots&\vdots&\ddots&\vdots&\vdots&\vdots\\
    0&0&0&0&0&\ldots&1&1&1\\
    0&0&0&0&0&\ldots&1&1&0\\
    0&0&0&0&0&\ldots&1&0&1
\end{pmatrix}.
\ee
$\hat{E}_6$, $\Gamma={2\text{T}}$:
\be \label{E6a}
a^{\bf 3}_{ij}=\begin{pmatrix}
    1&1&0&0&0&0&0\\
    1&1&1&0&0&0&0\\
    0&1&1&1&0&1&0\\
    0&0&1&1&1&0&0\\
    0&0&0&1&1&0&0\\
    0&0&1&0&0&1&1\\
    0&0&0&0&0&1&1
\end{pmatrix}.
\ee
$\hat{E}_7$, $\Gamma={2\text{O}}$:
\be
a^{\bf 3}_{ij}=\begin{pmatrix}
    1&1&0&0&0&0&0&0\\
    1&1&1&0&0&0&0&0\\
    0&1&1&1&0&0&0&0\\
    0&0&1&1&1&0&0&1\\
    0&0&0&1&1&1&0&0\\
    0&0&0&0&1&1&1&0\\
    0&0&0&0&0&1&1&0\\
    0&0&0&1&0&0&0&1
\end{pmatrix}.
\ee
$\hat{E}_8$, $\Gamma={2\text{I}}$:
\be
a^{\bf 3}_{ij}=\begin{pmatrix}
    1&1&0&0&0&0&0&0&0\\
    1&1&1&0&0&0&0&0&0\\
    0&1&1&1&0&0&0&0&0\\
    0&0&1&1&1&0&0&0&0\\
    0&0&0&1&1&1&0&0&0\\
    0&0&0&0&1&1&1&0&1\\
    0&0&0&0&0&1&1&1&0\\
    0&0&0&0&0&0&1&1&0\\
    0&0&0&0&0&1&0&0&1
\end{pmatrix}.
\ee

\section{Superspace Feynman Rules} \label{sec:SuperspaceFeynman}

In this appendix, we present the Feynman rules derived from the $\NN=2$ Lagrangian (\ref{eq:NN=2generalaction}). For more background and derivations of supergraph rules, see \cite{Gates:1983nr, Wess:1992cp}. We use Wick rotated rules i.e. we have transformed $e^{-iS}\to e^{S_E}$.  We decompose the superfields as follows:
\be 
        \Phi_i=(\Phi_A)_iT_i^A\;,\; Q_{ij}=(Q_A)_{ij}B_{ij}^A\;,\;\Qb_{ji}=(\Qb_A)_{ji}B_{ji}^A\;,
\ee
where here $\Phi_i=(V_i,c_i,c_i',\bar{c}_i,\bar{c}_i',Z_i,\Zb_i)$ stands for any adjoint field and $(T_i)^A$ and $(B_{ij})^A$ are the representation matrices of the $\SU(n_iN)\times \SU(n_iN)$ and $\SU(n_iN)\times\SU(n_jN)$ gauge groups respectively. These satisfy 
\begin{subequations}\label{eq:tracerelationships}
    \begin{align}
        \Tr_i(T^A_iT^B_i)=&\delta^{AB},\quad A,B=1,\ldots,(n_iN)^2-1,\\
        \Tr_i(B_{ij}^AB^B_{ji})=&\delta^{AB},\quad A,B=1,\ldots,n_in_jN^2.
    \end{align}
\end{subequations}
Since these matrices form a complete basis we have
\begin{subequations}
    \begin{align}
        (T_i^A)^{m}_n(T_i^A)^l_k=&\delta^{m}_k\delta^{l}_n-\frac{1}{n_iN}\delta^{m}_n\delta^{l}_k,\\
        (B^A_{ij})^m_n(B^A_{ji})^{l}_k=&\delta^{m}_{k}\delta^{l}_n.
    \end{align}
\end{subequations}
The matrices $T^A_i$ satisfy  the usual commutation relation, 
\begin{equation}
    [T^A_i,T^B_i]=if_i^{AB}\phan_CT_i^C,
\end{equation}
with the structure constants satisfying 
\begin{equation}
    f_i^{ACD}f_i^{BCD}=2n_iN\delta^{AB}.
\end{equation}
Now for $G_i=\SU(n_iN)$:
\begin{subequations}
    \begin{align}
    C(G_i)=&2n_iN,\quad C(\square_i)=C(\bar{\square}_i)=1,\\
    C_2(G_i)=&2n_iN,\quad C_2(\square_i)=C_2(\bar{\square}_i)=n_iN-\frac{1}{n_iN}.
    \end{align}
\end{subequations}

Let us now consider the propagators of the theory
\begin{equation}
    \begin{split}
        \begin{tikzpicture}
            \node[left] at (-1,0) {$\expval{(V_A)_i(V_B)_i}=$};
        \end{tikzpicture}&
        \begin{tikzpicture}
            \draw[decorate,decoration={coil,aspect=0,segment length=4.5,amplitude=0.7}] (-1,0)--(1,0);
            \node[right] at (1,0) {$-\frac{\delta_{AB}}{n_i}\frac{\delta^{(4)}(\theta_1-\theta_2)}{p^2},$};
        \end{tikzpicture}\\
        \begin{tikzpicture}
            \node[left] at (-1,0) {$\expval{(\Zb_A)_i(Z_B)_i}=$};
        \end{tikzpicture}&
        \begin{tikzpicture}
            \draw (-1,0)--(1,0);
            \node[right] at (1,0) {$\frac{\delta_{AB}}{n_i}\frac{\delta^{(4)}(\theta_1-\theta_2)}{p^2},$};
        \end{tikzpicture}\\
        \begin{tikzpicture}
            \node[left] at (-1,0) {$\expval{(\Qb_A)_{ji}(Q_{B})_{ij}}=$};
        \end{tikzpicture}&
        \begin{tikzpicture}
            \draw[dashed] (-1,0)--(1,0);
            \node[right] at (1,0) {$\frac{\delta_{AB}}{n_in_j}\frac{\delta^{(4)}(\theta_1-\theta_2)}{p^2},$};
        \end{tikzpicture}\\
        \begin{tikzpicture}
            \node[left] at (-1,0) {$\expval{(\bar{c}'_A)_i(c_B)_i}=-\expval{(c'_A)_i(\bar{c}_B)_i}=$};
        \end{tikzpicture}&
        \begin{tikzpicture}
            \draw[dotted,thick] (-1,0)--(1,0);
            \node[right] at (1,0) {$\frac{\delta_{AB}}{n_i}\frac{\delta^{(4)}(\theta_1-\theta_2)}{p^2}.$};
        \end{tikzpicture}
    \end{split}
\end{equation}
where we are working in Fermi-Feynman gauge $\xi_i=1+\mathcal{O}(\gymi^2)$ to avoid infrared problems \cite{Grisaru:1979wc}.

We now consider the vertices of the theory. Our convention is that the colour indices ($A,B,C$) are read counterclockwise starting with the leg to the left. 

The cubic gauge vertex is given by
\begin{equation}
    \begin{tikzpicture}[baseline=-0.72cm]
        \node[left] at (-0.75,0) {$\ver_{V_i^3}=\biggl($};
        \draw[decorate,decoration={coil,aspect=0,segment length=4.5,amplitude=0.7}] (-0.75,0)--(0,0);
        \draw[decorate,decoration={coil,aspect=0,segment length=4.5,amplitude=0.7}] (0,0)--(0.375,0.6495);
        \draw[decorate,decoration={coil,aspect=0,segment length=4.5,amplitude=0.7}] (0,0)--(0.375,-0.6495);
        \node[above,rotate=60,scale=0.7] at (0.1875,0.3248) {$\codif^\alpha$};
            \node[below,rotate=300,scale=0.7] at (0.1875,-0.3248) {$\bcodif^2\codif_\alpha$};
    \end{tikzpicture}
    \begin{tikzpicture}
        \node[left] at (-0.75,0) {$-$};
        \draw[decorate,decoration={coil,aspect=0,segment length=4.5,amplitude=0.7}] (-0.75,0)--(0,0);
        \draw[decorate,decoration={coil,aspect=0,segment length=4.5,amplitude=0.7}] (0,0)--(0.375,0.6495);
        \draw[decorate,decoration={coil,aspect=0,segment length=4.5,amplitude=0.7}] (0,0)--(0.375,-0.6495);
        \node[above,rotate=60,scale=0.7] at (0.1875,0.3248) {$\bcodif^2\codif_\alpha$};
        \node[below,rotate=300,scale=0.7] at (0.1875,-0.3248) {$\codif^\alpha$};
    \end{tikzpicture}
    \begin{tikzpicture}
        \node[left] at (-0.75,0) {$+$};
        \draw[decorate,decoration={coil,aspect=0,segment length=4.5,amplitude=0.7}] (-0.75,0)--(0,0);
        \draw[decorate,decoration={coil,aspect=0,segment length=4.5,amplitude=0.7}] (0,0)--(0.375,0.6495);
        \draw[decorate,decoration={coil,aspect=0,segment length=4.5,amplitude=0.7}] (0,0)--(0.375,-0.6495);
        \node[above,scale=0.7] at (-0.375,0) {$\codif_\alpha\bcodif^2$};
        \node[below,rotate=300,scale=0.7] at (0.1875,-0.3248) {$\codif^\alpha$};
    \end{tikzpicture}
    \begin{tikzpicture}
        \node[left] at (-0.75,0) {$-$};
        \draw[decorate,decoration={coil,aspect=0,segment length=4.5,amplitude=0.7}] (-0.75,0)--(0,0);
        \draw[decorate,decoration={coil,aspect=0,segment length=4.5,amplitude=0.7}] (0,0)--(0.375,0.6495);
        \draw[decorate,decoration={coil,aspect=0,segment length=4.5,amplitude=0.7}] (0,0)--(0.375,-0.6495);
        \node[below,scale=0.7] at (-0.375,0) {$\codif_\alpha\bcodif^2$};
        \node[above,rotate=60,scale=0.7] at (0.1875,0.3248) {$\codif^\alpha$};
    \end{tikzpicture}
    \begin{tikzpicture}
        \node[left] at (-0.75,0) {$+$};
        \draw[decorate,decoration={coil,aspect=0,segment length=4.5,amplitude=0.7}] (-0.75,0)--(0,0);
        \draw[decorate,decoration={coil,aspect=0,segment length=4.5,amplitude=0.7}] (0,0)--(0.375,0.6495);
        \draw[decorate,decoration={coil,aspect=0,segment length=4.5,amplitude=0.7}] (0,0)--(0.375,-0.6495);
         \node[below,scale=0.7] at (-0.375,0) {$\codif^\alpha$};
        \node[above,rotate=60,scale=0.7] at (0.1875,0.3248) {$\bcodif^2\codif_\alpha$};
    \end{tikzpicture}
    \begin{tikzpicture}
        \node[left] at (-0.75,0) {$-$};
        \draw[decorate,decoration={coil,aspect=0,segment length=4.5,amplitude=0.7}] (-0.75,0)--(0,0);
        \draw[decorate,decoration={coil,aspect=0,segment length=4.5,amplitude=0.7}] (0,0)--(0.375,0.6495);
        \draw[decorate,decoration={coil,aspect=0,segment length=4.5,amplitude=0.7}] (0,0)--(0.375,-0.6495);
        \node[above,scale=0.7] at (-0.375,0) {$\codif^\alpha$};
        \node[above,rotate=300,scale=0.7] at (0.1875,-0.3248) {$\bcodif^2\codif_\alpha$};
        \node[right] at (0.7,0) {$\biggr)\frac{\gym}{2} \kappa_in_i\Tr_i(T_i^A[T_i^B,T_i^C]).$};
    \end{tikzpicture}
\end{equation}

The cubic gauge-matter vertices are given by
\begin{equation}
    \begin{split}
        &\begin{tikzpicture}
            \node[left] at (-0.75,0) {$\ver_{\bar{Z}_iV_i Z_i}=$};
            \draw[decorate,decoration={coil,aspect=0,segment length=4.5,amplitude=0.7}] (-0.75,0)--(0,0);
            \draw (0,0)--(0.375,0.6495);
            \draw (0,0)--(0.375,-0.6495);
            \node[above,rotate=60,scale=0.7] at (0.1875,0.3248) {$\bcodif^2$};
            \node[below,rotate=300,scale=0.7] at (0.1875,-0.3248) {$\codif^2$};
            \node[right] at (0.7,0) {$\gym \kappa_in_i\Tr_i(T_i^A[T_i^B,T_i^C]),$};
        \end{tikzpicture}\\
        &\begin{tikzpicture}
            \node[left] at (-0.75,0) {$\ver_{\Qb_{ij}V_jQ_{ji}}=$};
            \draw[decorate,decoration={coil,aspect=0,segment length=4.5,amplitude=0.7}] (-0.75,0)--(0,0);
            \draw[dashed] (0,0)--(0.375,0.6495);
            \draw[dashed] (0,0)--(0.375,-0.6495);
            \node[above,rotate=60,scale=0.7] at (0.1875,0.3248) {$\bcodif^2$};
            \node[below,rotate=300,scale=0.7] at (0.1875,-0.3248) {$\codif^2$};
            \node[right] at (0.7,0) {$\gym\kappa_jn_in_j\Tr_j(T_j^AB^B_{ji}B_{ij}^C),$};
        \end{tikzpicture}\\
        &\begin{tikzpicture}
            \node[left] at (-0.75,0) {$\ver_{Q_{ji}V_i\Qb_{ij}}=$};
            \draw[decorate,decoration={coil,aspect=0,segment length=4.5,amplitude=0.7}] (-0.75,0)--(0,0);
            \draw[dashed] (0,0)--(0.375,0.6495);
            \draw[dashed] (0,0)--(0.375,-0.6495);
            \node[above,rotate=60,scale=0.7] at (0.1875,0.3248) {$\codif^2$};
            \node[below,rotate=300,scale=0.7] at (0.1875,-0.3248) {$\bcodif^2$};
            \node[right] at (0.7,0) {$-\gym\kappa_in_in_j\Tr_i(T_i^AB_{ij}^BB^C_{ji}),$};
        \end{tikzpicture}\\
        &\begin{tikzpicture}
            \node[left] at (-0.75,0) {$\ver_{V_ic_ic'_i}=$};
            \draw[decorate,decoration={coil,aspect=0,segment length=4.5,amplitude=0.7}] (-0.75,0)--(0,0);
            \draw[thick,dotted] (0,0)--(0.375,0.6495);
            \draw[thick,dotted] (0,0)--(0.375,-0.6495);
            \node[above,rotate=60,scale=0.7] at (0.1875,0.3248) {$\bcodif^2$};
            \node[below,rotate=300,scale=0.7] at (0.1875,-0.3248) {$\codif^2$};
            \node[right] at (0.7,0) {$\frac{\gym}{2} \kappa_in_i\Tr_i(T_i^A[T_i^B,T_i^C]),$};
        \end{tikzpicture}\\
        &\begin{tikzpicture}
            \node[left] at (-0.75,0) {$\text{V}_{V_ic_i\bar{c}'_i}=$};
            \draw[decorate,decoration={coil,aspect=0,segment length=4.5,amplitude=0.7}] (-0.75,0)--(0,0);
            \draw[thick,dotted] (0,0)--(0.375,0.6495);
            \draw[thick,dotted] (0,0)--(0.375,-0.6495);
            \node[above,rotate=60,scale=0.7] at (0.1875,0.3248) {$\bcodif^2$};
            \node[below,rotate=300,scale=0.7] at (0.1875,-0.3248) {$\codif^2$};
            \node[right] at (0.7,0) {$\frac{\gym}{2} \kappa_in_i\Tr_i(T_i^A[T_i^B,T_i^C]),$};
        \end{tikzpicture}\\
        &\begin{tikzpicture}
            \node[left] at (-0.75,0) {$\ver_{V_i\bar{c}_ic'_i}=$};
            \draw[decorate,decoration={coil,aspect=0,segment length=4.5,amplitude=0.7}] (-0.75,0)--(0,0);
            \draw[thick,dotted] (0,0)--(0.375,0.6495);
            \draw[thick,dotted] (0,0)--(0.375,-0.6495);
            \node[above,rotate=60,scale=0.7] at (0.1875,0.3248) {$\codif^2$};
            \node[below,rotate=300,scale=0.7] at (0.1875,-0.3248) {$\bcodif^2$};
            \node[right] at (0.7,0) {$\frac{\gym}{2} \kappa_in_i\Tr_i(T_i^A[T_i^B,T_i^C]),$};
        \end{tikzpicture}\\
        &\begin{tikzpicture}
            \node[left] at (-0.75,0) {$\ver_{V_i\bar{c}_i\bar{c}'_i}=$};
            \draw[decorate,decoration={coil,aspect=0,segment length=4.5,amplitude=0.7}] (-0.75,0)--(0,0);
            \draw[thick,dotted] (0,0)--(0.375,0.6495);
            \draw[thick,dotted] (0,0)--(0.375,-0.6495);
            \node[above,rotate=60,scale=0.7] at (0.1875,0.3248) {$\codif^2$};
            \node[below,rotate=300,scale=0.7] at (0.1875,-0.3248) {$\bcodif^2$};
            \node[right] at (0.7,0) {$\frac{\gym}{2} \kappa_in_i\Tr_i(T_i^A[T_i^B,T_i^C]).$};
        \end{tikzpicture}
    \end{split}
\end{equation}
\
The cubic matter vertices are given by
\begin{equation}
    \begin{split}
        &\begin{tikzpicture}
            \node[left] at (-0.75,0) {$\ver_{Q_{ji} Z_iQ_{ij}}=$};
            \draw (-0.75,0)--(0,0);
            \draw[dashed] (0,0)--(0.375,0.6495);
            \draw[dashed] (0,0)--(0.375,-0.6495);
            \node[above,rotate=60,scale=0.7] at (0.1875,0.3248) {$\bcodif^2$};
            \node[below,rotate=300,scale=0.7] at (0.1875,-0.3248) {$\bcodif^2$};
            \node[right] at (0.7,0) {$i\gym \kappa_id_{ji}\Tr_i(T_i^AB_{ij}^B,B_{ji}^C)$};
        \end{tikzpicture},\\
        &\begin{tikzpicture}
            \node[left] at (-0.75,0) {$\ver_{\Qb_{ji}\Zb_i\Qb_{ij}}=$};
            \draw (-0.75,0)--(0,0);
            \draw[dashed] (0,0)--(0.375,0.6495);
            \draw[dashed] (0,0)--(0.375,-0.6495);
            \node[above,rotate=60,scale=0.7] at (0.1875,0.3248) {$\codif^2$};
            \node[below,rotate=300,scale=0.7] at (0.1875,-0.3248) {$\codif^2$};
            \node[right] at (0.7,0) {$-i\gym \kappa_i\bar{d}_{ji}\Tr_i(T_i^AB_{ij}^B,B_{ji}^C)$};
        \end{tikzpicture}.
    \end{split}
\end{equation}
Finally, each ghost loop contributes a factor of $-1$.

\section{Finiteness of the ADE theories}\label{sec:betafunction}

As a consistency check of our ADE lagrangian (\ref{eq:NN=2generalaction}) and the resulting Feynman rules in Appendix \ref{sec:SuperspaceFeynman}, in this appendix we will verify the finiteness of the general ADE theories. We largely follow \cite{Grisaru:1979wc}, where the finiteness of $\Ncal=4$ SYM was demonstrated using superspace techniques. First of all, note that the $Q_{ij}Z_jQ_{ji}$ vertex coming from the superpotential is purely chiral and hence does not receive any perturbative corrections, and similarly for the pure vector vertices as they also arise from a chiral term. Hence to prove $\beta_i=0$ it is enough to show that the self-energies of the chiral fields vanish.

Given the 1-loop exactness of $\NN=2$ theories \cite{Seiberg:1988ur}, it is enough to show the vanishing of the one-loop beta function at a generic node $i$: 
\begin{equation}\label{eq:NSVZBeta}
    \beta_i=-\frac{g_{\text{YM},i}^3}{16\pi^2}\left[3C(G_i)-\sum_{j}C(R_j)(1-\gamma_j)\right],
\end{equation}
where $\gamma_j$ is the anomalous dimension of the matter fields coupled to the $i^\mathrm{th}$ gauge field.

Now let us focus on an arbitrary node, labelled by $i$, with corresponding gauge group $G_i=\SU(n_iN)$. First we should notice that we have one adjoint field $Z_i$ at this node, so we have $C(R_{Z_i})(1-\gamma_{Z_i})=C(G_i)(1-\gamma_{Z_i})$ with $C(G_i)$ given in Appendix \ref{sec:SuperspaceFeynman}. Let us now consider the bifundamentals $Q_{ij}$ and $Q_{ji}$ in the fundamental and antifundamental representation of $G_i=\SU(n_iN)$. From Appendix \ref{sec:SuperspaceFeynman}, we see that $C(\square_i)=C(\bar{\square}_i)=1$. However, we are treating $\SU(n_jN)$ as a flavour group, thus there are $n_jN$ copies of $Q_{ij}$ and $Q_{ji}$. Hence, we can write for the group that the $\beta$-function \eqref{eq:NSVZBeta} for the corresponding gauge coupling is given by:
\be
\begin{split}\label{eq:betafunctionN=2}
    \beta_i=&-\frac{\gym^3\kappa_i^3}{16\pi^2}\left[(2+\gamma_{\Phi_i})C(G_i)-\sum_{j=1}^Ma^{\mathbf{2}}_{ij}n_jN\biggl(C(\square_i)(1-\gamma_{Q_{ij}})+C(\bar{\square}_i)(1-\gamma_{Q_{ji}})\biggr)\right]\\
    =&-\frac{\gym^3\kappa_i^3}{16\pi^2}\left[2n_iN(2+\gamma_{\Phi_i})-\sum_{j=1}^Ma^{\mathbf{2}}_{ij}n_jN\bigl(2-\gamma_{Q_{ij}}-\gamma_{Q_{ji}}\bigr)\right].
\end{split}
\ee
We now compute the anomalous dimensions of the chiral superfields \begin{equation}\label{fig:vectorselfenergy}
    \begin{split}
        &\begin{tikzpicture}
    \node[left,scale=0.7] at (-2,0) {$(\Phib_A)_i(-p,\theta)$};
    \node[right,scale=0.7] at (2,0) {$(\Phi_B)_i(p,\theta)$};
    \draw (-2,0)--(2,0);
    \fill[blue!15,opacity=0.5]
        (-1,0) .. controls (-0.7,1) and (0.7,1) .. (1,0) --
        (1,0) -- (-1,0) -- cycle;
    \draw[decorate,decoration={coil,aspect=0,segment length=4.5,amplitude=0.7}]
        (-1,0).. controls (-0.7,1) and (0.7,1).. (1,0);
    \draw[->] (-0.2,0.85).. controls (-0.1,0.9) and (0.1,0.9).. (0.2,0.85);
    \node[above,scale=0.6] at (-1.3,0.1) {$(V_C)_i$};
    \node[above,scale=0.6] at (1.3,0.1) {$(V_C)_i$};
    \node[below,scale=0.6] at (-0.8,0) {$(\Phi_D)_i$};
    \node[below,scale=0.6] at (0.8,0) {$(\Phib_D)_i$};
    \draw[->] (0,0)--(-0.1,0);
    \node[below,scale=0.6] at (0,0) {$p-k$};
    \node[above,scale=0.6] at (0,0.9) {$k$};
\end{tikzpicture}
    \end{split}
\end{equation}
The contribution from \eqref{fig:vectorselfenergy} is
\begin{equation}
    \begin{split}
        \Gamma^{(2)}_{\text{vec}}[(\Phib_A)_i,(\Phi_B)_i]=&-\kappa_i^2\int\frac{d^{4}p}{(2\pi)^{4}}\int d^4\theta(\Phib^A)_i(\Phi^B)_i f^{ACD}f^{BCD}N^{-1}I(\lambda,\mu,\epsilon)\\
        =&-2\lambda \kappa_i^2n_i\frac{\Gamma(\epsilon)\Gamma(1-\epsilon)^2}{\Gamma(2-2\epsilon)}\int\frac{d^{4}p}{(2\pi)^{4}}\int d^4\theta(\Phib^A)_i(\Phi^A)_i\left[\frac{4\pi\mu^2}{p^2}\right]^\epsilon.
    \end{split}
\end{equation}
Next we consider the chiral self interaction 
\begin{equation}\label{eq:chiralselfenergy}
    \begin{split}
        &\begin{tikzpicture}
             \node[left,scale=0.7] at (-2,0) {$(\Phib_A)_i(-p,\theta)$};
             \node[right,scale=0.7] at (2,0) {$(\Phi_B)_i(p,\theta)$};
           \draw (-2,0)--(-1,0); 
           \draw (1,0)--(2,0);
           \draw[dashed,fill=blue!15,opacity=0.5] (0,0) circle (1);
           \node[left,scale=0.5] at (-1,0.2)  {$(\Qb_D)_{ij}$};
           \node[left,scale=0.5] at (-1,-0.2) {$(\Qb_C)_{ji}$};
           \node[right,scale=0.5] at (1,0.2)  {$(Q_D)_{ji}$};
           \node[right,scale=0.5] at (1,-0.2) {$(Q_C)_{ij}$};
           \draw[->] (0,-1)--(-0.1,-1);
           \draw[->] (0,1)--(0.1,1);
           \node[below,scale=0.6] at (0,-1) {$p-k$};
           \node[above,scale=0.6] at (0,1) {$k$};
        \end{tikzpicture}
    \end{split}
\end{equation}
where we sum over $j$. Let us first consider the matrix part of \eqref{eq:chiralselfenergy}
\begin{equation}
    \begin{split}
        (B_{ji}^C)^m_n&(T_i^A)^{n}_k(B_{ij}^D)^k_m(B^D_{ji})^{m'}_{n'}(T^B_i)^{n'}_{k'}(B_{ij}^C)^{k'}_{m'}\\
        =&\delta^m_{m'}\delta^{k'}_n\delta^{k}_{n'}\delta^{m'}_{m}(T_i^A)^n_k(T_i^B)^{n'}_{k'}\\
        =&\delta^m_m(T_i^A)^n_k(T_i^B)^k_n =n_jN\Tr(T^A_iT^B_i) =n_jN\delta^{AB}.
    \end{split}
\end{equation}
Next we consider
\begin{equation}
        \sum_{j=1}^M\frac{a^{\mathbf{2}}_{ij}\bar{d}_{ji}d_{ji}}{n_i^2n_j}
        =\sum_{j=1}^Ma^\mathbf{2}_{ij}n_j
        =2n_i \;.
    \end{equation}
Then we have
\begin{equation}
    \begin{split}
        \Gamma^{(2)}_{\text{chiral}}[(\Phib_A)_i,(\Phi_B)_i]=&2\kappa_i^2n_i\int\frac{d^4p}{(2\pi)^4}\int d^4\theta(\Phib^A)_i(\Phi^A)_iI(\lambda,\mu,\epsilon)\\
        =&2\lambda \kappa_i^2n_i\frac{\Gamma(\epsilon)\Gamma(1-\epsilon)^2}{\Gamma(2-2\epsilon)}\int\frac{d^4p}{(2\pi)^4}\int d^4\theta(\Phib_A)_i(\Phi_A)_i\left[\frac{4\pi\mu^2}{p^2}\right]^{\epsilon}.
    \end{split}
\end{equation}
Hence, the one-loop contribution to the effective action of the adjoint chiral superfields is
\begin{equation}
    \Gamma^{(2)}_{\text{one-loop}}[(\Phib_A)_i,(\Phi_B)_i]=\Gamma^{(2)}_{\text{vec}}[(\Phib_A)_i,(\Phi_B)_i]+\Gamma^{(2)}_{\text{chiral}}[(\Phib_A)_i,(\Phi_B)_i]=0,
\end{equation}
and it follows that
\be\label{eq:anomalousdimadj}
\gamma_{\Phi_i}=0.
\ee
Let us now compute the self-energy of the bifundamentals. The vector self-energy is given by
\begin{equation}
    \begin{split}
        &\begin{tikzpicture}
            \node[left,scale=0.7] at (-2,0) {$(\Qb_A)_{ij}(-p,\theta)$};
            \node[right,scale=0.7] at (2,0) {$(Q_B)_{ji}(p,\theta)$};
            \fill[blue!15,opacity=0.5]
        (-1,0) .. controls (-0.7,1) and (0.7,1) .. (1,0) --
        (1,0) -- (-1,0) -- cycle;
            \draw[dashed] (-2,0)--(2,0);
            \draw[decorate,decoration={coil,aspect=0,segment length=4.5,amplitude=0.7}] (-1,0).. controls (-0.7,1) and (0.7,1).. (1,0);
            \node[above,scale=0.6] at (-1.3,0.1) {$(V_C)_i$};
            \node[above,scale=0.6] at (1.3,0.1) {$(V_C)_i$};
            \node[below,scale=0.6] at (-0.8,0) {$(Q_D)_{ji}$};
            \node[below,scale=0.6] at (0.8,0) {$(\Qb_D)_{ij}$};
            \draw[->] (0,0)--(-0.1,0);
            \node[below,scale=0.6] at (0,0) {$p-k$};
            \draw[->] (-0.2,0.85).. controls (-0.1,0.9) and (0.1,0.9).. (0.2,0.85);
            \node[above,scale=0.6] at (0,0.9) {$k$};
        \end{tikzpicture}\\
        &\hspace{3.6cm}+\\
        &\begin{tikzpicture}
            \node[left,scale=0.7] at (-2,0) {$(\Qb_A)_{ij}(-p,\theta)$};
            \node[right,scale=0.7] at (2,0) {$(Q_B)_{ji}(p,\theta)$};
            \fill[blue!15,opacity=0.5]
        (-1,0) .. controls (-0.7,-1) and (0.7,-1) .. (1,0) --
        (1,0) -- (-1,0) -- cycle;
            \draw[dashed] (-2,0)--(2,0);
            \draw[decorate,decoration={coil,aspect=0,segment length=4.5,amplitude=0.7}] (-1,0).. controls (-0.7,-1) and (0.7,-1).. (1,0);
            \node[below,scale=0.6] at (-1.3,-0.1) {$(V_C)_j$};
            \node[below,scale=0.6] at (1.3,-0.1) {$(V_C)_j$};
            \node[above,scale=0.6] at (-0.8,0) {$(Q_D)_{ji}$};
            \node[above,scale=0.6] at (0.8,0) {$(\Qb_D)_{ij}$};
            \draw[->] (0,0)--(-0.1,0);
            \node[above,scale=0.6] at (0,0) {$p-k$};
            \draw[->] (-0.2,-0.85).. controls (-0.1,-0.9) and (0.1,-0.9).. (0.2,-0.85);
            \node[below,scale=0.6] at (0,-0.9) {$k$};
        \end{tikzpicture}
    \end{split}
\end{equation}
The chiral self-energy is given by
\begin{equation}
    \begin{split}
        &\begin{tikzpicture}
            \node[left,scale=0.7] at (-2,0) {$(\Qb_A)_{ij}(-p,\theta)$};
            \node[right,scale=0.7] at (2,0) {$(Q_B)_{ji}(p,\theta)$};
            \fill[blue!15,opacity=0.5]
        (-1,0) .. controls (-0.7,1) and (0.7,1) .. (1,0) --
        (1,0) -- (-1,0) -- cycle;
            \draw[dashed] (-2,0)--(2,0);
            \draw(-1,0).. controls (-0.7,1) and (0.7,1).. (1,0);
            \node[above,scale=0.6] at (-1.3,0.1) {$(\Phib_C)_i$};
            \node[above,scale=0.6] at (1.3,0.1) {$(\Phi_C)_i$};
            \node[below,scale=0.6] at (-0.8,0) {$(\Qb_D)_{ji}$};
            \node[below,scale=0.6] at (0.8,0) {$(Q_D)_{ij}$};
            \draw[->] (0,0)--(-0.1,0);
            \node[below,scale=0.6] at (0,0) {$p-k$};
            \draw[->] (-0.2,0.85).. controls (-0.1,0.9) and (0.1,0.9).. (0.2,0.85);
            \node[above,scale=0.6] at (0,0.9) {$k$};
        \end{tikzpicture}\\
        &\hspace{3.6cm}+\\
        &\begin{tikzpicture}
            \node[left,scale=0.7] at (-2,0) {$(\Qb_A)_{ij}(-p,\theta)$};
            \node[right,scale=0.7] at (2,0) {$(Q_B)_{ji}(p,\theta)$};
            \fill[blue!15,opacity=0.5]
        (-1,0) .. controls (-0.7,-1) and (0.7,-1) .. (1,0) --
        (1,0) -- (-1,0) -- cycle;
            \draw[dashed] (-2,0)--(2,0);
            \draw(-1,0).. controls (-0.7,-1) and (0.7,-1).. (1,0);
            \node[below,scale=0.6] at (-1.3,-0.1) {$(\Phib_C)_j$};
            \node[below,scale=0.6] at (1.3,-0.1) {$(\Phi_C)_j$};
            \node[above,scale=0.6] at (-0.8,0) {$(\Qb_D)_{ji}$};
            \node[above,scale=0.6] at (0.8,0) {$(Q_D)_{ij}$};
            \draw[->] (0,0)--(-0.1,0);
            \node[above,scale=0.6] at (0,0) {$p-k$};
            \draw[->] (-0.2,-0.85).. controls (-0.1,-0.9) and (0.1,-0.9).. (0.2,-0.85);
            \node[below,scale=0.6] at (0,-0.9) {$k$};
        \end{tikzpicture}
    \end{split}
\end{equation}
The vector self-energy contributes 
\begin{equation}
    \begin{split}
        \Gamma^{(2)}_{\text{vec}}&[(\Qb_A)_{ij},(Q_B)_{ji}]=-n_in_j(\kappa_i^2+\kappa_j^2)\int\frac{d^4p}{(2\pi)^4}\int d^4\theta(\Qb_A)_{ij}(Q_A)_{ji}I(\lambda,\mu,\epsilon)\\
        &\quad=-\lambda n_in_j(\kappa_i^2+\kappa_j^2)\frac{\Gamma(\epsilon)\Gamma(1-\epsilon)^2}{\Gamma(2-2\epsilon)}\int\frac{d^4p}{(2\pi)^4}\int d^4\theta(\Qb_A)_{ij}(Q_A)_{ji}\left[\frac{4\pi\mu^2}{p^2}\right]^{\epsilon}.
    \end{split}
\end{equation}
From the chiral vertex, we get the following contributions
\begin{equation}
    \begin{split}
        &\bar{d}_{ji}d_{ji}=(n_in_j)^2.
    \end{split}
\end{equation}
Hence the chiral self-energy contributes
\begin{equation}
    \begin{split}
        \Gamma^{(2)}_{\text{chiral}}&[(\Qb_A)_{ij},(Q_B)_{ji}]=n_in_j(\kappa_i^2+\kappa_j^2)\int\frac{d^4p}{(2\pi)^4}\int d^4\theta(\Qb_A)_{ij}(Q_A)_{ji}I(\lambda,\mu,\epsilon)\\
        &\quad=\lambda n_in_j(\kappa_i^2+\kappa_j^2)\frac{\Gamma(\epsilon)\Gamma(1-\epsilon)^2}{\Gamma(2-2\epsilon)}\int\frac{d^4p}{(2\pi)^4}\int d^4\theta(\Qb_A)_{ij}(Q_A)_{ji}\left[\frac{4\pi\mu^2}{p^2}\right]^{\epsilon}.
    \end{split}
\end{equation}
Thus, the one loop contribution to the bifundamental effective potential is 
\begin{equation}
    \Gamma^{(2)}_{\text{one-loop}}[(\Qb_A)_{ij},(Q_B)_{ji}]=\Gamma^{(2)}_{\text{vec}}[(\Qb_A)_{ij},(Q_B)_{ji}]+\Gamma^{(2)}_{\text{chiral}}[(\Qb_A)_{ij},(Q_B)_{ji}]=0.
\end{equation}
Hence, since this holds for all bifundamental fields, we have
\be\label{eq:anomalousdimbifun}
\gamma_{Q_{ij}}=\gamma_{Q_{ji}}=0.
\ee
Then from \eqref{eq:anomalousdimadj} and \eqref{eq:anomalousdimbifun}, we find \eqref{eq:betafunctionN=2} to be:
\be\label{eq:betaivanishing}
    \beta_i=-\frac{\gym^3\kappa_i^3N}{16\pi^2}\left[2n_i-\sum_{j=1}^Ma^{\mathbf{2}}_{ij}n_j\right]
    =-\frac{\gym^3\kappa_i^3N}{16\pi^2}\left[2n_i-2n_i\right]
    =0\;.
\ee
Since the gauge group, labelled by $i$, was arbitrary, \eqref{eq:betaivanishing} states that $\beta_i=0$ for $i=1,\ldots,M$ and hence the ADE theories are all finite.

\section{Index of Some Short Multiplets} \label{sec:Indexshort}

In this Appendix we provide some details on how the various terms in the superconformal index map to specific BPS multiplets and operators. In the diagrams below, each entry is of the form $R_{(j_1,j_2)}$ and operators with a negative sign come from equations of motion. We represent the action of $\Qcal$ as moving to the left and the action of $\bar{\Qcal}$ as moving to the right. The underlined entries satisfy $\bar{\delta}_{1\dot-}=0$, and contribute to the index. This material is standard, and we refer to \cite{Dolan:2002zh,Gadde:2009dj} and the reviews \cite{Pomoni:2019oib,Eberhardt:2020cxo} for more details.

\subsection{The $\bar{\mathcal{E}}_{r(0,0)}$ multiplet}\label{sec:Er(0,0)}

The highest weight state of the $\bar{\mathcal{E}}_{r(0,0)}$ multiplet obeys the shortening condition $\Delta=-r$, which follows from $\bar{\Qcal}_{i\dot{\alpha}}\ket{R,r}^\text{h.w.}=0$ for all $i$ and $\dot{\alpha}$. The highest weight state is materialised by $\Tr Z^\ell$, where $\ell=-r$. The case $\bar{\mathcal{E}}_{1(0,0)}$ is special as it describes the $\NN=2$ vector multiplet (together with its equations of motion and and the auxiliary field). The field content of the $\NN=2$ vector multiplet without the equations of motion is captured by $\bar{\mathcal{D}}_{0(0,0)}$. $\bar{\mathcal{E}}_{2(0,0)}$ is important as it contains the Lagrangian of the $\NN=2$ theories as a descendant. Schematically, we can write the Lagrangian as $\mathcal{Q}^4\Tr Z^2$. The highest weight operators of $\bar{\mathcal{E}}_{r(0,0)}$ parametrise the \emph{Coulomb branch} (the vacua described by $\expval{Z}=a$ and $\expval{Q}=0$). 
\be
\begin{array}{c|ccccc}
        \Delta & & & & & \\
        \ell & & & & &\underline{0_{(0,0)}}\\
        \ell+\frac{1}{2} & & & &\underline{\frac{1}{2}_{\left(\pm\frac{1}{2},0\right)}}& \\
        \ell+1 & & &0_{(\pm1,0)},\;\underline{1_{(0,0)}}& & \\
        \ell+\frac{3}{2} & &\frac{1}{2}_{\left(\pm\frac{1}{2},0\right)}& & & \\
        \ell+2 &0_{(0,0)}&\phantom{\frac{3}{2}} & & & \\\hline
        r &-\ell+2&-\ell+\frac{3}{2}&-\ell+1&-\ell+\frac{1}{2}&-\ell
    \end{array}
\ee
\begin{table}[h]
    \centering
        \renewcommand{\arraystretch}{1.3}
    \begin{tabular}{|c|c|c|}\hline
        $\Delta$ &  $R_{(j_1,j_2)}$&$\mathcal{I}(p,q,t)$\tabularnewline\hline
         $\ell$&$0_{\left(0,0\right)}$&$(pqt^{-1})^\ell$\tabularnewline\hline
         $\ell+\frac{1}{2}$&$\frac{1}{2}_{\left(\pm,0\right)}$&$-p^{\ell}q^{\ell}t^{1-\ell}(p^{-1}+q^{-1})$\tabularnewline\hline
         $\ell+1$&$1_{\left(0,0\right)}$&$p^{\ell-1}q^{\ell-1}t^{2-\ell}$\tabularnewline\hline
    \end{tabular}
    \caption{Operators with $\bar{\delta}_{1\dot{-}}=0$ in $\mathcal{E}_{\ell(0,0)}$.}
    \label{tab:IndexcontriutionEl(0,0)}
\end{table}
Hence, we have (dividing by $(1-p)(1-q)$ to take the BPS derivatives into account):
\be\label{eq:Erindex}
\mathcal{I}[\bar{\mathcal{E}}_{-\ell(0,0)}]=\,\frac{(pqt^{-1})^\ell\left(1-t(p^{-1}+q^{-1})+p^{-1}q^{-1}t^2\right)}{(1-p)(1-q)}\,.
\ee
Taking the various limits we find
\be
   \mathcal{I}_M[\bar{\mathcal{E}}_{-\ell(0,0)}]=
   \mathcal{I}_S[\bar{\mathcal{E}}_{-\ell(0,0)}]=
   \mathcal{I}_{HL}[\bar{\mathcal{E}}_{-\ell(0,0)}]=0\;\;,\;\;
   \mathcal{I}_C[\bar{\mathcal{E}}_{-\ell(0,0)}]=\,T^\ell\;.
\ee
For $\ell\ge2$, we sum the contribution of the operators from the above Table \ref{tab:IndexcontriutionEl(0,0)} and, dividing by $\left(1-p\right)\left(1-q\right)$ from the BPS derivatives, we obtain
\be
\begin{split}\label{eq:ErindexSum}
   \sum_{\ell=2}^\infty\mathcal{I}[\bar{\mathcal{E}}_{-\ell(0,0)}]=&\,\frac{1}{\left(1-p\right)\left(1-q\right)}\sum_{\ell=2}^\infty (pqt^{-1})^\ell\left(1-t(p^{-1}+q^{-1})+p^{-1}q^{-1}t^2\right)\\
   =&\,\frac{p^2q^2t^{-2}\left(1-t(p^{-1}+q^{-1})+p^{-1}q^{-1}t^2\right)}{\left(1-pqt^{-1}\right)\left(1-p\right)\left(1-q\right)}\,.
\end{split}
\ee
The various limits are 
\be\label{eq:ErlimitsSum}
    \sum_{\ell=2}^\infty\mathcal{I}_{M}[\bar{\mathcal{E}}_{-\ell(0,0)}]=
    \sum_{\ell=2}^\infty\mathcal{I}_{S}[\bar{\mathcal{E}}_{-\ell(0,0)}]=
    \sum_{\ell=2}^\infty\mathcal{I}_{HL}[\bar{\mathcal{E}}_{-\ell(0,0)}]=0\;\;,\;\;
    \sum_{\ell=2}^\infty\mathcal{I}_{C}[\bar{\mathcal{E}}_{-\ell(0,0)}]=\,\frac{T^2}{1-T}\,.
\ee

\subsection{The $\hat{\mathcal{B}}_R$ multiplet}\label{sec:BR}

The highest weight states of the $\hat{\mathcal{B}}_R$ multiplets satisfy the shortening condition $\Delta=2R$. The shortening condition of $\hat{\mathcal{B}}_R$ requires $j_1=j_2=r=0$.  The highest weight states of $\hat{\mathcal{B}}_R$ parametrise the \emph{Higgs branch} (the vacua described by $\expval{Z}=0$ and $\expval{Q}\ne0$).  
\be
\begin{array}{c|ccccc}
         \Delta& & & & & \\
         2R& & &\underline{R_{(0,0)}}& &\\
         2R+\frac{1}{2}& &(R-\frac{1}{2})_{\left(\frac{1}{2},0\right)}& &\underline{(R-\frac{1}{2})_{\left(0,\frac{1}{2}\right)}}& \\
         2R+1&(R-1)_{(0,0)}& &(R-1)_{\left(\frac{1}{2},\frac{1}{2}\right)}& &(R-1)_{(0,0)}\\
         2R+\frac{3}{2}& &(R-\frac{3}2)_{(0,\half)} & & (R-\frac{3}{2})_{(\half,0)}& \\
         2R+2& & &(R-2)_{(0,0)}&\phantom{\frac{3}{2}} & \\\hline
         r&1&\frac{1}{2}&0&-\frac{1}{2}&-1
\end{array}
\ee
\begin{table}[ht]
    \centering
        \renewcommand{\arraystretch}{1.3}
    \begin{tabular}{|c|c|c|}\hline
        $\Delta$ &  $R_{(j_1,j_2)}$&$\mathcal{I}(p,q,t)$\tabularnewline\hline
         $2R$&$R_{\left(0,0\right)}$&$t^{R}$\tabularnewline\hline
         $2R+\frac{1}{2}$&$(R+\frac{1}{2})_{\left(0,\half\right)}$&$-pqt^{R-1}$\tabularnewline\hline
         
    \end{tabular}
    \caption{Operators with $\bar\delta_{1\dot-}=0$ in $\hat{\mathcal{B}}_R$.}
    \label{tab:IndexcontriutionBR}
\end{table}
Summing the contributions from Table \ref{tab:IndexcontriutionBR} and dividing by by $\left(1-p\right)\left(1-q\right)$ to account for the BPS derivatives gives
\be\label{eq:SingleLetterBR}
\mathcal{I}[\hat{B}_R]=\frac{t^R(1-pqt^{-1})}{(1-p)(1-q)}\,.
\ee
Taking the limits we find
\be\label{eq:singleletterBRlimits}
   \mathcal{I}_{M}[\hat{B}_R]=\,\frac{t^R}{(1-q)}\;\;,\;\;
   \mathcal{I}_{S}[\hat{B}_R]=\,\frac{q^R}{(1-q)}\;\;,\;\;
   \mathcal{I}_{HL}[\hat{B}_R]=\,t^R\;\;,\;\;
   \mathcal{I}_{C}[\hat{B}_R]=\,0\,.\ee
   Note that for $R=\half$, we get the hypermultiplet. For $R=1$ we get $\Delta=2$ and the highest-weight state of $\hat{\mathcal{B}}_1$ is $\Mtrip$ of the $\NN=2$ chiral ring,  which is a triplet of $\SU(2)_R$. It also contains the flavour current as the vector field with $\Delta=3$, labelled by $0_{\left(\half,\half\right)}$. We note that although the $r=1$ and $r=-1$ $(0)_{(0,0)}$ states at $\Delta=3$ are scalar, corresponding to $\Qcal_1^2(XY)$ and $\bar{\Qcal}_2^2(XY)$ respectively, their lowest components are fermion bilinears with the pure scalar parts arising at the next order in the gauge coupling. Specifically, applying the $\Ncal=2$ transformations from Appendix \ref{sec:ExtendedSusyTransformations}, we find:
\be\begin{split}
\Qcal_1^2(XY)&=2\psi_X\psi_{Y}+\,i\gym([\Yb,\Zb]Y+X[\Xb,\Zb])\;,\\
\bar{\Qcal}_2^2(XY)&=2\bar{\psi}_{X}\bar{\psi}_{Y}+\,i\gym([\Yb,Z]Y+X[\Xb,Z])\;.
\end{split}\ee
As the pure scalar parts of these operators are at higher-loop order, we do not expect to see them as protected states in our one-loop spectrum (and indeed we do not). Finally, the $\Delta=4$ element of $\hat{\mathcal{B}}_1$, denoted $-0_{(0,0)}$, corresponds to the conservation of the flavour current.

\subsection{Index of Some Operators}\label{sec:IndexOperators}

Let us now focus on some specific gauge-invariant composite operators that contribute to the index. First we consider $\Tr_i Q_{ij}^k$ and $\Tr_j Q_{ji}^k$, with $k\in\Zset_{k\ge2}$ such that the strings $Q_{ij}^k$ and $Q_{ji}^k$ start and end on nodes $i$ and $j$ respectively. These operators are highest-weight states of the multiplet $\hat{\mathcal{B}}_{\frac{k}{2}}$. Their index is given by
\be\label{eq:Indexqk}
\mathcal{I}[Q_{ij}^k]=\mathcal{I}[Q_{ji}^k]=\,t^{\frac{k}{2}}\;,
\ee
and summing over all values of $\ell$ we find
\be\label{eq:Indexsumqk}
\sum_{\ell=1}^\infty\mathcal{I}[Q_{ij}^{\ell k}]=\sum_{\ell=1}^\infty\mathcal{I}[Q_{ji}^{\ell k}]=\,\frac{t^{\frac{k}{2}}}{1-t^{\frac{k}{2}}}\,.
\ee
Let us now consider the index of the highest weight state of the triplet $\Mtrip_\text{h.w.}$ (a.k.a the moment map), $\Tr Q_{ij}Q_{ji}$, which is the primary of the $\hat{\mathcal{B}}_1$ multiplet:
\be\label{eq:IndexMomentMap}
\mathcal{I}[\Mtrip]=\,t\;.
\ee
Comparing \eqref{eq:IndexMomentMap} to Table \ref{tab:n=2-multiplets} we can see that 
\be\label{eq:indexlambdamomentmamp}
\mathcal{I}[\bar{\lambda}_{Z\dot{+}}]=-\mathcal{I}[\Mtrip]\,.
\ee
Next, we consider the index of the states $\Tr(Q_{ij}Q_{jk})^R$, with $R\in\Zset_{\ge2}$, where $(Q_{ij}Q_{jk})^R$ is schematic notation for a string of $Q$'s starting and ending at node $i$. This operator is the primary of the multiplet $\hat{\mathcal{B}}_{R\ge2}$. Its index is given by
\be\label{eq:qtildeqR}
\mathcal{I}[(Q_{ij}Q_{jk})^R]=\,t^{R}\,.
\ee
Hence we have 
\be\label{eq:sumqtildeqR}
\sum_{\ell=1}^{\infty}\mathcal{I}[(Q_{ij}Q_{ji})^{R\ell}]=\,\frac{t^R}{1-t^R}\,.
\ee
Now we consider the index of the operator $\Tr\bar{\lambda}_{Z\dot{+}}(Q_{ij}Q_{jk})^R$, with $R\in\Zset_{\ge2}$ and the same schematic notation as above. This operator is the primary of the multiplet $\mathcal{D}_{R+\half(0,\half)}$ and its index is given by
\be\label{eq:indexlambdaqtildeqR}
\mathcal{I}[\bar{\lambda}_{Z\dot{+}}(Q_{ij}Q_{ji})^R]=-t^{R+1}\,,
\ee
and so
\be\label{eq:sumindexlambdaqtildeqR}
\sum_{\ell=1}^\infty\mathcal{I}[\bar{\lambda}_{Z\dot{+}}(Q_{ij}Q_{jk})^{R\ell}]=-\frac{t^{R+1}}{1-t^R}\,.
\ee
Finally, we consider the operator $\Tr Z^\ell$, with $\ell\in\Zset_{\ge2}$, which is the primary of $\bar{\mathcal{E}}_{-\ell(0,00)}$. Its index is 
\be\label{eq:indexphiell}
\mathcal{I}[Z^\ell]=\,\left(pqt^{-1}\right)^\ell\,.
\ee
We then find
\be\label{eq:sumindexphiell}
\sum_{\ell=2}^\infty\mathcal{I}[Z^\ell]=\,\frac{p^2q^2t^{-2}}{(1-pqt^{-1})}\,.
\ee
We can now consider how specific limits of the index pick out single operators from a multiplet. From  Appendix \ref{sec:Er(0,0)} and \eqref{eq:sumindexphiell}, we can see that the Coulomb-branch index has
\be\label{eq:CBEr(0,0)}
\sum_{\ell=2}^\infty\mathcal{I}_{C}[\bar{\mathcal{E}}_{-\ell(0,0)}]=\,\sum_{\ell=2}^\infty\mathcal{I}[Z^\ell]\,.
\ee
Next, we have from Appendix \ref{sec:BR} and \eqref{eq:indexlambdamomentmamp} that
\be\label{eq:HLB1}
\mathcal{I}_{HL}[\hat{\mathcal{B}}_1]=\,\mathcal{I}[\Mtrip]=\,-\mathcal{I}[\bar{\lambda}_{Z\dot{+}}]\,.
\ee
Finally, we have the index of the following (schematic) operators
\be\begin{split}
\sum_{\ell=1}^\infty\mathcal{I}[Q_{ij}^{\ell k}]=\sum_{\ell=1}^\infty\mathcal{I}[Q_{ji}^{\ell k}]=&\,\frac{t^{\frac{k}{2}}}{1-t^{\frac{k}{2}}}\;,\\
    \sum_{\ell=1}^\infty\mathcal{I}[(Q_{ij}Q_{ji})^{R\ell}]=&\,\frac{t^R}{1-t^R}\;,\\
    \sum_{\ell=1}^\infty\mathcal{I}[\bar{\lambda}_{Z\dot{+}}(Q_{ij}Q_{ji})^{R\ell}]=&\,-\frac{t^{R+1}}{1-t^R}\,.
\end{split}\ee

\section{Extended SUSY Transformations}\label{sec:ExtendedSusyTransformations}

In this Appendix we consider the $\Ncal=2$ supersymmetry transformations on the component fields. This will allow us to determine which operators belong to the same multiplets, which among other things will provide additional checks on the spectrum of the $\Ncal=2$ Hamiltonians. We start with the $\NN=4$ on-shell transformations of the component fields, as given e.g. in \cite{Eden:2004ua,Grant:2008sk}\footnote{Compared to those works, we use $\varphi_{AB}\ra \varphi^{ab}$. Note that our conventions differ slightly compared to those works in other places such as the normalisation of the superpotential.}:
\be\begin{split}\label{eq:N=4Susytransformations}
    \delta\varphi^{ab}=&\,\eta^{[a}\lambda^{b]}-\frac{1}{2}\epsilon^{abcd}\bar{\eta}_{[c}\bar{\lambda}_{d]} \;,\\
        \delta\lambda^a_{\alpha}=&\,2\eta^{a\beta}F_{\alpha\beta}-\frac{i\gym}{2}\eta^b_{\alpha}[\varphi^{ac},\bar{\varphi}_{bc}]+i\bar{\eta}_b^{\dot{\beta}}D_{\alpha\dot{\beta}}\varphi^{ba}\;,\\
        \delta\bar{\lambda}_{a\dot{\alpha}}=&\,2\bar{F}_{\dot{\alpha} \dot{\beta}}\bar{\eta}^{a\dot{\beta}}-\frac{i\gym}{2}\bar{\eta}_{\dot{\alpha} b}[\varphi^{cb},\bar{\varphi}_{ca}]-i\eta_{b}^\beta D_{\beta\dot{\alpha}}\bar{\varphi}_{ab}\;,\\
        \delta A_{\alpha\dot{\alpha}}=&\,\eta^1_{\alpha}\bar{\lambda}_{1\dot{\alpha}}-\lambda^1_{\alpha}\bar{\eta}_{1\dot{\alpha}}+\eta^2_{\alpha}\bar{\lambda}_{2\dot{\alpha}}-\lambda^2_{\alpha}\bar{\eta}_{2\dot{\alpha}}
        +\eta^3_{\alpha}\bar{\lambda}_{3\dot{\alpha}}-\lambda^3_{\alpha}\bar{\eta}_{3\dot{\alpha}}+\eta^4_{\alpha}\bar{\lambda}_{4\dot{\alpha}}-\lambda^4_{\alpha}\bar{\eta}_{4\dot{\alpha}}\;,\\
        \delta F_{\alpha\beta}=&\,\frac{i\gym}{2}\bigl(\eta^{1}_{\beta}([\lambda^2_{\alpha},\Zb]+[\lambda^3_{\alpha},\Xb]+[\lambda^4_\alpha,\Yb])+\eta^1_{\alpha}([\lambda^2_{\beta},\Zb]+[\lambda^3_{\beta},\Xb]+[\lambda^4_\beta,\Yb])\bigr)\\
        &\,+\eta^2_{\lbrace\beta}D^{\dot{\alpha}}_{\alpha\rbrace}\bar{\lambda}_{2\dot{\alpha}}+\eta^3_{\lbrace\beta}D^{\dot{\alpha}}_{\alpha\rbrace}\bar{\lambda}_{3\dot{\alpha}}+\eta^4_{\lbrace\beta}D^{\dot{\alpha}}_{\alpha\rbrace}\bar{\lambda}_{4\dot{\alpha}}\\
        &\,-\bar{\eta}_{1\dot{\alpha}}D_{\lbrace\alpha}^{\dot{\alpha}}\lambda^1_{\beta\rbrace}-\bar{\eta}_{2\dot{\alpha}}D_{\lbrace\alpha}^{\dot{\alpha}}\lambda^2_{\beta\rbrace}-\bar{\eta}_{3\dot{\alpha}}D_{\lbrace\alpha}^{\dot{\alpha}}\lambda^3_{\beta\rbrace}-\bar{\eta}_{4\dot{\alpha}}D_{\lbrace\alpha}^{\dot{\alpha}}\lambda^4_{\beta\rbrace}\;,\\
        \delta\bar{F}_{\dot{\alpha}\dot{\beta}}=&\,\frac{i\gym}{2}\bigl(\bar{\eta}_{1\dot{\beta}}([\bar{\lambda}_{2\dot{\alpha}},Z]+[\bar{\lambda}_{3\dot{\alpha}},X]+[\bar{\lambda}_{4\dot{\alpha}},Y])+\bar{\eta}_{1\dot{\alpha}}([\bar{\lambda}_{2\dot{\beta}},Z]+[\bar{\lambda}_{3\dot{\beta}},X]+[\bar{\lambda}_{4\dot{\beta}},Y])\bigr)\\
        &\,+\bar{\eta}_{2\lbrace\dot{\beta}}D_{\dot{\alpha}\rbrace}^\alpha\lambda_{\alpha}^2+\bar{\eta}_{3\lbrace\dot{\beta}}D_{\dot{\alpha}\rbrace}^\alpha\lambda_{\alpha}^3+\bar{\eta}_{4\lbrace\dot{\beta}}D_{\dot{\alpha}\rbrace}^\alpha\lambda_{\alpha}^4\\
        &\,-\eta^{1\alpha} D_{\alpha\lbrace\dot{\alpha}}\lambda_{1\dot{\beta}\rbrace}-\eta^{2\alpha} D_{\alpha\lbrace\dot{\alpha}}\lambda_{2\dot{\beta}\rbrace}-\eta^{3\alpha} D_{\alpha\lbrace\dot{\alpha}}\lambda_{3\dot{\beta}\rbrace}-\eta^{4\alpha} D_{\alpha\lbrace\dot{\alpha}}\lambda_{4\dot{\beta}\rbrace}\,.
\end{split}\ee
In the first equation, $[a, b]$ denotes antisymmetrisation with weight one. $F_{\alpha\beta}\equiv\frac{1}{2}(\sigma^{\mu\nu})_{\alpha\beta}F_{\mu\nu}$ and $\bar{F}_{\dot{\alpha}\dot{\beta}}\equiv\frac{1}{2}(\bar{\sigma}^{\mu\nu})_{\dot{\alpha}\dot{\beta}}F_{\mu\nu}$. In \eqref{eq:N=4Susytransformations} we have used the fermionic equations of motion
\be\begin{split}
    D_{\alpha\dot{\alpha}}\bar{\lambda}^{\dot{\alpha}}_1=&\,i\gym\bigl([\Zb,\lambda^2_{\alpha}]+[\Xb,\lambda^3_{\alpha}]+[\Yb,\lambda^4_{\alpha}]\bigr)\;,\\
    D_{\alpha\dot{\alpha}}\lambda^{1\alpha}=&\,i\gym\bigl([Z,\bar{\lambda}_{2\dot{\alpha}}]+[X,\bar{\lambda}_{3\dot{\alpha}}]+[Y,\bar{\lambda}_{4\dot{\alpha}}]\bigr)\,.
\end{split}\ee
Note that we use the notation 
\be\varphi^{ab}=\begin{pmatrix}
    0 & Z & X & Y \\
    -Z  & 0 & \Yb & -\Xb \\
     -X & -\Yb & 0 & \Zb \\
     -Y & \Xb & -\Zb & 0 
     \end{pmatrix} \;\text{and}
\;\;\bar{\varphi}_{ab}=\begin{pmatrix}
    0&\Zb&\Xb&\Yb\\
    -\Zb&0&Y&-X\\
    -\Xb&-Y&0&Z\\
    -\Yb&X&-Z&0
\end{pmatrix}.\ee
Then from \eqref{eq:N=4Susytransformations}, the scalars transform as 
\be\begin{split}\label{eq:N=4susyscalars}
    \delta X=&\eta^1\lambda^3-\eta^3\lambda^1+\bar{\eta}_4\bar{\lambda}_2-\bar{\eta}_{2}\bar{\lambda}_4\;,\\
    \delta Y=&\eta^1\lambda^4-\eta^{4}\lambda^{1}+\bar{\eta}_{2}\bar{\lambda}_{3}-\bar{\eta}_{3}\bar{\lambda}_{2}\;,\\
    \delta Z=&\eta^1\lambda^2-\eta^{2}\lambda^{1}+\bar{\eta}_{3}\bar{\lambda}_{4}-\bar{\eta}_{4}\bar{\lambda}_{3}\;,\\
    \delta\Xb=&\bar{\eta}_1\bar{\lambda}_3-\bar{\eta}_3\bar{\lambda}_1+\eta^4\lambda^2-\eta^2\lambda^4\;,\\
    \delta\Yb=&\bar{\eta}_1\bar{\lambda}_4-\bar{\eta}_4\bar{\lambda}_1+\eta^2\lambda^3-\eta^3\lambda^2\;,\\
    \delta\Zb=&\bar{\eta}_1\bar{\lambda}_2-\bar{\eta}^2\bar{\lambda}^1+\eta^3\lambda^4-\eta^4\lambda^3\;,
\end{split}\ee
while for the fermions we have
\be\begin{split}\label{eq:N=4susyfermions}
    \delta\lambda^1_{\alpha}=&2\eta^{1\beta}F_{\alpha\beta}-i\gym\bigl(\eta^2_{\alpha}[X,Y]+\eta^3_{\alpha}[Y,Z]+\eta^4_{\alpha}[Z,X]
    +\half\eta^1_{\alpha}([X,\Xb]+[Y,\Yb]+[Z,\Zb])\bigr)\\
    &-i\bar{\eta}_2^{\dot{\alpha}}D_{\alpha\dot{\alpha}}Z-i\bar{\eta}_3^{\dot{\alpha}}D_{\alpha\dot{\alpha}}X-i\bar{\eta}_4^{\dot{\alpha}}D_{\alpha\dot{\alpha}}Y\;,\\
    \delta\lambda^2_{\alpha}=&2\eta^{2\beta}F_{\alpha\beta}-i\gym(\eta^1_{\alpha}[\Yb,\Xb]+\eta^3_\alpha[Z,\Xb]+\eta^4_\alpha[Z,\Yb]
    +\half\eta^2_{\alpha}([X,\Xb]+[Y,\Yb]-[Z,\Zb]))\\
    &+i\bar{\eta}_1^{\dot{\alpha}}D_{\alpha\dot{\alpha}}Z-i\bar{\eta}_3^{\dot{\alpha}}D_{\alpha\dot{\alpha}}\Yb+i\bar{\eta}_4^{\dot{\alpha}}D_{\alpha\dot{\alpha}}\Xb\;,\\
    \delta\lambda^3_{\alpha}=&2\eta^{3\beta}F_{\alpha\beta}-i\gym(\eta^1_{\alpha}[\Zb,\Yb]+\eta^2_\alpha[X,\Zb]+\eta^4_\alpha[X,\Yb]
    +\half\eta^3_{\alpha}([Y,\Yb]+[Z,\Zb]-[X,\Xb]))\\
    &+i\bar{\eta}^{\dot{\alpha}}_1D_{\alpha\dot{\alpha}}X+i\bar{\eta}^{\dot{\alpha}}_2D_{\alpha\dot{\alpha}}\Yb-i\bar{\eta}^{4\dot{\alpha}}D_{\alpha\dot{\alpha}}\Zb\;,\\
    \delta\lambda^4_{\alpha}=&2\eta^{4\beta}F_{\alpha\beta}-i\gym(\eta^1_{\alpha}[\Xb,\Zb]+\eta^2_\alpha[Y,\Zb]+\eta^3_\alpha[X,\Yb]
    +\half\eta^4_{\alpha}([X,\Xb]+[Z,\Zb]-[Y,\Yb]))\\
    &+i\bar{\eta}_1^{\dot{\alpha}}D_{\alpha\dot{\alpha}}Y-i\bar{\eta}_2^{\dot{\alpha}}D_{\alpha\dot{\alpha}}\Xb+i\bar{\eta}_3^{\dot{\alpha}}D_{\alpha\dot{\alpha}}\Zb\;,\\
    \delta\bar{\lambda}_{1\dot{\alpha}}=&2\bar{F}_{\dot{\alpha}\dot{\beta}}\bar{\eta}^{\dot{\beta}}_1-i\gym\bigl(\bar{\eta}_{2\alpha}[\Xb,\Yb]+\bar{\eta}_{3\dot{\alpha}}[\Yb,\Zb]+\bar{\eta}_{4\dot{\alpha}}[\Zb,\Xb]
    +\half\bar{\eta}_{1\dot{\alpha}}([X,\Xb]+[Y,\Yb]+[Z,\Zb])\bigr)\\
    &+i\eta^{\alpha}_2D_{\alpha\dot{\alpha}}\Zb+i\eta^{\alpha}_3D_{\alpha\dot{\alpha}}\Xb+i\eta^{\alpha}_4D_{\alpha\dot{\alpha}}\Yb\;,\\
    \delta\bar{\lambda}_{2\dot{\alpha}}=&2\bar{F}_{\dot{\alpha}\dot{\beta}}\bar{\eta}^{\dot{\beta}}_2-i\gym(\bar{\eta}_{1\dot{\alpha}}[Y,X]+\bar{\eta}_{3\dot{\alpha}}[\Zb,X]+\bar{\eta}_{4\dot{\alpha}}[\Zb,Y]
    +\half\bar{\eta}_{2\dot{\alpha}}([X,\Xb]+[Y,\Yb]-[Z,\Zb]))\\
    &-i\eta^{1\alpha}D_{\alpha\dot{\alpha}}\Zb-i\eta^{3\alpha}D_{\alpha\dot{\alpha}}Y+i\eta^{4\alpha}D_{\alpha\dot{\alpha}}X\;,\\
    \delta\bar{\lambda}_{3\dot{\alpha}}=&2\bar{F}_{\dot{\alpha}\dot{\beta}}\bar{\eta}^{\dot{\beta}}_3-i\gym(\bar{\eta}_{1\dot{\alpha}}[Z,Y]+\bar{\eta}_{2\dot{\alpha}}[Z,\Xb]+\bar{\eta}_{4\dot{\alpha}}[\Xb,Y]
    +\half\bar{\eta}_{3\dot{\alpha}}([Y,\Yb]+[Z,\Zb]-[X,\Xb]))\\
    &-i\eta^{1\alpha}D_{\alpha\dot{\alpha}}\Xb-i\eta^{2\alpha}D_{\alpha\dot{\alpha}}Y+i\eta^{4\alpha}D_{\alpha\dot{\alpha}}Z\;,\\
    \delta\bar{\lambda}_{4\dot{\alpha}}=&2\bar{F}_{\dot{\alpha}\dot{\beta}}\bar{\eta}_4^{\dot{\beta}}-i\gym(\bar{\eta}_{1\dot{\alpha}}[X,Z]+\bar{\eta}_{2\dot{\alpha}}[\Yb,Z]+\bar{\eta}_{3\dot{\alpha}}[\Xb,Y]
    +\half\bar{\eta}_{4\dot{\alpha}}([X,\Xb]+[Z,\Zb]-[Y,\Yb]))\\
    &-i\eta^{1\alpha}D_{\alpha\dot{\alpha}}\Yb+i\eta^{2\alpha}D_{\alpha\dot{\alpha}}X-i\eta^{3\alpha}D_{\alpha\dot{\alpha}}Z\;.
\end{split}\ee
In our $\Ncal=2$ context, we will relabel $\lambda_1\equiv\lambda_V$, $\lambda_2\equiv\lambda_Z$ to be the fermions in the $\Ncal=2$ vector multiplet, while $\lambda_3\equiv\psi_X$ and $\lambda_4\equiv\psi_Y$ will be those in the hypermultiplet. From the above transformations, one can see that $\Qcal^1$ and $\Qcal^2$ (and their conjugates) act within the $\Ncal=2$ vector and hypermultiplets, while $\Qcal^3$ and $\Qcal^4$ are broken. One also notices that in the above conventions $\Qcal^1$ is the $\Ncal=1$ SYM supercharge acting within the $\Ncal=1$ multiplets.  

We then find that the $\NN=2$ on-shell transformations of the scalar fields are
\be\begin{split}\label{eq:N=2Scalar}
    \delta X=&\eta^1\psi_X-\bar{\eta}_{2}\bar{\psi}_Y\;,\\
    \delta Y=&\eta^1\psi_Y+\bar{\eta}_{2}\bar{\psi}_X\;,\\
    \delta Z=&\eta^1\lambda_Z-\eta^{2}\lambda_{V}\;,\\
    \delta\Xb=&\bar{\eta}_1\bar{\psi}_X-\eta_2\psi_Y\;,\\
    \delta\Yb=&\bar{\eta}_1\bar{\psi}_Y+\eta^2\psi_X\;,\\
    \delta\Zb=&\bar{\eta}_1\bar{\lambda}_Z-\bar{\eta}_2\bar{\lambda}_V\;.
\end{split}\ee
The $\Ncal=2$ transformations of the fermions are
\be\begin{split}\label{eq:N=2fermions}
    \delta\lambda_{V\alpha}=&2\eta^{1\beta}F_{\alpha\beta}-i\bar{\eta}_2^{\dot{\alpha}}D_{\alpha\dot{\alpha}}Z-i\gym\bigl(\eta^2_{\alpha}[X,Y]+\half\eta^1_{\alpha}([X,\Xb]+[Y,\Yb]+[Z,\Zb])\bigr)\;,\\
\delta\lambda_{Z\alpha}=&2\eta^{2\beta}F_{\alpha\beta}+i\bar{\eta}_1^{\dot{\alpha}}D_{\alpha\dot{\alpha}}Z-i\gym(\eta^1_{\alpha}[\Yb,\Xb]+\half\eta^2_{\alpha}([X,\Xb]+[Y,\Yb]-[Z,\Zb]))\;,\\
\delta\psi_{X\alpha}=&i\bar{\eta}^{\dot{\alpha}}_1D_{\alpha\dot{\alpha}}X+i\bar{\eta}^{\dot{\alpha}}_2D_{\alpha\dot{\alpha}}\Yb-i\gym(\eta^1_{\alpha}[\Zb,\Yb]+\eta^2_\alpha[X,\Zb])\;,\\
\delta\psi_{Y\alpha}=&i\bar{\eta}_1^{\dot{\alpha}}D_{\alpha\dot{\alpha}}Y-i\bar{\eta}_2^{\dot{\alpha}}D_{\alpha\dot{\alpha}}\Xb-i\gym(\eta^1_{\alpha}[\Xb,\Zb]+\eta^2_\alpha[Y,\Zb])\;,\\
\delta\bar{\lambda}_{V\dot{\alpha}}=&2\bar{F}_{\dot{\alpha}\dot{\beta}}\bar{\eta}^{\dot{\beta}}_1+i\eta^{\alpha}_2D_{\alpha\dot{\alpha}}\Zb-i\gym\bigl(\bar{\eta}_{2\alpha}[\Yb,\Xb]
+\half\bar{\eta}_{1\dot{\alpha}}([X,\Xb]+[Y,\Yb]+[Z,\Zb])\bigr)\;,\\
\delta\bar{\lambda}_{Z\dot{\alpha}}=&2\bar{F}_{\dot{\alpha}\dot{\beta}}\bar{\eta}^{\dot{\beta}}_2-i\eta^{1\alpha}D_{\alpha\dot{\alpha}}\Zb-i\gym(\bar{\eta}_{1\dot{\alpha}}[Y,X]
    +\half\bar{\eta}_{2\dot{\alpha}}([X,\Xb]+[Y,\Yb]-[Z,\Zb]))\;,\\
    \delta\bar{\psi}_{X\dot{\alpha}}=&-i\eta^{1\alpha}D_{\alpha\dot{\alpha}}\Xb-i\eta^{2\alpha}D_{\alpha\dot{\alpha}}Y-i\gym(\bar{\eta}_{1\dot{\alpha}}[Z,Y]+\bar{\eta}_{2\dot{\alpha}}[\Xb,Z])\;,\\
    \delta\bar{\psi}_{Y\dot{\alpha}}=&-i\eta^{1\alpha}D_{\alpha\dot{\alpha}}\Yb+i\eta^{2\alpha}D_{\alpha\dot{\alpha}}X-i\gym(\bar{\eta}_{1\dot{\alpha}}[X,Z]+\bar{\eta}_{2\dot{\alpha}}[\Yb,Z])\;.
\end{split}\ee
Finally, the $\NN=2$ transformations of the gauge field $A_{\alpha\dot{\alpha}}$ and the field strengths $F_{\alpha\beta}$ and $\bar{F}_{\dot{\alpha}\dot{\beta}}$ are 
\be\begin{split}\label{eq:N=2Gauge}
    A_{\alpha\dot{\alpha}}=&\,\eta^1_{\alpha}\bar{\lambda}_{V\dot{\alpha}}-\lambda_{V\alpha}\bar{\eta}_{1\dot{\alpha}}+\eta^2_{\alpha}\bar{\lambda}_{Z\dot{\alpha}}-\lambda_{Z\alpha}\bar{\eta}_{2\dot{\alpha}}\;,\\
        \delta F_{\alpha\beta}=&\,\frac{i\gym}{2}\bigl(\eta^{1}_{\beta}([\lambda_{Z\alpha},\Zb]+[\psi_{X\alpha},\Xb]+[\psi_{Y\alpha},\Yb])+\eta^1_{\alpha}([\lambda_{Z\beta},\Zb]+[\psi_{X\beta},\Xb]+[\psi_{Y\beta},\Yb])\bigr) \\
        &\,+\eta^2_{\lbrace\beta}D^{\dot{\alpha}}_{\alpha\rbrace}\bar{\lambda}_{Z\dot{\alpha}}-\bar{\eta}_{1\dot{\alpha}}D_{\lbrace\alpha}^{\dot{\alpha}}\lambda_{V\beta\rbrace}-\bar{\eta}_{2\dot{\alpha}}D_{\lbrace\alpha}^{\dot{\alpha}}\lambda_{Z\beta\rbrace} \;,\\
        \delta\bar{F}_{\dot{\alpha}\dot{\beta}}=&\,\frac{i\gym}{2}\bigl(\bar{\eta}_{1\dot{\beta}}([\bar{\lambda}_{Z\dot{\alpha}},Z]+[\bar{\psi}_{X\dot{\alpha}},X]+[\bar{\psi}_{Y\dot{\alpha}},Y])+\bar{\eta}_{1\dot{\alpha}}([\bar{\lambda}_{Z\dot{\beta}},Z]+[\bar{\psi}_{X\dot{\beta}},X]+[\bar{\psi}_{Y\dot{\beta}},Y])\bigr)\\
        &\,+\bar{\eta}_{2\lbrace\dot{\beta}}D_{\dot{\alpha}\rbrace}^\alpha\lambda_{Z\alpha}-\eta^{1\alpha} D_{\alpha\lbrace\dot{\alpha}}\lambda_{V\dot{\beta}\rbrace}-\eta^{2\alpha} D_{\alpha\lbrace\dot{\alpha}}\lambda_{Z\dot{\beta}\rbrace}\,.
\end{split}\ee
Note that we are only considering the transformations of the fields under the Poincar\'e supercharges $\Qcal$ and $\bar{\Qcal}$, not the special conformal supercharges $\Scal$ and $\bar{\Scal}$, as we would like to find the descendants of primary operators.

\section{Konishi Descendants} \label{sec:Konishi}

In studying the scalar spin-chain spectrum, one encounters states at different lengths which are superconformal descendants of each other. These states have classical dimensions differing by integers, but have the same anomalous dimensions. Confirming the presence of such states in the explicit spectrum provides a consistency check of the Hamiltonian, as well as of the supercharges used to derive the descendants.  In this appendix we consider the superconformal descendants of the Konishi operator, which classically takes the form
\be
\mathcal{K}_1=\Tr(\Xb X+\Yb Y+\Zb Z)\;,
\ee
where the subscript identifies it as an $\SU(4)_R$ singlet state. In the $\NN=4$ SYM literature, most discussions of the anomalous dimensions of the Konishi operator don’t work directly with $\mathcal{K}_1$, but instead study one of its descendants, in particular, its $L=4$ descendant in a holomorphic $\SU(2)$ sector, 
\be \label{K84}
\mathcal{K}_{84}=\Tr([X,Y][X,Y])\sim\Tr(XYXY-XXYY)\;,
\ee
(see e.g. \cite{Fiamberti:2007rj,Fiamberti:2008sh,Bajnok:2008bm}), or its descendant in the $\mathrm{SL}(2)$ sector (e.g. \cite{Gromov:2009zb}). The latter is not accessible to us as we work in the scalar sector, so we will focus on $\mathcal{K}_{84}$. Note that in $\Ncal=4$ SYM there are three holomorphic $\SU(2)$ sectors: The $XY$, $XZ$, and $YZ$ sectors. However, in the $\Ncal=2$ theories, these are not equivalent. As indicated in (\ref{K84}), we expect the unbroken supercharges to relate $\mathcal{K}_1$ to the corresponding $L=4$ operator in the $XY$ sector, and we will verify this below. 

Before we consider $\mathcal{K}_{84}$, let us consider the following $L=3$ descendant of Konishi in $\NN=4$ SYM:
\be
\bar{\Qcal}_1^2\mathcal{K}_1\equiv\,\gym\mathcal{K}_{10}\,.
\ee
As indicated,  $\mathcal{K}_{10}$ is the highest weight state of the $\mathbf{10}$ of $\SU(4)_R$. Applying \eqref{eq:N=4susyscalars} and \eqref{eq:N=4susyfermions}, we see that classically
\be\begin{split}
    \gym\mathcal{K}_{10}=&\,2\gym\Tr([Y,Z]X+[Z,X]Y+[X,Y]Z)=-3B_{\NN=4}\,,
\end{split}\ee
where $B_{\NN=4}$ \emph{classically} coincides with the superpotential. However, it is a composite operator and hence is not protected from renormalisation. We also recall that quantum effects (the Konishi anomaly) lead to a fermionic operator appearing on the right-hand side 
\be
\gym\mathcal{K}_{10}=-3(B_{\NN=4}+c\gym F^\text{anomaly})\,,
\ee
Referring to \cite{Bianchi:2001cm,Eden:2004ua} for more details on the quantum aspects, we will focus on classical descendants.

As indicated, $\mathcal{K}_{84}$ is the highest weight state of the $\mathbf{84}$ of $\SU(4)_R$. We can obtain the $XY$-sector $\mathcal{K}_{84}$ (\ref{K84}) from $\mathcal{K}_1$ by acting with the following combination of supercharges
\be\label{eq:KonishiL=4fromL=2}
\gym^2\mathcal{K}_{84}\sim\Qcal_2^2\bar{\Qcal}_1^2\mathcal{K}_1.
\ee
Since in our $\Ncal=2$ context the $\Qcal_1$ and $\Qcal_2$ supercharges are still present, we expect the same $\SU(4)\ra \SU(3)\ra \SU(2)$ sequence of descendants, where however the $\SU(2)$ is only the $XY$ sector. Note that we will continue to label these states by their corresponding $\SU(4)$ representations.  

In the following we will consider the action of the $\Ncal=2$ supercharges on the Konishi operator of the $\Zset_3$, $\Dfour$ and $\hat{E}_6$ quiver theories. We can confirm the presence of the descendants listed below, with the given coefficients (and of course the same anomalous dimensions), in the spectrum of the corresponding quiver theories, which provides an additional check on our Hamiltonians.

\subsection{Konishi descendants for $\Zset_3$}\label{Z3Konishi}

The orbifold-point $\Zset_3$ Konishi operator is 
\be\begin{split}\label{eq:Z3Konishi}
    \mathcal{K}_{\Zset_3}=&\Tr_1\Zb_1Z_1+\Tr_2\Zb_2Z_2+\Tr_3\Zb_3Z_3\\
    &+\Tr_1\Xb_{13}X_{31}+\Tr_2\Xb_{21}X_{12}+\Tr_3\Xb_{32}X_{23}\\
    &+\Tr_1\Yb_{12}Y_{21}+\Tr_2\Yb_{23}Y_{32}+\Tr_3\Yb_{31}Y_{13}.
\end{split}\ee
From \eqref{eq:N=2Scalar} we have 
\be\begin{split}\label{eq:Z3KonishiScalarfirst}
    \bar{\Qcal}_{1\dot{\alpha}}\Zb_i=&(\bar{\lambda}_i)_{Z\dot{\alpha}} \;,\\
    \bar{\Qcal}_{1\dot{\alpha}}\Xb_{ij}=&(\bar{\psi}_{ij})_{X\dot{\alpha}} \;,\\
    \bar{\Qcal}_{1\dot{\alpha}}\Yb_{ji}=&(\bar{\psi}_{ji})_{Y\dot{\alpha}}\;,\\
    \Qcal_{2\alpha} Z_i=&-(\lambda_i)_{V\alpha}\,.
\end{split}\ee
Then from \eqref{eq:N=2fermions} and \eqref{eq:Z3KonishiScalarfirst}, we can write 
\be\begin{split}\label{eq:Z3Q^2scalars}
    \bar{\Qcal}_1^2\Zb_i=&i\gym\kappa_i(Y_{ii-1}X_{i-1i}-X_{ii+1}Y_{i+1i})\;,\\
    \bar{\Qcal}_1^2\Xb_{ii-1}=&i\gym(\kappa_iZ_iY_{ii-1}-\kappa_{i-1}Y_{ii-1}Z_{i-1})\;,\\
    \bar{\Qcal}_1^2\Yb_{ii+1}=&i\gym(\kappa_{i+1}X_{ii+1}Z_{i+1}-\kappa_iZ_iX_{ii+1})\;,\\
    \Qcal_2^2Z_i=&-i\gym\kappa_i(Y_{ii-1}X_{i-1i}-X_{ii+1}Y_{i+1i})\;,
\end{split}\ee
where $i=1,2,3$, with $4\equiv1$ and $0\equiv3$.

As we move away from the orbifold point, the Konishi operator \eqref{eq:Z3Konishi} will be renormalised and in order to be an eigenstate of the deformed dilatation operator, $\Hcal\mathcal{K}_{\Zset_3}^\text{one-loop}$ $=\Delta(\kappa)\mathcal{K}_{\Zset_3}^\text{one-loop}$, it will acquire $(\kappa_1,\kappa_2,\kappa_3)$-dependent coefficients $a_i,b_i$ ($i=1,2,3$):
\be\begin{split}\label{eq:Z3eigenvalueKonishi}
    \mathcal{K}_{\Zset_3}^\text{one-loop}=&\,a_1\Tr_1Z_1\Zb_1+a_2\Tr_2Z_2\Zb_2+a_3\Tr_3Z_3\Zb_3\\
    &+b_1(\Tr_1X_{12}\Xb_{21}+\Tr_2Y_{21}\Yb_{12})+b_2(\Tr_2X_{23}\Xb_{32}+\Tr_3Y_{32}\Yb_{23})\\
    &+b_3(\Tr_3X_{31}\Xb_{13}+\Tr_1Y_{13}\Yb_{31})\,,
\end{split}\ee
such that  $a_i,b_i\to1$ and $\Delta(\kappa)\to12$ as $(\kappa_1,\kappa_2,\kappa_3)\to(1,1,1)$. The fact that the coefficients of $X\Xb$ and $Y\Yb$ are equal follows from the $\SU(2)_R$ symmetry. From \eqref{eq:Z3Q^2scalars} we can find that the $\SU(3)$-sector descendant of \eqref{eq:Z3eigenvalueKonishi},  $\gym\mathcal{K}^\text{one-loop}_{\Zset_3\;10}$ is given as follows:
\be\begin{split}
    \bar{\Qcal}_1^2\mathcal{K}^\text{one-loop}_{\Zset_3}=&-\gym\bigl[\kappa_1c_{1,1}\Tr_2Y_{21}Z_1X_{12}-\kappa_2c_{2,2}\Tr_1X_{12}Z_2Y_{21}\\
&+\kappa_2c_{2,2}\Tr_3Y_{32}Z_2X_{23}-\kappa_3c_{3,3}\Tr_2X_{23}Z_3Y_{32}\\
&+\kappa_3c_{3,3}\Tr_1Y_{13}Z_3X_{31}-\kappa_1c_{1,1}\Tr_3X_{31}Z_1Y_{13}\bigr]\\
=&\gym\mathcal{K}^\text{one-loop}_{\Zset_3\;10}\,,
\end{split}\ee
where $c_{i,j}\equiv\,a_i+2b_j\,$. The $XY$-sector descendant of \eqref{eq:Z3eigenvalueKonishi} is
\be\begin{split} \label{Z3L4Konishi}
    \Qcal_2^2\bar{\Qcal}_1^2\mathcal{K}_{\Zset_3}^\text{one-loop}
=&\gym^2\bigl[(\kappa_1^2c_{1,1}+\kappa_2^2c_{2,2})\Tr_1X_{12}Y_{21}X_{12}Y_{21}-2\kappa_2^2c_{2,2}\Tr_1X_{12}X_{23}Y_{32}Y_{21}\\
&+(\kappa_2^2c_{2,2}+\kappa_3^2c_{3,3})\Tr_2X_{23}Y_{32}X_{23}Y_{32}-2\kappa_3^2c_{3,3}\Tr_2X_{23}X_{31}Y_{13}Y_{32}\\
&+(\kappa_3^2c_{3,3}+\kappa_1^2c_{1,1})\Tr_3X_{31}Y_{13}X_{31}Y_{13}-2\kappa_1^2c_{1,1}\Tr_3X_{31}X_{12}Y_{21}Y_{13}\bigr]\\
=&\gym^2\mathcal{K}^\text{one-loop}_{\Zset_3\;84}\,.
\end{split}\ee

\subsection{Konishi descendants for $\Dfour$} \label{sec:D4Konishi}

In the $\Dfour$ case, the orbifold-point  Konishi operator is 
\be\begin{split}\label{eq:D4Konishi}
    \mathcal{K}_{\hat{\text{D}}_4}=&\Tr_1\Zb_1Z_1+\Tr_2\Zb_2Z_2+\Tr_3\Zb_3Z_3+\Tr_4\Zb_4Z_4+2\Tr_5\Zb_5Z_5\\
    &+2\Tr_5\Xb_{51}X_{15}+2\Tr_1\Yb_{15}Y_{51}+2\Tr_2\Xb_{25}X_{52}+2\Tr_5\Yb_{52}Y_{25}\\
    &+2\Tr_5\Xb_{53}X_{35}+2\Tr_3\Yb_{35}Y_{53}+2\Tr_4\Xb_{45}X_{54}+2\Tr_5\Yb_{54}Y_{45}\,.
\end{split}\ee
From \eqref{eq:N=2Scalar} we find
\be\begin{split}\label{eq:D4KonishiScalarfirst}
        \bar{\Qcal}_{1\dot{\alpha}}\Zb_i=&(\bar{\lambda}_i)_{Z\dot{\alpha}}\;,\\
    \bar{\Qcal}_{1\dot{\alpha}}\Yb_{i5}=&(\bar{\psi}_{i5})_{Y\dot{\alpha}}\;,\\
    \bar{\Qcal}_{1\dot{\alpha}}\Xb_{5i}=&(\bar\psi_{5i})_{X\dot{\alpha}}\;,\\
    \bar{\Qcal}_{1\dot{\alpha}}\Xb_{j5}=&(\bar{\psi}_{j5})_{X\dot{\alpha}}\;,\\
    \bar{\Qcal}_{1\dot{\alpha}}\Yb_{5j}=&(\bar{\psi}_{5j})_{Y\dot{\alpha}}\;,\\
    \Qcal_{2\alpha} Z_i=&-(\lambda_i)_{V\alpha}\,,
\end{split}\ee
while from \eqref{eq:N=2fermions} and \eqref{eq:D4KonishiScalarfirst}, we can write 
\be\begin{split}\label{eq:D4Q^2scalars}
    \Qcal_2^2Z_{i}=&2i\gym\kappa_iX_{i5}Y_{5i}\;,\quad i\text{ odd, }\;i\ne5\;,\\
    \Qcal_2^2Z_i=&-2i\gym\kappa_iY_{i5}X_{5i}\;,\quad i\text{ even}\;,\\
    \Qcal_2^2Z_5=&i\gym\kappa_5(X_{52}Y_{25}-Y_{51}X_{15}+X_{54}Y_{45}-Y_{53}X_{35})\;,\\
    \bar\Qcal_1^2\Zb_i=&-2i\gym\kappa_iX_{i5}Y_{5i}\;,\quad i\text{ odd, }\;i\ne5\;,\\
    \bar\Qcal_1^2\Zb_i=&2i\gym\kappa_iY_{i5}X_{5i}\;,\quad i\text{ even}\;,\\
    \bar\Qcal_1^2\Zb_5=&2i\gym\kappa_5(Y_{51}X_{15}-X_{52}Y_{25}+Y_{53}X_{35}-X_{54}Y_{45})\;,\\
    \bar\Qcal_1^2\Yb_{i5}=&i\gym(\kappa_5X_{i5}Z_5-\kappa_iZ_iX_{i5})\;,\\
    \bar\Qcal_1^2\Yb_{5i}=&i\gym(\kappa_iX_{5i}Z_i-\kappa Z_5X_{5i})\;,\\
    \bar\Qcal_1^2\Xb_{5i}=&i\gym(\kappa_5Z_{5}Y_{5i}-\kappa_iY_{5i}Z_i)\;,\\
    \bar\Qcal_1^2\Xb_{i5}=&i\gym(\kappa_iZ_iY_{i5}-\kappa_5Y_{i5}Z_5)\;.
\end{split}\ee
As we move away from the orbifold point, the renormalised Konishi operator \eqref{eq:D4Konishi}, satisfying $\Hcal\mathcal{K}_{\hat{\text{D}}_4}^\text{one-loop}=\Delta(\kappa)\mathcal{K}_{\hat{\text{D}}_4}^\text{one-loop}$,  will acquire coefficients $a_i,b_i$ that depend on $(\kappa_1,\kappa_2,\kappa_3,\kappa_4,\kappa_5)$:
\be\begin{split}\label{eq:D4eigenvalueKonishi}
    \mathcal{K}_{\hat{\text{D}}_4}^\text{one-loop}=&a_1\Tr_1Z_1\Zb_1+a_2\Tr_2Z_2\Zb_2+a_3\Tr_3Z_3\Zb_3+a_4\Tr_4Z_4\Zb_4+2a_5\Tr_5Z_5\Zb_5\\
    &+2b_1(\Tr_1X_{15}\Xb_{51}+\Tr_5Y_{51}\Yb_{15})+2b_2(\Tr_5X_{52}\Xb_{25}+\Tr_2Y_{25}\Yb_{52})\\
    &+2b_3(\Tr_3X_{35}\Xb_{53}+\Tr_5Y_{53}\Yb_{35})+2b_4(\Tr_5X_{54}\Xb_{45}+\Tr_4Y_{45}\Yb_{54})\,,
\end{split}\ee
with $a_i,b_i\to1$ and $\Delta(\kappa)\to12$ as $(\kappa_1,\kappa_2,\kappa_3,\kappa_4,\kappa_5)\to(1,1,1,1,1)$. From \eqref{eq:D4Q^2scalars}, we can find the descendant of \eqref{eq:D4eigenvalueKonishi} in the $\SU(3)$ sector as
\be\begin{split}\label{eq:K10one-loopD4}
    \bar{\Qcal}_1^2\mathcal{K}_{\hat{\text{D}}_4}^\text{one-loop}=&-2\gym\bigl[\kappa_1c_{1,1}\Tr_5Y_{51}Z_1X_{15}-\kappa_5c_{5,1}\Tr_1X_{15}Z_5Y_{51}\\
    &\qquad\quad+\kappa_5c_{5,2}\Tr_2Y_{25}Z_5X_{52}-\kappa_2c_{2,2}\Tr_5X_{52}Z_2Y_{25}\\
    &\qquad\quad+\kappa_3c_{3,3}\Tr_5Y_{53}Z_3X_{35}-\kappa_5c_{5,3}\Tr_3X_{35}Z_5Y_{53}\\
    &\qquad\quad+\kappa_5c_{5,4}\Tr_4Y_{45}Z_5X_{54}-\kappa_4c_{4,4}\Tr_5X_{54}Z_4Y_{45}\bigr]\\
    =&\gym\mathcal{K}^\text{one-loop}_{\hat{\text{D}}_4\;10}\;,
\end{split}\ee
where $c_{i,j}\equiv\,a_i+2b_j\,$. The next scalar descendant of \eqref{eq:D4eigenvalueKonishi} brings us to the $XY$ sector:
\be\begin{split}
    \Qcal_2^2\bar{\Qcal}_1^2\mathcal{K}_{\hat{\text{D}}_4}^\text{one-loop}=&2\gym^2\bigl[(2\kappa_1^2c_{1,1}+\kappa_5^2c_{5,1})\Tr_1X_{15}Y_{51}X_{15}Y_{51}\\
    &+(2\kappa_2^2c_{2,2}+\kappa_5^2c_{5,2})\Tr_5X_{52}Y_{25}X_{52}Y_{25}\\
    &+(2\kappa_3^2c_{3,3}+\kappa_5^2c_{5,3})\Tr_3X_{35}Y_{35}X_{35}Y_{53}\\
    &+(2\kappa_4^2c_{4,4}+\kappa_5^2c_{5,4})\Tr_5X_{54}Y_{45}X_{54}Y_{45}\\
    &+\kappa_5^2(c_{5,1}+c_{5,3})\Tr_1X_{15}Y_{53}X_{35}Y_{51}+\kappa_5^2(c_{5,2}+c_{5,4})\Tr_{5}X_{54}Y_{45}X_{52}Y_{25}\\
    &-\kappa_5^2(c_{5,1}+c_{5,2})\Tr_1X_{15}X_{52}Y_{25}Y_{51}-\kappa_5^2(c_{5,1}+c_{5,4})\Tr_1X_{15}X_{54}Y_{45}Y_{51}\\
    &-\kappa_5^2(c_{5,2}+c_{5,3})\Tr_3X_{35}X_{52}Y_{25}Y_{53}-\kappa_5^2(c_{5,3}+c_{5,4})\Tr_3X_{35}X_{54}Y_{45}Y_{53}\bigr]\\
    =&\gym^2\mathcal{K}^\text{one-loop}_{\hat{\text{D}}_4\;84}\,.
\end{split}\ee

\subsection{Konishi descendants for $\hat{E}_6$}  \label{sec:E6Konishi}

The orbifold-point $\hat{E}_6$ Konishi operator is given by
\be\begin{split}\label{2TKonishi}
    \mathcal{K}_{\hat{E}_6}=&\Tr_1\Zb_1Z_1+\Tr_3\Zb_3Z_3+\Tr_5\Zb_5Z_5+2[\Tr_2\Zb_2Z_2+\Tr_4\Zb_4Z_4+\Tr_6\Zb_6Z_6]+3\Tr_7\Zb_7Z_7\\
    &+2[\Tr_2\Xb_{21}X_{12}+\Tr_1\Yb_{12}Y_{21}+\Tr_4\Xb_{43}X_{34}+\Tr_3\Yb_{34}Y_{43}+\Tr_6\Xb_{65}X_{56}+\Tr_5\Yb_{56}Y_{65}]\\
    &+6[\Tr_7\Xb_{72}X_{27}+\Tr_2\Yb_{27}Y_{72}+\Tr_7\Xb_{74}X_{47}+\Tr_4\Yb_{47}Y_{74}+\Tr_7\Xb_{76}X_{67}+\Tr_6\Yb_{67}Y_{76}].
\end{split}\ee
From \eqref{eq:N=2Scalar}
\be\begin{split}\label{eq:2TKonishiScalarfirst}
    \bar{\Qcal}_{1\dot{\alpha}}\Zb_i=&\,(\bar\lambda_i)_{Z\dot{\alpha}}\;,\\
    \bar{\Qcal}_{1\dot{\alpha}}\Yb_{ij}=&\,(\bar{\psi}_{ij})_{Y\dot{\alpha}}\;,\\
    \bar{\Qcal}_{1\dot{\alpha}}\Xb_{ji}=&\,(\bar\psi_{ji})_{X\dot{\alpha}}\;,\\
    \Qcal_{2\alpha} Z_i=&\,-(\lambda_i)_{V\alpha}\,.
\end{split}\ee
Then from \eqref{eq:N=2fermions} and \eqref{eq:2TKonishiScalarfirst},we can write 
\be\begin{split}\label{eq:2TKonishiSecond}
    \Qcal_2^2Z_i=&-2i\gym\kappa_iX_{ii+1}Y_{i+1i}\;,\\
    \Qcal_2^2Z_{i+1}=&-i\gym\kappa_{i+1}(3X_{i+17}Y_{7i+1}-Y_{i+1i}X_{ii+1})\;,\\
    \Qcal_2^2Z_7=&2i\gym\kappa_7(Y_{72}X_{27}+Y_{74}X_{47}+Y_{76}X_{67})\;,\\
    \bar\Qcal_1^2\Zb_i=&2i\gym\kappa_iX_{ii+1}Y_{i+1i}\;,\\
    \bar\Qcal_1^2\Zb_{i+1}=&i\gym\kappa_{i+1}(3X_{i+17}Y_{7i+1}-Y_{i+1i}X_{ii+1})\;,\\
    \bar\Qcal_1^2\Zb_7=&-2i\gym\kappa_7(Y_{72}X_{27}+Y_{74}X_{47}+Y_{76}X_{67})\;,\\
    \bar\Qcal_1^2\Yb_{ii+1}=&i\gym(\kappa_{i+1}X_{ii+1}Z_{i+1}-\kappa_iZ_iX_{ii+1})\;,\\
    \bar\Qcal_1^2\Xb_{i+1i}=&i\gym(\kappa_{i+1}Z_{i+1}Y_{i+1i}-\kappa_iY_{i+1i}Z_i)\;,\\
    \bar\Qcal_1^2\Yb_{i+17}=&i\gym(\kappa_7X_{i+17}Z_7-\kappa_{i+1}Z_{i+1}X_{i+17})\;,\\
    \bar\Qcal_1^2\Xb_{7i+1}=&i\gym(\kappa_7Z_7Y_{7i+1}-\kappa_{i+1}Y_{7i+1}Z_{i+1})\;,
\end{split}\ee
where $i=1,3,5$.
As we move away from the orbifold point, the Konishi operator \eqref{2TKonishi} will be renormalised and in order to be an eigenstate of the dilatation operator, coefficients $a_i,b_i$ that depend on $(\kappa_1,\kappa_2,\kappa_3,\kappa_4,\kappa_5,\kappa_6,\kappa_7)$ will be introduced as follows
\be\begin{split}\label{eq:one-loop2TKonishi}
    \mathcal{K}_{\hat{E}_6}^\text{one-loop}=&a_1\Tr_1Z_1\Zb_1+a_3\Tr_3Z_3\Zb_3+a_5\Tr_5Z_5\Zb_5\\&+2[a_2\Tr_2Z_2\Zb_2+a_4\Tr_4Z_4\Zb_4+a_6\Tr_6Z_6\Zb_6]+3a_7\Tr_7\Zb_7Z_7\\
    &+2[b_1(\Tr_1X_{12}\Xb_{21}+\Tr_2Y_{21}\Yb_{12})+b_3(\Tr_3X_{34}\Xb_{43}+\Tr_4Y_{43}\Yb_{34})\\
    &+b_5(\Tr_5X_{56}\Xb_{65}+\Tr_6Y_{65}\Yb_{56})]\\
    &+6[b_2(\Tr_2X_{27}\Xb_{72}+\Tr_7Y_{72}\Yb_{27})+b_4(\Tr_4X_{47}\Xb_{74}+\Tr_7Y_{74}\Yb_{47})\\&+b_6(\Tr_6X_{67}\Xb_{76}+\Tr_7Y_{76}\Yb_{67})]\\
    =&\mathcal{K}_{\hat{E}_6\;10}^\text{one-loop}\,.
\end{split}\ee
such that $\Hcal\mathcal{K}_{\hat{E}_6}^\text{one-loop}=\Delta(\kappa)\mathcal{K}_{\hat{E}_6}^\text{one-loop}$, with $a_i,b_i\to1$ and $\Delta(\kappa)\to12$ as $(\kappa_1,\kappa_2,\kappa_3,\kappa_4,\kappa_5,\kappa_6,\kappa_7)$ $\to(1,1,1,1,1,1,1)$.
Then acting on \eqref{eq:one-loop2TKonishi} with \eqref{eq:2TKonishiSecond} we find
\be\begin{split}\label{eq:2TKone-loop10}
    \gym\mathcal{K}^\text{one-loop}_{\hat{\text{E}}_6\;10}=-2\gym&\bigl[\kappa_1c_{1,1}\Tr_2Y_{21}Z_1X_{12}-\kappa_2c_{2,1}\Tr_1X_{12}Z_2Y_{21}\\
&+3\kappa_2c_{2,2}\Tr_7Y_{72}Z_2X_{27}-3\kappa_7c_{7,2}\Tr_2X_{27}Z_7Y_{72}\\
&+3\kappa_4c_{4,4}\Tr_7Y_{74}Z_4X_{47}-3\kappa_7c_{7,4}\Tr_4X_{47}Z_7Y_{74}\\
&+\kappa_3c_{3,3}\Tr_4Y_{43}Z_3X_{34}-\kappa_4c_{4,3}\Tr_3X_{34}Z_4Y_{43}\\
&+3\kappa_6c_{6,6}\Tr_7Y_{76}Z_6X_{67}-3\kappa_7c_{7,6}\Tr_6X_{67}Z_7Y_{76}\\
    &\,+\kappa_5c_{5,5}\Tr_6Y_{65}Z_{5}X_{56}-\kappa_6c_{6,5}\Tr_5X_{56}Z_6Y_{65}\bigr]\,,
\end{split}\ee
where $c_{i,j}\equiv\,a_i+2b_j\,$.
The scalar descendant of \eqref{eq:one-loop2TKonishi} in the $XY$ sector is
\be\begin{split}
    \Qcal_2^2\bar{\Qcal}_1^2\mathcal{K}^\text{one-loop}_{\hat{E}_6}=&2\gym^2\bigl[(2\kappa_1^2c_{1,1}+\kappa_2^2c_{2,1})\Tr_1X_{12}Y_{21}X_{12}Y_{21}\\
    &+(2\kappa_3^2c_{3,3}+\kappa_4^2c_{4,3})\Tr_3X_{34}Y_{43}X_{34}Y_{43}\\
    &+(2\kappa_5^2c_{5,5}+\kappa_6^2c_{6,5})\Tr_5X_{56}Y_{65}X_{56}Y_{65}\\
    &+3(2\kappa_5^2(c_{7,2}+c_{7,4}+c_{7,6})+3\kappa_2^2c_{2,2})\Tr_2X_{27}Y_{72}X_{27}Y_{72}\\
    &-3\kappa_2^2(c_{2,1}+c_{2,2})\Tr_1X_{12}X_{27}Y_{72}Y_{21}\\
    &+3(2\kappa_5^2(c_{7,2}+c_{7,4}+c_{7,6})+3\kappa_4^2c_{4,4})\Tr_4X_{47}Y_{74}X_{47}Y_{74}\\
    &-3\kappa_4^2(c_{4,3}+c_{4,4})\Tr_3X_{34}X_{47}Y_{74}Y_{43}\\
    &+3(2\kappa_5^2(c_{7,2}+c_{7,4}+c_{7,6})+3\kappa_6^2c_{6,6})\Tr_6X_{67}Y_{76}X_{67}Y_{76}\\
    &-3\kappa_6^2(c_{6,5}+c_{6,6})\Tr_5X_{56}X_{67}Y_{76}Y_{65}\bigr]\\
=&\gym^2\mathcal{K}^\text{one-loop}_{\hat{\text{E}}_6\;84}\,.
\end{split}\ee

\bibliography{Quiver}

\providecommand{\href}[2]{#2}\begingroup\raggedright\begin{thebibliography}{100}

\bibitem{Minahan:2002ve}
J.~A. Minahan and K.~Zarembo, ``{The Bethe ansatz for N=4 superYang-Mills},''
  \href{http://dx.doi.org/10.1088/1126-6708/2003/03/013}{{\em JHEP} {\bfseries
  03} (2003) 013},
\href{http://arxiv.org/abs/hep-th/0212208}{{\ttfamily arXiv:hep-th/0212208
  [hep-th]}}.

\bibitem{Beisert:2004ry}
N.~Beisert, ``{The Dilatation operator of N=4 super Yang-Mills theory and
  integrability},'' \href{http://dx.doi.org/10.1016/j.physrep.2004.09.007}{{\em
  Phys. Rept.} {\bfseries 405} (2004) 1--202},
  \href{http://arxiv.org/abs/hep-th/0407277}{{\ttfamily arXiv:hep-th/0407277}}.

\bibitem{Serban:2010sr}
D.~Serban, ``{Integrability and the AdS/CFT correspondence},''
  \href{http://dx.doi.org/10.1088/1751-8113/44/12/124001}{{\em J. Phys. A}
  {\bfseries 44} (2011) 124001},
  \href{http://arxiv.org/abs/1003.4214}{{\ttfamily arXiv:1003.4214 [hep-th]}}.

\bibitem{Beisert:2010jr}
N.~Beisert {\em et~al.}, ``{Review of AdS/CFT Integrability: An Overview},''
  \href{http://dx.doi.org/10.1007/s11005-011-0529-2}{{\em Lett. Math. Phys.}
  {\bfseries 99} (2012) 3--32},
\href{http://arxiv.org/abs/1012.3982}{{\ttfamily arXiv:1012.3982 [hep-th]}}.

\bibitem{Bombardelli:2016rwb}
D.~Bombardelli, A.~Cagnazzo, R.~Frassek, F.~Levkovich-Maslyuk, F.~Loebbert,
  S.~Negro, I.~M. Szécsényi, A.~Sfondrini, S.~J. van Tongeren, and
  A.~Torrielli, ``{An integrability primer for the gauge-gravity
  correspondence: An introduction},''
  \href{http://dx.doi.org/10.1088/1751-8113/49/32/320301}{{\em J. Phys.}
  {\bfseries A49} no.~32, (2016) 320301},
\href{http://arxiv.org/abs/1606.02945}{{\ttfamily arXiv:1606.02945 [hep-th]}}.

\bibitem{Zoubos:2010kh}
K.~Zoubos, ``{Review of AdS/CFT Integrability, Chapter IV.2: Deformations,
  Orbifolds and Open Boundaries},''
  \href{http://dx.doi.org/10.1007/s11005-011-0515-8}{{\em Lett. Math. Phys.}
  {\bfseries 99} (2012) 375},
\href{http://arxiv.org/abs/1012.3998}{{\ttfamily arXiv:1012.3998 [hep-th]}}.

\bibitem{Wang:2003cu}
X.-J. Wang and Y.-S. Wu, ``{Integrable spin chain and operator mixing in N=1,2
  supersymmetric theories},''
  \href{http://dx.doi.org/10.1016/j.nuclphysb.2003.12.040}{{\em Nucl. Phys.}
  {\bfseries B683} (2004) 363--386},
\href{http://arxiv.org/abs/hep-th/0311073}{{\ttfamily arXiv:hep-th/0311073
  [hep-th]}}.

\bibitem{Ideguchi:2004wm}
K.~Ideguchi, ``{Semiclassical strings on AdS(5) x S(5)/Z(M) and operators in
  orbifold field theories},''
  \href{http://dx.doi.org/10.1088/1126-6708/2004/09/008}{{\em JHEP} {\bfseries
  09} (2004) 008}, \href{http://arxiv.org/abs/hep-th/0408014}{{\ttfamily
  arXiv:hep-th/0408014}}.

\bibitem{Beisert:2005he}
N.~Beisert and R.~Roiban, ``{The Bethe ansatz for Z(S) orbifolds of N=4 super
  Yang-Mills theory},''
  \href{http://dx.doi.org/10.1088/1126-6708/2005/11/037}{{\em JHEP} {\bfseries
  11} (2005) 037},
\href{http://arxiv.org/abs/hep-th/0510209}{{\ttfamily arXiv:hep-th/0510209
  [hep-th]}}.

\bibitem{Solovyov:2007pw}
A.~Solovyov, ``Bethe ansatz equations for general orbifolds of {N=4 SYM},''
  \href{http://dx.doi.org/10.1088/1126-6708/2008/04/013}{{\em JHEP} {\bfseries
  04} (2008) 013},
\href{http://arxiv.org/abs/0711.1697}{{\ttfamily arXiv:0711.1697 [hep-th]}}.

\bibitem{Beccaria:2011qd}
M.~Beccaria and G.~Macorini, ``{Y-system for $Z_S$ Orbifolds of N=4 SYM},''
  \href{http://dx.doi.org/10.1007/JHEP01(2012)112}{{\em JHEP} {\bfseries 06}
  (2011) 004}, \href{http://arxiv.org/abs/1104.0883}{{\ttfamily arXiv:1104.0883
  [hep-th]}}. [Erratum: JHEP 01, 112 (2012)].

\bibitem{deLeeuw:2012hp}
M.~de~Leeuw and S.~J. van Tongeren, ``{The spectral problem for strings on
  twisted AdS$_5 \times$ S$^5$},''
  \href{http://dx.doi.org/10.1016/j.nuclphysb.2012.03.004}{{\em Nucl. Phys. B}
  {\bfseries 860} (2012) 339--376},
  \href{http://arxiv.org/abs/1201.1451}{{\ttfamily arXiv:1201.1451 [hep-th]}}.

\bibitem{Skrzypek:2022cgg}
T.~Skrzypek, ``{Integrability treatment of AdS/CFT orbifolds},''
  \href{http://dx.doi.org/10.1088/1751-8121/ace947}{{\em J. Phys. A} {\bfseries
  56} no.~34, (2023) 345401}, \href{http://arxiv.org/abs/2211.03806}{{\ttfamily
  arXiv:2211.03806 [hep-th]}}.

\bibitem{Gadde:2009dj}
A.~Gadde, E.~Pomoni, and L.~Rastelli, ``The {V}eneziano limit of {N = 2}
  superconformal {QCD}: Towards the string dual of {N = 2 SU(N(c)) SYM} with
  {N(f) = 2 N(c)},'' \href{http://arxiv.org/abs/0912.4918}{{\ttfamily
  arXiv:0912.4918 [hep-th]}}.

\bibitem{Gadde:2010zi}
A.~Gadde, E.~Pomoni, and L.~Rastelli, ``Spin chains in {N=2} superconformal
  theories: From the {$Z_2$} quiver to superconformal {QCD},''
  \href{http://dx.doi.org/10.1007/JHEP06(2012)107}{{\em JHEP} {\bfseries 1206}
  (2012) 107},
\href{http://arxiv.org/abs/1006.0015}{{\ttfamily arXiv:1006.0015 [hep-th]}}.

\bibitem{Gadde:2010ku}
A.~Gadde and L.~Rastelli, ``Twisted magnons,''
  \href{http://dx.doi.org/10.1007/JHEP04(2012)053}{{\em JHEP} {\bfseries 04}
  (2012) 053}, \href{http://arxiv.org/abs/1012.2097}{{\ttfamily arXiv:1012.2097
  [hep-th]}}.

\bibitem{Gadde:2012rv}
A.~Gadde, P.~Liendo, L.~Rastelli, and W.~Yan, ``On the integrability of planar
  {$N=2$} superconformal gauge theories,''
  \href{http://dx.doi.org/10.1007/JHEP08(2013)015}{{\em JHEP} {\bfseries 08}
  (2013) 015}, \href{http://arxiv.org/abs/1211.0271}{{\ttfamily arXiv:1211.0271
  [hep-th]}}.

\bibitem{Veneziano:1976wm}
G.~Veneziano, ``{Some Aspects of a Unified Approach to Gauge, Dual and Gribov
  Theories},'' \href{http://dx.doi.org/10.1016/0550-3213(76)90412-0}{{\em Nucl.
  Phys. B} {\bfseries 117} (1976) 519--545}.

\bibitem{Pomoni:2013poa}
E.~Pomoni, ``{Integrability in N=2 superconformal gauge theories},''
\href{http://arxiv.org/abs/1310.5709}{{\ttfamily arXiv:1310.5709 [hep-th]}}.

\bibitem{Mitev:2014yba}
V.~Mitev and E.~Pomoni, ``{Exact effective couplings of four dimensional gauge
  theories with $\mathcal N=$ 2 supersymmetry},''
  \href{http://dx.doi.org/10.1103/PhysRevD.92.125034}{{\em Phys. Rev. D}
  {\bfseries 92} no.~12, (2015) 125034},
  \href{http://arxiv.org/abs/1406.3629}{{\ttfamily arXiv:1406.3629 [hep-th]}}.

\bibitem{Mitev:2015oty}
V.~Mitev and E.~Pomoni, ``{Exact Bremsstrahlung and Effective Couplings},''
  \href{http://dx.doi.org/10.1007/JHEP06(2016)078}{{\em JHEP} {\bfseries 06}
  (2016) 078}, \href{http://arxiv.org/abs/1511.02217}{{\ttfamily
  arXiv:1511.02217 [hep-th]}}.

\bibitem{DiVecchia:2004jw}
P.~Di~Vecchia and A.~Tanzini, ``{N=2 super Yang-Mills and the XXZ spin
  chain},'' \href{http://dx.doi.org/10.1016/j.geomphys.2004.09.001}{{\em J.
  Geom. Phys.} {\bfseries 54} (2005) 116--130},
  \href{http://arxiv.org/abs/hep-th/0405262}{{\ttfamily arXiv:hep-th/0405262}}.

\bibitem{Pomoni:2021pbj}
E.~Pomoni, R.~Rabe, and K.~Zoubos, ``{Dynamical spin chains in 4D $ \mathcal{N}
  $ = 2 SCFTs},'' \href{http://dx.doi.org/10.1007/JHEP08(2021)127}{{\em JHEP}
  {\bfseries 08} (2021) 127}, \href{http://arxiv.org/abs/2106.08449}{{\ttfamily
  arXiv:2106.08449 [hep-th]}}.

\bibitem{Felder:1994pb}
G.~Felder, ``{Conformal field theory and integrable systems associated to
  elliptic curves},''
\href{http://arxiv.org/abs/hep-th/9407154}{{\ttfamily arXiv:hep-th/9407154
  [hep-th]}}.

\bibitem{Felder:1994be}
G.~Felder, ``{Elliptic quantum groups},'' in {\em {11th International
  Conference on Mathematical Physics (ICMP-11) (Satellite colloquia: New
  Problems in the General Theory of Fields and Particles, Paris, France, 25-28
  Jul 1994)}}, pp.~211--218.
\newblock 7, 1994.
\newblock \href{http://arxiv.org/abs/hep-th/9412207}{{\ttfamily
  arXiv:hep-th/9412207}}.

\bibitem{Bozkurt:2024tpz}
D.~N. Bozkurt, J.~M. Nieto~Garc{\'\i}a, and E.~Pomoni, ``{Long-range to the
  rescue of Yang-Baxter},''
  \href{http://dx.doi.org/10.1016/j.nuclphysb.2026.117598}{{\em Nucl. Phys. B}
  {\bfseries 1030} (2026) 117598},
  \href{http://arxiv.org/abs/2408.03365}{{\ttfamily arXiv:2408.03365
  [hep-th]}}.

\bibitem{Bozkurt:2025exl}
D.~N. Bozkurt, J.~M. Nieto~Garc{\'\i}a, Z.~Kong, and E.~Pomoni, ``{Long-range
  to the Rescue of Yang-Baxter II},''
  \href{http://arxiv.org/abs/2507.08934}{{\ttfamily arXiv:2507.08934
  [hep-th]}}.

\bibitem{Bertle:2024djm}
H.~Bertle, E.~Pomoni, X.~Zhang, and K.~Zoubos, ``{Hidden symmetries of 4D $
  \mathcal{N} $ = 2 gauge theories},''
  \href{http://dx.doi.org/10.1007/JHEP02(2025)205}{{\em JHEP} {\bfseries 02}
  (2025) 205}, \href{http://arxiv.org/abs/2411.11612}{{\ttfamily
  arXiv:2411.11612 [hep-th]}}.

\bibitem{Minahan:2010js}
J.~A. Minahan, ``{Review of AdS/CFT Integrability, Chapter I.1: Spin Chains in
  N=4 Super Yang-Mills},''
  \href{http://dx.doi.org/10.1007/s11005-011-0522-9}{{\em Lett. Math. Phys.}
  {\bfseries 99} (2012) 33--58},
\href{http://arxiv.org/abs/1012.3983}{{\ttfamily arXiv:1012.3983 [hep-th]}}.

\bibitem{deLeeuw:2017cop}
M.~de~Leeuw, A.~C. Ipsen, C.~Kristjansen, and M.~Wilhelm, ``{Introduction to
  integrability and one-point functions in $\mathcal N=$ 4 supersymmetric
  Yang\textendash{}Mills theory and its defect cousin},''
  \href{http://arxiv.org/abs/1708.02525}{{\ttfamily arXiv:1708.02525
  [hep-th]}}.

\bibitem{Sieg:2010jt}
C.~Sieg, ``{Review of AdS/CFT Integrability, Chapter I.2: The spectrum from
  perturbative gauge theory},''
  \href{http://dx.doi.org/10.1007/s11005-011-0508-7}{{\em Lett. Math. Phys.}
  {\bfseries 99} (2012) 59--84},
  \href{http://arxiv.org/abs/1012.3984}{{\ttfamily arXiv:1012.3984 [hep-th]}}.

\bibitem{Sieg:2010tz}
C.~Sieg, ``{Superspace calculation of the three-loop dilatation operator of N=4
  SYM theory},'' \href{http://dx.doi.org/10.1103/PhysRevD.84.045014}{{\em Phys.
  Rev. D} {\bfseries 84} (2011) 045014},
  \href{http://arxiv.org/abs/1008.3351}{{\ttfamily arXiv:1008.3351 [hep-th]}}.

\bibitem{Pomoni:2011jj}
E.~Pomoni and C.~Sieg, ``{From N=4 gauge theory to N=2 conformal QCD:
  three-loop mixing of scalar composite operators},''
  \href{http://arxiv.org/abs/1105.3487}{{\ttfamily arXiv:1105.3487 [hep-th]}}.

\bibitem{Douglas:1996sw}
M.~R. Douglas and G.~W. Moore, ``{D-branes, quivers, and ALE instantons},''
  \href{http://arxiv.org/abs/hep-th/9603167}{{\ttfamily arXiv:hep-th/9603167}}.

\bibitem{Johnson:1996py}
C.~V. Johnson and R.~C. Myers, ``{Aspects of type IIB theory on ALE spaces},''
  \href{http://dx.doi.org/10.1103/PhysRevD.55.6382}{{\em Phys. Rev. D}
  {\bfseries 55} (1997) 6382--6393},
  \href{http://arxiv.org/abs/hep-th/9610140}{{\ttfamily arXiv:hep-th/9610140}}.

\bibitem{Kachru:1998ys}
S.~Kachru and E.~Silverstein, ``{4-D conformal theories and strings on
  orbifolds},'' \href{http://dx.doi.org/10.1103/PhysRevLett.80.4855}{{\em Phys.
  Rev. Lett.} {\bfseries 80} (1998) 4855--4858},
  \href{http://arxiv.org/abs/hep-th/9802183}{{\ttfamily arXiv:hep-th/9802183}}.

\bibitem{Lawrence:1998ja}
A.~E. Lawrence, N.~Nekrasov, and C.~Vafa, ``{On conformal field theories in
  four-dimensions},''
  \href{http://dx.doi.org/10.1016/S0550-3213(98)00495-7}{{\em Nucl. Phys. B}
  {\bfseries 533} (1998) 199--209},
  \href{http://arxiv.org/abs/hep-th/9803015}{{\ttfamily arXiv:hep-th/9803015}}.

\bibitem{Oz:1998hr}
Y.~Oz and J.~Terning, ``{Orbifolds of AdS(5) x S**5 and 4-d conformal field
  theories},'' \href{http://dx.doi.org/10.1016/S0550-3213(98)00454-4}{{\em
  Nucl. Phys. B} {\bfseries 532} (1998) 163--180},
  \href{http://arxiv.org/abs/hep-th/9803167}{{\ttfamily arXiv:hep-th/9803167}}.

\bibitem{Gukov:1998kk}
S.~Gukov, ``{Comments on N=2 AdS orbifolds},''
  \href{http://dx.doi.org/10.1016/S0370-2693(98)01005-3}{{\em Phys. Lett. B}
  {\bfseries 439} (1998) 23--28},
  \href{http://arxiv.org/abs/hep-th/9806180}{{\ttfamily arXiv:hep-th/9806180}}.

\bibitem{Lomont1959}
J.~S. Lomont, {\em Applications of Finite Groups}.
\newblock Academic Press, New York, 1959.

\bibitem{Sternberg94}
S.~Sternberg, {\em Group Theory and Physics}.
\newblock Cambridge University press, 1994.

\bibitem{Klein1884}
F.~Klein, {\em Vorlesungen über das Ikosaeder und die Auflösung der
  Gleichungen vom fünften Grade}.
\newblock Teubner, 1884.
\newblock \url{http://eudml.org/doc/203220}.

\bibitem{Benvenuti:2006qr}
S.~Benvenuti, B.~Feng, A.~Hanany, and Y.-H. He, ``{Counting BPS Operators in
  Gauge Theories: Quivers, Syzygies and Plethystics},''
  \href{http://dx.doi.org/10.1088/1126-6708/2007/11/050}{{\em JHEP} {\bfseries
  11} (2007) 050},
\href{http://arxiv.org/abs/hep-th/0608050}{{\ttfamily arXiv:hep-th/0608050
  [hep-th]}}.

\bibitem{McKay1980}
J.~McKay, ``Graphs, singularities, and finite groups,'' in {\em Proceedings of
  Symposia in Pure Mathematics}, vol.~37, pp.~183--186.
\newblock American Mathematical Society, 1980.

\bibitem{Fuchs:1992nq}
J.~Fuchs, {\em {Affine Lie algebras and quantum groups: An Introduction, with
  applications in conformal field theory}}.
\newblock Cambridge University Press, 3, 1995.

\bibitem{DiFrancesco}
P.~{Di Francesco}, P.~{Mathieu}, and D.~{S\'en\'echal}, {\em Conformal Field
  Theory}.
\newblock Springer-Verlag, 1997.

\bibitem{Dymarsky:2005uh}
A.~Dymarsky, I.~R. Klebanov, and R.~Roiban, ``{Perturbative search for fixed
  lines in large N gauge theories},''
  \href{http://dx.doi.org/10.1088/1126-6708/2005/08/011}{{\em JHEP} {\bfseries
  08} (2005) 011}, \href{http://arxiv.org/abs/hep-th/0505099}{{\ttfamily
  arXiv:hep-th/0505099}}.

\bibitem{Dymarsky:2005nc}
A.~Dymarsky, I.~R. Klebanov, and R.~Roiban, ``{Perturbative gauge theory and
  closed string tachyons},''
  \href{http://dx.doi.org/10.1088/1126-6708/2005/11/038}{{\em JHEP} {\bfseries
  11} (2005) 038}, \href{http://arxiv.org/abs/hep-th/0509132}{{\ttfamily
  arXiv:hep-th/0509132}}.

\bibitem{Kronheimer:1989zs}
P.~B. Kronheimer, ``The construction of {ALE} spaces as hyper-k{\"a}hler
  quotients,'' {\em J. Diff. Geom.} {\bfseries 29} no.~3, (1989) 665--683.

\bibitem{Intriligator:1996ex}
K.~A. Intriligator and N.~Seiberg, ``{Mirror symmetry in three-dimensional
  gauge theories},'' \href{http://dx.doi.org/10.1016/0370-2693(96)01088-X}{{\em
  Phys. Lett. B} {\bfseries 387} (1996) 513--519},
  \href{http://arxiv.org/abs/hep-th/9607207}{{\ttfamily arXiv:hep-th/9607207}}.

\bibitem{Minahan:2011dd}
J.~A. Minahan, ``{Supergraphs and the cubic Leigh-Strassler model},''
  \href{http://dx.doi.org/10.1007/JHEP12(2011)093}{{\em JHEP} {\bfseries 12}
  (2011) 093},
\href{http://arxiv.org/abs/1108.1583}{{\ttfamily arXiv:1108.1583 [hep-th]}}.

\bibitem{Pomoni:2019oib}
E.~Pomoni, ``{4D $\mathcal{N}=2$ SCFTs and spin chains},''
  \href{http://dx.doi.org/10.1088/1751-8121/ab7f66}{{\em J. Phys. A} {\bfseries
  53} no.~28, (2020) 283005}, \href{http://arxiv.org/abs/1912.00870}{{\ttfamily
  arXiv:1912.00870 [hep-th]}}.

\bibitem{Gates:1983nr}
S.~J. Gates, M.~T. Grisaru, M.~Rocek, and W.~Siegel, {\em {Superspace Or One
  Thousand and One Lessons in Supersymmetry}}, vol.~58 of {\em Frontiers in
  Physics}.
\newblock 1983.
\newblock \href{http://arxiv.org/abs/hep-th/0108200}{{\ttfamily
  arXiv:hep-th/0108200}}.

\bibitem{Beisert:2002bb}
N.~Beisert, C.~Kristjansen, J.~Plefka, G.~W. Semenoff, and M.~Staudacher,
  ``{BMN correlators and operator mixing in N=4 superYang-Mills theory},''
  \href{http://dx.doi.org/10.1016/S0550-3213(02)01025-8}{{\em Nucl. Phys. B}
  {\bfseries 650} (2003) 125--161},
  \href{http://arxiv.org/abs/hep-th/0208178}{{\ttfamily arXiv:hep-th/0208178}}.

\bibitem{Beisert:2003tq}
N.~Beisert, C.~Kristjansen, and M.~Staudacher, ``{The Dilatation operator of
  conformal N=4 superYang-Mills theory},''
  \href{http://dx.doi.org/10.1016/S0550-3213(03)00406-1}{{\em Nucl. Phys. B}
  {\bfseries 664} (2003) 131--184},
  \href{http://arxiv.org/abs/hep-th/0303060}{{\ttfamily arXiv:hep-th/0303060}}.

\bibitem{Wess:1992cp}
J.~Wess and J.~Bagger, {\em {Supersymmetry and supergravity}}.
\newblock Princeton University Press, Princeton, NJ, USA, 1992.

\bibitem{Romelsberger:2005eg}
C.~R\"omelsberger, ``{Counting chiral primaries in N = 1, d=4 superconformal
  field theories},''
  \href{http://dx.doi.org/10.1016/j.nuclphysb.2006.03.037}{{\em Nucl. Phys. B}
  {\bfseries 747} (2006) 329--353},
  \href{http://arxiv.org/abs/hep-th/0510060}{{\ttfamily arXiv:hep-th/0510060}}.

\bibitem{Kinney:2005ej}
J.~Kinney, J.~M. Maldacena, S.~Minwalla, and S.~Raju, ``{An Index for 4
  dimensional super conformal theories},''
  \href{http://dx.doi.org/10.1007/s00220-007-0258-7}{{\em Commun. Math. Phys.}
  {\bfseries 275} (2007) 209--254},
  \href{http://arxiv.org/abs/hep-th/0510251}{{\ttfamily arXiv:hep-th/0510251}}.

\bibitem{Dobrev:1985qv}
V.~K. Dobrev and V.~B. Petkova, ``{All Positive Energy Unitary Irreducible
  Representations of Extended Conformal Supersymmetry},''
  \href{http://dx.doi.org/10.1016/0370-2693(85)91073-1}{{\em Phys. Lett. B}
  {\bfseries 162} (1985) 127--132}.

\bibitem{Ferrara:1999ed}
S.~Ferrara and A.~Zaffaroni, ``{Superconformal field theories, multiplet
  shortening, and the AdS(5) / SCFT(4) correspondence},'' in {\em {Conference
  Moshe Flato}}, pp.~177--188.
\newblock 2000.
\newblock \href{http://arxiv.org/abs/hep-th/9908163}{{\ttfamily
  arXiv:hep-th/9908163}}.

\bibitem{Dolan:2002zh}
F.~A. Dolan and H.~Osborn, ``{On short and semi-short representations for
  four-dimensional superconformal symmetry},''
  \href{http://dx.doi.org/10.1016/S0003-4916(03)00074-5}{{\em Annals Phys.}
  {\bfseries 307} (2003) 41--89},
  \href{http://arxiv.org/abs/hep-th/0209056}{{\ttfamily arXiv:hep-th/0209056}}.

\bibitem{Cordova:2016emh}
C.~Cordova, T.~T. Dumitrescu, and K.~Intriligator, ``{Multiplets of
  Superconformal Symmetry in Diverse Dimensions},''
  \href{http://dx.doi.org/10.1007/JHEP03(2019)163}{{\em JHEP} {\bfseries 03}
  (2019) 163}, \href{http://arxiv.org/abs/1612.00809}{{\ttfamily
  arXiv:1612.00809 [hep-th]}}.

\bibitem{Eberhardt:2020cxo}
L.~Eberhardt, ``{Superconformal symmetry and representations},''
  \href{http://dx.doi.org/10.1088/1751-8121/abd7b3}{{\em J. Phys. A} {\bfseries
  54} no.~6, (2021) 063002}, \href{http://arxiv.org/abs/2006.13280}{{\ttfamily
  arXiv:2006.13280 [hep-th]}}.

\bibitem{Rastelli:2016tbz}
L.~Rastelli and S.~S. Razamat, ``{The supersymmetric index in four
  dimensions},'' \href{http://dx.doi.org/10.1088/1751-8121/aa76a6}{{\em J.
  Phys. A} {\bfseries 50} no.~44, (2017) 443013},
  \href{http://arxiv.org/abs/1608.02965}{{\ttfamily arXiv:1608.02965
  [hep-th]}}.

\bibitem{Gadde:2020yah}
A.~Gadde, ``{Lectures on the Superconformal Index},''
  \href{http://dx.doi.org/10.1088/1751-8121/ac42ac}{{\em J. Phys. A} {\bfseries
  55} no.~6, (2022) 063001}, \href{http://arxiv.org/abs/2006.13630}{{\ttfamily
  arXiv:2006.13630 [hep-th]}}.

\bibitem{Gadde:2011uv}
A.~Gadde, L.~Rastelli, S.~S. Razamat, and W.~Yan, ``{Gauge Theories and
  Macdonald Polynomials},''
  \href{http://dx.doi.org/10.1007/s00220-012-1607-8}{{\em Commun. Math. Phys.}
  {\bfseries 319} (2013) 147--193},
  \href{http://arxiv.org/abs/1110.3740}{{\ttfamily arXiv:1110.3740 [hep-th]}}.

\bibitem{Witten:1982df}
E.~Witten, ``{Constraints on Supersymmetry Breaking},''
  \href{http://dx.doi.org/10.1016/0550-3213(82)90071-2}{{\em Nucl. Phys. B}
  {\bfseries 202} (1982) 253}.

\bibitem{Nakayama:2005mf}
Y.~Nakayama, ``{Index for orbifold quiver gauge theories},''
  \href{http://dx.doi.org/10.1016/j.physletb.2006.03.045}{{\em Phys. Lett. B}
  {\bfseries 636} (2006) 132--136},
  \href{http://arxiv.org/abs/hep-th/0512280}{{\ttfamily arXiv:hep-th/0512280}}.

\bibitem{Gadde:2010en}
A.~Gadde, L.~Rastelli, S.~S. Razamat, and W.~Yan, ``{On the Superconformal
  Index of N=1 IR Fixed Points: A Holographic Check},''
  \href{http://dx.doi.org/10.1007/JHEP03(2011)041}{{\em JHEP} {\bfseries 03}
  (2011) 041}, \href{http://arxiv.org/abs/1011.5278}{{\ttfamily arXiv:1011.5278
  [hep-th]}}.

\bibitem{Eager:2012hx}
R.~Eager, J.~Schmude, and Y.~Tachikawa, ``{Superconformal Indices,
  Sasaki-Einstein Manifolds, and Cyclic Homologies},''
  \href{http://dx.doi.org/10.4310/ATMP.2014.v18.n1.a3}{{\em Adv. Theor. Math.
  Phys.} {\bfseries 18} no.~1, (2014) 129--175},
  \href{http://arxiv.org/abs/1207.0573}{{\ttfamily arXiv:1207.0573 [hep-th]}}.

\bibitem{Feng:2007ur}
B.~Feng, A.~Hanany, and Y.-H. He, ``{Counting gauge invariants: The Plethystic
  program},'' \href{http://dx.doi.org/10.1088/1126-6708/2007/03/090}{{\em JHEP}
  {\bfseries 03} (2007) 090},
\href{http://arxiv.org/abs/hep-th/0701063}{{\ttfamily arXiv:hep-th/0701063
  [hep-th]}}.

\bibitem{Ruijsenaars:1997aqs}
S.~N.~M. Ruijsenaars, ``{First order analytic difference equations and
  integrable quantum systems},'' \href{http://dx.doi.org/10.1063/1.531809}{{\em
  J. Math. Phys.} {\bfseries 38} no.~2, (1997) 1069}.

\bibitem{Beem:2013sza}
C.~Beem, M.~Lemos, P.~Liendo, W.~Peelaers, L.~Rastelli, and B.~C. van Rees,
  ``{Infinite Chiral Symmetry in Four Dimensions},''
  \href{http://dx.doi.org/10.1007/s00220-014-2272-x}{{\em Commun. Math. Phys.}
  {\bfseries 336} no.~3, (2015) 1359--1433},
  \href{http://arxiv.org/abs/1312.5344}{{\ttfamily arXiv:1312.5344 [hep-th]}}.

\bibitem{Banerjee:2023ddh}
A.~Banerjee and M.~Buican, ``{Nonperturbative explorations of chiral rings in
  4D N=2 SCFTs},'' \href{http://dx.doi.org/10.1103/PhysRevD.108.105010}{{\em
  Phys. Rev. D} {\bfseries 108} no.~10, (2023) 105010},
  \href{http://arxiv.org/abs/2306.12521}{{\ttfamily arXiv:2306.12521
  [hep-th]}}.

\bibitem{Cachazo:2002ry}
F.~Cachazo, M.~R. Douglas, N.~Seiberg, and E.~Witten, ``{Chiral rings and
  anomalies in supersymmetric gauge theory},''
  \href{http://dx.doi.org/10.1088/1126-6708/2002/12/071}{{\em JHEP} {\bfseries
  12} (2002) 071}, \href{http://arxiv.org/abs/hep-th/0211170}{{\ttfamily
  arXiv:hep-th/0211170}}.

\bibitem{Hanany:2006uc}
A.~Hanany and C.~Romelsberger, ``{Counting BPS operators in the chiral ring of
  N=2 supersymmetric gauge theories or N=2 braine surgery},''
  \href{http://dx.doi.org/10.4310/ATMP.2007.v11.n6.a4}{{\em Adv. Theor. Math.
  Phys.} {\bfseries 11} no.~6, (2007) 1091--1112},
  \href{http://arxiv.org/abs/hep-th/0611346}{{\ttfamily arXiv:hep-th/0611346}}.

\bibitem{Hayling:2017cva}
J.~Hayling, C.~Papageorgakis, E.~Pomoni, and D.~Rodr{\'\i}guez-G{\'o}mez,
  ``{Exact Deconstruction of the 6D (2,0) Theory},''
  \href{http://dx.doi.org/10.1007/JHEP06(2017)072}{{\em JHEP} {\bfseries 06}
  (2017) 072}, \href{http://arxiv.org/abs/1704.02986}{{\ttfamily
  arXiv:1704.02986 [hep-th]}}.

\bibitem{Arkani-Hamed:2001wsh}
N.~Arkani-Hamed, A.~G. Cohen, D.~B. Kaplan, A.~Karch, and L.~Motl,
  ``{Deconstructing (2,0) and little string theories},''
  \href{http://dx.doi.org/10.1088/1126-6708/2003/01/083}{{\em JHEP} {\bfseries
  01} (2003) 083}, \href{http://arxiv.org/abs/hep-th/0110146}{{\ttfamily
  arXiv:hep-th/0110146}}.

\bibitem{Arai:2019aou}
R.~Arai, S.~Fujiwara, Y.~Imamura, and T.~Mori, ``{Finite $N$ corrections to the
  superconformal index of toric quiver gauge theories},''
  \href{http://dx.doi.org/10.1093/ptep/ptaa023}{{\em PTEP} {\bfseries 2020}
  no.~4, (2020) 043B09}, \href{http://arxiv.org/abs/1911.10794}{{\ttfamily
  arXiv:1911.10794 [hep-th]}}.

\bibitem{Fujiwara:2023bdc}
S.~Fujiwara, Y.~Imamura, T.~Mori, S.~Murayama, and D.~Yokoyama, ``{Simple-Sum
  Giant Graviton Expansions for Orbifolds and Orientifolds},''
  \href{http://dx.doi.org/10.1093/ptep/ptae006}{{\em PTEP} {\bfseries 2024}
  no.~2, (2024) 023B02}, \href{http://arxiv.org/abs/2310.03332}{{\ttfamily
  arXiv:2310.03332 [hep-th]}}.

\bibitem{Bourdier:2015sga}
J.~Bourdier, N.~Drukker, and J.~Felix, ``{The $\mathcal{N}=2$ Schur index from
  free fermions},'' \href{http://dx.doi.org/10.1007/JHEP01(2016)167}{{\em JHEP}
  {\bfseries 01} (2016) 167}, \href{http://arxiv.org/abs/1510.07041}{{\ttfamily
  arXiv:1510.07041 [hep-th]}}.

\bibitem{Bourdier:2015wda}
J.~Bourdier, N.~Drukker, and J.~Felix, ``{The exact Schur index of
  $\mathcal{N}=4$ SYM},'' \href{http://dx.doi.org/10.1007/JHEP11(2015)210}{{\em
  JHEP} {\bfseries 11} (2015) 210},
  \href{http://arxiv.org/abs/1507.08659}{{\ttfamily arXiv:1507.08659
  [hep-th]}}.

\bibitem{Eleftheriou:2022kkv}
G.~Eleftheriou, ``{Root of unity asymptotics for Schur indices of 4d Lagrangian
  theories},'' \href{http://dx.doi.org/10.1007/JHEP01(2023)081}{{\em JHEP}
  {\bfseries 01} (2023) 081}, \href{http://arxiv.org/abs/2207.14271}{{\ttfamily
  arXiv:2207.14271 [hep-th]}}.

\bibitem{Bershadsky:1998mb}
M.~Bershadsky, Z.~Kakushadze, and C.~Vafa, ``{String expansion as large N
  expansion of gauge theories},''
  \href{http://dx.doi.org/10.1016/S0550-3213(98)00272-7}{{\em Nucl. Phys. B}
  {\bfseries 523} (1998) 59--72},
  \href{http://arxiv.org/abs/hep-th/9803076}{{\ttfamily arXiv:hep-th/9803076}}.

\bibitem{Bershadsky:1998cb}
M.~Bershadsky and A.~Johansen, ``{Large N limit of orbifold field theories},''
  \href{http://dx.doi.org/10.1016/S0550-3213(98)00526-4}{{\em Nucl. Phys. B}
  {\bfseries 536} (1998) 141--148},
  \href{http://arxiv.org/abs/hep-th/9803249}{{\ttfamily arXiv:hep-th/9803249}}.

\bibitem{Pestun:2016zxk}
V.~Pestun {\em et~al.}, ``{Localization techniques in quantum field
  theories},'' \href{http://dx.doi.org/10.1088/1751-8121/aa63c1}{{\em J. Phys.
  A} {\bfseries 50} no.~44, (2017) 440301},
  \href{http://arxiv.org/abs/1608.02952}{{\ttfamily arXiv:1608.02952
  [hep-th]}}.

\bibitem{Pini:2017ouj}
A.~Pini, D.~Rodriguez-Gomez, and J.~G. Russo, ``{Large $N$ correlation
  functions in $ \mathcal{N}=$ 2 superconformal quivers},''
  \href{http://dx.doi.org/10.1007/JHEP08(2017)066}{{\em JHEP} {\bfseries 08}
  (2017) 066}, \href{http://arxiv.org/abs/1701.02315}{{\ttfamily
  arXiv:1701.02315 [hep-th]}}.

\bibitem{Galvagno:2020cgq}
F.~Galvagno and M.~Preti, ``{Chiral correlators in $ \mathcal{N} $ = 2
  superconformal quivers},''
  \href{http://dx.doi.org/10.1007/JHEP05(2021)201}{{\em JHEP} {\bfseries 05}
  (2021) 201}, \href{http://arxiv.org/abs/2012.15792}{{\ttfamily
  arXiv:2012.15792 [hep-th]}}.

\bibitem{Preti:2022inu}
M.~Preti, ``{Correlators in superconformal quivers made QUICK},''
  \href{http://arxiv.org/abs/2212.14823}{{\ttfamily arXiv:2212.14823
  [hep-th]}}.

\bibitem{Pini:2024uia}
A.~Pini and P.~Vallarino, ``{Integrated correlators at strong coupling in an
  orbifold of $ \mathcal{N} $ = 4 SYM},''
  \href{http://dx.doi.org/10.1007/JHEP06(2024)170}{{\em JHEP} {\bfseries 06}
  (2024) 170}, \href{http://arxiv.org/abs/2404.03466}{{\ttfamily
  arXiv:2404.03466 [hep-th]}}.

\bibitem{Niarchos:2019onf}
V.~Niarchos, C.~Papageorgakis, and E.~Pomoni, ``{Type-B Anomaly Matching and
  the 6D (2,0) Theory},'' \href{http://dx.doi.org/10.1007/JHEP04(2020)048}{{\em
  JHEP} {\bfseries 04} (2020) 048},
  \href{http://arxiv.org/abs/1911.05827}{{\ttfamily arXiv:1911.05827
  [hep-th]}}.

\bibitem{Niarchos:2020nxk}
V.~Niarchos, C.~Papageorgakis, A.~Pini, and E.~Pomoni, ``{(Mis-)Matching Type-B
  Anomalies on the Higgs Branch},''
  \href{http://dx.doi.org/10.1007/JHEP01(2021)106}{{\em JHEP} {\bfseries 01}
  (2021) 106}, \href{http://arxiv.org/abs/2009.08375}{{\ttfamily
  arXiv:2009.08375 [hep-th]}}.

\bibitem{Billo:2021rdb}
M.~Billo, M.~Frau, F.~Galvagno, A.~Lerda, and A.~Pini, ``{Strong-coupling
  results for $ \mathcal{N} $ = 2 superconformal quivers and holography},''
  \href{http://dx.doi.org/10.1007/JHEP10(2021)161}{{\em JHEP} {\bfseries 10}
  (2021) 161}, \href{http://arxiv.org/abs/2109.00559}{{\ttfamily
  arXiv:2109.00559 [hep-th]}}.

\bibitem{Billo:2022gmq}
M.~Bill{\`o}, M.~Frau, A.~Lerda, A.~Pini, and P.~Vallarino, ``{Structure
  Constants in N=2 Superconformal Quiver Theories at Strong Coupling and
  Holography},'' \href{http://dx.doi.org/10.1103/PhysRevLett.129.031602}{{\em
  Phys. Rev. Lett.} {\bfseries 129} no.~3, (2022) 031602},
  \href{http://arxiv.org/abs/2206.13582}{{\ttfamily arXiv:2206.13582
  [hep-th]}}.

\bibitem{Billo:2022fnb}
M.~Billo, M.~Frau, A.~Lerda, A.~Pini, and P.~Vallarino, ``{Localization vs
  holography in 4d$ \mathcal{N} $ = 2 quiver theories},''
  \href{http://dx.doi.org/10.1007/JHEP10(2022)020}{{\em JHEP} {\bfseries 10}
  (2022) 020}, \href{http://arxiv.org/abs/2207.08846}{{\ttfamily
  arXiv:2207.08846 [hep-th]}}.

\bibitem{Billo:2022lrv}
M.~Billo, M.~Frau, A.~Lerda, A.~Pini, and P.~Vallarino, ``{Strong coupling
  expansions in $ \mathcal{N} $ = 2 quiver gauge theories},''
  \href{http://dx.doi.org/10.1007/JHEP01(2023)119}{{\em JHEP} {\bfseries 01}
  (2023) 119}, \href{http://arxiv.org/abs/2211.11795}{{\ttfamily
  arXiv:2211.11795 [hep-th]}}.

\bibitem{Skrzypek:2023fkr}
T.~Skrzypek and A.~A. Tseytlin, ``{On AdS/CFT duality in the twisted sector of
  string theory on AdS$_{5}${\texttimes}
  S$^{5}$/{\ensuremath{\mathbb{Z}}}$_{2}$ orbifold background},''
  \href{http://dx.doi.org/10.1007/JHEP03(2024)045}{{\em JHEP} {\bfseries 03}
  (2024) 045}, \href{http://arxiv.org/abs/2312.13850}{{\ttfamily
  arXiv:2312.13850 [hep-th]}}.

\bibitem{Martinez:2025jjq}
C.~B. Mart{\'\i}nez and T.~Skrzypek, ``{Exploring the twisted sector of
  {\ensuremath{\mathbb{Z}}}$_{L}$ orbifolds: matching
  {\ensuremath{\alpha}}'-corrections to localisation},''
  \href{http://dx.doi.org/10.1007/JHEP05(2026)157}{{\em JHEP} {\bfseries 05}
  (2026) 157}, \href{http://arxiv.org/abs/2512.16984}{{\ttfamily
  arXiv:2512.16984 [hep-th]}}.

\bibitem{Corcoran:2024ofo}
L.~Corcoran, M.~de~Leeuw, and B.~Pozsgay, ``{Integrable models on Rydberg atom
  chains},'' \href{http://dx.doi.org/10.21468/SciPostPhys.18.4.139}{{\em
  SciPost Phys.} {\bfseries 18} no.~4, (2025) 139},
  \href{http://arxiv.org/abs/2405.15848}{{\ttfamily arXiv:2405.15848
  [cond-mat.str-el]}}.

\bibitem{Trebst_2008}
S.~Trebst, M.~Troyer, Z.~Wang, and A.~W.~W. Ludwig, ``A short introduction to
  {F}ibonacci anyon models,''
  \href{http://dx.doi.org/10.1143/ptps.176.384}{{\em Progress of Theoretical
  Physics Supplement} {\bfseries 176} (2008) 384–407},
  \href{http://arxiv.org/abs/0902.3275}{{\ttfamily arXiv:0902.3275
  [cond-mat.stat-mech]}}.

\bibitem{Aasen:2020jwb}
D.~Aasen, P.~Fendley, and R.~S.~K. Mong, ``{Topological Defects on the Lattice:
  Dualities and Degeneracies},''
  \href{http://arxiv.org/abs/2008.08598}{{\ttfamily arXiv:2008.08598
  [cond-mat.stat-mech]}}.

\bibitem{Blakeney:2025ext}
M.~Blakeney, L.~Corcoran, M.~de~Leeuw, B.~Pozsgay, and E.~Vernier,
  ``{Temperley-Lieb integrable models and fusion categories},''
  \href{http://dx.doi.org/10.1007/JHEP04(2026)165}{{\em JHEP} {\bfseries 04}
  (2026) 165}, \href{http://arxiv.org/abs/2510.19902}{{\ttfamily
  arXiv:2510.19902 [cond-mat.str-el]}}.

\bibitem{Braylovskaya:2016btd}
N.~Braylovskaya, P.~E. Finch, and H.~Frahm, ``{Exact solution of the $D_3$
  non-Abelian anyon chain},''
  \href{http://dx.doi.org/10.1103/PhysRevB.94.085138}{{\em Phys. Rev. B}
  {\bfseries 94} no.~8, (2016) 085138},
  \href{http://arxiv.org/abs/1606.00793}{{\ttfamily arXiv:1606.00793
  [cond-mat.str-el]}}.

\bibitem{Andrews:1984af}
G.~Andrews, R.~Baxter, and P.~Forrester, ``{Eight vertex SOS model and
  generalized Rogers-Ramanujan type identities},''
  \href{http://dx.doi.org/10.1007/BF01014383}{{\em J. Statist. Phys.}
  {\bfseries 35} (1984) 193--266}.

\bibitem{Felder:2020tct}
G.~Felder and M.~Ren, ``{Quantum Groups for Restricted SOS Models},''
  \href{http://dx.doi.org/10.3842/sigma.2021.005}{{\em SIGMA} {\bfseries 17}
  (2021) 005}, \href{http://arxiv.org/abs/2010.01060}{{\ttfamily
  arXiv:2010.01060 [math.RT]}}.

\bibitem{Pasquier:1986jc}
V.~Pasquier, ``{Two-dimensional critical systems labelled by Dynkin
  diagrams},'' \href{http://dx.doi.org/10.1016/0550-3213(87)90332-4}{{\em Nucl.
  Phys. B} {\bfseries 285} (1987) 162--172}.

\bibitem{Pearce:1990ila}
P.~A. Pearce, ``{Temperley-Lieb operators and critical A-D-E lattice models},''
  \href{http://dx.doi.org/10.1142/S0217979290000371}{{\em Int. J. Mod. Phys. B}
  {\bfseries 4} no.~5, (1990) 715--734}.

\bibitem{Bianchi:2001cm}
M.~Bianchi, S.~Kovacs, G.~Rossi, and Y.~S. Stanev, ``{Properties of the Konishi
  multiplet in N=4 SYM theory},''
  \href{http://dx.doi.org/10.1088/1126-6708/2001/05/042}{{\em JHEP} {\bfseries
  05} (2001) 042}, \href{http://arxiv.org/abs/hep-th/0104016}{{\ttfamily
  arXiv:hep-th/0104016}}.

\bibitem{deLeeuw:2011rw}
M.~de~Leeuw and S.~J. van Tongeren, ``{Orbifolded Konishi from the Mirror
  TBA},'' \href{http://dx.doi.org/10.1088/1751-8113/44/32/325404}{{\em J. Phys.
  A} {\bfseries 44} (2011) 325404},
  \href{http://arxiv.org/abs/1103.5853}{{\ttfamily arXiv:1103.5853 [hep-th]}}.

\bibitem{vonNeumannWigner29}
J.~{von Neumann} and E.~Wigner, ``Über merkwürdige diskrete eigenwerte,''
  {\em Phys. Z.} {\bfseries 30} (1929) 467.

\bibitem{HaakeChaos}
F.~Haake, S.~Gnutzmann, and M.~Ku\'s, {\em Quantum Signatures of Chaos}.
\newblock Springer, 2018.

\bibitem{McLoughlin:2022jyt}
T.~McLoughlin and A.~Spiering, ``{Chaotic spin chains in AdS/CFT},''
  \href{http://dx.doi.org/10.1007/JHEP09(2022)240}{{\em JHEP} {\bfseries 09}
  (2022) 240}, \href{http://arxiv.org/abs/2202.12075}{{\ttfamily
  arXiv:2202.12075 [hep-th]}}.

\bibitem{Gomez96}
C.~{G\'omez}, M.~{Ruiz-Altaba}, and G.~{Sierra}, {\em Quantum Groups in
  Two-Dimensional Physics}.
\newblock Cambridge, 1996.

\bibitem{Berenstein:2002jq}
D.~E. Berenstein, J.~M. Maldacena, and H.~S. Nastase, ``{Strings in flat space
  and pp waves from N=4 superYang-Mills},''
  \href{http://dx.doi.org/10.1088/1126-6708/2002/04/013}{{\em JHEP} {\bfseries
  04} (2002) 013}, \href{http://arxiv.org/abs/hep-th/0202021}{{\ttfamily
  arXiv:hep-th/0202021}}.

\bibitem{Oz:2002wy}
Y.~Oz and T.~Sakai, ``{Exact anomalous dimensions for N=2 ADE SCFTs},''
  \href{http://dx.doi.org/10.1088/1126-6708/2003/01/044}{{\em JHEP} {\bfseries
  01} (2003) 044}, \href{http://arxiv.org/abs/hep-th/0208078}{{\ttfamily
  arXiv:hep-th/0208078}}.

\bibitem{Feng:2000af}
B.~Feng, A.~Hanany, Y.-H. He, and N.~Prezas, ``{Discrete torsion, non-Abelian
  orbifolds and the {S}chur multiplier},''
  \href{http://dx.doi.org/10.1088/1126-6708/2001/01/033}{{\em JHEP} {\bfseries
  01} (2001) 033}, \href{http://arxiv.org/abs/hep-th/0010023}{{\ttfamily
  arXiv:hep-th/0010023}}.

\bibitem{Nunez:2023loo}
C.~Nunez, L.~Santilli, and K.~Zarembo, ``{Linear Quivers at Large-N},''
  \href{http://dx.doi.org/10.1007/s00220-024-05186-1}{{\em Commun. Math. Phys.}
  {\bfseries 406} no.~1, (2025) 6},
  \href{http://arxiv.org/abs/2311.00024}{{\ttfamily arXiv:2311.00024
  [hep-th]}}.

\bibitem{Katz:1997eq}
S.~Katz, P.~Mayr, and C.~Vafa, ``{Mirror symmetry and exact solution of 4-D N=2
  gauge theories: 1.},''
  \href{http://dx.doi.org/10.4310/ATMP.1997.v1.n1.a2}{{\em Adv. Theor. Math.
  Phys.} {\bfseries 1} (1998) 53--114},
  \href{http://arxiv.org/abs/hep-th/9706110}{{\ttfamily arXiv:hep-th/9706110}}.

\bibitem{Bhardwaj2013}
L.~Bhardwaj and Y.~Tachikawa, ``{Classification of 4d N=2 gauge theories},''
  \href{http://dx.doi.org/10.1007/JHEP12(2013)100}{{\em JHEP} {\bfseries 12}
  (2013) 100}, \href{http://arxiv.org/abs/1309.5160}{{\ttfamily arXiv:1309.5160
  [hep-th]}}.

\bibitem{Ren:2023mtn}
M.~Ren, ``{Baxterization for the dynamical Yang-Baxter equation},''
  \href{http://arxiv.org/abs/2310.04728}{{\ttfamily arXiv:2310.04728
  [math.RT]}}.

\bibitem{Bundzik:2005zg}
D.~Bundzik and T.~{M\aa nsson}, ``{The General Leigh-Strassler deformation and
  integrability},'' \href{http://dx.doi.org/10.1088/1126-6708/2006/01/116}{{\em
  JHEP} {\bfseries 01} (2006) 116},
\href{http://arxiv.org/abs/hep-th/0512093}{{\ttfamily arXiv:hep-th/0512093
  [hep-th]}}.

\bibitem{Rey:2010ry}
S.-J. Rey and T.~Suyama, ``{Exact Results and Holography of Wilson Loops in N=2
  Superconformal (Quiver) Gauge Theories},''
  \href{http://dx.doi.org/10.1007/JHEP01(2011)136}{{\em JHEP} {\bfseries 01}
  (2011) 136}, \href{http://arxiv.org/abs/1001.0016}{{\ttfamily arXiv:1001.0016
  [hep-th]}}.

\bibitem{Fraser:2015xha}
B.~Fraser, ``{Higher rank Wilson loops in the ${\mathcal{N}}=2{SU}(N)\times
  {SU}(N)$ conformal quiver},''
  \href{http://dx.doi.org/10.1088/1751-8113/49/2/02LT03}{{\em J. Phys. A}
  {\bfseries 49} no.~2, (2016) 02LT03},
  \href{http://arxiv.org/abs/1503.05634}{{\ttfamily arXiv:1503.05634
  [hep-th]}}.

\bibitem{Zarembo:2020tpf}
K.~Zarembo, ``{Quiver CFT at strong coupling},''
  \href{http://dx.doi.org/10.1007/JHEP06(2020)055}{{\em JHEP} {\bfseries 06}
  (2020) 055}, \href{http://arxiv.org/abs/2003.00993}{{\ttfamily
  arXiv:2003.00993 [hep-th]}}.

\bibitem{Ouyang:2020hwd}
H.~Ouyang, ``{Wilson loops in circular quiver SCFTs at strong coupling},''
  \href{http://dx.doi.org/10.1007/JHEP02(2021)178}{{\em JHEP} {\bfseries 02}
  (2021) 178}, \href{http://arxiv.org/abs/2011.03531}{{\ttfamily
  arXiv:2011.03531 [hep-th]}}.

\bibitem{Fiol:2020ojn}
B.~Fiol, J.~Martfnez-Montoya, and A.~Rios~Fukelman, ``{The planar limit of $
  \mathcal{N} $ = 2 superconformal quiver theories},''
  \href{http://dx.doi.org/10.1007/JHEP08(2020)161}{{\em JHEP} {\bfseries 08}
  (2020) 161}, \href{http://arxiv.org/abs/2006.06379}{{\ttfamily
  arXiv:2006.06379 [hep-th]}}.

\bibitem{Beccaria:2021ksw}
M.~Beccaria and A.~A. Tseytlin, ``{$1/N$ expansion of circular Wilson loop in
  $\mathcal N=2$ superconformal $SU(N)\times SU(N)$ quiver},''
  \href{http://dx.doi.org/10.1007/JHEP04(2021)265}{{\em JHEP} {\bfseries 04}
  (2021) 265}, \href{http://arxiv.org/abs/2102.07696}{{\ttfamily
  arXiv:2102.07696 [hep-th]}}.

\bibitem{Galvagno:2021bbj}
F.~Galvagno and M.~Preti, ``{Wilson loop correlators in $ \mathcal{N} $ = 2
  superconformal quivers},''
  \href{http://dx.doi.org/10.1007/JHEP11(2021)023}{{\em JHEP} {\bfseries 11}
  (2021) 023}, \href{http://arxiv.org/abs/2105.00257}{{\ttfamily
  arXiv:2105.00257 [hep-th]}}.

\bibitem{Pini:2023lyo}
A.~Pini and P.~Vallarino, ``{Wilson loop correlators at strong coupling in $
  \mathcal{N} $ = 2 quiver gauge theories},''
  \href{http://dx.doi.org/10.1007/JHEP11(2023)003}{{\em JHEP} {\bfseries 11}
  (2023) 003}, \href{http://arxiv.org/abs/2308.03848}{{\ttfamily
  arXiv:2308.03848 [hep-th]}}.

\bibitem{Pini:2024zwi}
A.~Pini, ``{Integrated correlators with a Wilson line in a $ \mathcal{N} $ = 2
  quiver gauge theory at strong coupling},''
  \href{http://dx.doi.org/10.1007/JHEP01(2025)195}{{\em JHEP} {\bfseries 01}
  (2025) 195}, \href{http://arxiv.org/abs/2410.17342}{{\ttfamily
  arXiv:2410.17342 [hep-th]}}.

\bibitem{DeSmet:2025mbc}
P.-J. De~Smet, A.~Pini, and P.~Vallarino, ``{Analytical and numerical routes to
  strong coupling in $ \mathcal{N} $ = 2 SCFTs},''
  \href{http://dx.doi.org/10.1007/JHEP07(2025)011}{{\em JHEP} {\bfseries 07}
  (2025) 011}, \href{http://arxiv.org/abs/2505.02525}{{\ttfamily
  arXiv:2505.02525 [hep-th]}}.

\bibitem{DeLillo:2025eqg}
L.~De~Lillo and A.~Pini, ``{Integrated correlators of coincident Wilson lines
  in SU(N) gauge theories at strong coupling},''
  \href{http://dx.doi.org/10.1007/JHEP10(2025)089}{{\em JHEP} {\bfseries 10}
  (2025) 089}, \href{http://arxiv.org/abs/2506.16905}{{\ttfamily
  arXiv:2506.16905 [hep-th]}}.

\bibitem{Ferrando:2025qkr}
G.~Ferrando, S.~Komatsu, G.~Lefundes, and D.~Serban, ``{Exact three-point
  functions in $ \mathcal{N} $ = 2 superconformal field theories: integrability
  vs. localization},'' \href{http://dx.doi.org/10.1007/JHEP05(2026)289}{{\em
  JHEP} {\bfseries 05} (2026) 289},
  \href{http://arxiv.org/abs/2503.07295}{{\ttfamily arXiv:2503.07295
  [hep-th]}}.

\bibitem{lePlat:2025eod}
D.~le~Plat and T.~Skrzypek, ``{Three-point functions from integrability in
  $\mathcal{N}=2$ orbifold theories},''
  \href{http://dx.doi.org/10.1007/JHEP12(2025)172}{{\em JHEP} {\bfseries 12}
  (2025) 172}, \href{http://arxiv.org/abs/2506.21323}{{\ttfamily
  arXiv:2506.21323 [hep-th]}}.

\bibitem{Beccaria:2023qnu}
M.~Beccaria and G.~P. Korchemsky, ``{Four-dimensional $\mathcal{N}$ = 2
  superconformal long circular quivers},''
  \href{http://dx.doi.org/10.1007/JHEP04(2024)054}{{\em JHEP} {\bfseries 04}
  (2024) 054}, \href{http://arxiv.org/abs/2312.03836}{{\ttfamily
  arXiv:2312.03836 [hep-th]}}.

\bibitem{Korchemsky:2025eyc}
G.~P. Korchemsky and A.~Testa, ``{Correlation functions in four-dimensional
  superconformal long circular quivers},''
  \href{http://dx.doi.org/10.1007/JHEP07(2025)223}{{\em JHEP} {\bfseries 07}
  (2025) 223}, \href{http://arxiv.org/abs/2501.17223}{{\ttfamily
  arXiv:2501.17223 [hep-th]}}.

\bibitem{Sobko:2025zci}
E.~Sobko, ``{Continuous quiver gauge theories},''
  \href{http://dx.doi.org/10.1103/PhysRevD.111.046022}{{\em Phys. Rev. D}
  {\bfseries 111} no.~4, (2025) 046022}.

\bibitem{Beccaria:2023kbl}
M.~Beccaria, G.~P. Korchemsky, and A.~A. Tseytlin, ``{Non-planar corrections in
  orbifold/orientifold $ \mathcal{N} $ = 2 superconformal theories from
  localization},'' \href{http://dx.doi.org/10.1007/JHEP05(2023)165}{{\em JHEP}
  {\bfseries 05} (2023) 165}, \href{http://arxiv.org/abs/2303.16305}{{\ttfamily
  arXiv:2303.16305 [hep-th]}}.

\bibitem{Gaberdiel:2022iot}
M.~R. Gaberdiel and F.~Galvagno, ``{Worldsheet dual of free $ \mathcal{N} $ = 2
  quiver gauge theories},''
  \href{http://dx.doi.org/10.1007/JHEP10(2022)077}{{\em JHEP} {\bfseries 10}
  (2022) 077}, \href{http://arxiv.org/abs/2206.08795}{{\ttfamily
  arXiv:2206.08795 [hep-th]}}.

\bibitem{Baume:2020ure}
F.~Baume, J.~J. Heckman, and C.~Lawrie, ``{6D SCFTs, 4D SCFTs, Conformal
  Matter, and Spin Chains},''
  \href{http://dx.doi.org/10.1016/j.nuclphysb.2021.115401}{{\em Nucl. Phys. B}
  {\bfseries 967} (2021) 115401},
  \href{http://arxiv.org/abs/2007.07262}{{\ttfamily arXiv:2007.07262
  [hep-th]}}.

\bibitem{Heckman:2020otd}
J.~J. Heckman, ``{Qubit Construction in 6D SCFTs},''
  \href{http://dx.doi.org/10.1016/j.physletb.2020.135891}{{\em Phys. Lett. B}
  {\bfseries 811} (2020) 135891},
  \href{http://arxiv.org/abs/2007.08545}{{\ttfamily arXiv:2007.08545
  [hep-th]}}.

\bibitem{Baume:2022cot}
F.~Baume, J.~J. Heckman, and C.~Lawrie, ``{Super-spin chains for 6D SCFTs},''
  \href{http://dx.doi.org/10.1016/j.nuclphysb.2023.116250}{{\em Nucl. Phys. B}
  {\bfseries 992} (2023) 116250},
  \href{http://arxiv.org/abs/2208.02272}{{\ttfamily arXiv:2208.02272
  [hep-th]}}.

\bibitem{Cherkis:2003wk}
S.~A. Cherkis and N.~J. Hitchin, ``{Gravitational instantons of type D(k)},''
  \href{http://dx.doi.org/10.1007/s00220-005-1404-8}{{\em Commun. Math. Phys.}
  {\bfseries 260} (2005) 299--317},
  \href{http://arxiv.org/abs/hep-th/0310084}{{\ttfamily arXiv:hep-th/0310084}}.

\bibitem{Ionas:2016oxr}
R.~A. Ionas, ``{Gravitational instantons of type $D_k$ and a generalization of
  the Gibbons-Hawking Ansatz},''
  \href{http://dx.doi.org/10.1007/s00220-017-2852-7}{{\em Commun. Math. Phys.}
  {\bfseries 355} no.~2, (2017) 691--740},
  \href{http://arxiv.org/abs/1611.09939}{{\ttfamily arXiv:1611.09939
  [math.DG]}}.

\bibitem{Anselmi:1993sm}
D.~Anselmi, M.~Billo, P.~Fre, L.~Girardello, and A.~Zaffaroni, ``{ALE manifolds
  and conformal field theories},''
  \href{http://dx.doi.org/10.1142/S0217751X94001199}{{\em Int. J. Mod. Phys. A}
  {\bfseries 9} (1994) 3007--3058},
  \href{http://arxiv.org/abs/hep-th/9304135}{{\ttfamily arXiv:hep-th/9304135}}.

\bibitem{Hanany:1998sd}
A.~Hanany and Y.-H. He, ``{Non-Abelian finite gauge theories},''
  \href{http://dx.doi.org/10.1088/1126-6708/1999/02/013}{{\em JHEP} {\bfseries
  02} (1999) 013}, \href{http://arxiv.org/abs/hep-th/9811183}{{\ttfamily
  arXiv:hep-th/9811183}}.

\bibitem{Grisaru:1979wc}
M.~T. Grisaru, W.~Siegel, and M.~Ro\v{c}ek, ``{Improved Methods for
  Supergraphs},'' \href{http://dx.doi.org/10.1016/0550-3213(79)90344-4}{{\em
  Nucl. Phys. B} {\bfseries 159} (1979) 429}.

\bibitem{Seiberg:1988ur}
N.~Seiberg, ``{Supersymmetry and Nonperturbative beta Functions},''
  \href{http://dx.doi.org/10.1016/0370-2693(88)91265-8}{{\em Phys. Lett. B}
  {\bfseries 206} (1988) 75--80}.

\bibitem{Eden:2004ua}
B.~Eden, C.~Jarczak, and E.~Sokatchev, ``{A Three-loop test of the dilatation
  operator in N = 4 SYM},''
  \href{http://dx.doi.org/10.1016/j.nuclphysb.2005.01.036}{{\em Nucl. Phys. B}
  {\bfseries 712} (2005) 157--195},
  \href{http://arxiv.org/abs/hep-th/0409009}{{\ttfamily arXiv:hep-th/0409009}}.

\bibitem{Grant:2008sk}
L.~Grant, P.~A. Grassi, S.~Kim, and S.~Minwalla, ``{Comments on 1/16 BPS
  Quantum States and Classical Configurations},''
  \href{http://dx.doi.org/10.1088/1126-6708/2008/05/049}{{\em JHEP} {\bfseries
  05} (2008) 049}, \href{http://arxiv.org/abs/0803.4183}{{\ttfamily
  arXiv:0803.4183 [hep-th]}}.

\bibitem{Fiamberti:2007rj}
F.~Fiamberti, A.~Santambrogio, C.~Sieg, and D.~Zanon, ``{Wrapping at four loops
  in N=4 SYM},'' \href{http://dx.doi.org/10.1016/j.physletb.2008.06.061}{{\em
  Phys. Lett. B} {\bfseries 666} (2008) 100--105},
  \href{http://arxiv.org/abs/0712.3522}{{\ttfamily arXiv:0712.3522 [hep-th]}}.

\bibitem{Fiamberti:2008sh}
F.~Fiamberti, A.~Santambrogio, C.~Sieg, and D.~Zanon, ``{Anomalous dimension
  with wrapping at four loops in N=4 SYM},''
  \href{http://dx.doi.org/10.1016/j.nuclphysb.2008.07.014}{{\em Nucl. Phys. B}
  {\bfseries 805} (2008) 231--266},
  \href{http://arxiv.org/abs/0806.2095}{{\ttfamily arXiv:0806.2095 [hep-th]}}.

\bibitem{Bajnok:2008bm}
Z.~Bajnok and R.~A. Janik, ``{Four-loop perturbative Konishi from strings and
  finite size effects for multiparticle states},''
  \href{http://dx.doi.org/10.1016/j.nuclphysb.2008.08.020}{{\em Nucl. Phys. B}
  {\bfseries 807} (2009) 625--650},
  \href{http://arxiv.org/abs/0807.0399}{{\ttfamily arXiv:0807.0399 [hep-th]}}.

\bibitem{Gromov:2009zb}
N.~Gromov, V.~Kazakov, and P.~Vieira, ``{Exact Spectrum of Planar ${\cal N}=4$
  Supersymmetric Yang-Mills Theory: Konishi Dimension at Any Coupling},''
  \href{http://dx.doi.org/10.1103/PhysRevLett.104.211601}{{\em Phys. Rev.
  Lett.} {\bfseries 104} (2010) 211601},
  \href{http://arxiv.org/abs/0906.4240}{{\ttfamily arXiv:0906.4240 [hep-th]}}.

\end{thebibliography}\endgroup
\bibliographystyle{utphys}

\end{document}